\input harvmac
\input rotate
\input epsf
\input xyv2
\input glshortspare.defs

\font\teneurm=eurm10 \font\seveneurm=eurm7 \font\fiveeurm=eurm5
\newfam\eurmfam
\textfont\eurmfam=\teneurm \scriptfont\eurmfam=\seveneurm
\scriptscriptfont\eurmfam=\fiveeurm
\def\eurm#1{{\fam\eurmfam\relax#1}}
 \font\teneusm=eusm10 \font\seveneusm=eusm7 \font\fiveeusm=eusm5
\newfam\eusmfam
\textfont\eusmfam=\teneusm \scriptfont\eusmfam=\seveneusm
\scriptscriptfont\eusmfam=\fiveeusm
\def\eusm#1{{\fam\eusmfam\relax#1}}
\font\tencmmib=cmmib10 \skewchar\tencmmib='177
\font\sevencmmib=cmmib7 \skewchar\sevencmmib='177
\font\fivecmmib=cmmib5 \skewchar\fivecmmib='177
\newfam\cmmibfam
\textfont\cmmibfam=\tencmmib \scriptfont\cmmibfam=\sevencmmib
\scriptscriptfont\cmmibfam=\fivecmmib
\def\cmmib#1{{\fam\cmmibfam\relax#1}}
\writedefs

\noblackbox\input rotate
\let\includefigures=\iftrue
\includefigures
\message{If you do not have epsf.tex (to include figures),}
\message{change the option at the top of the tex file.}
\def\figin{\epsfcheck\figin}\def\figins{\epsfcheck\figins}
\def\epsfcheck{\ifx\epsfbox\UnDeFiNeD
\message{(NO epsf.tex, FIGURES WILL BE IGNORED)}
\gdef\figin##1{\vskip2in}\gdef\figins##1{\hskip.5in}
\else\message{(FIGURES WILL BE INCLUDED)}%
\gdef\figin##1{##1}\gdef\figins##1{##1}\fi}
\def\DefWarn#1{}

\def\underarrow#1{\vbox{\ialign{##\crcr$\hfil\displaystyle
{#1}\hfil$\crcr\noalign{\kern1pt\nointerlineskip}\rightarrowfill\crcr}}}

\def\EUV{\eusm V}
\def\EUT{\eusm T}
\def\EUW{\eusm W}
\def\EUS{\eusm S}
\def\EUX{\eusm X}
\def\EUZ{\eusm Z}
\def\EUY{\eusm Y}
\def\EUQ{\eusm Q}
\def\EUA{\eusm A}

\def\EUC{\eusm C}
\def\EUD{\eusm D}
\def\EUE{\eusm E}
\def\EUF{\eusm F}
\def\EUH{\eusm H}
\def\EUI{\eusm I}
\def\EUJ{\eusm J}
\def\EUK{\eusm K}
\def\EUM{\eusm M}
\def\EUN{\eusm N}
\def\EUO{\eusm O}
\def\EUP{\eusm P}
\def\EUQ{\eusm Q}
\def\EUR{\eusm R}

\def\cA{{\cal A}}
\def\cD{{\cal D}}
\def\cF{{\cal F}}
\def\cO{{\cal O}}

\def\ot{\otimes}
\def\End{{\rm End}}
\def\ra{\rightarrow}
\def\figinsert{\goodbreak\midinsert}
\def\ifig#1#2#3{\DefWarn#1\xdef#1{fig.~\the\figno}
\writedef{#1\leftbracket fig.\noexpand~\the\figno}%
\figinsert\figin{\centerline{#3}}\medskip\centerline{\vbox{\baselineskip12pt
\advance\hsize by -1truein\noindent\footnotefont{\bf
Fig.~\the\figno:} #2}}
\bigskip\endinsert\global\advance\figno by1}
\else
\def\ifig#1#2#3{\xdef#1{fig.~\the\figno}
\writedef{#1\leftbracket fig.\noexpand~\the\figno}%
\global\advance\figno by1} \fi
\noblackbox
\input amssym.tex
\def\hat{\widehat}
%
\overfullrule=0pt
\def\G{{\Gamma}}
\def\M{{\EUM}}
\def\MH{{{\EUM}_H}}
\def\CZ{{\cal Z}}
\def\MHAD{\MH(G_{\rm ad})}
\def\tilde{\widetilde}
\def\bar{\overline}
\def\Gad{{G_{\rm ad}}}
%
\def\EUBB{\cmmib B}
\def\EUWW{\eurm W}
\def\tilde{\widetilde}
\def\bar{\overline}

\def\bz{{\bar z}}

\def\bt{{\bar t}}

\def\bz{{\bar z}}
\def\bpartial{{\bar\partial}}
\def\End{{\rm End}}
\def\Tr{{\rm Tr}}
\def\eps{\epsilon}
\def\RR{{\Bbb{R}}}
\def\cL{{\cal L}}

\def\CC{{\Bbb{C}}}

\def\LG{{^L\negthinspace G}}

\def\CZ{{\cal Z}}

\def\G{{\Gamma}}
\def\M{{\EUM}}
\def\MH{{{\EUM}_H}}
%
\def\tilde{\widetilde}
\def\bar{\overline}

\font\zfont = cmss10 
\font\litfont = cmr6 
\def\bigone{\hbox{1\kern -.23em {\rm l}}}
\def\ZZ{\hbox{\zfont Z\kern-.4emZ}}
\def\half{{\litfont {1 \over 2}}}

\def\CM{{\EUM}}
\def\Re{{\rm Re ~}}
\def\Im{{\rm Im ~}}


\def\bz{{\bar z}}
\def\bt{{\bar t}}

\def\cM{{\EUM}}

\def\LG{{^L\negthinspace G}}

\def\RR{{\Bbb{R}}}
\def\CC{{\Bbb{C}}}
\def\ZZ{{\Bbb{Z}}}
\def\eps{\epsilon}
\def\ra{\rightarrow}

\def\ad{\rm {ad}}

\font\zfont = cmss10 
\font\litfont = cmr6 
\def\bigone{\hbox{1\kern -.23em {\rm l}}}
\def\ZZ{\hbox{\zfont Z\kern-.4emZ}}
\def\half{{\litfont {1 \over 2}}}

\def\CM{{\EUM}}

\font\zfont = cmss10 
\font\litfont = cmr6 
\def\bigone{\hbox{1\kern -.23em {\rm l}}}
\def\ZZ{\hbox{\zfont Z\kern-.4emZ}}
\def\half{{\litfont {1 \over 2}}}

\def\CM{{\EUM}}
\def\Re{{\rm Re ~}}
\def\Im{{\rm Im ~}}

\def\g{{\frak g}}
\def\G{{\Gamma}}
\def\M{{\EUM}}
\def\MH{{{\EUM}_H}}
%
\def\tilde{\widetilde}
\def\bar{\overline}

\font\zfont = cmss10 
\font\litfont = cmr6 
\def\bigone{\hbox{1\kern -.23em {\rm l}}}
\def\ZZ{\hbox{\zfont Z\kern-.4emZ}}
\def\half{{\litfont {1 \over 2}}}

\def\neg{\negthinspace}

\def\CM{{\EUM}}
\def\Re{{\rm Re ~}}
\def\Im{{\rm Im ~}}


\def\bz{{\bar z}}

\let\includefigures=\iftrue
\def\g{{\frak g}}

\def\HG{{\widehat \Gamma}}
\def\rr{r}
\def\bz{{\bar z}}
\def\bpartial{{\bar\partial}}
\def\End{{\rm End}}
\def\Tr{{\rm Tr}}
\def\eps{\epsilon}
\def\RR{{\Bbb{R}}}
\def\cL{{\cal L}}

\def\CC{{\Bbb{C}}}

\def\LG{{^L\negthinspace G}}

\def\CZ{{\cal Z}}
 \noindent
 \Title{hep-th/0604151} {\vbox{ \centerline{
Electric-Magnetic Duality}
\bigskip
\centerline{And The Geometric Langlands Program}}}
\smallskip
\centerline{Anton Kapustin}
\smallskip
\centerline{\it{Department of Physics, California Institute of
Technology}} \centerline{\it{Pasadena, CA 91125}}
\medskip
\centerline{and}
\medskip
\centerline{Edward Witten}
\smallskip
\centerline{\it{School of Natural Sciences, Institute for Advanced
Study}} \centerline{\it{Princeton, New Jersey 08540}}
\bigskip\bigskip
\noindent The geometric Langlands program can be described in a
natural way by compactifying on a Riemann surface $C$ a twisted
version of ${\EUN}=4$ super Yang-Mills theory in four dimensions.
The key ingredients are electric-magnetic duality of gauge theory,
mirror symmetry of sigma-models, branes, Wilson and 't Hooft
operators, and topological field theory. Seemingly esoteric
notions of the geometric Langlands program, such as Hecke
eigensheaves and ${\cal D}$-modules, arise naturally from the
physics.

\noindent\Date{April, 2006} \listtoc \writetoc

\newsec{Introduction}
\seclab\intro

 The Langlands program for number fields \ref\lang{R. Langlands,
``Problems In The Theory Of Automorphic Forms,'' in Lect. Notes in
Math. {\bf 170} (Springer-Verlag, 1970), pp. 18-61; ``Where Stands
Functoriality Today?'' in {\it Representation Theory And
Automorphic Forms}, Proc. Symp. Pure Math. {\bf 61} (American
Mathematical Society, 1997), pp. 457-471.} unifies many classical
and contemporary results in number theory and is a vast area of
research. \nref\drinfeld{V. Drinfeld, ``Langlands Conjecture For
$GL(2)$ Over Function Field,'' Proc. Int. Congress. Math.
(Helsinki, 1978), pp. 565-574; ``Two-Dimensional $l$-Adic
Representations Of The Fundamental Group Of A Curve Over A Finite
Field And Automorphic Forms
On $GL(2)$,'' Amer. J. Math. {\bf 105} (1983) 85-114.}%
\nref\lafforgue{L. Lafforgue, ``Chtoucas de Drinfeld et
Correspondance de Langlands,'' Invent. Math. {\bf 147} (2002) 1-241.}%
It has an analog for curves over a finite field, which has also
been the subject of much celebrated work \refs{\drinfeld,
\lafforgue}. In addition, a geometric version of the Langlands
program for curves has been much developed \nref\laumon{G. Laumon,
``Correspondance de Langlands Geometrique Pour Les Corps
de Fonctions,'' Duke. Math. J. {\bf 54} (1987) 309-359.}%
\nref\bd{A. Beilinson and V. Drinfeld, ``Quantization Of Hitchin's
Integrable System And Hecke Eigensheaves,'' preprint (ca. 1995),
http://www.math.uchicago.edu/~arinkin/langlands/.}%
\nref\fgv{E. Frenkel, D. Gaitsgory and K. Vilonen, ``On The
Geometric Langlands Conjecture,'' Journal of the AMS {\bf 15}
(2001) 367-417.}%
\nref\gaitsg{D. Gaitsgory, ``On A Vanishing Conjecture Appearing
In The Geometric Langlands Correspondence,'' Ann. of Math. (2)
{\bf 160} (2004) 617-182,
math.AG/0204081.}%
 \nref\brav{A. Braverman and R.
Bezrukavnikov, ``Geometric Langlands Correspondence For ${\cal
D}$-Modules in Prime
Characteristic: The $GL(n)$ Case,'' math.AG/0602255.}%
\refs{\laumon - \brav}, both for curves over a field of
characteristic $p$ and for ordinary complex Riemann surfaces. For
a survey that is relatively readable for physicists, with numerous
references, see \ref\frenkel{E. Frenkel, ``Lectures On The
Langlands Program And Conformal Field Theory,''
arXiv:hep-th/0512172.}.

Our focus in the present paper is on the geometric Langlands
program for complex Riemann surfaces.  We aim to show how this
program can be understood as a chapter in quantum field theory. No
prior familiarity with the Langlands program is assumed; instead,
we assume a familiarity with subjects such as supersymmetric gauge
theories, electric-magnetic duality, sigma-models, mirror
symmetry, branes, and topological field theory. The theme of the
paper is to show that when these familiar physical ingredients are
applied to just the right problem, the geometric Langlands program
arises naturally. Seemingly esoteric notions such as Hecke
eigensheaves, ${\cal D}$-modules, and so on, appear spontaneously
in the physics, with new insights about their properties.

The first hints of a connection between the Langlands program and
quantum field theory came from the work of Goddard, Nuyts, and
Olive (GNO), who showed in 1976 \ref\gno{P. Goddard, J. Nuyts, and
D. I. Olive, ``Gauge Theories And Magnetic Charge,'' Nucl. Phys.
{\bf B125} (1977) 1-28.} that in gauge theories, though electric
charge takes values in the weight lattice of the gauge group,
magnetic charge takes values in the weight lattice of a dual
group. Magnetic charges for general compact Lie groups had been
first analyzed by Englert and Windey \ref\engwin{F. Englert and P.
Windey, ``Quantization Condition For 't Hooft Monopoles In Compact
Simple Lie Groups,'' Phys. Rev. {\bf D14} (1976) 2728-2731.}. The
GNO analysis motivated the Montonen-Olive electric-magnetic
duality conjecture \ref\montolive{C. Montonen and D. I. Olive,
``Magnetic Monopoles As Gauge Particles?'' Phys. Lett. {\bf B72}
(1977) 117-120.} according to which a specific gauge theory based
on a given gauge group is equivalent to a similar theory with the
coupling constant inverted and the gauge group replaced by its
dual.

For a gauge group $G$, the GNO dual group is actually the same as
the Langlands dual group $^L\negthinspace G$, which plays an
important role in formulating the Langlands conjectures. (For some
examples of the correspondence between $G$ and $^L\negthinspace
G$, see the table.) This was observed by Atiyah, who suggested to
the second author at the end of 1977 that the Langlands program is
related to quantum field theory and recommended the two papers
\refs{\gno,\montolive}. There resulted a further development
\ref\wo{D. I. Olive and E. Witten, ``Supersymmetry Algebras That
Include Topological Charges,'' Phys. Lett. {\bf B78} (1978)
97-101.} in which it was understood that Montonen-Olive duality is
more natural in supersymmetric gauge theory.  It was later
understood that ${\EUN}=4$ supersymmetry (i.e., the maximal
possible supersymmetry in four dimensions) is the right context
for this duality \ref\whog{H. Osborn, ``Topological Charges For
${\EUN}=4$ Supersymmetric Gauge Theories And Monopoles Of Spin
1,'' Phys. Lett. {\bf B83} (1979) 321-326.}, and that the
$\Bbb{Z}_2$ duality originally proposed has a natural extension to
$SL(2,\Bbb{Z})$ \nref\sltwoz{J. Cardy and E. Rabinovici, ``Phase
Structure Of $\Bbb{Z}_p$ Models In The Presence Of A Theta
Parameter,'' Nucl. Phys. {\bf B205} (1982) 1-16; J. Cardy,
``Duality And The Theta Parameter In Abelian Lattice Models,''
Nucl. Phys. {\bf B205} (1982) 17-26.}%
\nref\slmore{A. Shapere and F. Wilczek, ``Selfdual Models With
Theta Terms,''
Nucl. Phys. {\bf B320} (1989) 669-695.}%
\refs{\sltwoz,\slmore} when the theta angle of the gauge theory is
included.\bigskip \centerline{
\vbox{\hsize=2truein\offinterlineskip\halign{\tabskip=2em plus2em
minus2em\hfil#\hfil&\vrule#&\hfil#\hfil\tabskip=0pt\cr $G$&depth
6pt&$^L\negthinspace G$\cr \noalign{\hrule}
$U(N)$&height14pt&$U(N)$\cr
$SU(N)$&height14pt&$PSU(N)=SU(N)/\Bbb{Z}_N$\cr
$Spin(2n)$&height14pt&$SO(2n)/\Bbb{Z}_2$\cr
$Sp(n)$&height14pt&$SO(2n+1)$\cr
$Spin(2n+1)$&height14pt&$Sp(n)/\Bbb{Z}_2$\cr
$G_2$&height14pt&$G_2$\cr $E_8$&height14pt&$E_8$
\cr\noalign{\medskip} }}} \nobreak \centerline{
\vbox{\hsize=3truein \noindent Table 1. Examples of the
correspondence between a Lie group $G$ and its Langlands or GNO
dual $^L\negthinspace G$.}}\bigskip

\nref\senschwarz{J. H. Schwarz and A. Sen, ``Duality Symmetries Of
$4-D$ Heterotic Strings,'' Phys. Lett. {\bf 312B} (1993) 105-114,
``Duality Symmetric Actions,'' Nucl. Phys. {\bf B411} (1994)
35-63, arXiv:hep-th/9304154. }%
 \nref\vafa{M. Bershadsky, A. Johansen, V.
Sadov, and C. Vafa, ``Topological Reduction Of $4-D$ SYM  To $2-D$
Sigma Models,'' Nucl. Phys. {\bf B448} (1995) 166-186,
arXiv:hep-th/9501096.}%
\nref\hm{J. A. Harvey, G. W. Moore, and A. Strominger, ``Reducing $S$ Duality To
$T$ Duality,'' Phys. Rev. {\bf D52} (1995) 7161-7167.}%
 In the early 1990's, extensions of
Montonen-Olive duality to string theory were conjectured
\senschwarz. Subsequently, Montonen-Olive duality and its
generalizations were studied from many new points of view and were
recognized as a crucial, though still mysterious, ingredient in
understanding field theory and string theory.  These developments
were far too extensive to be reviewed here, but one observation of
that period, though a sideline at the time, is a starting point
for the present paper. Compactification of ${\EUN}=4$ super
Yang-Mills theory from four dimensions to two dimensions was
studied \refs{\vafa,\hm} and was shown to lead at low energies to
a two-dimensional supersymmetric sigma-model in which the target
space is a hyper-Kahler manifold that is Hitchin's moduli space
${\EUM}_H$ of stable Higgs bundles
 \ref\hitchin{N. Hitchin,
``The Self-Duality Equations On A Riemann Surface,'' Proc. London
Math. Soc. (3) {\bf 55} (1987) 59-126}.
 Electric-magnetic duality in four dimensions
reduces in two dimensions to $T$-duality of the sigma-model.  This
particular $T$-duality was subsequently used mathematically
\ref\thadhau{T. Hausel and M. Thaddeus, ``Mirror Symmetry,
Langlands Duality, And The Hitchin System,'' Invent. Math. {\bf
153} (2003) 197-229, math.AG/0205236.} to show (for $SU(N)$) that
the Hodge numbers of the Higgs bundle moduli space of a gauge
group $G$ are equal to those of $^L\negthinspace G$. The geometry
underlying this $T$-duality was investigated in \ref\donga{R.
Donagi and D. Gaitsgory, ``The Gerbe Of Higgs Bundles,''
Transform. Groups {\bf 7} (2002) 109-153, math.AG/0005132.} and
subsequently in \ref\dopa{R. Donagi and T. Pantev, ``Torus
Fibrations, Gerbes, and Duality,'' math.AG/0306213.} for any
semi-simple Lie group $G$.

Other clues about the relation of the geometric Langlands program
to quantum field theory have come from relatively recent
mathematical work. The approach of Beilinson and Drinfeld to the
geometric Langlands program is based on quantization of $\MH$, as
the title of their paper implies \bd. The $T$-duality of
${\EUM}_H$, understood mathematically as a Fourier-Mukai
transform, has been interpreted as a sort of semiclassical
approximation to the geometric Langlands program. This point of
view underlies the paper \ref\arinkin{D. Arinkin, ``Moduli Of
Connections With A Small Parameter On A Curve,''
math.AG/0409373.}. We understand that there have also been
important unpublished contributions by other mathematicians,
including Donagi and Pantev. The second author learned of this
interpretation of the $T$-duality of $\MH$ from a lecture by D.
Ben-Zvi at a conference on the geometric Langlands program, held
at the IAS in the spring of 2004. This was an extremely strong
hint that it must be possible to understand the geometric
Langlands program using four-dimensional electric-magnetic duality
(which leads to this particular $T$-duality) and branes (the
natural quantum field theory setting for interpreting $T$-duality
as a Fourier-Mukai transform).  This hint was the starting point
for the present paper.

\nref\eww{E. Witten, ``Topological Quantum Field Theory,'' Commun.
Math. Phys. {\bf 117} (1988) 353-386.}%
To summarize this paper in the briefest possible terms, we will
develop six main ideas.   The first is that from a certain twisted
version of ${\EUN}=4$ supersymmetric Yang-Mills theory in four
dimensions, one can construct a family of four-dimensional
topological field theories in four dimensions. After reviewing
some of the background in section \review, we explain this
construction in section \tqftfour. The twisting procedure is
formally just analogous to the construction by which Donaldson
theory can be obtained \eww\ from ${\EUN}=2$ super Yang-Mills
theory. The second main idea, developed in sections
\compacthitchin\ and \tqfttwo, is that, extending the insights of
\refs{\vafa,\hm}, compactification on a Riemann surface $C$ gives
in two dimensions a family of topological sigma-models, with
target ${\EUM}_H$, which are ``generalized $B$-models.'' Moreover,
for a special value of the parameter, four-dimensional $S$-duality
acts as two-dimensional mirror symmetry.  The third main idea,
developed in section \loopop, is that Wilson and 't Hooft line
operators are topological operators that act on the branes of the
two-dimensional sigma-model in a natural fashion. Here we consider
an operator that maps one brane to another (or roughly speaking,
one theory to another), not the more familiar sort of operator in
Hilbert space that maps one state to another. A brane that is
mapped by the Wilson or 't Hooft operators to, roughly speaking, a
multiple of itself is what we call an electric or magnetic
eigenbrane. $S$-duality will automatically exchange the electric
and magnetic eigenbranes. The fourth main idea, explained in
section \electricbranes, is that, in the right context, electric
eigenbranes are in natural correspondence with homomorphisms of
$\pi_1(C)$ to the complexification of the Langlands dual group
$^L\negthinspace G_{\Bbb{C}}$. The fifth main idea, developed in
sections \thooftheckeop\ and \bogeqheck, is that 't Hooft
operators correspond naturally to geometric Hecke operators
similar to those of the geometric Langlands program but acting on
Higgs bundles instead of ordinary $G$-bundles. It takes one more
important idea, developed in section \abranes, to make contact
with the usual formulation of the geometric Langlands
correspondence. We show that, because of the existence of a
canonical coisotropic brane on ${\EUM}_H$, the magnetic
eigenbranes in our sense are naturally associated to ${\cal
D}$-modules (modules for the sheaf of differential operators) on
the moduli space ${\EUM}(G,C)$ of holomorphic $G$-bundles on $C$.

Sections 7, 8, 9, and 10 of this paper and some expository
portions of other sections are primarily adapted from a
forthcoming book \ref\witfur{E. Witten, {\it Gauge Theory And The
Geometric Langlands Program}, to appear.} which will also contain
some additional results. This material is included here to make
the paper more comprehensible.

\nref\frthree{E. Frenkel, ``Affine Algebras, Langlands Duality,
And Bethe Ansatz,'' in the Proceedings of the XIth International
Congress of Mathematical Physics, Paris, 1994, ed. D. Iagolnitzer
(International Press, 1995) 606-642, q-alg/9506003.}%
 \nref\bdtwo{A.
Beilinson And V. Drinfeld, {\it Chiral Algebras}, American
Mathematical Society Colloquium Publications {\bf 51}
(American Mathematical Society, 2004).}%
 \nref\frenzvi{E. Frenkel
and D. Ben-Zvi, {\it Vertex Algebras And Algebraic Curves},
Mathematical Surveys And Monographs {\bf 88},
second edition (American Mathematical Society, 2004).}%
\nref\gaitfren{E. Frenkel and D. Gaitsgory, ``Local Geometric
Langlands Correspondence And Affine Kac-Moody Algebras,''
math.RT/0508382.}

\nref\witold{E. Witten, ``Free Fermions On An Algebraic Curve,''
in {\it The Mathematical Heritage Of Hermann Weyl}, ed. R. Wells
(American Mathematical Society, Durham, 1988) pp. 329-344,
``Quantum Field Theory, Grassmannians, And Algebraic Curves,''
Commun. Math. Phys. {\bf 113} (1988) 529-600.} \nref\zed{A. Neveu and J. Scherk,
``Parameter-Free Regularization Of One-Loop Unitary Dual Diagram,'' Phys. Rev. {\bf D1} (1970)
2355-2359.}%
\nref\qwed{D. J. Gross, A. Neveu, J. Scherk, and J. H. Schwarz,
``Renormalization And Unitarity In The Dual-Resonance Model,''
Phys. Rev. {\bf D2} (1970) 697-710.}%
 \nref\add{P. Goddard and D. Olive, ``Algebras,
Lattices, And Strings,'' in {\it Vertex Operators In Mathematics
And Physics} (1985), ed. J. Lepowsky et. al.} \nref\bdd{J. Harvey
and G. Moore, ``Algebras, BPS States, And Strings,'' Nucl. Phys.
{\bf B463} (1996) 315-368, arXiv:hep-th/9510182, ``Exact
Gravitational Threshold Correction in the FHSV Model,'' Phys. Rev.
{\bf D57} (1998) 2329-2336, arXiv:hep-th/9611176. } \nref\cdd{R.
Dijkgraaf, E. Verlinde, and H. Verlinde, ``Counting Dyons In
${\EUN}=4$ String Theory,'' Nucl. Phys. {\bf B484} (1997) 543-561,
arXiv:hep-th/9607026.} \nref\greens{M. B. Green and S. Sethi,
``Supersymmetry Constraints On Type IIB Supergravity,'' Phys. Rev.
{\bf D59} (1999) 046006,
hep-th/9808061.}%
\nref\whoto{H. Nicolai,  B. Pioline, J. Plefka and A. Waldron,
``$R^4$ Couplings, the Fundamental Membrane and Exceptional Theta
Correspondences," JHEP {\bf 03} (2001) 036, hep-th/0102123.}%
 \nref\ddd{M. Gunaydin, A. Neitzke, B. Pioline,
and A. Waldron, ``BPS Black Holes, Quantum Attractor Flows, And
Automorphic Forms,'' arXiv:hep-th/0512296.}%
 \nref\edd{R.
Schimmrigk, ``The Langlands Program and String Modular K3
Surfaces,'' arXiv:hep-th/0603234.}%
 One obvious gap in our analysis is
that we consider only the unramified case of the geometric
Langlands correspondence.  We expect that it is possible to apply
somewhat similar ideas to the ramified case.  A second major gap
is that we do not shed light on the utility of two-dimensional
conformal field theory for the geometric Langlands program
\refs{\bd,\frenkel,\frthree-\gaitfren}. The last of these
references applies conformal field theory to the ramified case.
Hopefully it will prove possible to deduce the conformal field
theory approach from the gauge theory approach of this paper. In
fact, there is an analogy even at a naive level \witold\ between
conformal field theory and the theory of automorphic
representations, which is the basis of the Langlands program.
Finally, though we have nothing to contribute about this here, an
additional clue about the relation of the Langlands program with
physics presumably comes from the diverse ways that automorphic
forms enter string theory. For a tiny sampling of this, see
\refs{\zed-\edd}.

A.K. would like to thank D. Arinkin, R. Bezrukavnikov, and D.
Orlov for useful conversations. In particular, Orlov's
explanations in 2002 about the abelian case of the geometric
Langlands program partially motivated the paper \ref\okap{A.
Kapustin, ``$A$-Branes And Noncommutative Geometry,''
arXiv:hep-th/0502212.}, which will enter our story in section 11.

E.W. would like to thank the many mathematicians who over the
years have explained matters relevant to the Langlands program,
including A. Beilinson, P. Deligne,  V. Drinfeld,  and K. Vilonen,
and especially M. F. Atiyah,   D. Ben-Zvi, R. Donagi, E. Frenkel,
and D. Kazhdan, and most recently M. Goresky and R. MacPherson. In
addition, among others,  T. Hausel, N. Hitchin, M. Hopkins, P.
Kronheimer, L. Jeffrey,  J. Morgan, G. Moore, D. Morrison, N.
Nekrasov, M. Thaddeus, C. Vafa, and E. J. Weinberg clarified some
points relevant to the present paper, and many of the physicists
at the IAS, including S. Hellerman, K. Intriligator, J. Maldacena,
N. Seiberg, and J. Walcher, made helpful comments.

 \nref\brink{L. Brink, J. H. Schwarz, and J.
Scherk, ``Supersymmetric Yang-Mills Theories,'' Nucl. Phys. {\bf
B121}
(1977) 77-92.}%

\newsec{${\EUN}=4$ Super Yang-Mills Theory And $S$-Duality}

\seclab\review

 \nref\nahm{W. Nahm, ``Supersymmetries And Their
Representations,'' Nucl. Phys.
{\bf B135} (1978) 149-166.}%

 In this section, we recall a few properties of
${\EUN}=4$ super Yang-Mills theory and its $S$-duality.

\subsec{Review Of ${\EUN}=4$ Super Yang-Mills Theory}
\subseclab\reviewsym

As in the original work \brink, ${\EUN}=4$ super Yang-Mills is
most easily constructed by dimensional reduction from ten
dimensions. Ten dimensions is the maximum possible dimension for
supersymmetric Yang-Mills theory  by virtue of Nahm's theorem
\nahm, and for given gauge group $G$, there is a very simple
supersymmetric Lagrangian which moreover is  unique up to the
choice of a few coupling parameters if we ask for a Lagrangian
quadratic in the curvature. In this paper, we always assume $G$ to
be compact and denote its complexification as $G_{\Bbb{C}}$.  This
differs from most other expositions of the geometric Langlands
program.

\lref\maldaads{J.~M.~Maldacena, ``The large N limit of
superconformal field theories and supergravity,''
  Adv.\ Theor.\ Math.\ Phys.\  {\bf 2}, 231 (1998),
  arXiv:hep-th/9711200.}
  \lref\wittenads{E.~Witten, ``Anti-de Sitter Space and Holography,''
  Adv.\ Theor.\ Math.\ Phys.\  {\bf 2}, 253-291 (1998),
  arXiv:hep-th/9802150.}
  \lref\gkpads{S.~S.~Gubser, I.~R.~Klebanov and A.~M.~Polyakov,
  ``Gauge Theory Correlators From Non-critical String Theory,''
  Phys.\ Lett.\  {\bf B428}  (1998) 105-114,
  arXiv:hep-th/9802109.}
  \lref\wittenbaryon{E.~Witten,
  ``Baryons and Branes in Anti de Sitter Space,''
  JHEP {\bf 9807}, 006 (1998),
  arXiv:hep-th/9805112.}

  \lref\vafageom{C.~Vafa,
  ``Geometric Origin of Montonen-Olive Duality,''
  Adv.\ Theor.\ Math.\ Phys.\  {\bf 1}, 158-166 (1997),
  arXiv:hep-th/9707131.}

\lref\british{N.~Dorey, C.~Fraser, T.~J.~Hollowood and
M.~A.~C.~Kneipp, ``$S$-Duality in ${\EUN}=4$ Supersymmetric Gauge
Theories With Arbitrary Gauge Group,'' Phys.\ Lett.\ B {\bf 383},
422-428 (1996), arXiv:hep-th/9605069. }

\lref\AKS{P. Argyres, A. Kapustin, N. Seiberg, ``On $S$-Duality
For Non-Simply-Laced Gauge Groups,'' arXiv:hep-th/0603048.}

\bigskip\noindent{\it Spacetime Conventions}

 \lref\vafawitten{C. Vafa and E. Witten, ``A Strong
Coupling Test Of $S$ Duality,'' Nucl. Phys. {\bf B431} (1994) 3-77,
arXiv:hep-th/9408074.}%

We begin by describing some conventions. We will work with Lorentz
signature $-++\dots +$ or Euclidean signature $+++\dots +$.
Basically, when  emphasizing questions of topological field
theory, we will use Euclidean signature, but when we want to
stress that the constructions are natural in physically sensible,
unitary quantum field theory, we use Lorentz signature.

We write the metric of ten-dimensional Minkowski space
$\Bbb{R}^{1,9}$ or Euclidean space $\Bbb{R}^{10}$ as
$ds^2=\sum_{I,J=0}^9g_{IJ}\,dx^I\,dx^J=\mp (dx^0)^2+(dx^1)^2+\dots
+(dx^9)^2$. These have symmetry groups $SO(1,9)$ or $SO(10)$,
whose spin representations ${\EUS}^+$ and ${\EUS}^-$ are of rank
16. They are real (and dual to one another) in Lorentz signature,
while in Euclidean signature they are complex conjugate
representations.

The gamma matrices $\Gamma_I,\,I=0,\dots,9$ (which in Lorentz
signature are real) reverse the chirality, mapping ${\EUS}^\pm$ to
${\EUS}^\mp$, and obey the Clifford algebra
$\{\Gamma_I,\Gamma_J\}=2g_{IJ}$. Moreover, the operator
\eqn\thop{\bar\Gamma=\Gamma_0\Gamma_1\cdots\Gamma_9} acts on
${\EUS}^+$ or ${\EUS}^-$ as multiplication by 1 or $-1$. Because
${\EUS}^+$ and ${\EUS}^-$ are dual, we can regard the $\Gamma$'s
not as maps from ${\EUS}^\pm$ to ${\EUS}^{\mp}$ but as bilinear
maps $\Gamma_I:{\EUS}^+\otimes {\EUS}^+\to \Bbb{R}$ or
$\Gamma_I:{\EUS}^-\otimes{\EUS}^-\to \Bbb{R}$. (In this paragraph,
we assume Lorentz signature; in Euclidean signature, all such maps
are to $\Bbb{C}$, since the spinors are complex.) If
$\beta,\gamma\in {\EUS}^+$, it is conventional to write $\bar\beta
\Gamma_I\gamma$ for the bilinear map $\Gamma_I(\beta,\gamma)$,
which can also be written in components as
$\sum_{a,b=1}^{16}\Gamma_{I\,ab}\beta^a\gamma^b$. (The bar in
$\bar \beta$ is conventional for spinors and should perhaps be
read as transpose, not complex conjugation.) A standard convention
is to define $\Gamma_{I_1I_2\dots I_k}$ to be zero if the indices
$I_1,I_2,\dots,I_k$ are not pairwise distinct and otherwise to
equal the product $\Gamma_{I_1}\Gamma_{I_2}\cdots \Gamma_{I_k}$.
So $\Gamma_{I_1I_2\dots I_k}$ reverses the chirality if $k$ is
odd, and again can be regarded as a bilinear map ${\EUS}^+\otimes
{\EUS}^+\to \Bbb{R}$ (or ${\EUS}^-\otimes {\EUS}^-\to \Bbb{R}$).
These maps are symmetric for $k=1,5,9$ and antisymmetric for
$k=3,7$.  For $\beta,\gamma\in {\EUS}^+$, we again write
$\bar\beta\Gamma_{I_1\dots I_k}\gamma$ for these bilinear maps
$\Gamma_{I_1\dots I_k}(\beta,\gamma)$. For $k$ even, we have
$\Gamma_{I_1I_2\dots I_k}:{\EUS}^\pm\to{\EUS}^\pm$, or
equivalently we have  bilinear maps $\Gamma_{I_1I_2\dots
I_k}:{\EUS}^+\otimes {\EUS}^-\to \Bbb{R}$, again denoted
$\bar\beta\Gamma_{I_1\dots I_k}\gamma$.

\bigskip\noindent{\it Fields, Transformation Laws, And Lagrangian}

The fields of ten-dimensional super Yang-Mills theory are the
gauge field $A$, which is a connection on a $G$-bundle $E$, and a
fermion field $\lambda$ that is a section of ${\EUS}^+\otimes {\rm
ad}(E)$; in other words, $\lambda$ is a positive chirality spinor
field with values in the adjoint representation of $G$. In Lorentz
signature, $\lambda$ is real, since the bundles ${\EUS}^+$ and
${\EUS}^-$ are real.  In Euclidean signature, $\lambda$ is not
real but its complex conjugate does not enter the formalism. The
covariant derivative is $D=d+A$ and the curvature of $A$ is
$F=D^2=dA+A\wedge A$.

We consider $A$ and $\lambda$ to take values in the real Lie
algebra of $G$, which has real structure constants.  This means
that, in a unitary representation of $G$, $A$ and $\lambda$ take
values in antihermitian matrices. This is opposite to the usual
physics convention, but is in accord with the math literature.
(Taking the fields to be antihermitian may look unnatural for
$G=U(1)$, which is the reason for the usual physics convention,
but it avoids unnatural factors of $i$ for general $G$.)  In a
standard set of physics conventions (see p. 4 of \ref\weinberg{S.
Weinberg, {\it Quantum Theory of Fields, II} (Cambridge University
Press, 1996).}), the covariant derivative is $D=d-iA'$ with a
hermitian gauge field $A'$.  The relation between our
antihermitian $A$ and this hermitian gauge field $A'$ is thus
\eqn\boloko{A=-iA'.}  The curvature of $A'$ is defined as
$F'=i(D)^2=dA'-i A'\wedge A'$, so \eqn\obolko{F=-iF'.}

We use the symbol ``$\Tr$'' to denote an invariant and negative
definite quadratic form on the Lie algebra of $G$.  As we assume
$G$ to be simple, such a quadratic form is unique up to a constant
multiple; we  normalize it so that for $M=S^4$, the characteristic
class ${1\over 8\pi^2}\int_M\Tr\, F\wedge F$ takes arbitrary
integer values. For $G=SU(N)$, this quadratic form can be obtained
as a trace in the $N$-dimensional representation, which motivates
denoting it as $\Tr$. Since our Lie algebras are generated by
antihermitian matrices, $\Tr$ is {\it negative}-definite.

With this understood, the supersymmetry transformation laws and
the Lagrangian of ten-dimensional super Yang-Mills theory can be
described as follows.  The generator of supersymmetry is a
constant spinor $\epsilon$ that takes values in ${\EUS}^+$, and
hence obeys \eqn\goggo{\bar\Gamma\epsilon=\epsilon.} ($\bar
\Gamma$ was defined in \thop; we take $\epsilon$ to be bosonic.)
The supersymmetry transformation generated by $\epsilon$ is
\eqn\hygo{\eqalign{\delta_S A_I & = i\bar\epsilon\Gamma_I\lambda
\cr
                   \delta_S\lambda & = {1\over 2}
                   \Gamma^{IJ}F_{IJ}\epsilon.\cr}}
For any field $\Phi$, the symbol $\delta_S \Phi$ is short
for\foot{In a $\Bbb{Z}_2$-graded algebra, the symbol $[A,B\}$
denotes $AB-(-1)^{AB}BA$.} $\sum_{a=1}^{16}[\epsilon^aQ_a,\Phi\}$,
where $Q_a$, taking values in ${\EUS}^-$, are the sixteen
supersymmetries. The symbol $\delta_S$ stands for supersymmetric
variation. The invariant action is\foot{This is written in Lorentz
signature, with the usual convention in which the kinetic energy
is positive. For Euclidean signature, one must change the overall
sign of the Lagrangian to make the bosonic part of the action
positive definite.} \eqn\ygo{{{\cmmib I}}_{10}={1\over e^2}\int
d^{10}x\,\Tr\left({1\over
2}F_{IJ}F^{IJ}-i\bar\lambda\Gamma^ID_I\lambda\right),} with an
arbitrary constant $e$, the gauge coupling.  The verification of
supersymmetry is described in \brink. Finally, the conserved
supercurrent that generates the supersymmetries \hygo\ is
\eqn\bygo{J^I=\half\Tr\, \Gamma^{JK}F_{JK}\Gamma^I\lambda.}

The bosonic symmetry of this theory is not just $SO(1,9)$ but the
``Poincar\'e group'' $\EUP$, which is an extension of $SO(1,9)$ by
the ``translation'' group of $\Bbb{R}^{1,9}$.  This translation
group is isomorphic to $\Bbb{R}^{1,9}$ itself (regarded as an
abelian group), and ${\EUP}$ is an extension \eqn\ugy{0\to
\Bbb{R}^{1,9}\to \EUP\to SO(1,9)\to 1.}  The generators of
$\Bbb{R}^{1,9}$ are called the ``momentum operators'' $P_I$.  The
algebra obeyed by the conserved supercharges $Q_a$, $a=1,\dots,16$
is \eqn\realify{\{Q_a,Q_b\}=\sum_{I=1}^{10}\Gamma^I_{ab}P_I.} In
addition, the $Q_a$ commute with $P_I$ and transform under
$SO(1,9)$ as ${\EUS}^-$.

\bigskip\noindent{\it Dimensional Reduction To Four Dimensions}

To reduce to four dimensions, we simply take all fields to be
independent of the coordinates $x^4,\dots,x^9$.  The components
$A_I$, $I=0,\dots,3$ describe the four-dimensional gauge field
$A=\sum_{\mu=0}^3A_\mu dx^\mu$, while the components $A_I$, $I\geq
4$, become four-dimensional scalar fields $\phi_i=A_{i+4}$,
$i=0,\dots,5$. The ten-dimensional curvature $F_{IJ}$ has three
types of contribution; depending on whether the number of indices
$I,J$ in the range $4,\dots, 9$ is zero, one, or two, we get a
four-dimensional curvature $F_{\mu\nu}$, a derivative
$D_\mu\phi_i$ of a scalar field, or a commutator $[\phi_i,\phi_j]$
of scalar fields.

The bosonic part of the action, in four dimensions, has all three
types of contribution and becomes \eqn\loopy{{{\cmmib
I}}_4={1\over e^2}\int d^4x\,\Tr\,\left({1\over
2}\sum_{\mu,\nu=0}^3F_{\mu\nu}F^{\mu\nu}+\sum_{\mu=0}^3\sum_{i=1}^6D_\mu\phi_i
D^\mu\phi_i +{1\over 2}\sum_{i,j=1}^6[\phi_i,\phi_j]^2\right).}
Together with the part of the action involving fermions, which can
be similarly written in four-dimensional terms, though we will not
do so, this is the essentially unique four-dimensional gauge
theory with the maximal possible supersymmetry.  If $G$ is simple
and if we want a Lagrangian quadratic in derivatives (the case
that leads to a sensible quantum theory), the action is unique
except for the choice of parameter $e$ and, in four dimensions,
another possible parameter that measures the topology of the
$G$-bundle $E$: \eqn\oopy{{{\cmmib I}}_{\theta}=-{\theta\over
8\pi^2}\int \Tr \,F\wedge F.} This last term is $\theta$ times the
second Chern class or instanton number of the bundle.  The
parameters $e$ and $\theta$ combine into a complex parameter
\eqn\ingo{\tau={\theta\over 2\pi}+{4\pi i\over e^2}.} As long as
we are on $\Bbb{R}^4$ or any four-manifold $M$ with
$H^2(M,\Bbb{Z})=0$ (for the generalization, in which $\pi_1(G)$
comes into play, see \refs{\vafawitten,\witfur}), there is an
elementary symmetry $\tau\to\tau+1$, which expresses the fact that
$(1/8\pi^2)\int\Tr F\wedge F$ is integer-valued, and that in
quantum mechanics one only cares about the action modulo an
integer multiple of $2\pi$. Equivalently, $\theta$ is an angular
variable, with $\theta\cong\theta+2\pi$.

The $SO(1,9)$ (or $SO(10)$) symmetry in ten dimensions becomes,
after dimensional reduction to four dimensions, $SO(1,3)\times
SO(6)$ (or $SO(4)\times SO(6)$). Allowing for the presence of
spinors in the theory, the symmetry is really $Spin(1,9)$ reduced
to $Spin(1,3)\times Spin(6)$ (or $Spin(10)$ reduced to
$Spin(4)\times Spin(6)$).  The group $Spin(6)$ is isomorphic to
$SU(4)$ and is known as the ``$R$ symmetry group'' of the theory.
We will call it $SU(4)_{{\cal R}}$.

\def\btwo{{\bf 2}}
\def\bone{{\bf 1}}
\def\b4{{\bf 4}}
\def\bb4{{\bf \bar 4}}
 The chirality condition $\bar\Gamma\epsilon=\epsilon$ in ten
dimensions becomes in four dimensions \eqn\yug{\hat\Gamma
\Gamma'\epsilon=\epsilon,} where
$\hat\Gamma=\Gamma_0\Gamma_1\Gamma_2\Gamma_3$ measures the
$Spin(1,3)$ chirality and $\Gamma'=\Gamma_4\Gamma_5\dots\Gamma_9$
measures the $Spin(6)$ chirality. $\hat \Gamma$ and $\Gamma'$ have
eigenvalues $\pm i$; \yug\ means that the eigenvalue of $\Gamma'$
is minus that of $\hat \Gamma$. The two eigenvalues of
$\hat\Gamma$ distinguish the two spin representations  of
$Spin(1,3)$, while the eigenvalues of $\Gamma'$ similarly label
the spin representations of $Spin(6)$. The complexification of
$Spin(1,3)$ is $SL(2)\times SL(2)$ and the two spin
representations correspond to the representations $({\bf 2},{\bf
1})$ and $({\bf 1},{\bf 2})$ of $SL(2)\times SL(2)$ (here $({\bf
2},{\bf 1})$ is the two-dimensional representation of the first
$SL(2)$ tensored with the trivial one-dimensional representation
of the second $SL(2)$, and vice-versa for $(\bone,\btwo)$). The
spin representations of $Spin(6)$ are the defining
four-dimensional representation of $SU(4)_{{\cal R}}$ and its
dual; we denote them as ${\bf 4}$ and ${\bf \bar 4}$. We pick
orientations so that $\hat \Gamma$ acts as $i^{-1}$ or $i$ on the
$({\bf 2},{\bf 1})$ and $({\bf 1},{\bf 2})$, respectively, and
$\Gamma'$ acts as $i$ and $i^{-1}$ on the $\bar {\bf 4}$ and ${\bf
4}$.
 So
\yug\ means that the four-dimensional supersymmetries transform
under $Spin(1,3)\times Spin(6)\sim SL(2)\times SL(2)\times
Spin(6)$ as \eqn\jugo{(\btwo,\bone,\bb4)\oplus (\bone,\btwo,\b4).}
The fermion fields $\lambda$ transform the same way under
$Spin(1,3)\times Spin(6)$.

If we write $\bar Q_{AX}$, $A=1,2$, $X=1,\dots, 4$ for the
supersymmetries of type $({\bf 2},\bone,{\bf \bar 4})$, and
similarly $ Q_{\dot A }^Y$, $\dot A=1,2$, $Y=1,\dots,4$ for those
of type $(\bone,{\bf 2},{\bf  4})$, then the algebra generated by
the supersymmetries is the reduction of \realify,
\eqn\ugo{\eqalign{\{\bar Q_{AX}, Q_{\dot A}^
Y\}&=\delta_X{}^Y\sum_{\mu=0}^3\Gamma^\mu_{A\dot A}P_\mu\cr
\{Q,Q\}=\{\bar Q,\bar Q\}&=0,\cr}} where now the four momentum
operators $P_\mu$ generate the translations of $\Bbb{R}^{1,3}$.
With suitable boundary conditions, additional terms appear  \wo\
on the right hand side of \ugo\ (they are related to the extra six
momenta that were dropped in going to four dimensions, and their
magnetic duals).  They will make a brief appearance at the end of
this section.

\subsec{$S$-Duality}

\subseclab\reviewsd

A review of $S$-duality is unfortunately beyond our scope in this
paper.  We will just mention a few relevant facts.

Since its imaginary part is positive, $\tau=\theta/2\pi +4\pi
i/e^2$ defines a point in the upper half plane ${\EUH}$.  The
group $SL(2,\Bbb{R})$ acts on ${\EUH}$ in the standard fashion
$\tau\to (a\tau+b)/(c\tau+d)$, with $ad-bc=1$.\foot{The group that
acts faithfully on ${\EUH}$ is the quotient $PSL(2,\Bbb{R})$, but
in application to four-dimensional gauge theory, one really needs
the double cover.} The transformation $T:\tau\to\tau+1$ is simply
a $2\pi$ shift of the angle $\theta$ and thus a classical symmetry
of the theory on $\Bbb{R}^4$, for any gauge group $G$. The
$S$-duality conjecture asserts that there exists an additional
quantum symmetry that inverts $\tau$, exchanges $G$ with
$^L\negthinspace G$, and exchanges electric and magnetic charges.
Moreover, this symmetry, which we will call $S$, combines with the
classical symmetry $T:\tau\to\tau+1$ to generate an infinite
discrete subgroup $\Gamma$ of $SL(2,\Bbb{R})$.

The most familiar case is the case that $G$ is simply-laced.  Then
\eqn\bivolo{S=\left(\matrix{ 0 & 1\cr -1 & 0\cr}\right)} acts as
$\tau\to -1/\tau$, and together with $T$ generates the group
$SL(2,\Bbb{Z})$.

If $G$ is not simply-laced, then the $S$-transformation is not
$\tau\to -1/\tau$.  Rather, it is $\tau\to -1/n_{\frak g}\tau$,
where $n_{\frak g}$ is $2$ for $F_4$ and 3 for $G_2$.  This
transformation can be achieved by the $SL(2,\Bbb{R})$
transformation \eqn\bivolo{S=\left(\matrix{ 0 & 1/\sqrt {n_{\frak
g}}\cr -\sqrt {n_{\frak g}} & 0\cr}\right).} This presence of a
factor of 2 or 3 in the action of $S$ on $\tau$ can be seen
\refs{\british,\AKS} by examining the BPS mass formulas for
electric and magnetic charges
 and reflects the relation between roots and
coroots for these groups. It also can be extracted from a
string-theoretic explanation of $S$-duality for non-simply-laced
groups \vafageom. The factor of 2 or 3 means that the duality
groups for $G_2$ or $F_4$ are not simply $SL(2,\Bbb{Z})$, but
certain infinite discrete subgroups of $SL(2,\Bbb{R})$ that are
known as Hecke groups.

The remaining simple Lie groups $Sp(k)$ and $Spin(2k+1)$ (and
their respective quotients by $\Bbb{Z}_2$) require some special
comment, because these are the only simple Lie groups such that
$G$ and $^L\negthinspace G$ have non-isomorphic Lie algebras.  The
Yang-Mills Lagrangian and therefore the definition of the $\tau$
parameter depend only on the Lie algebra $\frak g$ of the gauge
group. Hence, whenever $G$ and $^L\negthinspace G$ have the same
Lie algebra, we can discuss how the $S$-transformation acts on
$\tau$ without distinguishing whether we have in mind $\tau$ of a
theory with gauge group $G$ or $\tau$ of a theory with gauge group
$^L\negthinspace G$. This indeed is what we have implicitly done
so far.

For the pair $Sp(k)$ and $Spin(2k+1)$, however, there is no
equally direct way to compare the two $\tau$ parameters.  Hence,
one may introduce separate gauge coupling parameters, say $\tau$
for $Sp(k)$ or $Sp(k)/\Bbb{Z}_2$ gauge theory, and $\tau'$ for
$Spin(2k+1)$ or $Spin(2k+1)/\Bbb{Z}_2=SO(2k+1)$ gauge theory. If
one normalizes the respective $\tau$ parameters so that the
respective $T$-transformations act by $\tau\to \tau+1$ and
$\tau'\to\tau'+1$, then $S$ acts by $\tau=-1/2\tau'$, $\tau'=
-1/2\tau$, just as for $F_4$.  (Thus we set $n_{\frak g}=2$ for
$Spin(2k+1)$ and $Sp(k)$.)

This normalization is natural, since it leads to the most uniform
gauge theory formulas for arbitrary gauge groups. A slight
complication is that there is for the same groups a second
normalization that one might also consider natural.  A useful
string theory realization of the $Sp(k)/Spin(2k+1)$ duality,
involving orientifold threeplanes \wittenbaryon, actually
motivates a different normalization.  In this normalization, the
$Sp(k)$ theory is described by a coupling parameter
$\tilde\tau=2\tau$, so that $S$ acts simply by $\tilde \tau =
-1/\tau'$, but instead $T$ acts by $\tilde\tau\to \tilde \tau+2$.
(In figure 3 of \wittenbaryon, the $Sp(k)$ theory appears twice,
precisely because  the $Sp(k)$ coupling parameter $\tilde\tau$ is
normalized in that paper so that the $T$-symmetry acts by
$\tilde\tau\to\tilde\tau+2$.)

A final comment on this is that the assertion that Montonen-Olive
duality exchanges $G$ and $^L\negthinspace G$ is not quite the
whole story.  On $\Bbb{R}^4$, this is an adequate description, but
on a general four-manifold, the full story is somewhat more
elaborate.  A $G$-bundle on a four-manifold has a characteristic
class $\xi\in H^2(M,\pi_1(G))$, studied in this paper in section
\centertop. A path integral can be defined for each value of
$\xi$, and the resulting partition functions $Z_\xi$ transform as
a unitary representation of the duality group
\refs{\vafawitten,\witfur}.

How much of this is important for the geometric Langlands program?
The basic geometric Langlands duality is the transformation $S$
that acts as $\tau\to -1/n_{\frak g}\tau$.  It acts on the
canonical parameter $\Psi$ (introduced in section \canonpar) by
$\Psi\to -1/n_{\frak g}\Psi$.  The basic geometric Langlands
duality involves comparing $\Psi=\infty$ to $\Psi=0$.  These are
exchanged by $S$ regardless of the value of $n_{\frak g}$.  So the
value of $n_{\frak g}$ is not very important for the basic
geometric Langlands duality.  In section \genotics, we come to a
generalization of the geometric Langlands duality to arbitrary
$\Psi$.  Here the precise duality group is important and,
therefore, the value of $n_{\frak g}$ does play a role.

\bigskip\noindent{\it Transformation of Supersymmetries}

There is one question about Montonen-Olive duality that actually
will play a bigger role in this paper: How does it act on the
supersymmetries?

 We cannot the answer this question by inspection
of the classical Lagrangian, because, apart from the subgroup
generated by $\tau\to \tau+1$, $\Gamma$ does not consist of
symmetries of the classical theory. This after all is what makes
$S$-duality interesting.  So obtaining the answer will require a
more subtle reasoning.

Before determining the answer, let us ask to what extent the
answer is unique.  Consider an element $\gamma=\left(\matrix{a & b
\cr c & d \cr}\right)$ of the duality group $\Gamma$ generated by
$S$ and $T$. Such an element acts on $\tau$ by $\tau\to
(a\tau+b)/(c\tau+d)$ and on the supersymmetry algebra by an
automorphism. This automorphism is not uniquely defined, since it
could be combined with a symmetry of the ${\EUN}=4$ super
Yang-Mills theory (at a fixed value of $\tau$). An important
simplification is that, according to the Montonen-Olive
conjecture, $\gamma$ commutes with the Poincar\'e group. Moreover,
one can define it to commute with  the global $R$-symmetry group
$SU(4)_{{\cal R}}$.\foot{To show this, we first observe that
conjugation by $\gamma$ generates an automorphism of $SU(4)_{{\cal
R}}$.  This automorphism is necessarily inner, as the classical
theory has no symmetry that acts trivially on spacetime and by an
outer automorphism of $SU(4)_{{\cal R}}$. Finally, given that
$\gamma$ generates an inner automorphism of $SU(4)_{{\cal R}}$, we
can combine it with an $SU(4)_{{\cal R}}$ element that generates
the inverse automorphism to get a duality symmetry that commutes
with $SU(4)_{{\cal R}}$. } Combining these facts, it follows that
$\gamma$ acts as a scalar multiplication $\exp(i\hat\phi)$ on the
supersymmetries that transform as the ${\bf \bar 4}$ (that is, the
$\bar Q_{AX}$) and as $\exp(-i\hat\phi)$ on those that transform
as ${\bf 4}$ (the $ Q_{\dot A}^Y$). Moreover, $\hat\phi$ must be
real to preserve the real structure of the algebra (with respect
to which, in Lorentz signature, $P$ is hermitian and $\bar Q$ is
the hermitian adjoint of $Q$).  We will call these symmetries
$U(1)$ chiral rotations. The action of $\gamma$ is defined up to
an element of the center of $SU(4)_{{\cal R}}$. The center is
generated by an element ${\EUI}$ that acts as $i$ on the ${\bf 4}$
of $SU(4)_{{\cal R}}$ and $-i$ on the $\bf{ \bar 4}$. Thus
$\hat\phi(\gamma)$ is defined up to
$\hat\phi(\gamma)\to\hat\phi(\gamma)+\pi/2$.

To determine $\hat\phi(\gamma)$, we can compute the action of
$\gamma$ in any convenient state. We choose to perform the
computation on the Coulomb branch of the theory, where the gauge
group is broken to $U(1)^r$ and the supersymmetry algebra is
centrally extended by electric and magnetic charges.  For an
abelian gauge group such as $U(1)$ or $U(1)^r$, one can calculate
everything explicitly and thereby determine how $\gamma$ acts on
the supersymmetries.  Or one can use the realization of ${\cal
N}=4$ super Yang-Mills theory as the gauge theory on a $D3$-brane
of Type IIB superstring theory; this gives a geometrical way to
determine the action of $\gamma$ on supercharges for $G=U(1)$ (and
this approach also extends directly to classical groups such as
$U(N)$).

\nref\witteneffect{
 E.~Witten,
 ``Dyons Of Charge $e\theta/ 2 \pi$,''
 Phys.\ Lett.\ {\bf B86} (1979) 283-287.}

We will follow the alternative approach of determining the result
by examining the mass formula for BPS states. To simplify
notation, we focus on an ${\EUN}=2$ subalgebra of the
supersymmetry algebra, which has a pair of right-handed
supercharges $Q_{\dot A}^i$, $i=1,2$ and a single central charge
$Z$. They satisfy \eqn\hilj{ \{Q_{\dot A}^i,Q_{\dot
B}^j\}=\eps_{\dot A\dot B}\eps^{ij} Z. } The complex scalar which
is the ${\EUN}=2$ superpartner of the massless gauge fields takes
values in the Cartan subalgebra ${\frak t}$ of the Lie algebra
${\frak g}$. We normalize this scalar so that its kinetic term is
$\tau$-independent and denote its expectation value by
$\vec{\phi}$.  The electric and magnetic charges, denoted by
$\vec{\eurm n}$, $\vec{\eurm m},$ take values in the weight and
coweight lattices of $\frak g$. The vectors $\vec{\phi},\vec{\eurm
n},\vec{\eurm m}$ are defined up to an action of the Weyl group.

To make our conventions clear, we will give the explicit
definition of $\eurm n$ and $\eurm m$ for $G=U(1)$.  Physically,
one usually describes $U(1)$ gauge theory with a real connection
$A'$ of curvature $F'=dA'$.  We write Minkowski spacetime as
$\Bbb{R}^{1,3}=\Bbb{R}\times \Bbb{R}^3$ where the first factor
parametrizes time and the second parametrizes space.  To agree
with Maxwell's equations (see, for example, p. 42 of
\ref\weintwo{S. Weinberg, {\it Gravitation And Cosmology} (WIley,
1972).}), one defines the electric and magnetic fields by
$E_i=F'{}^{0i}$ and $B_i=\half\epsilon_{ijk}F'_{jk}$ (here
$\epsilon_{ijk}$ is the antisymmetric tensor on $\Bbb{R}^3$ with
$\epsilon_{123}=1$). Then $\eurm n$ and $\eurm m$ are defined by
\eqn\eltro{\eqalign{{\eurm n} & = {1\over 4\pi}\int_Y d\Omega\,
\vec n\cdot \vec E\cr {\eurm m}&={1\over 4\pi}\int_Yd\Omega\,\vec
n\cdot \vec B\cr}} where $Y$ is a large sphere at infinity in
$\Bbb{R}^3$ with volume-form $d\Omega$ and normal vector $\vec n$.
In terms of the antihermitian gauge field $A$ related to $A'$ by
$A=-iA'$ and $F=-iF'$ (recall eqns. \boloko\ and \obolko), we have
\eqn\feltro{\eqalign{{\eurm n} & = {i\over 4\pi}\int_Y d\Omega\,
n^i F^{0i}\cr {\eurm m}&={i\over 4\pi}\int_Yd\Omega\,\half
\epsilon_{ijk}n^i F^{jk}.\cr}} In terms of differential forms, the
second formula is \eqn\ofeltro{{\eurm m}={i\over 2\pi}\int_YF
=\int_Y c_1({\cal L}),} where ${\cal L}$ is a line bundle on which
$A$ is a connection.

The central charge $Z$ is given by \eqn\pikf{ Z=\sqrt{2\over {\rm
Im}\,\tau}\,\left(\vec{\eurm m}~\vec{\eurm n}\right)\cdot
\left(\matrix{\tau\vec \phi \cr\vec \phi \cr}\right)=\sqrt {2\over
{\rm Im}\,\tau}\,\vec \phi\cdot (\vec{\eurm n}+\tau \vec{\eurm m})
 }  For simply-laced $G$, the low energy abelian gauge theory on
 the Coulomb branch admits an $SL(2,\Bbb{Z})$ group of duality
 symmetries such that a general element
$\eurm M=\left(\matrix{a&b\cr c&d\cr}\right)$, which  acts on
$\tau$ by $\tau\to (a\tau+b)/(c\tau+d)$, acts on $\vec{\phi},\,
\vec{{\eurm m}},\, \vec{{\eurm n}} $ by
\eqn\xilj{\eqalign{\vec{\phi} & \to
\vec{\phi}\cr\left(\matrix{\vec{{\eurm m}} & \vec{{\eurm
n}}}\right)&\to \left(\matrix{\vec{{\eurm m}} & \vec{{\eurm
n}}}\right){\eurm M}^{-1}=\left(\matrix{ \vec{{\eurm m}} &
\vec{{\eurm n}}}\right)\left(\matrix{d & -b  \cr -c & a
\cr}\right).} } The $S$-duality conjecture asserts that this
symmetry of the low energy theory actually extends to a symmetry
of the full theory, necessarily acting in the same way. That the
generator $T:\tau\to\tau+1$ of the duality group $\Gamma$ extends
to the full theory is clear (since it is just a $2\pi$ shift in
the $\theta$ angle), so the conjecture really is that $S:\tau\to
-1/\tau$ similarly extends to the full theory. The action of $T$
claimed in \xilj\ can be seen in the full theory by a direct
computation \witteneffect.  This direct computation is valid for
all $G$, simply-laced or not; a variant of it will be presented in
section \tophooft. From \pikf\ and \xilj, one deduces the
transformation of the central charge: \eqn\mujk{ Z\to
{|c\tau+d|\over {c\tau+d}}\, Z.} Then the centrally-extended
supersymmetry algebra implies that the right-handed
supersymmetries get multiplied by
\eqn\tokio{\exp(-i\hat\phi(\gamma))=\left({|c\tau+d|\over
{c\tau+d}}\right)^{1/2}.}

The square root means that the group that acts on the
supersymmetries is a double cover of the duality group $\Gamma$.
The extension is by a symmetry $(-1)^F$ of order two that changes
the sign of all fermions. This symmetry is the {\it square} of the
generator ${\EUI} $ of the center of $SU(4)_{\cal R}$. We avoided
getting an extension of $\Gamma$ by the full center by taking
$\vec{\phi}$ to be invariant under $\Gamma$ rather than invariant
up to sign.

In case $G$ is not simply-laced, this argument needs to be phrased
more carefully.  The duality group is generated by $T:\tau\to
\tau+1$ and $S:\tau\to -1/n_{\frak g}\tau$.  $T$ leaves fixed the
supersymmetries, so $\hat\phi=0$ for $T$, in keeping with \tokio.
As for $S$, its action is now \eqn\xiljo{\eqalign{\vec{\phi} &
\mapsto {\frak R}\cdot\vec{\phi}\cr\left(\matrix{\vec{{\eurm m}} &
\vec{{\eurm n}}}\right)&\mapsto \left(\matrix{{\frak R}\cdot
\vec{{\eurm m}} & {\frak R}\cdot\vec{{\eurm
n}}}\right)\left(\matrix{0 & -1/\sqrt n_{\frak g} \cr \sqrt
n_{\frak g} & 0 \cr}\right).} } Here $\frak R$ is an orthogonal
transformation of $\frak t$ which for simply-laced groups can be
taken to be trivial, and for non-simply-laced groups is described
in \AKS. The square of $\frak R$ belongs to the Weyl group, so the
square of $S$ acts trivially on the moduli space. $\frak R$ does
not affect the computation of the transformation of the central
charge, and a small computation shows that $n_{\frak g}$ cancels
out, so \tokio\ remains valid in the general case.

\newsec{Topological Field Theory From ${\EUN}=4$ Super
Yang-Mills Theory} \seclab\tqftfour

Our next goal is to describe how to construct a four-dimensional
topological quantum field theory (TQFT) on any
four-manifold\foot{We generally assume $M$ to be oriented, as the
construction is more interesting in this case.  However, if we
specialize to the case that the theta angle of eqn. \oopy\
vanishes and the canonical parameter $\Psi$ of section \canonpar\
equals 0 or $\infty$, then the construction makes sense even for
unorientable $M$.} $M$ by twisting of ${\EUN}=4$ super Yang-Mills
theory. In constructing TQFT's, it is most natural to use
Euclidean signature. Topological field theory is most naturally
related to Euclidean signature, and in any event the twisting we
use does not work well in Lorentz signature. When we specialize to
$M=\Bbb{R}\times W$ or $\Bbb{R}^2\times C$ for a three-manifold
$W$ or a two-manifold $C$, we can usefully return to Lorentz
signature by taking Lorentz signature on $\Bbb{R}$ or $\Bbb{R}^2$.

One important change in going to Euclidean signature is that the
spin representations $(\btwo,\bone)$ and $(\bone,\btwo)$ of
$Spin(4)$ are pseudoreal, while for $Spin(1,3)$ these
representations are complex conjugates of each other.
Correspondingly,
 the operator
$\Gamma_0\Gamma_1\Gamma_2\Gamma_3$ that distinguishes the two spin
representations now squares to $+1$; to avoid confusion, we will
call it $\hat\Gamma_E$ when using Euclidean signature and restrict
$\hat\Gamma$ for the Lorentz signature case.  Similarly, the two
spin representations ${\EUS}^\pm$ are complex conjugates of each
other in Euclidean signature, and the operator
$\Gamma_0\Gamma_1\dots\Gamma_9$ that distinguishes them, which we
now call $\bar\Gamma_E$, has eigenvalues $\mp i$ on ${\EUS}^\pm$.

To determine the sign, letting a subscript $L$ or $E$ refer to
Lorentz or Euclidean signature, we make the Wick rotation from
Lorentz to Euclidean signature by $ix^0_L=x^0_E$, so
$\partial/\partial x^0_L=i\partial/\partial x^0_E$ and
$\Gamma_{0\,L}=i\Gamma_{0\,E}$.  Hence, $\bar\Gamma_L =
i\bar\Gamma_E$, and as $\bar\Gamma=\bar\Gamma_L$ acts on
$\EUS^\pm$ as multiplication by $\pm 1$, $\bar\Gamma_E$ acts by
multiplication by $\mp i$.  In particular, if $\epsilon$ is a
supersymmetry generator, then
\eqn\nohinks{\bar\Gamma_E\epsilon=-i\epsilon.}

\subsec{Twisting ${\EUN}=4$ Super Yang-Mills}

\subseclab\twisting

\bigskip\noindent{\it General Idea Of Twisting}

\nref\yamron{J. P. Yamron, ``Topological Actions {}From Twisted
Supersymmetric Theories,'' Phys. Lett. {\bf B213} (1988) 325-330.}
\nref\marcus{N. Marcus, ``The Other Topological Twisting Of ${\cal
N}=4$ Yang-Mills,'' Nucl. Phys. {\bf B452} (1995) 331-345,
arXiv:hep-th/9506002.}%
\nref\bt{M. Blau and G. Thompson, ``Aspects Of $N_T\geq 2$
Topological Gauge Theories And $D$-Branes,'' Nucl. Phys. {\bf
B492} (1997) 545-590, arXiv:hep-th/9612143.}%
\nref\llozano{ J. M. F. Labastida and C. Lozano, ``Mathai-Quillen
Formulation Of
 Twisted ${\EUN}=4$ Supersymmetric Gauge Theories In Four Dimensions,'' Nucl. Phys.
 {\bf B502} (1997) 741-790.}%
 \nref\lozano{ C. Lozano,
``Duality In Topological Quantum Field Theories,''
arXiv:hep-th/9907123.} \nref\labloz{J. M. F. Labastida and C.
Lozano, ``Duality In Twisted ${\EUN}=4$ Supersymmetric Gauge
Theories In Four
Dimensions,'' Nucl. Phys. {\bf B537} (1999) 203-242.}%

 ${\EUN}=2$ super Yang-Mills theory can be twisted to make a
quantum field theory realization of Donaldson theory \eww.
Similarly \vafawitten, ${\EUN}=4$ super Yang-Mills theory can be
twisted in three ways to make a topological field theory.  Two of
the twisted theories, including one that was investigated in
detail in \vafawitten, are closely analogous to Donaldson theory
in the sense that they lead to instanton invariants which, like
the Donaldson invariants of four-manifolds, can be expressed in
terms of the Seiberg-Witten invariants.   The third twist, which
was mentioned in \yamron\ and has been investigated in
\refs{\marcus-\labloz}, has had no applications until now. It
turns out to be the twist relevant to the geometric Langlands
program, and we will call it the GL twist.

To give an inevitably very brief explanation of the notion of
twisting, we first take $M=\Bbb{R}^4$. This has rotational
symmetry group $Spin(4)$, of course, while the ${\EUN}=4$ theory
has the larger symmetry $Spin(4)\times Spin(6)$.
 ``Twisting'' means replacing
$Spin(4)$ by a different subgroup $Spin'(4)$ of $Spin(4)\times
Spin(6)$ that is isomorphic to $Spin(4)$ and acts on $\Bbb{R}^4$
the same way that $ Spin(4)$ does, but acts differently on the
${\EUN}=4$ gauge theory. To accomplish this, we pick a
homomorphism $\varkappa:Spin(4)\to Spin(6)$ and set
$Spin'(4)=(1\times \varkappa)(Spin(4))\subset Spin(4)\times
Spin(6)$.

We also want to pick $\varkappa $ such that the action of
$Spin'(4)$ on ${\EUS}^+$ has a non-zero invariant vector.  Since
the supersymmetry generator $\epsilon$ takes values in  $
{\EUS}^+$, a choice of  an invariant vector in ${\EUS}^+$ will
give us a $Spin'(4)$-invariant supersymmetry that we will call
$Q$.  It will automatically obey $Q^2=0$, by virtue of \ugo.
(There is no $Spin'(4)$ invariant operator on the right hand side
of \ugo, so as $Q^2=\half\{Q,Q\}$ is $Spin'(4)$-invariant, it must
vanish.) Once $Q$ is chosen, we change the physical interpretation
of the theory to say that we are only interested in $Q$-invariant
path integrals, operators, and states, and that we consider
anything of the form $[Q,{\EUO}\}$, for any operator ${\EUO}$, to
be trivial. So henceforth, the interesting operators or states lie
in suitable cohomology groups of $Q$.

It turns out that theories obtained this way are, loosely
speaking, topological field theories. (They may depend on the
smooth structure of $M$, which goes into the definition of the
quantum field theory, before or after twisting.) This is proved by
showing that the definition of the theory on flat $\Bbb{R}^4$ can
be extended to any four-manifold in such a way that the $Q$
symmetry is retained and the choice of metric is irrelevant modulo
$Q$-exact terms.
 For the GL twist of ${\EUN}=4$ super Yang-Mills theory,
 one approach to this can be found in section 7 of \lozano. In
the present paper, we will take another route; once we have worked
out the topological equations and the vanishing theorems, we will
see directly, in section \toplag, what action on a curved
four-manifold has the right properties.

\bigskip\noindent{\it Description Of Twist}

Now we will describe the GL twist. The homomorphism
$\varkappa:Spin(4)\to Spin(6)$ is chosen so that the ${\bf 4}$ of
$Spin(6)=SU(4)_{{\cal R}}$ transforms as $({\bf 2},\bone)\oplus
(\bone,{\bf  2})$ of $Spin(4)=SU(2)\times SU(2)$, which we will
refer to as $SU(2)_\ell\times SU(2)_\rr$.\foot{ The ``$\ell$'' and
``$r$'' in $SU(2)_\ell$ and $SU(2)_\rr$ are usually read as
``left'' and ``right,'' referring to the spin of physical massless
particles.} (The ${\bf \bar 4}$ of $Spin(6)$, which is the complex
conjugate of the ${\bf 4}$, transforms the same way under
$Spin(4)$, since the $({\bf 2},\bone)$ and $(\bone,{\bf 2})$ of
$Spin(4)$ are pseudoreal.)

This choice of $\varkappa$ amounts to embedding $SU(2)_\ell\times
SU(2)_\rr$ in $Spin(6)=SU(4)_{{\cal R}}$ in the following way:
\eqn\norml{\left(\matrix{SU(2)_\ell &  0 \cr
                                                  0    &
                                                  SU(2)_\rr\cr}\right).}
This embedding obviously commutes with an additional $U(1)$
group,\foot{The global structure of the combined  group is
actually not a product $SU(2)_\ell\times SU(2)_\rr\times U(1)$,
but rather the quotient $(SU(2)_\ell\times SU(2)_\rr\times
U(1))/\Bbb{Z}_2$.} whose generator ${\cal K}$ we can take to have
the form \eqn\orml{{\cal K}=i\left(\matrix{1&  0 \cr
                                                  0    &
                                                  -1\cr}\right)}
in $2\times 2$ blocks.  So our embedding is such that the ${\bf
4}$ of $Spin(6)$ transforms under $SU(2)_\ell\times
SU(2)_\rr\times U(1)$ as $({\bf 2},\bone)^{1}\oplus (\bone,{\bf
2})^{-1}$, where now the exponent is the eigenvalue of $-i{\cal
K}$, which we will call ${\EUK}$. The ${\bf \bar 4}$ transforms as
the complex conjugate of this, or $({\bf 2},\bone)^{-1}\oplus
(\bone,{\bf 2})^{1}$.

This twist can also be conveniently described in terms of $SO$
groups rather than $Spin$ groups.  To do so, we use the fact that
the fundamental, six-dimensional vector representation ${\bf 6_v}$
of $SO(6)$ is, in terms of $Spin(6)=SU(4)_{{\cal R}}$, the same as
$\wedge^2 {\bf 4}$, the skew-symmetric part of ${\bf 4}\otimes{\bf
4}$. So ${\bf 6_v}=\wedge^2(({\bf 2},\bone)^{1}\oplus (\bone,{\bf
2})^{-1})=(\btwo,\btwo)^0\oplus (\bone,\bone)^2\oplus
(\bone,\bone)^{-2}$.  Here $(\btwo,\btwo)$ is the same as the
vector representation ${\bf 4_v}$ of $SO(4)$. So the ${\bf 6_v}$
is the sum of a vector ${\bf 4_v}$ of $SO(4)$ with ${\EUK}=0$, and
two $SO(4)$  scalars with ${\EUK}=\pm 2$. This corresponds to the
obvious homomorphism of $SO(4)\times U(1)=SO(4)\times SO(2)$ to
$SO(6)$:\foot{The $U(1)$ in $(SU(2)\times SU(2)\times
U(1))/\Bbb{Z}_2\subset SU(4)$ is a double cover of the $U(1)$ in
$SO(4)\times U(1)\subset SO(6)$.} \eqn\tomo{\left(\matrix{SO(4) &
0 \cr 0 & SO(2)\cr}\right).}

The six spin zero fields $\phi_i$, $i=0,\dots,5$ of ${\EUN}=4$
super Yang-Mills theory transform as ${\bf 6_v}$ of $SO(6)$, so
this analysis applies to them.  We can pick coordinates so the
first four fields $\phi_0,\dots,\phi_3$ form a ${\bf 4_v}$ of
$SO(4)$, while $\phi_4$ and $\phi_5$ are the $SO(4)$ scalars,
which are rotated by $SO(2)$.  Moreover, we can label the scalars
so that the linear combinations $\sigma=(\phi_4-i\phi_5)/\sqrt 2$
and $\bar\sigma=(\phi_4+i\phi_5)/\sqrt 2$ have ${\EUK}=2$ and
${\EUK}=-2$. The fields $\phi_0,\dots,\phi_3$ can then be
interpreted geometrically as the components of an adjoint-valued
one-form, while $\sigma$ is a scalar field or zero-form with
values in the complexification of the Lie algebra.  So the bosonic
fields of the theory are a gauge field, which locally is an
adjoint-valued one-form $A=A_\mu \,dx^\mu$, along with a second
adjoint-valued one-form $\phi=\phi_\mu\,dx^\mu$, and the complex
scalars $\sigma$ and $\bar\sigma$.

\bigskip\noindent{\it Transformation Of The Supersymmetries}

We can likewise analyze how the supersymmetries transform under
$Spin'(4)$. The ${\bf \bar 4}$ of $Spin(6)$ transforms as
$(\btwo,\bone)^{-1}\oplus (\bone,\btwo)^{1}$ of $Spin'(4)\times
U(1)$, and the ${\bf  4}$ as $(\btwo,\bone)^{1}\oplus
(\bone,\btwo)^{-1}$. So, using \jugo, the supersymmetries that
transform as $(\btwo,\bone)$ of $Spin(4)$ transform under
$Spin'(4)\times U(1)$ as \eqn\lollypop{(\btwo,\bone)^0\otimes
((\btwo,\bone)^{-1}\oplus (\bone,\btwo)^{1}) =
(\bone,\bone)^{-1}\oplus ({\bf 3},\bone)^{-1}
\oplus(\btwo,\btwo)^{1}.} And the supersymmetries that transform
as $(\bone,\btwo)$ of $Spin(4)$ transform under $Spin'(4)\times
U(1)$ as \eqn\tollypop{(\bone,\btwo)^0\otimes
((\btwo,\bone)^{1}\oplus (\bone,\btwo)^{-1}) = (\bone,\bone)^{-1}
\oplus (\bone,{\bf 3})^{-1} \oplus(\btwo,\btwo)^{1}.}

Next, we can find the $Spin'(4)$-invariant supersymmetries. {}From
\lollypop\ and \tollypop, there is one invariant supersymmetry
generator $\epsilon_\ell$ that is left-handed in the
four-dimensional sense (transforms as $({\bf 2},\bone)$ under the
original $Spin(4)$), and one such generator $\epsilon_\rr$ that is
right-handed (transforms as $(\bone,{\bf 2})$). With our
conventions, they are distinguished by
\eqn\onog{\eqalign{\hat\Gamma_E\epsilon_\ell& = -\epsilon_\ell\cr
\hat\Gamma_E\epsilon_\rr& = \epsilon_\rr.\cr}} A choice of
$\epsilon_\ell$ determines a natural choice of $\epsilon_\rr$,
namely \eqn\huy{\epsilon_\rr=N\epsilon_\ell,} where $N={1\over
4}\sum_{\mu=0}^3\Gamma_{\mu+4}\Gamma_{\mu}$. This idea here is
that $N$ commutes with $Spin'(4)$ but anticommutes with
$\hat\Gamma_E$, so the definition \huy\ makes sense.  We have
normalized $N$ so that $N^2\epsilon=-\epsilon$ if $\epsilon$ is
$Spin'(4)$-invariant. Hence we have also
\eqn\uy{\epsilon_\ell=-N\epsilon_\rr.} By exploiting the
$Spin'(4)$ symmetry, one can show that for $\mu=0,\dots,3$, we
have \eqn\uzu{\Gamma_{\mu+4}\epsilon_\ell =
-\Gamma_\mu\epsilon_r,~\Gamma_{\mu+4}\epsilon_r=\Gamma_\mu\epsilon_\ell.}

\ifig\zunko{\bigskip  A family of topological field theories
parametrized by the variable $t$. }
{\epsfxsize=3.5in\epsfbox{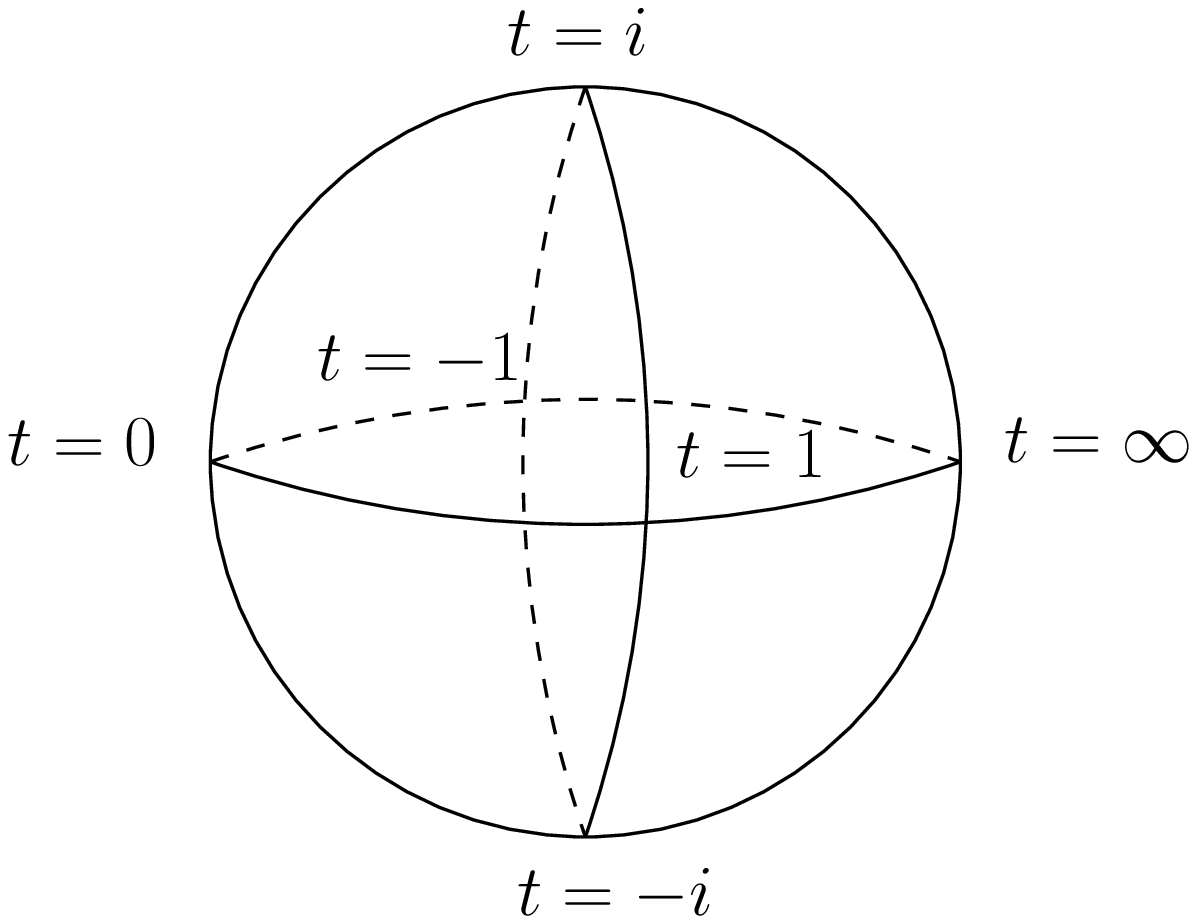}}

The generators of $Spin'(4)$ are
$\Gamma_{\mu\nu}+\Gamma_{\mu+4,\nu+4}$, for $\mu,\nu=0,\dots,3$,
so a $Spin'(4)$-invariant spinor $\epsilon$ obeys
$(\Gamma_{\mu\nu}+\Gamma_{\mu+4,\nu+4})\epsilon=0$.  Setting
$\mu,\nu=0,1$, we learn that $\Gamma_{0145}\epsilon=\epsilon$, and
setting $\mu,\nu=2,3$, we learn that
$\Gamma_{2367}\epsilon=\epsilon$.  So
$\Gamma_0\Gamma_1\dots\Gamma_7\epsilon=\epsilon$, and using also
the chirality condition
$\Gamma_0\Gamma_1\dots\Gamma_9\epsilon=-i\epsilon$ of eqn.
\nohinks, we learn that
\eqn\banother{\Gamma_{89}\epsilon=-i\epsilon.} It follows in
particular that ${\cal K}=\Gamma_{89}$ acts with the same
eigenvalue $\EUK=-i{\cal K}=-1$ on $\epsilon_\ell $ and
$\epsilon_r$.  It also follows that $\Gamma_{8-i9}\epsilon=0$,
where $\Gamma_{8-i9}=(\Gamma_8-i\Gamma_9)/\sqrt 2$.  These facts
are helpful in verifying the supersymmetry transformation laws
presented below.

We now can see why this construction gives a family of topological
field theories parametrized by $\Bbb{CP}^1$ (\zunko).  We pick any
nonzero complex linear combination of $\epsilon_\ell$ and
$\epsilon_\rr$ \eqn\utu{\epsilon=u\epsilon_\ell+v\epsilon_\rr,}
and take this to be the generator of the topological symmetry.  An
overall scaling of $\epsilon$ would not matter, so the possible
choices for the topological symmetry are parametrized by
$\Bbb{CP}^1$. Because there is a natural choice of $\epsilon_\rr$
once $\epsilon_\ell$ is given, this has much more structure than
an abstract copy of $\Bbb{CP}^1$; it has a natural affine
parameter $t=v/u$.

We can use \tokio\ to determine the transformation of $u, v$, and
$t$ under $S$-duality.  Under an element $\gamma$ of the duality
group $\Gamma$, we have $\epsilon_\ell\to
\exp(i\hat\phi(\gamma))\epsilon_\ell$, $\epsilon_\rr\to
\exp(-i\hat\phi(\gamma))\epsilon_\rr$.  Hence, to ensure
invariance of $\epsilon=u\epsilon_\ell+v\epsilon_\rr$, we have
$u\to \exp(-i\hat\phi(\gamma))u$, $v\to \exp(i\hat\phi(\gamma))v$,
and finally the transformation of $t=v/u$ is \eqn\duco{t\to
\exp(2i\hat\phi(\gamma))t={{c\tau+d}\over\ {|c\tau+d|}}t.}
Similarly, we can determine the action on $t$ of the center of
$SU(4)_{{\cal R}}$. The generator ${\EUI}$ of the center acts by
$\epsilon_\ell\to i\epsilon_\ell$, $\epsilon_\rr\to
i^{-1}\epsilon_\rr$, so by a similar reasoning to the above, we
have \eqn\ungu{{\EUI}(t)=-t.}

Thus, the topological field theories with parameter $t$ or $-t$
are equivalent. This is a trivial equivalence in the sense that it
follows from a symmetry of the classical action -- as opposed to a
non-trivial $S$-duality.  The $\Bbb{Z}_2$ symmetry \ungu\ is the
only trivial equivalence among the twisted topological field
theories parametrized by $t$. In making this assertion, it is
essential that $\epsilon_\ell$ and $\epsilon_\rr$ have the same
eigenvalue of ${\cal K}$. Otherwise, a transformation in the group
$\Bbb{C}^*$ generated by ${\cal K}$ would act non-trivially on $t$
and would cause the whole $\Bbb{CP}^1$ family of topological field
theories to be trivially equivalent.  Of course, it is only
because there are (almost) no trivial equivalences that it is
possible for the non-trivial equivalences coming from $S$-duality
to generate something interesting like the geometric Langlands
program.

We write $Q_\ell$ and $Q_\rr$ for the supersymmetries generated by
$\epsilon_\ell$ and $\epsilon_\rr$. The supersymmetry generated by
the linear combination $\epsilon=u\epsilon_\ell+v\epsilon_\rr$ is
then \eqn\hut{Q=uQ_\ell+vQ_\rr,} and this is the topological
symmetry of the twisted theory. It is also convenient to write
\eqn\nut{\delta_T=\epsilon_\ell\delta_\ell+\epsilon_\rr\delta_\rr,}
where for any field $\Phi$,
\eqn\but{\delta_T\Phi=[Q,\Phi\},~~~\delta_\ell\Phi=
[Q_\ell,\Phi\},~~~\delta_\rr\Phi=[Q_\rr,\Phi\}.} ($T$  stands for
twisted or topological.)  Since $\epsilon_\ell$ and $\epsilon_r$
have $\EUK=-1$, it follows that $Q_\ell$, $Q_\rr$, and $Q$ all
have $\EUK=1$. Indeed, the supersymmetry transformation laws
presented below show explicitly that $\delta_T$ increases $\EUK$
by 1 for every field.  Thus in mathematical language, the space of
gauge-invariant operators or states is a complex, $\Bbb{Z}$-graded
by $\EUK$, with differential $Q$.

\bigskip\noindent{\it The Fermion Fields And The Supersymmetry
Transformations}

The fermion fields $\lambda$ transform the same way as the
supersymmetry generators $\epsilon$, so using \lollypop\ and
\tollypop, we can identify them in four-dimensional terms. The
fields with ${\EUK}=1$ are two copies of $(\btwo,\btwo)^1$,
corresponding to two one-forms $\psi$ and $\tilde\psi$. The fields
transforming as $({\bf 3},{\bf 1})^{-1}$ and $(\bone,{\bf
3})^{-1}$ are the selfdual and anti-selfdual parts $\chi^\pm$ of a
two-form $\chi$ of ${\EUK}=-1$. Finally, there are two zero-forms
$\eta$ and $\tilde\eta$ of ${\EUK}=-1$.

It is helpful to define these four-dimensional fields more
precisely by specifying the following expansion of the
ten-dimensional spinor field $\lambda$ in terms of
four-dimensional fermion fields:\foot{Since $\lambda$,
$\epsilon_\ell$, and $\epsilon_\rr$ are all of positive chirality,
this expansion must be made using only even elements of the
Clifford algebra. To ensure this, certain factors of
$\Gamma_{8+i9}={1\over \sqrt 2}\left(\Gamma_8+ i \Gamma_9\right)$
are included in the expansion.}
 \eqn\iko{\eqalign{\lambda=&\left(\eta
+\sum_\mu\Gamma^\mu\psi_\mu \Gamma_{8+i9}
+\sum_{\mu<\nu}\Gamma^{\mu\nu}\chi^+_{\mu\nu}\right)\epsilon_\ell
\cr+& \left(\tilde\eta +\sum_{\mu}\Gamma^\mu\tilde\psi_\mu
\Gamma_{8+i9}+\sum_{
\mu<\nu}\Gamma^{\mu\nu}\chi^-_{\mu\nu}\right)\epsilon_\rr.\cr}}
This formula uniquely determines sixteen fields ($\eta$ and
$\tilde\eta$, four components each of $\psi$ and $\tilde\psi$, and
six components of $\chi$) in terms of the sixteen components of
$\lambda$.  Note that, because
$\hat\Gamma_E\epsilon_\ell=-\epsilon_\ell$,
$\Gamma_{\mu\nu}\epsilon_\ell$ is selfdual, so
$\Gamma^{\mu\nu}\chi_{\mu\nu}\epsilon_\ell=\Gamma^{\mu\nu}\chi^+_{\mu\nu}\epsilon_\ell$.
Likewise
$\Gamma^{\mu\nu}\chi_{\mu\nu}\epsilon_\rr=\Gamma^{\mu\nu}\chi^-_{\mu\nu}\epsilon_\rr$.
So one could equivalently replace $\chi^\pm$ by $\chi$ in \iko.

It is convenient to normalize $\epsilon_\ell$ up to sign so that
\eqn\zomop{\bar\epsilon_\ell \Gamma_{8+i9}\epsilon_\ell = 1.} Then
it follows from $\epsilon_r=N\epsilon_\ell$ that likewise
\eqn\omop{\bar\epsilon_r\Gamma_{8+i9}\epsilon_r=1.} {}From
considerations of four-dimensional chirality, it follows that
\eqn\tomop{\bar\epsilon_\ell\Gamma_{8+i9}\epsilon_r
=\bar\epsilon_r\Gamma_{8+i9}\epsilon_\ell=0.}

The next step is to describe how the fields transform under the
topological symmetries, that is, the supersymmetries generated by
$\epsilon_\ell$ and $\epsilon_\rr$.   For example, to transform
the bosons, we start with $\delta_S A_I=i\bar\epsilon
\Gamma_I\lambda$. Upon setting $\epsilon =
u\epsilon_\ell+v\epsilon_\rr$, we get for the four-dimensional
gauge field \eqn\kono{\delta_T A_\mu = iu\bar \epsilon_\ell
\Gamma_\mu \lambda +iv\bar\epsilon_\rr\Gamma_\mu\lambda.} After
some $\Gamma$-matrix gymnastics, one finds the transformation of
the four-dimensional gauge field under the topological
supersymmetry: \eqn\bono{\delta_T A_\mu=iu
\psi_\mu+iv\tilde\psi_\mu.} Similarly, for the other bosonic
fields we find \eqn\lono{\delta_T\phi_\mu=iv\psi_\mu -
iu\tilde\psi_\mu.} \eqn\otono{\eqalign{\delta_T\sigma & = 0 \cr
                       \delta_T\bar\sigma & = iu\eta+iv\tilde\eta.\cr}}
The vanishing of $\delta_T\sigma$ actually follows merely from the
fact that $\sigma$ has ${\EUK}=2$, and there is no appropriate
field of ${\EUK}=3$.

Likewise, we can find the transformations of the four-dimensional
fermionic fields starting with the underlying transformation law
\eqn\invo{\delta_T\lambda=\half\sum_{I,J=0}^9F_{IJ}\Gamma^{IJ}\epsilon,}
where $\epsilon=u\epsilon_\ell+v\epsilon_\rr$. The fermionic
fields with ${\EUK}=-1$ transform by
\eqn\helpme{\eqalign{\delta_T\chi^+ & = u(F-\phi\wedge\phi)^+
+v(D\phi)^+\cr \delta_T\chi^-& = v(F-\phi\wedge\phi)^- -
u(D\phi)^-\cr \delta_T\eta& = vD^*\phi+u[\bar\sigma,\sigma] \cr
\delta_T\tilde\eta& = -uD^* \phi+v[\bar\sigma,\sigma].\cr}} Those
of ${\EUK}=1$ transform by \eqn\mimo{\eqalign{\delta_T\psi& = u\,
D\sigma +v[\phi,\sigma]\cr \delta_T\tilde\psi & = v\, D\sigma - u
[\phi,\sigma].\cr}} In these formulas, $D=d+[A,\,\cdot \,]$ is the
covariant derivative with respect to the connection $A$; a Lie
bracket is implicit in the adjoint-valued two-form
$\phi\wedge\phi$; and $D^*\phi=\star D \star \phi =
D_\mu\phi^\mu$, with $\star$ the Hodge star.

\subsec{A Family Of $A$-Models}

\subseclab\familya

We are in the following general situation.  We have a quantum
theory with a fermionic symmetry $Q=uQ_\ell+vQ_\rr$ that obeys
$Q^2=0$ (modulo a gauge transformation). In such a situation, in
general, the path integral localizes on fields that are invariant
under $Q$. When the space $\eurm Y$ of $Q$-invariant fields is
smooth and the gauge group acts freely on it, the path integral
can be evaluated by a Gaussian or one-loop approximation,
expanding around $\eurm Y$. More generally, one might have to go
to higher order, but the path integral is always determined by the
structure of a finite order infinitesimal neighborhood of $\eurm
Y$.

It suffices to identify the bosonic fields that are invariant
under $Q$, since fermions are in any case infinitesimal, and
automatically are incorporated in perturbation theory.  The
condition for a bosonic field to be invariant under $Q$ is that
$\delta_T\Psi$ vanishes for every fermionic field $\Psi$.  In
other words, $\eurm Y$ is defined by setting to zero the right
hand sides of \helpme\ and \mimo.

For these expressions to vanish is a condition that is invariant
under scaling of $u$ and $v$ and so only depends on $t=v/u$.
Vanishing of the right hand side of \helpme\ gives\foot{To get the
third equation, we take a $t$-dependent linear combination of  the
conditions $\delta_T\eta=0$ and $\delta_T\tilde\eta=0$, if
$t\not=\pm i$.  What happens if $t=\pm i$ is discussed below.}
\eqn\loopme{\eqalign{(F-\phi\wedge\phi+tD\phi)^+ & = 0
\cr
                      (F-\phi\wedge\phi-t^{-1}D\phi)^-& = 0 \cr
                      D^* \phi & = 0.\cr}}

 The most elementary interpretation of these equations is
for $-t^{-1}=t$, or equivalently $t=\pm i$.  For these values of
$t$, which will be important for the geometric Langlands program,
let ${\cal A}$ be the ${G}_{\Bbb{C}}$-valued connection ${\cal
A}=A+i\phi$, with curvature ${\cal F}=d{\cal A}+{\cal A}\wedge
{\cal A}$.  Then the first two equations in \loopme\ are
equivalent to ${\cal F}=0$, so that a solution of \loopme\ at
$t=\pm i$ determines a complex-valued flat connection and hence a
 homomorphism $\pi_1(M)\to G_{{\Bbb{C}}}$. Eqn. \loopme\ instructs
us to impose the further condition $D^*\phi=0$ and to divide by
$G$-valued gauge transformations. The combined operation
constructs the moduli space ${\cal Y}$ of homomorphisms
$\vartheta:\pi_1(M)\to {{{G}}}_{\Bbb{C}}$ (we refer to this space
as ${\cal Y}_M$ or ${\cal Y}_M(G)$ if greater clarity is needed).
This statement depends on a theorem of Corlette \ref\corlette{K.
Corlette, ``Flat $G$-Bundles With Canonical Metrics,'' J. Diff.
Geom. {\bf 28} (1988)361-382.} (the two-dimensional case was also
proved by Donaldson in the appendix to \hitchin): dividing by
$G$-valued gauge transformations on the pair $A,\phi$, requiring
that ${\cal F}=0$ and imposing $D^*\phi=0$ is equivalent to simply
dropping that last equation, imposing $\CF=0$ and a condition of
stability\foot{ As is usual for moduli problems involving
non-compact symmetry groups such as $G_{\Bbb{C}}$, the moduli
space ${\cal Y}_M(G)$ does not literally parametrize all conjugacy
classes of homomorphisms $\pi_1(M)\to G_{\Bbb{C}}$. To get a good
moduli space, one must drop certain ``unstable orbits,''
corresponding in this problem to homomorphisms that can be reduced
to a triangular form but not to a direct sum.  That setting $D^*
\phi=0$ and dividing by $G$-valued gauge transformations has the
effect of dropping such unstable orbits was shown in \corlette.
We explain the role of stability  more fully in section
\complexstr.} and dividing by the group of $G_{\Bbb{C}}$-valued
gauge transformations.

More generally, if $t$ is not real, then the equations \loopme\
for fields $A,\phi$ valued in the real Lie algebra of $G$ are
overdetermined, rather than elliptic.  Setting to zero separately
the hermitian and antihermitian parts of these equations, we learn
that they imply ${\cal F}=0$ for any non-real $t$.

We want to understand, however, what happens when $t$ is real.
(The most important values of $t$ for the geometric Langlands
program are $t=\pm i$ and $t=\pm 1$.) In this case, the equations
are real and we get a family of real elliptic equations
parametrized by $t$.  As written, these equations are regular for
$t\not= 0,\infty$.  However, as is clear from the homogeneous
expressions on the right hand side of \helpme, the equations can
be extended over $t=0$ or $t=\infty$, by multiplying the second
equation by $t$ or the first by $t^{-1}$.  Thus, at $t=0$, we
should replace the second equation by $(D\phi)^-=0$, and at
$t=\infty$, we replace the first by $(D\phi)^+=0$.  So we get a
family of real elliptic equations parametrized by $\Bbb{RP}^1$.

We can similarly understand the  conditions for unbroken
supersymmetry which come from \mimo:
 \eqn\toopme{\eqalign{D\sigma + t[\phi,\sigma]& = 0\cr
                     D\sigma - t^{-1}[\phi,\sigma]& = 0.\cr
                     }}
For any $t$ other than $\pm i$, these equations imply that
$D\sigma=[\phi,\sigma]=0$, a condition which means that the gauge
symmetry generated by $\sigma$ leaves invariant the given solution
of \loopme.
In addition, for $t\not=\pm i$, a linear combination of the equations $\delta_T\eta=0$
and $\delta_T\tilde \eta=0$ gives
\eqn\woopem{[\sigma,\bar\sigma]=0.}
As long as we only consider
supersymmetric fields that are irreducible -- or more generally,
those for which the automorphism group is finite -- \toopme\
implies that $\sigma=0$. In this paper, we generally consider only
supersymmetric fields that are irreducible. Some simple remarks on
the  reducible case will appear in \witfur.

At $t=\pm i$, the conclusion is similar although the details are different.  In this case,
$\toopme$ says that $\sigma$ generates a symmetry of the flat $G_\Bbb{C}$-valued connection
$\CA$ or $\bar\CA$.  Again, as long as we only consider supersymmetric fields that are irreducible,
we can assume that $\sigma=0$.  (For $t=\pm i$, in case of a reducible supersymmetric configuration
with $\sigma\not=0$, we do not get separate conditions $D^*\phi=[\sigma,\bar\sigma]=0$, but only
the linear combination $D^*\phi\pm i[\sigma,\bar\sigma]=0$.)

For real $t$, the theory discussed here fits in the general
framework of cohomological field theories \ref\witco{E. Witten,
``Introduction To Cohomological Field Theories,'' Int. J. Mod.
Phys.{\bf A6} (1991) 2775-2792.}, like Donaldson theory in four
dimensions or $A$-models in two dimensions.  The fields are the
bosonic fields $A,\phi$ of ${\EUK}=0$; the equations are \loopme;
and the symmetries are simply the gauge symmetries. As in any such
theory, the path integral in the absence of operator insertions
can be evaluated by counting, with signs, the number of solutions
of the equations, as long as those solutions are isolated and
irreducible.  The sign with which a given solution contributes is
simply  the sign of the corresponding one-loop fermion
determinant. A smooth compact family of irreducible solutions,
parametrized by a space ${\eurm Y}_0$ (which is a component of
${\eurm Y}$), makes a contribution that is plus or minus the Euler
characteristic of ${\eurm Y}_0$, with the sign again coming from
the fermion determinant. Contributions of singularities in ${\eurm
Y}$ and of reducible solutions -- such as the trivial solution
with $A=\phi=0$ -- are more subtle to evaluate and will not be
considered here.

Since the number of solutions, weighted by sign, is invariant
under continuous deformation of a family of elliptic equations,
the partition function on a closed four-manifold without operator
insertions must be independent of $t$, at least for real $t$. By
holomorphy, this is also true for complex $t$.  That the partition
function on a closed four-manifold is independent of $t$ can be
seen more directly from the path integral, as we explain later.
{}From a mathematical point of view, the vanishing theorems of
section \vanishing\ will give a strong statement about
$t$-independence: the space of solutions of \loopme\ is actually
independent of $t$ except at $t=0,\infty$.

The importance of the parameter $t$ is that it is possible to
introduce operators or, in case $M$ has a boundary, boundary
conditions, that preserve the topological symmetry only for a
particular value of $t$. Like Donaldson theory, the TQFT
considered here has local operators that preserve the topological
symmetry. They have been discussed in \refs{\marcus-\labloz} and
will be further analyzed elsewhere \witfur.  However, they are not
the most important operators for the  geometric Langlands program.
The important operators, introduced in section \loopop, will be
the Wilson and 't Hooft line operators, which have no obvious
close analogs in more familiar cohomological field theories.

If our theory for real $t$ is somewhat analogous to a
two-dimensional $A$-model, what does it correspond to for complex
$t$?  The theory at $t=\pm i$ is analogous to a two-dimensional
$B$-model.
 In the $B$-model that describes  maps $\Phi:\Sigma\to X$ from a
Riemann surface $\Sigma$ to a complex manifold $X$,  the
supersymmetric fields are the constant maps of $\Sigma$ to $X$.
The obvious similarity between the two cases is that the constant
maps or the flat $G_{\Bbb{C}}$-valued connections are relatively
elementary to describe  by using over-determined rather than
elliptic equations (namely $d\Phi=0$ in one case, ${\cal F}=0$ in
the other) in which the complex structure is obvious. A deeper
analogy will be apparent when we consider dimensional reduction in
section \compacthitchin.

\subsec{Vanishing Theorems}

\subseclab\vanishing

Let $\EUV^+(t)= (F-\phi\wedge\phi+tD\phi)^+$, $\EUV^-(t)=
(F-\phi\wedge\phi-t^{-1}D\phi)^-$, and $\EUV_0= D^* \phi$, so the
supersymmetric equations are $\EUV^+(t)=\EUV^-(t)=\EUV_0=0$.  In
this section, we concentrate on the case of real $t$, so that the
supersymmetric equations are elliptic.

Like many first order equations associated with supersymmetry,
such as the equations for a holomorphic curve, the Yang-Mills
instanton equations, the Seiberg-Witten equations, or the
equations studied in \vafawitten, these are subject to unusual
vanishing theorems. In the case at hand, we have:

\bigskip
\noindent {\it Vanishing Theorem 1~}  Let $M$ be a compact
four-manifold  without boundary, and $E$ a $G$-bundle over $M$
with
 nonzero Pontryagin class, $\int_M
\Tr\,F\wedge F\not=0$.  Then for any $t\not=0,\infty$, the
supersymmetric equations $\EUV^+(t)=\EUV^-(t)=\EUV_0=0$ have no
solutions.

\bigskip \noindent {\it Vanishing Theorem 2~} On such an $M$, any
field that obeys those equations
 for one value of $t$ other than $0,\infty$ obeys them for all $t$
 and hence is given by a flat $G_\Bbb{C}$-valued connection.

\bigskip

The shortest proof of the first vanishing theorem comes from the
following identity: \eqn\hombo{\int_M\Tr\,F\wedge
F=\int_M\Tr\left(\EUV^+(t)\wedge \EUV^+(-t^{-1})+\EUV^-(t)\wedge
\EUV^-(-t^{-1})\right).} The proof of the identity is
straightforward and uses integration by parts. Clearly, this
identity implies that if for some $t$, $\EUV^\pm(t)=0$, then
$\int_M\Tr\, F\wedge F=0$.

This proof fails if $t=0,\infty$ because of the poles in
$\EUV^\pm$. In fact, the result is false for $t=0,\infty$; the
equations at $t=0$ or $t=\infty$ have at least the obvious
solutions given by an instanton or anti-instanton with $\phi=0$.

For the second vanishing theorem, we need a similar but more
intricate identity: \eqn\ormob{\eqalign{&-\int_Md^4x\sqrt
g\,\Tr\,\left({t^{-1}\over
t+t^{-1}}\EUV^+(t)_{\mu\nu}\EUV^+(t)^{\mu\nu}+{t\over
t+t^{-1}}\EUV^-(t)_{\mu\nu}\EUV^-(t)^{\mu\nu}+\EUV_0^2\right)\cr
&=-\int d^4x\sqrt g\,\Tr\,\left({1\over 2}F_{\mu\nu}
F^{\mu\nu}+D_\mu\phi_\nu D^\mu\phi^\nu
+R_{\mu\nu}\phi^\mu\phi^\nu+ {1\over
2}[\phi_\mu,\phi_\nu]^2\right)\cr&~~~~~~~+{t-t^{-1}\over
(t+t^{-1})}\int_M\Tr\, F\wedge F. \cr} } What is surprising about
this identity is that apart from the topological invariant
proportional to $\int\,\Tr\,F\wedge F$, the right hand side is
independent of $t$.  The proof is similar to the proof of \hombo,
but more elaborate. $R_{\mu\nu}$ is the Ricci tensor of $M$, which
enters in integrating by parts to relate $\int_Md^4x\sqrt g
D_\mu\phi_\nu D^\nu\phi^\mu$ to $\int d^4x\sqrt g
(D_\mu\phi^\mu)^2$.  Unlike \hombo, \ormob\ has a natural limit as
$t\to 0,\infty$.  One simply replaces $t (\EUV^-)^2/(t+t^{-1})$ by
$\lim_{t\to 0}(t\EUV^-)^2$ or $t^{-1}(\EUV^+)^2/(t+t^{-1})$ by
$\lim_{t\to \infty}(t^{-1}\EUV^+)^2$.  So arguments based on
\ormob\ are valid at $t=0,\infty$.

In section \toplag, we will understand the physical meaning of
\ormob. For now, we simply use it to complete the proof of the
vanishing theorems.

To prove the second vanishing theorem, we must show that if the
bundle $E$ admits a pair $A,\phi$ for which $\EUV^\pm(t)=\EUV_0=0$
for some value of $t$ other than $0,\infty$, then these equations
hold for all $t$.  We already know that in such a solution,
$\int_M\Tr\,F\wedge F=0$, whence the right hand side of \ormob\ is
independent of $t$.  So if the left hand side vanishes for one
value of $t$, it vanishes for all $t$. Moreover, the left hand
side is a sum of squares, so it vanishes for given $t$ if and only
if $\EUV^\pm(t)=\EUV_0=0$. Moreover, as we have already noted, the
vanishing of $\EUV^\pm(t)$ and $\EUV_0$ for all $t$ is equivalent
to the vanishing of the curvature ${\cal F}=d{\cal A}+{\cal
A}\wedge {\cal A}$ of the $G_{\Bbb{C}}$-valued connection ${\cal
A}=A+i\phi$.  This completes the argument.

Finally, we can use \ormob\ to prove a slightly stronger version
of the first vanishing theorem:

\bigskip\noindent{\it Vanishing Theorem 1$'$~} For $\int_M\Tr\,
F\wedge F>0$, the supersymmetric equations have no solutions
except at $t=0$, and for $\int_M\Tr\,F\wedge F<0$, they have no
solutions except at $t=\infty$.\bigskip

Let ${\eurm B}(t)$ denote the left hand side of \ormob, and let
$f(t)=(t-t^{-1})/(t+t^{-1})$.  We have for any $t,u$
\eqn\toto{{\eurm B}(u)-{\eurm B}(t)=(f(u)-f(t))\int_M\Tr \,F\wedge
F.} If $\EUV^\pm(t)=\EUV_0=0$, then ${\eurm B}(t)=0$, and as the
function ${\eurm B}$ is positive semidefinite, the left hand side
of \toto\ is positive semidefinite. To make the right hand side
positive semidefinite, $t$ must be a minimum of the function $f$
if $\int_M\Tr\, F\wedge F>0$, and a maximum if $\int_M\Tr\,F\wedge
F<0$.  But the only minimum is at $t=0$, and the only maximum is
at $t=\infty$.

\bigskip\noindent{\it A Note On Compactification}

A special case of the vanishing theorems  is relevant to
compactification.  We take $M=\Sigma\times C$, where $\Sigma$ and
$C$ are two compact Riemann surfaces.  In our applications, $C$ is
the Riemann surface on which we study the geometric Langlands
program, and we think of $\Sigma$ as being much larger than $C$.
However, in topological field theory, the metrics on $\Sigma$ and
$C$ are inessential.

For $t\not=0,\infty$, the second vanishing theorem says that any
solution of the supersymmetric equations on $M$ is given by a flat
$G_{\Bbb{C}}$-valued connection.  A flat bundle $E\to
M=\Sigma\times C$ has a decomposition\foot{In this discussion, for
simplicity, we use the language of vector bundles, as if
$G=SU(N)$. To extend the argument to any $G$, one simply
formulates it in terms of subgroups of $G$ rather than subbundles
of $E$.} $E=\oplus_{i=1}^n E'_i\otimes E''_i$, where the $E'_i$
are flat bundles pulled back from  $\Sigma$ and the $E''_i$ are
flat bundles pulled back from $C$.

If we restrict to a point $x\in \Sigma$, then the $E'_i$ become
trivial. If $d_i$ is the rank of $E_i'$, then $E$ reduces to
$E_x=\oplus_{i=1}^n d_iE''_i$, a direct sum of flat bundles pulled
back from $C$. $E_x$ is a flat bundle over $C$, and as such
defines a point in Hitchin's moduli space $\MH$, which we describe
more thoroughly in section \compacthitchin. For now, we can just think of $\MH$ as the
moduli space of flat $G_\Bbb{C}$-bundles over $C$.  Thus any
supersymmetric field on $\Sigma\times C$ is given by a map from
$\Sigma$ to $\MH$, namely $x\to E_x$.

The moduli space $\MH$ has singularities that correspond to
reducible flat bundles.  In the present construction, the bundle
$E_x$ is reducible if $n>1$. For $n=1$, we have $E=E'\otimes E''$,
where the two factors are pullbacks from $\Sigma $ and $C$,
respectively.  $E_x$ is still reducible unless $E'$ has rank 1, in
which case, for $G=SU(N)$, it must be trivial.  Thus, the case in
which the singularities of $\MH$ are avoided is precisely the case
that $E$ is the pullback of an irreducible flat bundle $E''\to C$.

 The supersymmetric fields are
also the configurations that minimize the action; this will become
clear in section \toplag, when we construct the supersymmetric
Lagrangian.   In the above reasoning, we took $\Sigma$ and $C$ to
be compact.  In our applications, we are also interested in the
case that $\Sigma$ is a complete but noncompact two-manifold, such
as $\Bbb{R}^2$, $\Bbb{R}^2_+$, or $\Bbb{R}\times I$ ($\Bbb{R}^2_+$
is a halfspace and $I$ is a closed interval). Given appropriate
boundary conditions and asymptotic behavior at infinity to justify
the proof of the second vanishing theorem, one still has in these
cases the decomposition of supersymmetric fields
$E=\oplus_{i=1}^nE'_i\otimes E''_i$. (For $\Sigma$ flat and
simply-connected, by $E'_i$ we mean a bundle with trivial
connection and covariantly constant Higgs field.) This leads to
the same conclusion that we had for compact $\Sigma$: a
supersymmetric field that avoids singularities of $\MH$ is a
pullback from $C$.

\subsec{The Topological Lagrangian}

\subseclab\toplag

Our goal now is to find a Lagrangian that possesses the
topological symmetry for any value of $t$ and reduces when $M$ is
flat to the Lagrangian of the underlying ${\EUN}=4$ super
Yang-Mills theory.  We could do this by hand, starting with \ygo\
for flat $M$ and asking what curvature dependent terms are needed
to maintain the topological symmetry when $M$ is not flat.  We
will follow a different approach.

A very useful first step is to compute the algebra generated on
the fields by the the supersymmetries $Q_\ell$ and $Q_\rr$, or
equivalently by $Q=uQ_\ell+vQ_\rr$.  For the fields $
A,\phi,\psi$, $\tilde\psi$, and $\sigma$, we compute
\eqn\ibo{\eqalign{\delta_T{}^2 A & = -i(u^2+v^2)(-D\sigma)\cr
\delta_T{}^2\phi & = -i(u^2+v^2)[\sigma,\phi]\cr \delta_T{}^2 \psi
& = -i(u^2+v^2)[\sigma,\psi]\cr \delta_T{}^2\tilde\psi& =
-i(u^2+v^2)[\sigma,\tilde\psi]\cr \delta_T{}^2\sigma& = 0.\cr}}
These results can be summarized by saying that if $\Phi$ is any
field of $\EUK\geq 0$, then
$\delta_T{}^2\Phi=-i(u^2+v^2){{\lie}}_\sigma(\Phi)$, where
${{\lie}}_\sigma(\Phi)$ is the first order change in $\Phi$ in a
gauge transformation generated by $\sigma$ (so $\lie$ stands for
Lie derivative). For example, ${{\lie}}_\sigma(A)=-D\sigma$, and
for the other fields here, ${{\lie}}_\sigma(\Phi)=[\sigma,\Phi]$.
The relation $\delta_T{}^2\Phi=-i(u^2+v^2){{\lie}}_\sigma(\Phi)$
can be expanded in powers of $u$ and $v$ and is equivalent to the
following:
\eqn\uvu{\eqalign{\delta_\ell^2\Phi=\delta_r^2\Phi&=-i{{\lie}}_\sigma(\Phi)\cr
\{\delta_\ell,\delta_\rr\}\Phi& = 0.\cr}}

If, however, one computes $\delta_T{}^2\Phi$ for a field of
$\EUK<0$, one does not get $-i(u^2+v^2){{\lie}}_\sigma(\Phi)$ in
an obvious way. In fact, the relation
$\delta_T{}^2\Phi=-i(u^2+v^2){{\lie}}_\sigma(\Phi)$ does hold for
all $\Phi$ in the twisted ${\EUN}=4$ super Yang-Mills theory, but
for some fields, it only holds upon using the equations of motion.

Construction of supersymmetric Lagrangians is generally much
easier if one can introduce ``auxiliary fields,'' which are extra
fields that can ultimately be eliminated by their equations of
motion if one so chooses, such that the algebra, in the present
case $\delta_T{}^2\Phi=-i(u^2+v^2){{\lie}}_\sigma(\Phi)$, is
satisfied without using equations of motion.   In theories with a
great deal of supersymmetry, such as ${\EUN}=4$ super Yang-Mills
theory, it can be very difficult to find suitable auxiliary
fields.   The present example, as first noted in \marcus, is a
case in which this problem arises.

However, the problem can be greatly alleviated by introducing an
auxiliary field $P$, a zero-form with values in the Lie algebra,
so as to close the algebra for $\eta$ and $\tilde \eta $ (but not
$\chi$). One takes the transformation laws of
$\bar\sigma,\eta,\tilde\eta$, and $P$ to be
\eqn\yfog{\eqalign{\delta_T\bar\sigma & = iu\eta +i v\tilde\eta
\cr
                    \delta_T\eta & = vP+u[\bar\sigma,\sigma]\cr
                    \delta_T\tilde\eta & =
                    -uP+v[\bar\sigma,\sigma]\cr
                    \delta_T P & =
                    -iv[\sigma,\eta]+i u[\sigma,\tilde\eta].\cr}}
These equations will reduce to those in section \twisting\ once we
impose the equations of motion -- which notably will set
$P=D^*\phi$.

Once we have a closed algebra on a set of fields -- in this case
all fields in the theory including $P$ but excluding $\chi$, since
we have not closed the algebra on $\chi$ -- it is generally a
simple matter to find the possible invariant Lagrangians.  In the
present case, we would like to find a partial action ${{\cmmib
I}}_0$ for the the fields on which we have closed the algebra
which obeys $\delta_\ell{{\cmmib I}}_0=\delta_\rr{{\cmmib
I}}_0=0$, and hence, for any $u,v$, obeys $\delta_T {{\cmmib
I}}_0=0$ where $\delta_T=u\delta_\ell+v\delta_\rr$. These
properties will hold if ${{\cmmib I}}_0=\delta_\ell\delta_\rr V$
for some gauge-invariant $V$. Indeed, if $V$ is gauge-invariant,
then ${{\lie}}_\sigma(V)=0$ and \uvu\ reduces to
$\delta_\ell^2V=\{\delta_\ell,\delta_\rr\}V=\delta_\rr^2V=0$.  So
$\delta_\ell\delta_\rr V$ will automatically be annihilated by
$\delta$.

In addition, we want to pick $V$ so that the equation of motion
for $P$ is $P=D^*\phi$, so as to be compatible with the formulas
of section \twisting.  So we take \eqn\obo{V={2\over e^2}\int_M
d^4x\sqrt g\left(-{1\over 2}\Tr\,\eta\tilde\eta -i\Tr\,\bar\sigma
D^*\phi\right),} and compute that ${\cmmib
I}_0=\delta_\ell\delta_\rr V$ is \eqn\nobo{\eqalign{{{\cmmib
I}}_0= {2\over e^2}\int_Md^4x\sqrt g\Tr& \left({1\over 2}P^2
-PD^*\phi+{1\over 2}[\bar\sigma,\sigma]^2-
D_\mu\bar\sigma\,D^\mu\sigma
-[\phi_\mu,\sigma][\phi^\mu,\bar\sigma]\right.\cr & + i\tilde\eta
D_\mu\tilde\psi^\mu+i\eta
D_\mu\psi^\mu-i\tilde\eta[\psi_\mu,\phi^\mu]+i\eta[\tilde\psi_\mu,\phi^\mu]
\cr &\left.-{i\over 2}[\sigma,\tilde\eta]\tilde\eta -{i\over
2}[\sigma,\eta]\eta+i[\bar\sigma,\psi_\mu]\psi^\mu
+i[\bar\sigma,\tilde\psi_\mu]\tilde\psi^\mu\right).\cr}} The
Euler-Lagrange equation for $P$ is $P=D^*\phi$, as desired, and
one can make the replacement \eqn\olgog{\Tr\left(
P^2-2PD^*\phi\right)\to-\Tr (D^*\phi)^2,} if one wishes.

${{\cmmib I}}_0$ manifestly possesses the topological symmetry for
any value of $t=v/u$.  Moreover, as it is of the form
$\delta_\ell\delta_\rr V$, where the metric of $M$ does not enter
in the definition of $\delta_\ell$ and $\delta_\rr $ but only in
the choice of $V$, it also has the key property that leads to a
topological field theory: its dependence on the metric of $M$ is
of the form $\delta_\ell\delta_\rr (\dots)$.

The only reason that ${{\cmmib I}}_0$ is not a satisfactory action
is that it is degenerate; it does not contain the kinetic energy
for the gauge fields or any terms containing $\chi$. It does not
seem to be possible to complete the construction of the action
with a construction as simple as that above.  Though it is
possible to add auxiliary fields so as to close the algebra on
$\chi$, it does not seem to be possible to do this in a way that
incorporates both $\delta_\ell$ and $\delta_\rr$ and is useful for
understanding the appropriate twisted ${\EUN}=4$ action. (The fact
that a $t$-dependent topological invariant occurs on the right
hand side of \ormob\ appears to be an obstruction to this.)
Instead, we will fix a particular value of $t$ and consider only
the differential $\delta_t=\delta_\ell+t\delta_\rr$. We add an
auxiliary field $H$, which is to be a two-form with values in the
Lie algebra, and postulate \eqn\hco{\eqalign{\delta_t\chi& = H\cr
                           \delta_t H & =
                           -i(1+t^2)[\sigma,\chi].}}
This is compatible with
\eqn\kino{\delta_t^2(\Phi)=-i(1+t^2){{\lie}}_\sigma(\Phi),} for
$\Phi=\chi,H$; this is the specialization of
$\delta_T^2=-i(u^2+v^2)\lie_\sigma$ for $(u,v)=(1,t)$. For \hco\
to agree with the transformation of $\chi$ found in section
\twisting, the equations of motion will have to give
$H^+=\EUV^+(t)$, $H^-=t\EUV^-(t)$. We will construct the action to
ensure this.

As before, an action annihilated by $\delta_t$ can be ${\cmmib
I}_1=\delta_t V_1$, for any gauge-invariant $V_1$.  We pick
\eqn\usefuld{V_1={2\over e^2}\int_Md^4x \sqrt g{1\over
(1+t^2)}\Tr\left(\chi^+_{\mu\nu}\left({1\over
2}H^{+\mu\nu}-\EUV^+(t)^{\mu\nu}\right)
+\chi^-_{\mu\nu}\left({1\over
2}H^{-\mu\nu}-t\EUV^-(t)^{\mu\nu}\right)\right).} Then
\eqn\usefuldud{\eqalign{{{{\cmmib I}}}_1= {1\over e^2}\int_Md^4x
\sqrt g&\left({2\over (1+t^2)}\Tr\left({1\over
2}H^+_{\mu\nu}H^{+\mu\nu}-H^+_{\mu\nu}\EUV^+(t)^{\mu\nu}\right)\right.\cr
& + {2\over (1+t^2)}\Tr\left({1\over
2}H^-_{\mu\nu}H^{-\mu\nu}-tH^-_{\mu\nu}\EUV^-(t)^{\mu\nu}\right)\cr
&+
2\Tr\left(\chi^+_{\mu\nu}\left(iD\psi+i[\tilde\psi,\phi]\right)^{\mu\nu}
+\chi^-_{\mu\nu}\left(iD\tilde\psi-i[\psi,\phi]\right)^{\mu\nu}\right)\cr
&+\Tr\left(i\chi^+_{\mu\nu}[\sigma,\chi^{+\mu\nu}]+
i\chi^-_{\mu\nu}[\sigma,\chi^{-\mu\nu}] \right)\biggr).\cr}} Upon
integrating $H^{\pm}$ out of this, we can write the equivalent
action \eqn\usefuldudd{\eqalign{{{\cmmib I}}_1= {1\over
e^2}\int_Md^4x \sqrt g&\left(-{t^{-1}\over
(t+t^{-1})}\Tr\,\EUV^+(t)_{\mu\nu}\EUV^+(t)^{\mu\nu}-{t\over
(t+t^{-1})}\Tr\,\EUV^-(t)_{\mu\nu}\EUV^-(t)^{\mu\nu}\right)\cr &+
2\Tr\left(\chi^+_{\mu\nu}\left(iD\psi+i[\tilde\psi,\phi]\right)^{\mu\nu}
+\chi^-_{\mu\nu}\left(iD\tilde\psi-i[\psi,\phi]\right)^{\mu\nu}\right)\cr
&+\Tr\bigl(i\chi^+_{\mu\nu}[\sigma,\chi^{+\mu\nu}]+
i\chi^-_{\mu\nu}[\sigma,\chi^{-\mu\nu}] \bigr)\biggr).\cr}}

Finally,  the useful identity \ormob\ enables us to write
\eqn\nuseful{\eqalign{{{\cmmib I}}_1=-{1\over e^2}\int_Md^4x \sqrt
g&\left(\Tr\left({1\over 2}F_{\mu\nu} F^{\mu\nu}+D_\mu\phi_\nu
D^\mu\phi^\nu +R_{\mu\nu}\phi^\mu\phi^\nu+ {1\over
2}[\phi_\mu,\phi_\nu]^2-(D^*\phi)^2
  \right)\right.\cr &-
2\Tr\left(\chi^+_{\mu\nu}\left(iD\psi+i[\tilde\psi,\phi]\right)^{\mu\nu}
+\chi^-_{\mu\nu}\left(iD\tilde\psi-i[\psi,\phi]\right)^{\mu\nu}\right)\cr
&-\Tr\left(i\chi^+_{\mu\nu}[\sigma,\chi^{+\mu\nu}]+
i\chi^-_{\mu\nu}[\sigma,\chi^{-\mu\nu}] \right)\biggr)\cr &
~~~~~~~+{t-t^{-1}\over e^2(t+t^{-1})}\int_M\Tr\, F\wedge F.\cr}}
Apart from the topological term, the part of ${{\cmmib
I}}_0+{{\cmmib I}}_1$ that depends only on $A$ and $\phi$ can also
\marcus\ be written \eqn\wingo{ {{\cmmib I}}^{(A,\phi)}=-{1\over
e^2}\int d^4x\,\sqrt g\, \Tr\left({1\over 2}
\cF_{\mu\nu}\bar\cF^{\mu\nu}+(D^*\phi)^2\right),} as one can show
with some integration by parts similar to what is used in proving
the vanishing theorems.  The analogous terms involving $\sigma$
can be found in \nobo\ and are \eqn\uwingo{{\cmmib
I}^\sigma={2\over e^2}\int d^4x \sqrt g\,\Tr\left( {1\over
2}[\bar\sigma,\sigma]^2- D_\mu\bar\sigma\,D^\mu\sigma
-[\phi_\mu,\sigma][\phi^\mu,\bar\sigma]\right).}

The key point -- which we already exploited in proving the
vanishing theorems -- is that ${{\cmmib I}}_1$ is independent of
$t$ except for the last term.  But that last term, being a
topological invariant, is automatically annihilated by $\delta_t$,
and indeed by both $\delta_\ell$ and $\delta_\rr$, all by itself.
So without spoiling the topological symmetry, we can add another
term to the action, \eqn\ucno{{{\cmmib I}}_2=-\left({t-t^{-1}\over
e^2(t+t^{-1})}-i{\theta\over 8\pi^2}\right)\int_M\Tr\,F\wedge F.}
We have simply chosen the coefficient to cancel the $t$-dependence
in ${{\cmmib I}}_1$, leaving us with the standard $\theta$
parameter multiplying this term in the action.

Finally, then, the overall  action is ${{\cmmib I}}={\cmmib
I}_0+{{\cmmib I}}_1+{{\cmmib I}}_2$. The construction makes it
manifest that ${{\cmmib I}}$ is annihilated by $\delta_t$ for the
specific value of $t$ used in constructing ${{\cmmib I}}_1$.  But
since in fact ${{\cmmib I}}$ is independent of $t$,
it is annihilated by $\delta_t$ for all $t$.

Moreover, ${{\cmmib I}}$ is the action of a topological field
theory. The change in ${{\cmmib I}}$ in a change in the metric of
$M$ is of the form $\delta_t(\dots)$, since ${{\cmmib I}}_0$ and
${{\cmmib I}}_1$ are of this form even before varying the metric,
and ${{\cmmib I}}_2$ does not depend on the metric of $M$ at all.

We have accomplished our goal of constructing an action that has
the full ${\Bbb{ CP}}^1$ family of topological symmetries --
making it clear, among other things, that the partition function
of this theory on a closed four-manifold $M$ without operator
insertions is independent of $t$.  One can readily verify that
(after eliminating $P$ via \olgog, whereupon the $(D^*\phi)^2$
terms cancel), the bosonic part of ${{\cmmib I}}$ reduces for
$M=\Bbb{R}^4$ to \loopy.  The topological symmetry implies that
the fermionic terms also agree. So this theory is a generalization
of ${\EUN}=4$ super Yang-Mills theory to a curved four-manifold.

\subsec{The Canonical Parameter}

\subseclab\canonpar

So far, we have obtained a family of topological field theories
that depend on a coupling parameter $\tau=\theta/2\pi+4\pi i/e^2$,
and a twisting parameter $t$.  But these parameters are not really
all independent.  We will now show that the theory really only
depends on a certain combination $\Psi$ of $t$ and $\tau$, which
we will call the canonical parameter.

Let us write the coefficient of the topological term in ${\cmmib
I}_2$, which takes the form $-\left({(t-t^{-1})/
e^2(t+t^{-1})}-i{\theta/8\pi^2}\right)$, as $-\Psi/4\pi i$, where
\eqn\yorgo{\Psi = {\tau+\bar\tau\over 2}+{\tau-\bar\tau\over
2}\left({t-t^{-1}\over t+t^{-1}}\right).} We will call $\Psi$ the
canonical parameter.

We claim that any observable ${\EUO}(\tau,\bar\tau,t)$ really
depends only on the canonical parameter $\Psi$. Indeed, the action
of the theory is of the form \eqn\juty{{{\cmmib I}}=\delta_tV
+{i\Psi\over 4\pi }\int_M\Tr \,F\wedge F.} To the extent that the
term $\delta_tV$ is irrelevant, the theory naturally only depends
on the coefficient $\Psi$.

At first sight, there is a fallacy in this argument.  In addition
to whatever $t$-dependence may be explicit in ${{\cmmib I}}$ (we
have arranged so that there is none, though individual terms are
separately $t$-dependent), there might be $t$-dependence coming
from the fact that the definition of $\delta_t$ depends on $t$.
The definition of the theory, after all, depends on the
topological symmetry $\delta_t$ as well as on the action. But in
fact, the $t$-dependence of $\delta_t$ can be eliminated (as long
as $t\not=\pm i$) by redefining the fields. One rescales
$\delta_t$ to $\delta'_t=\delta_t/\sqrt{1+t^2}$, so that the
algebra becomes $(\delta'_t)^2=-i{{\lie}}_\sigma$, independent of
$t$. Then,  with the auxiliary fields $P$ and $H$ included, after
some small redefinitions of the fermions, every multiplet takes
the form $\delta'_t U=V$, $\delta'_tV=-i{{\lie}}_\sigma(U)$, for
some pair of fields $U$, $V$.  (This is a standard type of
multiplet in cohomological field theories, related mathematically
to the Cartan model of equivariant cohomology.)  Once this is
done, the transformation laws of the multiplets are completely
independent of $t$.

\bigskip\noindent{\it Some Special Values}

Some special cases of the relation between $\Psi$, $t$, and $\tau$
are worthy of note. We focus especially on values that have a
simple behavior under $S$-duality and a simple interpretation
after compactification to two dimensions.

If $t=\pm 1$, we have $\Psi={\rm Re}\,\tau$.  Thus, any real value
of $\Psi$ can be achieved at $t=\pm 1$, simply by varying the
$\theta$ angle of the gauge theory, but only real values are
possible. More generally, if $|t|=1$, then $\Psi$ is real.

To make $\Psi$ complex, we must get away from $|t|=1$.  All values
of $\Psi$ in the upper or lower half plane are possible,
respectively, for $t$ real and of modulus greater than or less
than 1.

If $t=\pm i$, then $\Psi=\infty$, independent of $\tau$.  Thus, at
$t=\pm i$, the value of $\tau$ is completely irrelevant.  This has
essentially been demonstrated directly by Marcus \marcus.

A final observation is that all values of $\Psi$ can be realized
at arbitrarily weak coupling, that is arbitrarily large ${\rm
Im}\,\tau$.  Indeed, inverting the definition of $\Psi$, we have
\eqn\helmo{t^2=-{\Psi-\bar\tau\over \Psi-\tau},} showing that if
we keep $\Psi$ and ${\rm Re}\,\tau$ fixed and take ${\rm
Im}\,\tau\to\infty$, then $t\to \pm 1$.  The limit ${\rm
Im}\,\tau\to\infty$  is the weak coupling limit where the theory
should be understandable.

\bigskip\noindent{\it Action Of $S$-Duality On The Canonical Parameter}

Now we want to determine how $\Psi$ transforms under $S$-duality.
We claim that a transformation that maps $\tau$ by $\tau\to
(a\tau+b)/(c\tau+d)$ acts likewise on $\Psi$: \eqn\yrto{\Psi\to
{a\Psi+b\over c\Psi+d}.}

Clearly, under $T:\tau\to\tau+1$, which leaves $t$ invariant, we
have $\Psi\to \Psi+1$.  We also have to check the behavior under
$S:\tau\to -1/n_{\frak g}\tau$.  This mapping transforms $t$ to
$\pm t\tau/|\tau|$ according to \duco. (The sign depends on how
$S$ is lifted to $SL(2,\Bbb{R})$, and does not matter as $\Psi$ is
invariant under $t\to -t$.) So $\Psi$ transforms by
\eqn\nelmo{\Psi\to -{1\over 2}\left({1\over n_{\frak
g}\tau}+{1\over n_{\frak g}\bar \tau}\right)-{1\over
2}\left({1\over n_{\frak g}\tau}-{1\over n_{\frak
g}\bar\tau}\right)\left({t^2(\tau/ \bar\tau)-1\over t^2(
\tau/\bar\tau)+1}\right)=-{1\over n_{\frak g} \Psi},} where
\helmo\ has been used.

\bigskip\noindent{\it Orientation Reversal}

It is a little unusual to find an action of a discrete group such
as $SL(2,\Bbb{Z})$ on a complex parameter $\Psi$ which, as here,
takes values in $\Bbb{CP}^1$, not just in the upper half of the
complex plane.  This being the case, it is conceivable to extend
$SL(2,\Bbb{Z})$ to $GL(2,\Bbb{Z})$ by adding an extra symmetry
that we can take to be \eqn\ysnon{{\EUT}=\left(\matrix{-1 & 0 \cr
0 & 1}\right).}  Such an element will act on $\Psi$ by $\Psi\to
-\Psi$, exchanging the upper half plane with the lower half plane.

In fact, such a symmetry exists precisely when the four-manifold
$M$ possesses a symmetry  ${\EUT}:M\to M$  that reverses its
orientation. Such a transformation leaves fixed the gauge coupling
$e^2$ and hence leaves fixed ${\rm Im}\,\tau=4\pi/e^2$. On the
other hand, it reverses the sign of the instanton number and hence
reverses the sign of ${\rm Re}\,\tau$.  Looking back at the
definition of the canonical parameter in \yorgo, we see that we
will get $\Psi\to-\Psi$ if ${\EUT}$ acts by \eqn\zongor{t\to
{1\over t}.} The points $t=\pm 1$ are fixed points of this
transformation.

We will have a symmetry of the topological equations \loopme\ if
we take ${\EUT}$ to act on the bosons $(A,\phi)$ by
\eqn\yoro{(A,\phi)\to ({\EUT}^*A,-{\EUT}^*\phi).} The action on
fermions is \eqn\zoro{\eqalign{(\psi,\tilde\psi)&\to
({\EUT}^*\tilde\psi,{\EUT}^*\psi)\cr (\chi^+,\chi^-)&\to
({\EUT}^*\chi^-,{\EUT}^*\chi^+)\cr (\eta,\tilde\eta)&\to
({\EUT}^*\tilde\eta,{\EUT}^*\eta).\cr}}  This gives a manifest
symmetry of the action.  The supersymmetry generators transform by
$(u,v)\to(v,u)$ and hence $\delta_t\to t\delta_{t^{-1}}$.

An important application, as we will see, is to the case
$M=\Bbb{R}\times W$, with ${\EUT}$ being a ``time-reversal''
symmetry that acts trivially on $W$ and acts on $\Bbb{R}$ by
multiplication by $-1$.

\newsec{Compactification And The Geometry Of Hitchin's Moduli Space}
\def\MH{{\EUM}_H}

\seclab\compacthitchin

We now consider compactification to two dimensions.  We take our
four-manifold to be $M=\Sigma\times C$, a product of two
two-manifolds.  We take $C$ to be a compact Riemann surface on
which we will study the geometric Langlands program.  Generally,
we assume that the genus of $C$ is greater than one, so that for
simple gauge group $G$, the generic flat connection on $C$ is
irreducible. $\Sigma$ may be either a complete but noncompact
two-manifold such as $\Bbb{R}^2$, or a second compact Riemann
surface.   In the latter case, we assume that the size of $\Sigma$
is much larger than that of $C$. We aim to reduce the
four-dimensional gauge theory on $M=\Sigma\times C$ to an
alternative description by an effective field theory on $\Sigma$.

To find the effective physics on $\Sigma$, we must find the
configurations on $M$ that minimize or nearly minimize the action
(Euclidean signature) or the energy (Lorentz signature).  Either
way, one arrives at the conclusion obtained in \vafa, with closely
related results in \hm: the four-dimensional twisted ${\EUN}=4$
supersymmetric gauge theory reduces on $\Sigma$ to a sigma-model
of maps $\Phi:\Sigma\to\MH$, where $\MH$ is the moduli space of
solutions on $C$ of Hitchin's equations  with gauge group $G$.

This  follows from our results in section \toplag.  Apart from the
topological term, which does not affect the equations of motion or
the energy of a classical configuration, the bosonic action of the
twisted ${\EUN}=4$ super Yang-Mills theory is a sum of squares. As
we see from \wingo\ and \uwingo, it is minimized precisely if the
topological equations are obeyed for all $t$, or in other words if
${\cal F}=D^*\phi=0$, along with
$D\sigma=[\phi,\sigma]=[\sigma,\bar\sigma]=0$. One simple type of
solution of these equations is obtained by taking $A$ and $\phi$
to be pullbacks from $C$, with $\sigma=0$. In fact, the equations
${\cal F}=D^*\phi=0$ for a connection $A$ and an adjoint-valued
one-form $\phi$ on the Riemann surface $C$ are one way to present
Hitchin's equations \hitchin.  By taking the real and imaginary
parts of ${\cal F}$, these equations can alternatively be written
\eqn\zongo{\eqalign{ F-\phi\wedge\phi & = 0\cr D\phi=D^*\phi&=0
.\cr}}  We will write $\MH$, or in more detail $\MH(G,C)$, for the
moduli space of solutions of these equations on $C$ up to gauge
transformations.  If the genus $g$ of $C$ is greater than one,
then $\MH(G,C)$ is a complex manifold of dimension $(2g-2){\rm
dim}\,G$.  It has singularities corresponding to reducible
solutions of Hitchin's equations, that is, solutions with
continuous unbroken gauge symmetries.

As we explained at the end of section \vanishing, any solution of
the equations ${\cal F}=D^*\phi=0$ on $M$ is associated with a map
from $\Sigma$ to $\MH$.  Moreover, if the singularities of $\MH$
are avoided, the solution on $M$ is actually a pullback from a
solution on $C$.  In this case, the conditions on $\sigma$ require
$\sigma=0$.
 So a  field of zero
energy or action that avoids singularities of $\MH$ is a constant
map from $\Sigma$ to a smooth point in  $\M_H$.

Accordingly, a field of almost zero energy or action comes from a
slowly varying map from $\Sigma$ to $\M_H$. (A map is slowly
varying if it varies significantly only over a length scale in
$\Sigma$ that is much greater than the size of $C$.)   So, as long
as we do not encounter fields on $C$ with continuous gauge
symmetries, the effective physics at long distances on $\Sigma$ is
given by a supersymmetric sigma-model in which the target space is
$\M_H(G,C)$.

The importance of the assumption that $\Sigma$ is very large
compared to $C$ is that  it ensures  that nonconstant but slowly
varying maps $\Sigma\to \MH(G,C)$ can exist, and moreover that a
pair of fields $(A,\phi)$ that determine a slowly varying map
$\Phi:\Sigma\to \MH(G,C)$ has very small action.\foot{There is
also a converse, which we explain in section \wilbranes.  Given a
slowly varying map $\Phi:\Sigma\to \MH(G,C)$, one can reconstruct
a pair of fields $(A,\phi)$ on $M$ that determine this map and
have very small action. } If alternatively we assume that $C$ is
much larger than $\Sigma$, a pair $(A,\phi)$ representing a
generic nonconstant map $\Phi:\Sigma\to \MH(G,C)$ has large
action; instead, we could get a good description involving slowly
varying maps $\tilde\Phi:C\to \MH(G,\Sigma)$.

The above analysis assumes that $G$ is semi-simple. If $G=U(1)$ or
$U(N)$,  some modification is required, because every solution of
Hitchin's equations has a $U(1)$ group of symmetries. As a result,
the low energy description involves the product of a sigma-model
and a supersymmetric $U(1)$ gauge theory. Moreover, for any $G$,
the low energy description breaks down at singularities of $\M_H$
corresponding to solutions of Hitchin's equations that have
continuous gauge symmetries (or extra continuous gauge symmetries
for $G=U(1)$ or $U(N)$). At such points, additional degrees of
freedom come into play: a field of zero energy or action can have
$\sigma\not=0$ and need not be a pullback from $C$.  (It has a
more general decomposition $E=\oplus_{i=1}^nE'_i\otimes E''_i$
described at the end of section \vanishing.)  For semi-simple $G$,
the generic solution of Hitchin's equations is irreducible if the
genus of $C$ is at least 2.  In this paper, we will keep away from
the singularities of $\M_H$ (by considering the Langlands
correspondence for irreducible flat $^L\negthinspace G_{\Bbb{C}}$
bundles, for example).  Some simple generalizations will be
considered in \witfur.

The sigma-model with target $\MH(G,C)$ will prove to be a powerful
tool for understanding the geometric Langlands program.  But its
relation to the underlying gauge theory is important.  For one
thing, key actors in the story are the  Wilson and 't Hooft line
operators, which are most naturally constructed in the underlying
four-dimensional gauge theory. In addition, the underlying gauge
theory is an important tool when the sigma-model breaks down
because of singularities of $\MH$. But many things can be
understood directly in the sigma-model.

The rest of this section is devoted to a review of the classical
geometry of $\MH$, focussing on just those topics that we need for
the present paper.

\subsec{$\M_H$ As A Hyper-Kahler Quotient}

\subseclab\mhhyper

For us the fact of central importance is that $\M_H$ can be
regarded as a hyper-Kahler quotient and therefore is a
hyper-Kahler manifold. Our first goal is to describe how this
comes about.

We consider a problem in gauge theory on a Riemann surface $C$,
the fields being a gauge field $ A$ and a one-form $\phi$ that
takes values in the adjoint representation.  (These arise by
restricting to $C$ the fields $A,\phi$ that appeared in section
\tqftfour\ in constructing the twisted gauge theory on a general
four-manifold $M$.)  We think of the space of all fields $A,\phi$
as an infinite-dimensional affine space $\EUW$.  We write $\delta$
for the exterior derivative on $\EUW$ (to denote this as $d$ would
invite confusion with the exterior derivative on $C$; hopefully,
there will be no confusion with the use of the symbol $\delta_S$
in section \tqftfour\ to denote supersymmetric variation).

We define a flat metric on $\EUW$ by \eqn\igob{ds^2=-{1\over
4\pi}\int_C|d^2z|\,\Tr\left(\delta A_z\otimes \delta A_{\bar
z}+\delta A_{\bar z}\otimes \delta A_z +\delta \phi_z\otimes
\delta \phi_{\bar z}+\delta \phi_{\bar z}\otimes \delta
\phi_z\right).} Here for a local complex coordinate $z$ on $C$, we
write $A=dz\,A_z+d\bar z A_{\bar z}$, $\phi = dz\,\phi_z+d\bar
z\,\phi_{\bar z}$, and $|d^2z|$ is the measure corresponding to
the $(1,1)$-form $i dz\wedge d\bar z$.  This metric on $\EUW$, in
which $\EUW$ is an infinite-dimensional Euclidean space, is
determined by the complex structure on $C$ (and does not depend on
the local coordinate $z$ that we used in writing it).

We want to extend this flat metric on $\EUW$ to a flat
hyper-Kahler structure.  To do this, we introduce three complex
structures $I,J,$ and $K$ on $\EUW$ obeying certain axioms.  They
will all be translation-invariant on the affine space $\EUW$, so
they can be defined by (constant) linear transformations of the
space of one-forms. Any linear transformation of square $-1$ will
define a complex structure; integrability is automatic.

\ifig\xunko{\bigskip  A family of complex structures on the
hyper-Kahler manifold $\M_H$. }
{\epsfxsize=3.5in\epsfbox{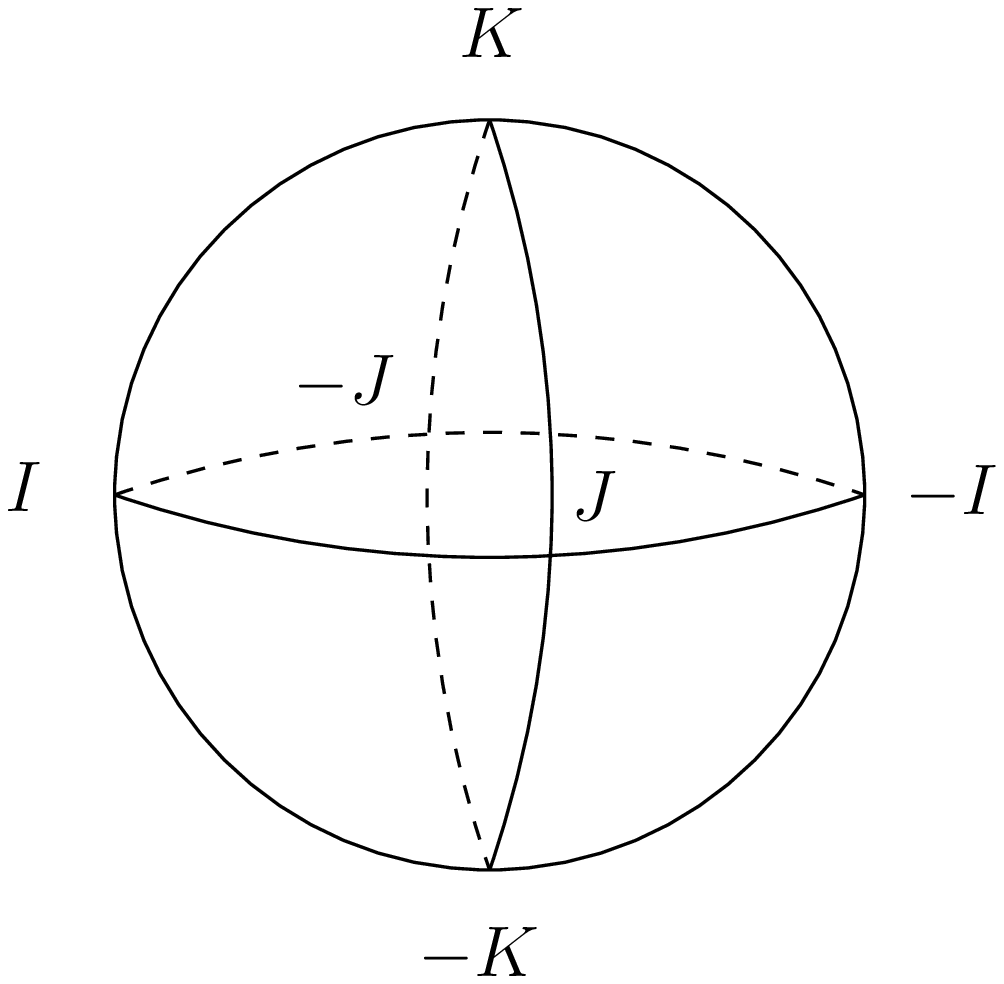}} A complex structure such as
$I, J,$ or $K$ is usually defined as a linear transformation of
the tangent space, but instead we will write down the transposed
matrices $I^t$, $J^t$, and $K^t$ acting on the cotangent space.
 We define $I^t$ by the formulas \eqn\uttu{\eqalign{ I^t(\delta A_{\bar z} )& =
i\delta A_{\bar z} \cr I^t(\delta \phi_{z}) & = i\delta \phi_{ z}
\cr I^t(\delta A_{ z}) & = -i\delta A_{ z} \cr I^t(\delta
\phi_{\bar z}) & = -i\delta \phi_{\bar z}, \cr}} which do ensure
$I^2=(I^t)^2=-1$. A translation-invariant complex structure on an
affine space $\EUW$ can also be characterized by saying which
linear functions on $\EUW$ are holomorphic.  For complex structure
$I$, the linear holomorphic functions are $A_{\bar z}$ and
$\phi_z$, evaluated at any point on $C$.

Similarly, we define a complex structure $J$ by
\eqn\buttu{\eqalign{J^t(\delta A_{\bar z}) & = -\delta \phi_{\bar
z}\cr J^t(\delta A_z) & = -\delta \phi_z \cr  J^t(\delta\phi_{\bar
z}) &  = \delta A_{\bar z} \cr J^t(\delta \phi_{ z}) & = \delta
A_z.\cr}}  The linear holomorphic functions are $\CA_{\bar
z}=A_{\bar z}+i\phi_{\bar z}$ and $\CA_z=A_{z}+i\phi_z$, or more
succinctly $\CA=A+i\phi$.

Finally, we define complex structure $K$ by
\eqn\bluttu{\eqalign{K^t(\delta A_{\bar z} )& = -
i\delta\phi_{\bar z} \cr K^t(\delta \phi_{\bar z}) & = - i\delta
A_{\bar z} \cr K^t(\delta A_{ z} )& =  i\delta\phi_{ z} \cr
K^t(\delta \phi_{ z}) & = i\delta A_{ z}. \cr}} The linear
holomorphic functions are $A_{\bar z}-\phi_{\bar z}$ and
$A_z+\phi_z$.

These linear transformations of the cotangent space of $\EUW$ obey
$I^tJ^t=-J^tI^t=-K^t$, and cyclic permutations thereof. Taking the
transpose, we get the usual form of the quaternion relations
$IJ=-JI=K$, etc.

Beyond the quaternion relations, the definition of a hyper-Kahler
structure requires that the metric $ds^2$ should be of type
$(1,1)$ in each complex structure.  Equivalently, we want
$ds^2=I^t\otimes I^t(ds^2)=J^t\otimes J^t(ds^2)=K^t\otimes
K^t(ds^2).$ This can be readily verified.  Because $\EUW$ is an
affine space and our complex structures are linear, there is
nothing else to check; we have obtained a flat hyper-Kahler
structure on $\EUW.$

A hyper-Kahler manifold has (\xunko) a family of complex
structures parametrized by $\Bbb{CP}^1$, obtained by taking linear
combinations of $I,$ $J$, and $K$.  In the case at hand, we can
describe this family simply.  For any $w\in \Bbb{C}^*$, we
consider the complex structure $I_w$ in which $A_{\bar
z}-w\phi_{\bar z}$ and $A_z+w^{-1}\phi_z$ are holomorphic. Thus,
$w=-i$ corresponds to complex structure $J$, and $w=1$ corresponds
to $K$.  We extend the definition to include $w=0,\infty$ by
setting $I_0=I$, and $I_\infty=-I$ (here $-I$ is the complex
structure opposite to $I$, in which $A_z$ and $\phi_{\bar z}$ are
holomorphic), as these are the limits of $I_w$ for $w\to
0,\infty$. So we get the expected family of complex structures,
parametrized by a copy of $\Bbb{CP}^1$ that we will call
$\Bbb{CP}^1_h$ (see \zunko). An explicit formula for $I_w$ is
\eqn\wyre{I_w={1-\bar w w\over 1+\bar w w}I+{i(w-\bar w)\over
1+\bar w w}J+{w+\bar w\over 1+\bar w w}K.} Of course, this is of
the form $aI+bJ+cK$ with $a^2+b^2+c^2=1$.

\def\CU{{\cal U}}
We introduce a group $\CU_1\cong U(1)$ that acts on $\phi$ by
$\phi_z\to \lambda\phi_z$, $\phi_{\bar z}\to
\lambda^{-1}\phi_{\bar z}$, $|\lambda|=1$, while leaving $A$
fixed. $\CU_1$ preserves the hyper-Kahler metric and acts on
$\Bbb{CP}^1_h$ by \eqn\zungo{I_w\to I_{\lambda^{-1}w};}
 it leaves
fixed the two points $w=0,\infty$.  We explain at the end of
section \complexstr\ in what sense the action of $\CU_1$ extends
to an action of its complexification $\CU\cong \Bbb{C}^*$.

Most of the complex structures on $\EUW$ in this family depend on
the complex structure of $C$.  The exceptions are $J$, which can
be characterized as the complex structure in which $\CA=A+i\phi$
is holomorphic, and $-J$, in which $\bar\CA=A-i\phi$ is
holomorphic.  The other complex structures on $ \EUW$ treat
differently the $z$ and $\bar z$ components of $A$ and $\phi$, a
distinction that depends on the complex structure of $C$.

Does the symmetry group ${\cal U}_1$  come from a symmetry of
four-dimensional Yang-Mills theory?  The answer to this question
is a little subtle.  The twisted ${\EUN}=4$ super Yang-Mills
theory compactified on $C$ does have the symmetry group $\CU_1$.
It acts by rotation of the $\phi_2-\phi_3$ plane; this is a
subgroup of $SU(4)_{\cal R}$ that is unbroken in this particular
compactification. However, this does not give a symmetry of the
underlying family of four-dimensional TQFT's, because it does not
preserve the family of supersymmetry generators
$\epsilon=u\epsilon_\ell+v\epsilon_\rr$ (it maps this family to a
larger family, whose interpretation we discuss in section
\twistedtft, of supersymmetries that lack the full $Spin'(4)$
symmetry).

But the element $-1\in {\cal U}_1$ does arise from a symmetry of
the four-dimensional TQFT. This element acts on $\MH$ just like
the central element ${\EUI}\in SU(4)_{\cal R}$, which we
introduced in section \reviewsd. Indeed, ${\EUI}$ acts as $i$ in
the ${\bf 4}$ of $SU(4)_{{\cal R}}$, so it acts as $-1$ on ${\bf
6}=\wedge^2{\bf 4}$, which is the representation that contains
$\phi$.  (Note that ${\EUI}$ also reverses the sign of untwisted
scalars.) The transformation $\phi\to-\phi$ is holomorphic in
complex structure $I$, but antiholomorphic in structures $J$ and
$K$.  The obvious component of the fixed point set of this
transformation is defined by $\phi=0$ (giving, as we discuss
later, an embedding of $\M$, the moduli space of stable bundles,
in $\MH$).  There also are \hitchin\ other components of the fixed
point set in which $\phi$ is gauge-equivalent to $-\phi$.

\bigskip\noindent{\it Symplectic Structures}

The three symplectic structures associated with the hyper-Kahler
structure on $\EUW$ can be defined as $\omega_I = I^t\otimes
1(ds^2)$, and similarly  $\omega_J=J^t\otimes 1(ds^2) $ and
$\omega_K=K^t\otimes 1(ds^2)$. We get \eqn\tomog{\eqalign{\omega_I
& = -{i\over 2\pi}\int_C|d^2z|\,\Tr\left(\delta A_{\bar z}\wedge
\delta A_z-\delta \phi_{\bar z}\wedge \delta\phi_z\right)\cr
&=-{1\over 4\pi}\int_C\,\Tr\left(\delta A\wedge \delta A
-\delta\phi\wedge\delta \phi\right) \cr
                      \omega_J & ={1\over 2\pi}\int_C |d^2z|\,\Tr\left(
                      \delta\phi_{\bar z}\wedge \delta A_z+\delta\phi_z\wedge
                      \delta A_{\bar z} \right)\cr
                      \omega_K & =
                      {i\over 2\pi}\int_C|d^2z|\,\Tr\left(\delta\phi_{\bar z}
                      \wedge\delta A_z-\delta\phi_z\wedge\delta A_{\bar z}\right)\cr
                      & = {1\over 2\pi}\int_C \,\Tr\,\delta\phi
                      \wedge\delta A.\cr}}
Here, $\omega_J$ depends on the complex structure of $C$, while
$\omega_I$ and $\omega_K$ do not.

Moreover, $\omega_I$ is of type $(1,1)$ with respect to complex
structure $I$, and similarly for $\omega_J$ and $\omega_K$.  We
also define $\Omega_I=\omega_J+i\omega_K$, along with cyclic
permutations $\Omega_J=\omega_K+i\omega_I$,
$\Omega_K=\omega_I+i\omega_J$:
 \eqn\ninto{\eqalign{\Omega_I & =
{1\over \pi}\int_C|d^2z|\,\Tr\,\delta\phi_z\wedge \delta A_{\bar
z} \cr
        \Omega_J & =-{i\over 4\pi}\int_C\,\Tr\,\delta \CA\wedge\delta \CA          \cr
        \Omega_K & = -{i\over 2\pi}\int_C |d^2z|\,\Tr\,\left( \delta A_{\bar z}\wedge
\delta A_z-\delta \phi_{\bar z}\wedge \delta\phi_z -
\delta\phi_{\bar z}\wedge \delta A_z-\delta\phi_z\wedge
                      \delta A_{\bar z}\right).\cr}}
$\Omega_I$ is of type $(2,0)$ in complex structure $I$, and
similarly $\Omega_J$ and $\Omega_K$ are of type $(2,0)$ in their
respective complex structures. $\Omega_I$ and $\Omega_K$ depend on
the complex structure of $C$, but $\Omega_J$ does not.

The closed two-forms $\omega_J$ and $\omega_K$ are actually exact.
Indeed, we can write very explicitly $\omega_J=\delta\lambda_J$
with $\lambda_J={1\over 2\pi}\int_C |d^2z|\,\Tr\,\left(\phi_{\bar
z}\delta A_z+\phi_z\delta A_{\bar z}\right)$. $\lambda_J$ is
gauge-invariant, because $\phi$ and $\delta A$ (unlike $A$ itself)
both transform in the adjoint representation of $G$.  Similarly
$\omega_K=\delta\lambda_K$ with $\lambda_K={1\over
2\pi}\int_C\,\Tr\,\phi\wedge \delta A$. Among other things, this
means that a topologically trivial line bundle over $\MH$ can be
endowed with a  connection whose  curvature is any desired linear
combination of $\omega_J$ and $\omega_K$.  One simply picks the
connection to be a suitable linear combination of $\lambda_J$ and
$\lambda_K$.

 \lref\quillen{D.~Quillen, ``Determinants Of Cauchy-Riemann Operators
  Over A Riemann Surface,'' Funct.\ Anal.\ Appl.\
  {\bf 19}, 31-34 (1985).}
On the other hand, $\omega_I$ is topologically non-trivial.  To
write $\omega_I=\delta\lambda_I$ along the above lines, one would
need in $\lambda_I$ a term proportional to $\int \Tr \,A\wedge
\delta A$, but this is not gauge-invariant because of the
inhomogeneous transformation of $A$ under gauge transformations.
However, we can write $\omega_I=\omega_I'+\delta\lambda_I$, where
$\lambda_I=(1/4\pi)\int_C\Tr\,\phi\wedge\delta\phi$ and
\eqn\frox{\omega_I'=-{1\over 4\pi}\int_C\Tr\,\delta A\wedge\delta
A.} So $\omega_I$ and $\omega_I'$ are cohomologous, and in
particular the cohomology class of $\omega_I$ is a pullback from
the subspace with $\phi=0$.  For simply-connected $G$, the Picard
group of line bundles over $\M$ is $\Bbb{Z}$, generated by a line
bundle ${{\frak L}}$  whose first Chern class is represented in de
Rham cohomology by $\omega_I'/2\pi$.  We will loosely call
${{\frak L}}$ the determinant line bundle, since for classical $G$
it can be defined \quillen\ as the determinant of a $\bar\partial$
operator.\foot{For $G=SU(N)$, if $E\to C$ is a holomorphic vector
bundle of rank $N$, the fiber ${\frak L}_E$ at $E$ of the
determinant line bundle $\frak L$ is ${\frak
L}_E=\det\,H^0(C,E)^{-1}\otimes \det\, H^1(C,E)$, where $\det\,V$
denotes the highest exterior power of a vector space $V$.  The
line bundle $\frak L\to \M$ so defined is ample.  Some authors
write ${\frak L}^{-1}$ for what we call $\frak L$.}

\bigskip\noindent{\it Hitchin Moduli Space as a Hyper-Kahler Quotient}

We now consider the group of gauge transformations acting on
$\EUW$ in an obvious way. It manifestly preserves the metric
\igob\ and the three symplectic structures \tomog . A short
computation shows that the Hitchin equations \zongo\ are
equivalent to the vanishing of the three moment maps corresponding
to the symplectic structures \tomog. Explicitly, let $\epsilon$
generate an infinitesimal gauge transformation  (thus, $\epsilon$
is a zero-form with values in ${\rm ad}(E)$), and let
$V(\epsilon)$ be the corresponding vector field on $\EUW$, which
acts by $\delta A=-D\epsilon$, $\delta\phi=[\epsilon,\phi]$. Then
the moment map $\mu$ for symplectic structure $\omega$ is found by
solving the condition $\iota_{V(\epsilon)}\omega=\delta\mu$ (here
$\iota_{V(\epsilon)}$ is the operator of contraction with
$V(\epsilon)$). Taking $\omega$ to be $\omega_I$, $\omega_J$, or
$\omega_K$, the corresponding moment maps turn out to be
\eqn\joget{\eqalign{\mu_I & =-{1\over
2\pi}\int_C\,\Tr\,\epsilon(F-\phi\wedge \phi),\cr \mu_J &
=-{1\over 2\pi}\int_C|d^2z|\,\Tr\,\epsilon\left(D_z\phi_{\bar
z}+D_{\bar z}\phi_z\right), \cr \mu_K& = -{i\over
2\pi}\int_C|d^2z|\,\Tr\,\epsilon\left(D_z\phi_{\bar z}-D_{\bar
z}\phi_z\right).\cr}} In general, if $X$ is a hyper-Kahler
manifold on which a Lie group $H$ acts with moment maps
$\vec\mu=(\mu_I,\mu_J,\mu_K)$, then the quotient by $H$ of the
solution set of the equations $\vec\mu=0$ is a hyper-Kahler
manifold \ref\rochitchin{N. J. Hitchin, A. Karlhede, U. Lindstrom,
and M. Rocek, ``Hyperkahler Metrics and Supersymmetry,'' Commun.
Math. Phys. {\bf 108} (1987) 535-589.}, called the hyper-Kahler
quotient $X/\negthinspace/\negthinspace/H$ of $X$ by $H$. So in
particular $\M_H$ is the hyper-Kahler quotient of the
infinite-dimensional space ${\EUW}$ by the group of gauge
transformations, and is a hyper-Kahler manifold.

It is possible to give a more ``physical'' explanation of why the
target of the sigma-model is the hyper-Kahler quotient of $\EUW$.
In general, in a two-dimensional gauge theory with $(4,4)$
supersymmetry and gauge group $H$, the hypermultiplets parametrize
a hyper-Kahler manifold $X$, and the low energy physics (assuming
for simplicity that $H$ acts freely on $X$) is a sigma-model with
target the hyper-Kahler quotient
$X/\negthinspace/\negthinspace/H$. (This is the context in which
the hyper-Kahler quotient construction was originally discovered
\rochitchin.) In the present example, compactification of the
twisted topological gauge theory in four dimensions give in two
dimensions a theory with $(4,4)$ supersymmetry.  (The twisting and
compactification preserve eight of the original sixteen
supersymmetries, of which four have one two-dimensional chirality
and four have the other.) The gauge group of this two-dimensional
theory is the group ${\cal G}$ of maps of $C$ to $G$, and the
hypermultiplets parametrize $\EUW$. This is precisely the setup
that leads to $\MH$ as a hyper-Kahler quotient.  So the low energy
physics is a sigma-model with target the hyper-Kahler quotient of
$\EUW$.

\bigskip\noindent{\it More Precise Account Of Dimensional
Reduction}

We can now be more precise about the dimensional reduction from
twisted four-dimensional super  Yang-Mills theory to the
sigma-model with target $\MH$.  Such a sigma-model has a metric
and a $B$-field.  We want to determine what they are.

We consider four-dimensional super Yang-Mills on $M=\Sigma\times
C$, where the radius of $\Sigma $ is much greater than the radius
of $C$.  The kinetic energy of the gauge fields is
\eqn\kinen{{{\cmmib I}}_{kin}=-{1\over 2e^2}\int_{\Sigma\times C}
|d^2y|\,|d^2z|\,\Tr\,F_{\mu\nu}F^{\mu\nu},} where $|d^2y|$ and
$|d^2z|$ are the Riemannian measures on $\Sigma$ and $C$,
respectively. This has contributions of type $(p,q)$, $p+q=2$,
where $p$ is the number of indices $\mu,\,\nu$ that are tangent to
$\Sigma$ and $q$ is the number tangent to $C$. It is useful to
write the gauge field on $\Sigma\times C$ as $A=A_\Sigma+A_C$,
where $A_\Sigma$ is tangent to $\Sigma$ and $A_C$ is tangent to
$C$.

The terms of type $(0,2)$ are part of the potential energy for
$A_C$ that causes the theory to be localized on slowly varying
maps  $\Phi:\Sigma\to \MH$. The terms of type $(2,0)$ give a
kinetic energy for $A_\Sigma$, which away from singularities of
$\MH$ is massive, because the gauge group is completely broken,
and can be integrated out in the infrared (as we describe
explicitly in section \wilbranes).

The kinetic energy for the sigma-model comes from the terms of
type $(1,1)$.  These terms are \eqn\binen{-{1\over e^2}\int
|d^2y|\,|d^2z|\,\sum_{\mu=0,1}\sum_{\nu=2,3}\,\Tr\,F_{\mu\nu}F^{\mu\nu}.}
We want to compare this to a general sigma-model of maps
$\Phi:\Sigma\to X$, where the target space $X$ has metric
$ds^2_X=g_{IJ}(X)\,dX^I\,dX^J$ in terms of local coordinates
$X^I$.  The kinetic energy of the sigma-model is then
\eqn\zinen{\int |d^2y|\, g_{IJ}\sum_{\mu=0,1}\partial_\mu
X^I\partial^\mu X^J.} In comparing \binen\ and \zinen, the role of
the $X^I$ is played by $A_C$, and $F_{ij}$ corresponds to
$\partial_i X^I$.  If we compare \binen\ and \zinen\ to our
original definition \igob\ of the hyper-Kahler metric $ds^2$ on
${\EUW}$ (which induces   on the hyper-Kahler quotient $\MH$ a
metric that we also call $ds^2$), we see that the metric $ds^2_g$
that is induced by the four-dimensional gauge theory is related to
$ds^2$ by \eqn\inog{ds^2_g={4\pi\over e^2}\,ds^2=({\rm
Im}\,\tau)\,ds^2.}

We similarly want to reduce the $\theta$ term of the gauge theory,
which in Euclidean signature is \eqn\bloko{{\cmmib
I}_\theta={i\theta\over 8\pi^2}\int_{\Sigma\times C}\,\Tr\,
F\wedge F.} Again, we evaluate this coupling for a gauge field
that represents a slowly varying map to $\MH$.  Because of the
topological invariance of ${{\cmmib I}}_\theta$, in evaluating
\bloko\ we can deform to a situation in which the restriction of
$F$ to $C$ vanishes (that is, $F_{23}=0$).  The deformation to
$F|_C=0$ can be achieved via a complex gauge transformation or
using Hitchin's second fibration $\MH\to\M$ (these are both
notions that we explain later).  Given this, in evaluating
${\cmmib I}_\theta$, we can replace $F={1\over
2}\sum_{\mu,\nu=0}^3dx^\mu\wedge dx^\nu \,F_{\mu\nu}$ by
$F^{1,1}=\sum_{\mu=0,1}\sum_{\nu=2,3}dx^\mu\wedge dx^\nu
\,F_{\mu\nu}$. In the sigma-model limit, $F^{1,1}$ corresponds to
$\delta A_C=\sum_{\mu=0,1}dx^\mu \,D_\mu A_C$, and ${\cmmib
I}_\theta$ reduces to\foot{A minus sign enters because in defining
$\omega_I'$, $\delta$, the exterior derivative on the space of
connections $A_C$, is taken to commute with a one-form on $C$ such
as $dx^ j$, $j=2,3$, which does not depend on $A_C$. For
differential forms on $\Sigma\times C$, in terms of which ${\cmmib
I}_\theta$ is defined, they would anticommute.}
\eqn\ino{-{i\theta\over 8\pi^2}\int_\Sigma \int_C \Tr\,\delta
A_C\wedge \delta A_C.}

Comparing to the definition \frox\ of $\omega_I'$, we see that
this is equivalent to \eqn\zinok{{i\theta\over
2\pi}\int_\Sigma\Phi^*(\omega'_{I})={i\theta\over 2\pi}\int_\Sigma
\Phi^*(\omega_I),} where the last step uses the fact that
$\omega_{I}-\omega'_{I}$ is exact.  In the general sigma-model,
the $B$-field is a closed two-form on the target space $X$; it
appears in the action via a term $-i\int_\Sigma\,\Phi^*(B)$. So
the $B$-field of the effective low energy sigma-model is
\eqn\palooka{B=-{\theta\over 2\pi}\omega_I=-({\rm
Re}\,\tau)\,\omega_I.}

We can describe this more conceptually.  For simply-connected $G$,
the instanton number $(1/8\pi^2)\int_M \Tr\,F\wedge F$ is
integer-valued for any four-manifold $M$.  In addition, for
simply-connected $G$, $\MH$ is simply-connected, and hence
$H^2(\MH,\Bbb{Z})=\pi_2(\MH)$.  The form $\omega_I$ has been
normalized so that $\omega_I/2\pi$ is the image in de Rham
cohomology of a generator of $H^2(\MH,\Bbb{Z})$. These facts imply
that for any $\Sigma$ and $C$, there is a map $\Phi:\Sigma\to \MH$
with $\int_\Sigma\Phi^*(\omega_I/2\pi)=1$ (and this is the smallest
achievable value).  Such a map corresponds to a gauge field on
$M=\Sigma\times C$ that has instanton number $\pm 1$ depending on
one's conventions.\foot{In the physics literature, instantons in
four dimensions are generally taken to be selfdual gauge fields,
obeying $F^-=0$, and in the math literature, they are generally
taken to be anti-selfdual, obeying $F^+=0$.  If an instanton in
the sigma model is understood as a holomorphic map to $\MH$, and
the usual physics convention is followed in four dimensions, then
a sigma model field of instanton number 1 comes from a gauge
theory instanton of instanton number $-1$.  The usual math
convention  avoids this minus sign.} More briefly, the theta
angles are the same in two and four dimensions because the
instantons are the same. If $G$ is not simply-connected, things
are a little more complicated; the instanton number is not always
an integer, and there are corresponding complications on the
sigma-model side.

\subsec{Complex Structures Of $\MH$}

\subseclab\complexstr

Our next task is to understand more concretely the complex
structures on $\MH$.

\bigskip\noindent{\it Description By Holomorphic Data}

Let us look at the hyper-Kahler quotient of $\EUW$ from the
standpoint of complex structure $I$. The combination
$\nu_I=\mu_J+i\mu_K$ is holomorphic in this complex structure:
\eqn\uttup{\nu_I=-{1\over \pi}\int_C|d^2z|\,\Tr\,\epsilon\,D_{\bar
z}\phi_z.} $\nu_I$ is holomorphic in complex structure $I$ because
it is the moment map derived from $\Omega_I$, which is of type
$(2,0)$ in complex structure $I$ (explicitly $\nu_I$ is
holomorphic because it depends only on $A_{\bar z}$ and $\phi_z$).
In such problems, it is often convenient \rochitchin\ to consider
separately the vanishing of the holomorphic moment map $\nu_I$ and
the real moment map $\mu_I$.

Any  connection $A$ on a smooth $G$-bundle $E$ over a Riemann
surface $C$ automatically turns $E$ into a holomorphic $G$-bundle
(which we will denote by the same name, hoping this causes no
confusion). One simply defines the $\bar\partial$ operator as
$\bar D=d\bar z\,D_{\bar z}$, which in complex dimension one
trivially obeys $\bar D^2=0$. The vanishing of the holomorphic
moment map $\nu_I$ tells us that $\bar D\phi=0$; in other words,
$\varphi=\phi_z\,dz$ is a holomorphic section of $K_C\otimes {\rm
ad}(E)$, where $K_C$ is the canonical line bundle of $C$.
Differently put, $\varphi$ is a holomorphic one-form valued in the
adjoint representation, that is in the adjoint bundle of $E$. We
will call a pair $(E,\varphi)$ consisting of a $G$-bundle $E$ and
a holomorphic section $\varphi$ of $K_C\otimes {\rm ad}(E)$ a
Hitchin pair.  $\varphi$ is often called the Higgs field, and the
bundle $E$ endowed with the choice of $\varphi$ is also called a
Higgs bundle.

To obtain the moduli space $\MH$, we must also set to zero the
real moment map $\mu_I$ and divide by the group of $G$-valued
gauge transformations.  However, as exploited in \hitchin\ (and as
is often the case in moduli problems) there is a simpler way to
understand these combined steps.  The space $\EUW_0$ of zeroes of
the holomorphic moment map $\nu_I$ admits the action not just of
ordinary $G$-valued gauge transformations, but of gauge
transformations valued in the complexification $G_{\Bbb{C}}$ of
$G$.  This is manifest from the holomorphy of $\nu_I$. Thus  we
can perform on $\EUW_0$ either of the following two operations:

{\it (I)} Restrict to the subspace of $\EUW_0$ with $\mu_I=0$ and
divide by $G$-valued gauge transformations.

 {\it (II)} Divide $\EUW_0$ by $G_{\Bbb{C}}$-valued gauge
transformations.

\noindent Operation {\it (I)} gives the desired moduli space
$\MH$, but operation {\it (II)} is much easier to understand and
nearly coincides with it.  (The reason they nearly coincide is
that almost every orbit of complex gauge transformations contains
a unique orbit of ordinary gauge transformations on which
$\mu_I=0$.)  The easiest way to understand operation {\it (I)} is
often to first understand operation {\it (II)} and then understand
its relation to operation {\it (I)}.

In the present case, the result of operation {\it (II)} is easy to
describe. It means that we do not care about the particular choice
of $A_{\bar z}, $ $\phi_z$, but only about the holomorphic
$G$-bundle $E$ that is determined by $A_{\bar z}$, and the
associated holomorphic section $\varphi$ of $K_C\otimes {\ad}(E)$.
Thus, in the present example, operation {\it (II)} gives us the
set of equivalence classes of Hitchin pairs $(E,\varphi)$.

This set is for many purposes a very good approximation to $\MH$;
for example, they differ in rather  high codimension if $C$ has
genus at least 2. To be more precise (most of the present paper
does not depend on the details),  we need the notion of stability.
For $G=SU(2)$, we interpret $E$ as a rank two holomorphic vector
bundle over $C$, and $\varphi$ as a holomorphic map $\varphi:E\to
E\otimes K_C$. A line bundle ${\cal L}\subset E$ is called
$\varphi$-invariant if $\varphi({\cal L})\subset {\cal L}\otimes
K_C$. A Hitchin pair $(E,\varphi)$ is called stable if every
$\varphi$-invariant line bundle ${\cal L}\subset E$ has negative
first Chern class. It is called semistable if each such ${\cal L}$
has non-positive first Chern class. For general $G$, one must
consider $\varphi$-invariant reductions of the structure group of
$E$ to a maximal parabolic subgroup $P$ of $G_{\Bbb{C}}$. The
bundle $E$ with such a reduction has a natural first Chern class
(because $P$ has a distinguished $U(1)$), and the pair
$(E,\varphi)$ is called stable (or semistable) if for every such
reduction the first Chern class is negative (or nonpositive). A
pair that is semistable but not stable is called strictly
semistable.

Stability is a mild condition in the sense that, for example, if a
pair $(E,\varphi)$ is stable, then so is every nearby pair.
Moreover, the pairs that are not stable have high codimension (if
the genus of $C$ is greater than 1).

Now we can go back to the question of comparing operations {\it
(I)} and {\it (II)}, or equivalently, describing in holomorphic
terms the moduli space $\MH$ of solutions of Hitchin's equations.
The result proved in \hitchin\ is that $\MH$ is the same as the
``moduli space of stable pairs $(E,\varphi)$,'' i.e., stable
Hitchin pairs. We get this moduli space by throwing away unstable
Hitchin pairs, imposing a certain equivalence relation on the
semistable ones, and then dividing by the complex-valued gauge
transformations.  This slightly modified   version of operation
{\it (II)} -- call it operation {\it (II)$'$} -- agrees precisely
with operation {\it (I)}.

In sum, $\MH$ parametrizes the Hitchin pairs $(E,\varphi)$ that
are stable, as well as certain equivalence classes of strictly
semistable Hitchin pairs.  The stable pairs correspond to smooth
points in $\MH$ as well as (for some $G$) certain orbifold
singularities. The strictly semistable pairs generally lead to
singularities in $\MH$ that are worse than orbifold singularities.

\bigskip\noindent{\it Analog For Complex Structure $J$}

All this has an analog in complex structure $J$. The holomorphic
moment map in complex structure $J$ is $\nu_J=\mu_K+i\mu_I$:
\eqn\nuJ{\nu_J =-{i\over 2\pi}\int_C\,\Tr\,\epsilon{\cal F}} To
construct $\MH$, we first impose the vanishing of $\nu_J$. A zero
of $\nu_J$ is simply a pair $A,\phi$ such that the curvature
${\cal F}$ of the $G_{\Bbb{C}}$-valued connection $\CA=A+i\phi$ is
equal to zero. If therefore we were to divide the zero set of
$\nu_J$ by the group of $G_{\Bbb{C}}$-valued gauge
transformations, we would simply get the set ${\cal Y}_0$ of all
homomorphisms $\vartheta:\pi_1(C)\to G_{\Bbb{C}}$, up to
conjugation. This is operation {\it (II)}.

Instead, to get $\MH$, we are supposed to carry out operation {\it
(I)}.  In other words, we are supposed to set $\mu_J=0$, i.e., to
impose the condition $D^*\phi=0$, and divide only by the
$G$-valued gauge transformations.  As  in complex structure $I$,
in comparing these two operations, a notion of stability intrudes.
For $G=SU(2)$, a homomorphism $\vartheta:\pi_1(C)\to G_{\Bbb{C}}$
is considered stable if the monodromies cannot be simultaneously
reduced to the triangular form \eqn\yto{\left(\matrix{ \alpha &
\beta \cr 0 & \alpha^{-1}}\right).}   If $\vartheta$ has such a
reduction, we call it strictly semistable.  (In complex structure
$J$, there is no notion of an unstable representation.) We
consider two strictly semistable representations to be equivalent
if they have the same diagonal monodromy elements, that is, the
same $\alpha$'s.  Note that each such equivalence class has a
distinguished representative with $\beta=0$.

For any $G$, the analog of putting the monodromies in triangular
form is to conjugate them into a parabolic subgroup $P$ of $G$.  A
representation is stable if it cannot be so conjugated, and
otherwise strictly semistable.  The equivalence relation on
strictly semistable representations has a natural analog for any
$G$ (two strictly semistable representations that both reduce to
$P$ are equivalent if the two flat $P$ bundles become equivalent
when projected to the maximal reductive subgroup of $P$).

A theorem of Corlette \corlette\ and Donaldson  (see the appendix
to \hitchin) is that $\MH$, in complex structure $J$, is  the
moduli space ${{\cal Y}}$ that parametrizes stable homomorphisms
$\vartheta:\pi_1(C)\to G_{\Bbb{C}}$, as well as the equivalence
classes of semistable ones.

\bigskip\noindent{\it Analog For $I_w$}

What we have just said has a direct analog in each of the complex
structure $I_w$, $w\not= 0,\infty$.  The holomorphic  variables in
complex structure $I_w$ are $A_{\bar z}-w\phi_{\bar z}$,
$A_z+w^{-1}\phi_z$.  Two of Hitchin's equations combine to the
holomorphic condition $[D_{\bar z}-w\phi_{\bar
z},D_z+w^{-1}\phi_z]=0$, and the third is a moment map condition.
The holomorphic condition says that the complex-valued connection
with components $(A_{\bar z}-w\phi_{\bar z},A_z+w^{-1}\phi_z)$ is
flat.  Setting to zero the moment map and dividing by gauge
transformations gives the same moduli space ${{\cal Y}}$ that we
found in complex structure $J$.  Thus all the complex structures
$I_w$, $w\not=0,\infty$ are equivalent.

In fact, in complex structure $I=I_{w=0}$, we can define a
symmetry group ${\cal U}\cong \Bbb{C}^*$ that acts on a Higgs
bundle by $(E,\varphi)\to (E,\lambda\varphi)$,
$\lambda\in\Bbb{C}^*$. (The stability condition is invariant under
this transformation.) For $|\lambda|=1$, this reduces to the group
${\cal U}_1$, described in section \mhhyper, which is a symmetry
group of the hyper-Kahler metric of $\MH$. For $|\lambda|\not=1$,
we do not get a symmetry of the hyper-Kahler metric of $\MH$, but
as is shown on pp. 107-8 of \hitchin, we get a group of
diffeomorphisms of $\MH$ that preserves complex structure $I$ and
transforms the other complex structures by \eqn\trymodo{I_w\to
I_{\lambda^{-1}w},} generalizing eqn. \zungo\ to
$|\lambda|\not=1$.

\subsec{Hitchin's Fibrations}

\subseclab\hitchfib

Now we will discuss in more detail the geometry of $\MH$.

If $E$ is a stable $G$-bundle, then the pair $(E,0)$ is a stable
Hitchin pair.  So we get a natural embedding of $\M$, the moduli
space of stable $G$-bundles on $C$, into the Hitchin moduli space
$\MH$.  $\M$ is a holomorphic submanifold of $\MH$ in complex
structure $I$, since it is defined by the equations $\phi_z=0$,
which are holomorphic in complex structure $I$.  In complex
structures $J$ and $K$, $\M$ is not holomorphic. Instead, it is
Lagrangian, since $\phi=0$ implies $\delta\phi=0$ and hence
$\mu_J=\mu_K=0$.

If $E$ is stable, then the Hitchin pair $(E,\varphi)$ is stable
for every $\varphi\in H^0(C,K_C\otimes\ad(E))$.  Moreover, the
tangent bundle to ${\EUM}$ at the point $E$ is $H^1(C,\ad(E))$
(which parametrizes first order deformations of $E$), so by Serre
duality, the cotangent space to $\M$ at the point $E$ is
$H^0(C,K_C\otimes\ad(E))$. So $\varphi$ takes values in this
cotangent space, and the space of all $(E,\varphi)$ with stable
$E$ is the cotangent bundle $T^*\M$. We thus actually get an
embedding of $T^*\M$ in $\MH$.  The holomorphic symplectic form
$\Omega_I$ of $\MH$ in complex structure $I$ restricts on $T^*\M$
to its natural symplectic structure as a holomorphic cotangent
bundle (that is, for any local holomorphic coordinates $q^\alpha$
on $\M$, we have $\Omega_I=\sum_\alpha dp_\alpha\wedge dq^\alpha$
for some holomorphic functions $p_\alpha$ that are linear on the
fibers of $T^*\M\to \M$).\foot{This statement is true because the
analogous statement is actually true on the space $\EUW$ of all
pairs $(A,\phi)$ before taking the hyper-Kahler quotient.  $\EUW$
can be understood as the cotangent space of the space ${\EUA}$ of
connections, with $A$ parameterizing ${\EUA}$ and $\phi$
parameterizing the cotangent space to ${\EUA}$; $\Omega_I$ is the
natural holomorphic symplectic form of ${\EUW}$ regarded as
$T^*{\EUA}$.  In general, this type of structure persists in
taking a complex symplectic quotient (in the present case, setting
$\nu_I=0$ and dividing by $G_{\Bbb{C}}$-valued gauge
transformations).}

The image of $T^*\M$ in $\MH$ is not all of $\MH$ because a
Hitchin pair $(E,\varphi)$ may be stable even if $E$ is unstable.
However, the stable Hitchin pairs $(E,\varphi)$ for which $E$ is
unstable are a set of very high codimension. Upon throwing away
this set, $\MH$ becomes isomorphic to $T^*\M$, and has a natural
map to $\M$ by forgetting $\varphi$. Even though it is only
defined away from a set of very high codimension, this map is
extremely useful.  We will call it Hitchin's second fibration (the
first one being another map that we introduce presently).

\bigskip\noindent{\it The Hitchin Fibration}

Another natural operation in complex structure $I$, apart from
mapping $\varphi$ to zero, is to calculate the gauge-invariant
polynomials in $\varphi$. For $G=SU(2)$, this simply means that we
consider the quadratic Casimir operator $w=\Tr\,\varphi^2$. Since
$\bar D\varphi=0$, we have $\bar\partial w=0$, so $w$ is a
holomorphic quadratic differential, taking values in
${\EUBB}=H^0(C,K_C^2)\cong \Bbb{C}^{3g-3}$. The Hitchin fibration,
as it is most commonly called, is the map $\pi:\MH\to {\EUBB}$
obtained by mapping the pair $(E,\varphi)$ to $w=\Tr\,\varphi^2$.

For any $G$, the Hitchin fibration is defined similarly, except
that one characterizes $\varphi$ by all of its independent
Casimirs (that is, all of the independent $G$-invariant
polynomials in $\varphi$), not just the quadratic one. For
example, for $G=SU(N)$, we  define $w_n=\Tr\,\varphi^n$,
$n=2,\dots,N$, and let ${\EUBB}=\oplus_{n=2}^NH^0(C,K_C^n)$. The
Hitchin fibration is then the map that takes $(E,\varphi)$ to the
point $(w_2,w_3,\dots,w_n)\in {\EUBB}$. For any $G$, one considers
instead of $\Tr\,\varphi^n$ the appropriate independent
homogeneous $G$-invariant polynomials ${\cal P}_i$. The number of
these polynomials equals the rank $r$ of $G$, and their degrees
$d_i$ obey the identity \eqn\yty{\sum_{i=1}^r(2d_i-1)={\rm
dim}(G).} For example, for $G=SU(N)$, the $d_i$ are $2,3,\dots,N$,
whence $\sum_i(2d_i-1)=N^2-1={\rm dim}(G)$.  And finally in
general we define ${\EUBB}=\oplus_i H^0(C,K_C^{d_i})$.

\def\CMF{{\cmmib F}}
Since $\dim \,H^0(C,K_C^n)=(2n-1)(g-1)$, it follows from \yty\
that the complex dimension of ${\EUBB}$ is $(g-1){\rm dim}(G)$,
which equals the dimension of $\M$, and  one half of the dimension
of $\M_H$. The Hitchin fibration $\pi:\MH\to {\EUBB}$ is
surjective, as we will discuss momentarily.   As the base
${\EUBB}$ of the Hitchin fibration $\pi:\MH\to {\EUBB}$ has one
half the dimension of $\MH$, it follows that the dimension of a
typical fiber $\CMF$ of $\pi$ is also half the dimension of $\MH$
and so equal to the dimension of ${\EUBB}$:
\eqn\yrof{\dim\,\CMF=\dim\,{\EUBB}={1\over
2}\dim\,\MH=(g-1)\dim\,G.}

Let us explain qualitatively why the Hitchin fibration is
surjective. For example, take $G=SU(2)$. Pick a stable $SU(2)$
bundle $E$. Consider the equations $\Tr\,\varphi^2=w$, where
$\varphi $ varies in the $(3g-3)$-dimensional space
$H^0(C,K_C\otimes \ad(E))$ and $w$ is a fixed element of the
$(3g-3)$-dimensional space ${\EUBB}=H^0(C,K_C^2)$. This is a
system of $3g-3$ quadratic equations for $3g-3$ complex variables.
The number of solutions is generically $2^{3g-3}$.  A similar
counting can be made for other $G$.

\bigskip\noindent{\it Complete Integrability}

\nref\duke{N. Hitchin, ``Stable Bundles And Integrable Systems,''
Duke Math. J. {\bf 54} (1987) 91-114.}%
 \nref\nekrasov{A. Gorsky
and N. Nekrasov, ``Elliptic Calogero-Moser System From
Two-Dimensional Current Algebra,'' arXiv:hep-th/9401021;
``Relativistic Calogero-Moser Model As Gauged WZW
Theory,'' Nucl. Phys. {\bf B436} (1995) 582-608.}%
 \nref\markdon{R.
Donagi and E. Markman, ``Spectral Covers, Algebraically Completely
Integrable Hamiltonian Systems, And Moduli Of Bundles,'' in {\it
Integrable Systems And Quantum Groups}, Lecture Notes in Math.
{\bf 1620} (Springer-Verlag, 1996) pp. 1-119,
alg-geom/9507017.}%

 We now want to explain one of Hitchin's main
results \duke: $\MH$ is a completely integrable Hamiltonian system
in the complex structure $I$.  We can find ${1\over 2}{\rm
dim}\MH$ functions $H_a$ on $\MH$ that are holomorphic in complex
structure $I$, are algebraically independent, and are
Poisson-commuting using the Poisson brackets obtained from the
holomorphic symplectic form $\Omega_I$.\foot{The reader may be
unaccustomed to completely integrable systems in this holomorphic
sense.  {}From such systems, one can extract completely integrable
Hamiltonian systems in the ordinary real sense (and moreover,
interesting and significant constructions arise in this way; see
\refs{\nekrasov,\markdon} for some examples). We pick $C$ to admit
a real structure -- that is, an involution that reverses its
complex structure. This induces real structures on $\M$ and $\MH$.
By specializing to a real slice in $\MH$, one gets then completely
integrable Hamiltonian systems in which the phase space
coordinates, the symplectic structure, and the commuting
Hamiltonians are all real.}

In fact, we can take the $H_a$ to be linear functions on
${\EUBB}$, since the dimension of ${\EUBB}$ is the same as the
desired number of functions.   We will explain the construction
first for $G=SU(2)$. We begin by picking  a basis $\alpha_a$,
$a=1,\dots,3g-3$ of the $(3g-3)$-dimensional space $ H^1(C,T_C)$,
which is dual to $H^0(C, K_C^2)\cong {\EUBB}$. (Here $T_C$ is the
holomorphic tangent bundle to $C$.) We represent $\alpha_a$ by
$(0,1)$-forms valued in $T_C$, which we call by the same name, and
we define \eqn\miro{H_a=\int_C\alpha_a \wedge \Tr\,\varphi^2.} We
claim that these functions are Poisson-commuting with respect to
the holomorphic symplectic form $\Omega_I$.

A natural proof uses the fact that the definition of the $H_a$
makes sense on the infinite-dimensional space $\EUW$, before
taking the hyper-Kahler quotient.  Using the symplectic structure
$\Omega_I$ on $\EUW$ to define Poisson brackets, the $H_a$ are
obviously Poisson-commuting. For in these Poisson brackets, given
the form \ninto\ of $\Omega_I$, $\varphi_z$ is conjugate to
$A_{\bar z}$ and commutes with itself. But the $H_a$ are functions
of $\varphi_z$ only, not $A_{\bar z}$.

The functions $H_a$ can be restricted to the locus with $\nu_I=0$,
and then, because they are invariant under the
$G_{\Bbb{C}}$-valued gauge transformations, they descend to
holomorphic functions on $\MH$. A general property of symplectic
reduction (in which one sets to zero a moment map, in this case
$\nu_I$, and then divides by the corresponding group, in this case
the group of $G_{\Bbb{C}}$-valued gauge transformations) is that
it maps Poisson-commuting functions to Poisson-commuting
functions.  So the $H_a$ are Poisson-commuting as functions on
$\MH$.  There are enough of them to establish the complete
integrability of $\MH$.

One point to note in this construction is that the $H_a$, as
functions on $\EUW$, depend on the particular choice of $T$-valued
$(0,1)$-forms $\alpha_a$ that represent the cohomology classes.
Any choice will do, but we have to make a choice.  But after
restricting and descending to $\MH$, the functions we get on $\MH$
depend only on the cohomology classes of the $\alpha_a$. In fact,
once we have $\bar D\varphi=0$ and hence
$\bar\partial\Tr\,\varphi^2=0$, the $H_a$ are clearly invariant
under $\alpha_a\to\alpha_a+\bar\partial\epsilon_a$.

The Poisson-commuting functions $H_a$ generate commuting flows on
$\MH$ that are holomorphic in complex structure $I$.  Complex tori
admit commuting flows, and one might surmise that the orbits
generated by the $H_a$ are complex tori at least generically. This
follows from general results about holomorphic symplectic
structures and compactness of the fibers of the Hitchin fibration
and can also be demonstrated  more directly, using the technique
of the spectral cover \refs{\hitchin,\duke}. This technique has
many applications in further development of this subject, as will
be explained elsewhere \witfur. In this paper, we will get as far
as we can without using the spectral cover construction.

The analog of the above construction for any $G$ is to replace
$\Tr\,\varphi^2$ by a general gauge-invariant  polynomial ${\cal
P}_i$ of degree $d_i$. The associated commuting Hamiltonians take
the form $H_{\alpha,i}=\int_C \alpha \,{\cal P}_i(\varphi)$, for
$\alpha\in H^1(C,K_C^{1-d_i})$. A simple dimension counting, using
the dimension formula \yty, shows that these Hamiltonians are
precisely sufficient in number to establish the complete
integrability of $\MH$.  By using the fact that the fibers of the
Hitchin fibration are compact, so that a holomorphic function must
be a pullback from the base of this fibration, one can show that
the commuting Hamiltonians generate the ring of holomorphic
functions on $\MH$.

One easy and important consequence of complete integrability is
that the fibers of the Hitchin fibration are Lagrangian
submanifolds in the holomorphic symplectic structure $\Omega_I$ or
equivalently in the real symplectic structures $\omega_J$ and
$\omega_K$.  Indeed, a fiber of this fibration is defined by
equations $H_k-h_k=0$, where $H_k$ are the commuting Hamiltonians
and $h_k$ are complex constants.  In general, the zero set of a
middle-dimensional collection of Poisson-commuting functions, such
as $H_k-h_k$ in the present case, is Lagrangian.

\newsec{Topological Field Theory In Two Dimensions}
\seclab\tqfttwo

In section \tqftfour, we analyzed a twisted version of ${\cal
N}=4$ super Yang-Mills theory on a general four-manifold $M$.  The
twisting preserved those supersymmetries that are invariant under
a subgroup $Spin'(4)\subset Spin(4)\times Spin(6)$.  The condition
for a supersymmetry generator to be invariant under $Spin'(4)$ is
\eqn\tory{\left(\Gamma_{\mu\nu}+\Gamma_{4+\mu,4+\nu}\right)\epsilon=0,
~~\mu,\nu=0,\dots,3.}  In addition, all 16 supersymmetry
generators obey a ten-dimensional chirality condition,
\eqn\nory{\Gamma_0\Gamma_1\dots\Gamma_9\epsilon=\cases{\epsilon&
~\text{in Lorentz signature}\cr-i\epsilon&~\text{in Euclidean
signature.}\cr}} In the twisted theory, four of the six scalar
fields $\phi_0,\dots,\phi_5$ of ${\EUN}=4$ super Yang-Mills theory
are reinterpreted as a one-form
$\phi=\sum_{\mu=0}^3\phi_\mu\,dx^\mu$. Two scalar fields, $\phi_4$
and $\phi_5$, are untwisted, and are rotated by a group
$SO(2)_{\cal R}$ (whose generator we called ${\cal K}$ in section
\twisting). Allowing for the fermions, the symmetry group of the
twisted theory is $Spin(2)_{\cal R}$, the double cover of
$SO(2)_{\cal R}$.

The twisted theory is a formal construction in the sense that
twisting violates unitarity and only works in Euclidean
signature.\foot{To construct a twisted theory in Lorentz
signature, we would have needed a suitable homomorphism
$Spin(1,3)\to Spin(6)$.  Because of the compactness of $Spin(6)$,
a non-trivial homomorphism does not exist. If one replaces
$Spin(6)$ by $Spin(1,5)$ to work around this, the couplings to
fermions will no longer be hermitian and the energy is no longer
bounded from below.} Suppose, however, that we split off the time
direction and take $M=\Bbb{R}\times W$, where $\Bbb{R}$
parametrizes the time, $W$ is a three-manifold, and the metric on
$M$ is  a product. Then time is not involved in the twisting, and
the twisted theory makes sense with Lorentz or Euclidean signature
and is unitary and physically sensible. In this case, as the
holonomy of $M$ reduces to $SO(3)$, unbroken supersymmetries need
only be invariant under a subgroup $Spin'(3)\subset Spin'(4)$;
equivalently, we need only impose \tory\ for $\mu,\nu=1,\dots,3$.
There are four unbroken supersymmetry generators, which obey
\tory\ for $\mu,\nu=1,\dots 3$. Three scalars are untwisted,
namely $\phi_0$, $\phi_4$ and $\phi_5$, and are rotated by a
$Spin(3)_{\cal R}$ symmetry that extends the $Spin(2)_{\cal R}$
symmetry that is present on any $M$.

Let us specialize this further to the case $M=\Bbb{R}^{1,1}\times
C$ (or $\Bbb{R}^2\times C$), for $C$ a Riemann surface.  The
holonomy group of $M$ reduces to $SO(2)$, and \tory\ collapses to
the single condition \eqn\yty{\Gamma_{2367}\epsilon=\epsilon,}
leaving an eight-dimensional space of unbroken supersymmetries.
Four scalars $\phi_0,\phi_1,\phi_4$, and $\phi_5$ are untwisted,
and the symmetry group that rotates the untwisted scalars is now
extended to $Spin(4)_{\cal R}$.

In fact, the structure in two dimensions is precisely that of
$(4,4)$ supersymmetry. If we combine \yty\ and \nory, we find (in
Lorentz signature) \eqn\byty{\Gamma_{01}\epsilon
=\Gamma_{4589}\epsilon.} Here $\Gamma_{01}$ is the operator that
distinguishes the two spinor representations of the Lorentz group
$SO(1,1)$ of $\Bbb{R}^{1,1}$.
 Similarly, $\Gamma_{4589}$ distinguishes the two spin
representations of $Spin(4)_{\cal R}=SU(2)'\times SU(2)''$. These
two spin representations transform under $SU(2)'\times SU(2)''$ as
$({\bf 2},{\bf 1})$ and $({\bf 1},{\bf 2})$, respectively. Eqn.
\byty\ says that the eigenvalues of $\Gamma_{01}$ and
$\Gamma_{4589}$ are equal for all two-dimensional supersymmetries.

So with a suitable choice of orientation of $\Bbb{R}^{1,1}$, the
supersymmetries transform as $({\bf 2},{\bf 1})^-\oplus({\bf
1},{\bf 2})^+$, where the superscript labels the eigenvalue of
$\Gamma_{01}$. But this is the usual structure of $(4,4)$
supersymmetry in two dimensions: one $SU(2)$ group of
$R$-symmetries -- namely $SU(2)'$ -- acts on the supersymmetries
of negative $Spin(1,1)$ chirality, while a second such group --
$SU(2)''$ --  acts on supersymmetries of positive chirality. In
fact, in section \compacthitchin, we have already identified the
effective low energy theory as  a sigma-model in which the target
space is the hyper-Kahler manifold $\MH$.  In general,
sigma-models with hyper-Kahler targets have $(4,4)$ supersymmetry
\ref\alvfr{L. Alvarez-Gaum\'e and D. Z. Freedman, ``Geometrical
Structure And Ultraviolet Finiteness In The Supersymmetric Sigma
Model,'' Commun. Math. Phys. {\bf 80} (181) 443.}.

In addition to the $Spin(4)_{\cal R}$ global symmetry, the twisted
gauge theory on $\Bbb{R}^{1,1}\times C$ has an additional symmetry
that appears because the holonomy group of $C$, by which we twist,
is abelian.  Even after twisting $\phi_2$ and $\phi_3$, it is
still possible to make a rotation of the $\phi_2-\phi_3$ plane.
This is the symmetry group, called ${\cal U}_1$ in section
\mhhyper, that acts on a Hitchin pair $(E,\varphi)$ by rotating
the phase of $\varphi$.

\subsec{Twisted Topological Field Theories}

\subseclab\twistedtft

We now want to study the sigma-model of target $\MH$ from the
viewpoint of topological field theory, so we replace
$\Bbb{R}^{1,1}$ by a general two-manifold $\Sigma$ that we
eventually will take to have Euclidean signature. We want to
discuss what twisted topological field theories on $\Sigma$ can be
constructed from the sigma-model of target $\MH$, and which of
these actually arise by compactifying on the Riemann surface $C$
the four-dimensional topological field theories that we
constructed in section \tqftfour.

 \nref\rock{S. J. Gates, C. M.
Hull, and M. Rocek, ``Twisted Multiplets And New Supersymmetric
Nonlinear Sigma Models,'' Nucl. Phys. {\bf B248} (1984) 157-186.}
 \nref\gualtieri{M. Gualtieri, ``Generalized Complex Geometry,'' D.
Phil. thesis, Oxford University, math.DG/0401221.}%
\nref\pestun{V. Pestun, ``Topological Strings In Generalized
Complex Space,'' arXiv:hep-th/0603145.}%

 \nref\hhitchin{N. Hitchin,
``Generalized Calabi-Yau Manifolds,'' Q.
J. Math. {\bf 54} (2003) 281-308, math.DG/0209099.}%
\nref\witog{E. Witten, ``Mirror Manifolds And Topological Field
Theory,'' in {\it Mirror Symmetry I}, S. T. Yau, ed. (American
Mathematical Society, 1998) pp. 121-160, arXiv:hep-th/9112056.}%
 \nref\kap{A. Kapustin, ``Topological
Strings
 On Noncommutative Manifolds,'' Int. J. Geom. Meth. Mod. Phys. {\bf 1} (2004) 49-81,
 arXiv:hep-th/0310057.}
 \nref\kli{ A. Kapustin and Y. Li, `` Topological Sigma Models With $H$ Flux
 And Twisted Generalized Complex Manifolds,'' arXiv:hep-th/0407249.}%

To construct a topological field theory from a sigma-model with a
target $X$, one picks a {\it pair} of complex structures
$(J_+,J_-)$ on $X$ that obey certain conditions of $(2,2)$
supersymmetry \rock.\foot{Generalized complex geometry leads to an
extension of this construction, which we will not need here, in
which only one of the two generalized complex structures that can
be formed \gualtieri\ from the pair $(J_+,J_-)$ is integrable.
One then aims as in \pestun\ to define a topological field theory using
this  integrable generalized complex structure.}
These conditions have recently
been reinterpreted \gualtieri\ in terms of generalized complex
geometry \hhitchin. Once one has $(2,2)$ supersymmetry, there is a
standard recipe \witog, via the twisting procedure that we
reviewed in section \twisting, for constructing a topological
field theory.

If $X$ is hyper-Kahler, things simplify. The sigma-model with
target $X$ has $(4,4)$ supersymmetry, and a structure of $(2,2)$
supersymmetry can be obtained by picking a $(2,2)$ subalgebra of
the $(4,4)$ supersymmetry algebra.\foot{Two $(2,2)$ subalgebras
that differ by conjugation by $Spin(4)_{\cal R}$ are considered
equivalent.} In terms of choosing complex structures, this amounts
to the following.  $X$ has a family of complex structures
parametrized by a copy of $\Bbb{CP}^1$ that we call
$\Bbb{CP}^1_h$, and a $(2,2)$ structure can be defined by picking
a pair of points $(J_+,J_-)\in \Bbb{CP}_h^1$. The conditions in
\rock\ are automatically obeyed for such a pair. In the case of
$\MH$, this gives the family of topological field theories that is
important for the geometric Langlands program.

We can conveniently characterize as follows the topological field
theory associated with the pair $(J_+,J_-)$.  Let $\Phi: \Sigma\to
\MH$ be the bosonic field of the sigma-model.   Then the equations
of unbroken supersymmetry read
\eqn\hson{(1-iJ_+)\bar\partial\Phi=0,} and
\eqn\hsono{(1-iJ_-)\partial \Phi=0.}  Here $\partial$ and
$\bar\partial$ are the usual $(1,0)$ and $(0,1)$ parts of the
exterior derivative on $\Sigma$; and $\partial\Phi$ and
$\bar\partial \Phi$ are understood as one-forms on $\Sigma$ with
values in the pullback of the tangent bundle $TX$ (on which $J_+$
and $J_-$ act). The first condition says that the map
$\Phi:\Sigma\to X$ is holomorphic in complex structure $J_+$; the
second says that it is antiholomorphic in complex structure $J_-$.
Analogously to how we found the conditions of unbroken
supersymmetry in section \familya, these equations arise because
there are fermi fields $\chi_+$, $\chi_-$ with
$\{Q,\chi_+\}=(1-iJ_+)\bar\partial\Phi$,
$\{Q,\chi_-\}=(1-iJ_-)\partial\Phi$.

There is much more information in the equations \hson\ and \hsono\
than in their solutions. To discuss the solutions of the
equations, one usually specializes to $\Sigma$ of Euclidean
signature (the most natural case for topological field theory).
Then for generic $J_+$, $J_-$,  they imply that the map
$\Phi:\Sigma\to X$ must be constant, a result that does not depend
on $J_+$ and $J_-$.  The equations themselves, understood
algebraically, and without fixing a real structure on the
two-manifold $\Sigma$, depend on $J_+$ and $J_-$.

An important example is the case $J_+=J_-=\hat J$ (where, in our
applications, $\hat J$ will be one of the complex structures $I_w$
of $\MH$, as described in eqn. \wyre). The equations combine to
$(1-i\hat J\,)d\Phi = 0$, which imply that $d\Phi=0$. The model is
called the $B$-model in complex structure $\hat J$. If instead
$J_+=-J_-=\hat J$, we get the $A$-model in complex structure $\hat
J$. (This model really depends on the symplectic form
$\omega_{\hat J}$ associated with complex structure $\hat J$
rather than  the complex structure {\it per se}. But calling it
the $A$-model in complex structure $\hat J$ is sometimes a
convenient shorthand.) For $\Sigma$ of Euclidean signature, the
equations of the $A$-model are redundant, as eqn. \hsono\ is the
complex conjugate of \hson. Either one of them asserts that the
map $\Phi:\Sigma\to \MH$ is holomorphic in complex structure $\hat
J$. If $J_-\not= \pm J_+$, we get a model that is neither an
$A$-model nor a $B$-model, but can be reduced to an $A$-model
using generalized complex geometry, as we discuss in section
\gencpx.

We recall from eqn. \wyre\ the explicit form of the family of
complex structures on $\MH$ parametrized by $\Bbb{CP}_h^1$.  In
terms of an affine parameter $w$ for $\Bbb{CP}_h^1$, we defined a
complex structure $I_w$ in which the holomorphic coordinates are
$A_{\bar z}-w\phi_{\bar z}$ and $A_z+w^{-1}\phi_z$. Explicitly, we
found \eqn\wyrre{I_w={1-\bar w w\over 1+\bar w w}I+{i(w-\bar
w)\over 1+\bar w w}J+{w+\bar w\over 1+\bar w w}K.}
 The pair
$(J_+,J_-)$ corresponds in this parametrization to a pair
$(w_+,w_-)$.

\bigskip\noindent{\it Reduction From Four Dimensions}

Now let us compare this to what we get by reduction from four
dimensions.  We learned in section \tqftfour\ that gauge theory
leads to a family of topological quantum field theories (TQFT's)
defined on any four-manifold $M$ and parametrized by a copy of
$\Bbb{CP}^1$ that we will call $\Bbb{CP}^1_g$.   This family
certainly reduces for $M=\Sigma\times C$ to a subfamily of the
family $\Bbb{CP}_h^1\times \Bbb{CP}_h^1$ that is natural from the
two-dimensional point of view. Both families come from a twisting
procedure applied to ${\EUN}=4$ super Yang-Mills theory on a
four-manifold $M$. In this procedure, a topological field theory
is obtained from the cohomology of a suitable linear combination
$Q$ of the supersymmetries.  To get the family $\Bbb{CP}^1_g$, we
require $Q$ to be $Spin'(4)$-invariant so that the construction
works for all $M$. But to get the sigma-model of target $\MH$, we
specialize to the manifold $M=\Sigma\times C$ of reduced holonomy,
and then, as we discuss more explicitly later, a weaker condition
on $Q$ suffices, leading to the larger family $\Bbb{CP}_h^1\times
\Bbb{CP}_h^1$.

We aim to identify the four-dimensional topological field theory
family $\Bbb{CP}^1_g$ as a subfamily of the sigma-model family
$\Bbb{CP}_h^1\times \Bbb{CP}_h^1$. The embedding of $\Bbb{CP}^1_g$
in $\Bbb{CP}_h^1\times \Bbb{CP}_h^1$ can be described by functions
$w_+(t)$, $w_-(t)$. We can compute these functions by specializing
to a convenient configuration or physical state.

A very convenient way to proceed is to simply abelianize the
problem, working in a vacuum in which $G$ is broken to its maximal
torus by the expectation values of some of the untwisted scalar
fields.   In the abelian case, from \loopme, the equations of
unbroken supersymmetry in four dimensions are
\eqn\onso{\eqalign{(F+t\,d\phi)^+&=0\cr(F-t^{-1}d\phi)^-&=0.\cr}}
Here, the one-form $\phi$ is a section of $T^*M=T^*\Sigma\oplus
T^*C$, and the gauge field $A$ is locally a one-form or section of
$T^*M$. To compare to a sigma-model on $\Sigma$ with target
$\MH(G,C)$, we must take $\Sigma $ to have radius much greater
than that of $C$.  In that limit, the parts of $A$ and $\phi$ that
take values in $T^*\Sigma$ are ``massive'' and can be dropped; so
$A$ and $\phi$ can be interpreted as sections of $T^*C$, slowly
varying on $\Sigma$.

We write $y$ for a local complex coordinate on $\Sigma$ and $z$
for a local complex coordinate on $C$.  The two-form $d\bar
y\wedge d\bar z$ is selfdual, while $dy\wedge d\bar z$ is
anti-selfdual. So the first equation in \onso\ leads to
\eqn\igot{\partial_{\bar y}\left(A_{\bar z}+t \phi_{\bar
z}\right)=0,} and the second leads to
\eqn\nigot{\partial_y\left(A_{\bar z}-t^{-1}\phi_{\bar
z}\right)=0.}

By comparing these results to eqns. \hson\ and \hsono, we can read
off the functions $w_+(t)$ and $w_-(t)$.   Eqn. \igot\ asserts
that the map $\Sigma\to \MH(G,C)$ is holomorphic if we take on
$\MH(G,C)$  the complex structure in which $A_{\bar z}+t\phi_{\bar
z}$ is holomorphic.  This agrees with eqn. \hson, which asserts
such holomorphy in complex structure $I_{w_+}$, if and only if
$w_+=-t$. Similarly, eqn. \nigot\ asserts that the map $\Sigma\to
\MH(G,C)$ is antiholomorphic if we take on $\MH(G,C)$ the complex
structure in which $A_{\bar z}-t^{-1} \phi_{\bar z}$ is
holomorphic.  This agrees with eqn. \hsono, which asserts
antiholomorphy in complex structure $I_{w_-}$, if and only if
$w_-=t^{-1}$.

So the embedding of $\Bbb{CP}^1_g$ in $\Bbb{CP}_h^1\times
\Bbb{CP}_h^1$ is defined by \eqn\tyroy{\eqalign{w_+ & = -t\cr
                                              w_- & = t^{-1}.\cr}}

\bigskip\noindent{\it Some Simple Considerations}

Let us now make a few simple simple observations about this
result.

First of all, when do we get a $B$-model?  For a $B$-model, we
want $w_+=w_-$.  So the condition is $t^{-1}=-t$, which occurs
precisely for $t=\pm i$.  Since the complex structures $I_w$ for
$w=\mp i$ coincide with $J$ and $-J$, we get this way the
$B$-model in complex structure $J$ or $-J$.

When do we get an $A$-model?  For an $A$-model, the complex
structures $I_{w_+}$ and $I_{w_-}$ should be opposite.  The map
$w\to -1/\bar w$ maps $I_w$ to its opposite (this is clear from
\wyrre), so we get an $A$-model in two dimensions if $w_-=-1/\bar
w_+$, which works out to $t=\bar t$.  In other words, precisely
for real $t$, we get an $A$-model.  For example, for $t=1$ or
$-1$, we get the $A$-model in complex structure $K$ or $-K$.  For
$t=0$ or $\infty$, we get the $ A$-model in complex structure $I$
or $-I$.

Now, let us compare this family to the more complete family
$\Bbb{CP}_h^1\times \Bbb{CP}_h^1$ of the sigma-model.  At the end
of section \complexstr, we described the group ${\cal U}\cong
\Bbb{C}^*$ of diffeomorphisms of $\MH(G,C)$; it acts on the family
of complex structures $I_w$ by $w\to \lambda^{-1} w$, $\lambda\in
\Bbb{C}^*$.   The topological field theory determined by a
pair\foot{This topological field theory depends also on the
hyper-Kahler metric and $B$-field of $\MH$.  The full dependence
on all variables is determined in section \gencpx\ using
generalized complex geometry.} $(w_+,w_-)$ is the same as that
determined by $(\lambda^{-1} w_+,\lambda^{-1} w_-)$. How many
really inequivalent TQFT's can we define from the two-dimensional
point of view?  The only invariant we can form from the pair
$(w_+,w_-)$ is the ratio $q=w_+/w_-$. We see that
\eqn\yto{q=-t^2,} so all values of $q$ can be achieved, but not
quite uniquely.  The points $t$ and $-t $ on $\Bbb{CP}^1_g$ lead
to equivalent theories in two dimensions. This equivalence
reflects the action of the center of $SU(4)_{\cal R}$ (recall eqn.
\ungu).

Although all values of the invariant $q$ do come from
four-dimensional TQFT's, it is not quite true that all $\Bbb{C}^*$
orbits on $\Bbb{CP}_h^1\times \Bbb{CP}_h^1$ have representatives
with such an origin.  The missing orbits are the
$\Bbb{C}^*$-invariant points  $(0,0)$ and $(\infty,\infty)$, and
also the orbits in which $w_+$ or $w_-$, but not both, is 0 or
$\infty$. Particularly notable is the fact that the points $(0,0)$
and $(\infty,\infty)$ are not equivalent to theories that
originate in four-dimensional TQFT's. These points correspond to
the $B$-models in complex structures $I$ and $-I$.

The $B$-model in complex structure $I$ has been the starting point
in mathematical efforts -- briefly surveyed in the introduction --
to interpret the geometric Langlands program in terms of the
geometry of $\MH$.  Because the Hitchin fibration is holomorphic
in complex structure $I$, the $T$-duality on the fibers of the
Hitchin fibration (whose relation to four-dimensional $S$-duality
we review soon) maps the $B$-model of complex structure $I$ to
itself, acting  on $D$-branes via the Fourier-Mukai transform that
is the starting point in the mathematical description.  Although
the point $(0,0)$ corresponding to this $B$-model is not in the
family $\Bbb{CP}_g^1$ that comes from four-dimensional TQFT's, it
is interesting that it can be infinitesimally perturbed to give
points on the $\Bbb{C}^*$ orbits corresponding to almost any point
on $\Bbb{CP}^1_g$.  We simply perturb $(0,0)$ to $(\alpha,\beta)$
for infinitesimal $\alpha$, $\beta$; the invariant $q=w_+/w_-$ is
then $q=\alpha/\beta$, and can take any value for arbitrarily
small $\alpha,\beta$.  Perhaps this fact will lead eventually to
an understanding of the geometric Langlands program based on
perturbing from the $B$-model in complex structure $I$.  Our
approach, however, will rely on the family $\Bbb{CP}^1_g$ that
comes directly from four dimensions.

Although certain $\Bbb{C}^*$ orbits, such as the point $(0,0)$, do
not arise by specializing a four-dimensional TQFT to
$M=\Sigma\times C$, this does not meant that they cannot be
described in four-dimensional gauge theory.  In section \twisting,
we obtained the family $\Bbb{CP}^1_g$ using the supersymmetry
generator \eqn\klop{\epsilon=u\epsilon_\ell+v\epsilon_\rr} that
was constrained to be $Spin'(4)$-invariant.  If we specialize to
$M=\Sigma\times C$, there is no need to ask for $Spin'(4)$
invariance.  The holonomy group of $\Sigma\times C$ is
$SO(2)\times SO(2)\subset SO(4)$, and $Spin'(2)\times Spin'(2)$
invariance is enough.  This means that we can generalize \klop\ to
\eqn\blop{\hat\epsilon=\left(u+\tilde u\,
\Gamma^*\right)\epsilon_\ell
                 +\left(v+\tilde v\,\Gamma^*\right)\epsilon_\rr,}
                 where
\eqn\ujbyty{\Gamma^*=\cases{\Gamma_{01}&\text{in Lorentz
signature}\cr i\Gamma_{01}&\text{in Euclidean signature}\cr}} is
the operator that distinguishes the two chiralities of
two-dimensional spinors.  We will adopt Euclidean signature here,
as this is more natural for topological field theory.

The supersymmetry generators in \blop\ are precisely the ones that
obey the following conditions:
\eqn\zonop{\eqalign{\Gamma_{2367}\epsilon& = \epsilon \cr
\Gamma_{0145}\epsilon & = \epsilon.\cr}} The first is eqn. \yty,
which says that the supersymmetry generated by $\epsilon$ is
unbroken by the curvature of $C$; the second says that it is
similarly unbroken by the curvature of $\Sigma$. With this more
general starting point, it is possible to get the whole family
$\Bbb{CP}_h^1\times \Bbb{CP}_h^1$ of two-dimensional TQFT's from
four-dimensional gauge theory, though not from a four-dimensional
TQFT.

It actually is convenient to rewrite \blop\ in an eigenbasis of
$\Gamma^*$:
\eqn\bilop{\hat\epsilon=\half\left(1-\Gamma^*\right)(u'\epsilon_\ell+v'\epsilon_r)
+\half\left(1+\Gamma^*\right)(u''\epsilon_\ell+v''\epsilon_r).}
Here $(u',v')$ and $(u'',v'')$ are, respectively, homogeneous
coordinates for the two factors of $\Bbb{CP}^1_h\times
\Bbb{CP}^1_h$.  Let $\hat\delta_T$ be the extended topological
symmetry generated by $\hat \epsilon$.  To determine the two
complex structures $(J_+,J_-)$, or equivalently, to determine the
pair $(w_+,w_-)$, we just need to compute
$\hat\delta_T\chi^+_{\bar y\,\bar z}$ and
$\hat\delta_T\chi^-_{y\bar z}$. Setting these to zero will give
the generalization of \onso. To determine the generalization of
the usual formulas from eqn. \helpme, all we need to know is that
if $\tilde\chi^+=\Gamma^{\bar y\,\bar z}\chi^+_{\bar y\,\bar z}$
and $\tilde\chi^-=\Gamma^{y\bar z}\chi^-_{y\bar z}$, then
$\tilde\chi^+\Gamma^*=\tilde\chi^+$  and
$\tilde\chi^-\Gamma^*=-\tilde\chi^-$. Using this, we get
\eqn\qengo{\eqalign{\hat\delta_T\chi^+_{\bar y\,\bar
z}&=u'(F-\phi\wedge\phi)_{\bar y\,\bar z}+v'(D\phi)_{\bar y\,\bar
z}\cr \hat\delta_T\chi^-_{ y \bar z}&=v''(F-\phi\wedge\phi)_{y
z}-u''(D\phi)_{yz}.\cr}} If, therefore, $t'=v'/u'$ and
$t''=v''/u''$, then the generalizations of eqns. \igot\ and
\nigot\ are \eqn\pokodo{\eqalign{\partial_{\bar y}(A_{\bar
z}+t'\phi_{\bar z})&=0\cr \partial_y(A_{\bar
z}-(t'')^{-1}\phi_{\bar z})&=0 .\cr}} This determines the two
complex structures: \eqn\plombob{(w_+,w_-)=(-t',(t'')^{-1}).}

Now we can determine how $S$-duality acts on the full family
$\Bbb{CP}^1_h\times \Bbb{CP}^1_h$.  We learned in eqn. \duco\ that
a duality transformation $\left(\matrix{a & b \cr c & d}\right)\in
SL(2,\Bbb{Z})$ (or its analog for a gauge group that is not simply-laced)
acts on $t$ by \eqn\utlo{t\to t{c\tau+d\over
|c\tau+d|}.}  The same reasoning shows that $t'$ and $t''$
transform in precisely the same way.  {}From this and \plombob, it
follows that the action of $S$-duality on $\Bbb{CP}^1_h\times
\Bbb{CP}^1_h$ is \eqn\yfog{\eqalign{w_+ & \to w_+{c \tau+d\over
|c\tau+d|}\cr
                   w_-  & \to w_-{|c\tau+d|\over c
                   \tau+d}.\cr}}

An important special case is that the $B$-model in complex
structure $I$, which corresponds to $(w_+,w_-)=(0,0)$, is
completely invariant under duality transformations. The $A$-model
in complex structure $I$, which corresponds to
$(w_+,w_-)=(0,\infty)$, is likewise invariant. This and other
statements about two-dimensional TQFT's that are not on the
distinguished family $\Bbb{CP}^1_g$ will be useful auxiliary
tools, but as will become clear, the geometric Langlands program
is really a statement or collection of statements about the
distinguished family.

\bigskip\noindent{\it Dependence On The Metric}

In general, a two-dimensional TQFT with target $\MH(G,C)$ will
depend on the complex structure of $C$, because this influences
the hyper-Kahler structure of $\MH(G,C)$.  However, if a
two-dimensional TQFT, such as those parametrized by
$\Bbb{CP}^1_g$,  descends from a four-dimensional one, it must be
independent of the complex structure on $C$.  After all, the
four-dimensional TQFT on $M=\Sigma\times C$ is independent of the
metric on $M$ and in particular on $C$.

In some cases, we can see directly that these models are
independent of the complex structure of $C$.   At $t=\pm i$, we
get the $B$-model in complex structure $\pm J$; this complex
structure is independent of the metric of $C$, as we observed in
section \mhhyper. For real $t$, we get the $A$-model in a complex
structure that is a linear combination of complex structures $I$
and $K$. The $A$-model is determined by the corresponding
symplectic structure, which is a linear combination of symplectic
structures $\omega_I$ and $\omega_K$; we  found these in section
\mhhyper\ to be independent of the metric of $C$.

To go farther, we will use generalized complex geometry.

\subsec{The Role Of Generalized Complex Geometry}

\subseclab\gencpx

Generalized complex geometry is a natural framework for describing
the topological field theories that can be constructed by twisting
two-dimensional sigma-models, such as the sigma-model with target
$\MH$.  Our goal here is to use generalized complex geometry to
clarify a few questions from a two-dimensional point of view. What
is the family $\Bbb{CP}^1_g$ and why is it independent of the
complex structure of $C$?  How, from a two-dimensional point of
view, can we understand the canonical parameter $\Psi$ introduced
in section \canonpar?   And what geometry of $\MH$ is really
needed in these constructions?

 Let $TX$ and $T^*X$ be the tangent
and cotangent bundles of a manifold $X$. We let $\hat T=TX\oplus
T^*X$ and we write a section of $\hat T$ as $\left(\matrix{v\cr
\xi\cr}\right)$, where $v$ is a section of $TX$ and $\xi$ is a
section of $T^*X$.  $\hat T$ has a natural indefinite metric, in
which $TX$ and $T^*X$ are both null and the inner product of $TX$
and $T^*X$ is the natural one, in which the inner product  of
$\left(\matrix{v\cr \xi\cr}\right)$ and $\left(\matrix{v'\cr
\xi'\cr}\right)$ is $v^i\xi'_i+v'{}^i\xi_i$.

\nref\lindst{  U.~Lindstrom, R.~Minasian, A.~Tomasiello and M.~Zabzine,
  ``Generalized Complex Manifolds and Supersymmetry,''
  Commun.\ Math.\ Phys.\  {\bf 257} (2005) 235, hep-th/0405085.}%
  \nref\lindstwo{   A.~Bredthauer, U.~Lindstrom, J.~Persson and M.~Zabzine,
  ``Generalized Kahler Geometry from Supersymmetric Sigma Models,''
  hep-th/0603130.      }%
 \nref\toma{M. Grana, R. Minasian, M. Petrini, and A. Tomasiello,
``Supersymmetric Backgrounds From Generalized Calabi-Yau
Manifolds,'' JHEP 0408:046 (2004), arXiv:hep-th/0406137; ``Type II
Strings And Generalized Calabi-Yau Manifolds,'' Comptes Rendus
Physique {\bf 5} (2004) 979-986,
arXiv:hep-th/0409176.}%
\nref\zoma{M. Grana, R. Minasian, M. Petrini, and A. Tomasiello,
``Generalized Structures Of ${\EUN}=1$ Vacua,'' JHEP 0511:020
(2005), arXiv:hep-th/0505212.}%
\nref\woma{M. Grana, J. Louis, and D. Waldram, JHEP 0601:008
(2006), arXiv:hep-th/0505264.}%
 A generalized
almost complex structure on a manifold $X$ is a linear
transformation ${\cal I}$ of $\widehat T=TX\oplus T^*X$ that
preserves the metric and obeys ${\cal I}^2=-1$.  If a certain
integrability condition is obeyed, it is called a generalized
complex structure \hhitchin.  (For more detail, see Gualtieri's
thesis \gualtieri, as well as \refs{\kap-\lindstwo,\pestun} for applications
to sigma-models and \refs{\toma-\woma} for applications to
supergravity.) One basic example of a generalized complex
structure is \eqn\momo{{\cal I}_{\hat J}=\left(\matrix{ {\hat J} &
0 \cr 0 & -{\hat J}\,^t\cr}\right),} where ${\hat J}:TX\to TX$ is
an ordinary complex structure, and ${\hat J}\,^t:T^*X\to T^*X$ is
its transpose.  If $\omega$ is a symplectic structure, then a
second basic example of generalized complex structure is given by
\eqn\zomo{{\cal I}_\omega=\left(\matrix{0 & -\omega^{-1}\cr \omega
& 0\cr}\right).} Here we regard $\omega$ as a linear map from $TX$
to $T^*X$; $\omega^{-1}$ is the inverse map from $T^*X$ to $TX$.
The topological field theory associated with ${\cal I}_{\hat J}$
is the $B$-model in complex structure ${\hat J}$, and the one
associated with ${\cal I}_\omega$ is the $A$-model with symplectic
structure $\omega$.\foot{The $A$-model is most commonly considered
on a Kahler manifold, and then $\omega$ is the Kahler form.
Because the $A$-model makes sense on symplectic manifolds more
generally, we simply refer to $\omega$ as the symplectic form.}

In each of these cases, the $B$-field vanishes.  A $B$-field can
be turned on as follows.  For any closed two-form $B_0$, let
\eqn\hyt{{\cal M}(B_0)=\left(\matrix{1 & 0 \cr
                                   B_0 & 1 \cr}\right).}
Then if ${\cal I}$ is an integrable generalized complex structure,
so is \eqn\nyt{{\cal I}(B_0)={\cal M}(B_0){\cal I}{\cal
M}(B_0)^{-1}.} The transformation ${\cal I}\to {\cal I}(B_0)$ has
the effect of shifting the $B$-field by $B_0$.  So in particular
the topological field theory derived from ${\cal I}_{\hat J}(B_0)$
is the $B$-model with complex structure ${\hat J}$ and $B$-field
$B_0$, and ${\cal I}_\omega(B_0)$ is similarly related to the
$A$-model with symplectic structure $\omega$ and $B$-field $B_0$.
Conjugation by ${\cal M}(B_0)$ is called the $B$-field transform.

It is shown in chapter 6 of \gualtieri\ that the conditions \rock\
for a sigma model to have $(2,2)$ supersymmetry are equivalent to the existence of a pair
of generalized complex structures obeying a certain compatibility
condition.  If the $(2,2)$ model has anomaly-free $R$-symmetries, then it can be twisted
in two ways to make a topological field theory.  It is believed that, in general, each of these
topological field theories is determined by one of the two generalized complex structures and independent
of the second.
(For example, if the target is a Kahler manifold and the $B$-field is flat, then the two
generalized complex structures are ${\cal I}_{\hat J}$ and ${\cal I}_\omega$, as described above;
the two topological field theories are the $B$-model, which is determined by the first, and the
$A$-model, which is determined by the second.)  Arguments supporting this claim can be found in
\refs{\kli,\pestun}, though a complete proof is not yet known.

In the case relevant to us that the $(2,2)$ model
comes from a hyper-Kahler metric $g$ with a pair of points
$J_+,J_-\in \Bbb{CP}^1_h$, there is a slight simplification in the formulas
of \gualtieri,
because the $B$-field can be taken to vanish (and restored later
by a $B$-field transform).  We let $\omega_\pm$ be the two
symplectic structures $\omega_\pm = gJ_\pm$.  One of the two
generalized complex structures determined by the pair $J_+$, $J_-$
with the hyper-Kahler metric $g$ is then, according to eqn. 6.3 of
\gualtieri, \eqn\ugg{{\cal J}={1\over 2}\left(\matrix{J_++J_- &
-(\omega_+^{-1}-\omega_-^{-1})\cr \omega_+-\omega_- &
-(J_+^t+J_-^t)\cr}\right).} The second one, not relevant for us,
is obtained by reversing the sign of $J_-$ and $\omega_-$.

Using \wyrre\ and $(w_+,w_-)=(-t,t^{-1})$, we have \eqn\ytro{
\eqalign{ J_+ & = {1-\bar t t\over 1+\bar t t}I-{i(t-\bar t)\over
1+\bar t t}J-{t+\bar t\over 1+\bar t t}K  \cr J_- & = -{1-\bar
tt\over 1+\bar t t}I-{i(t-\bar t)\over 1+\bar t t}J+{t+\bar t\over
1+\bar t t}K  .\cr}} The associated symplectic structures are
 \eqn\bytro{
\eqalign{ \omega_+ & ={\rm Im}\,\tau\left( {1-\bar t t\over 1+\bar
t t}\omega_I-{i(t-\bar t)\over 1+\bar t t}\omega_J-{t+\bar t\over
1+\bar t t}\omega_K\right) \cr \omega_- & = {\rm
Im}\,\tau\left(-{1-\bar tt\over 1+\bar t t}\omega_I-{i(t-\bar
t)\over 1+\bar t t}\omega_J+{t+\bar t\over 1+\bar t
t}\omega_K\right) .\cr}} (The factor of ${\rm Im}\,\tau$ was
obtained in \inog.) So we compute that the generalized complex
structure determined by this data is \eqn\islo{{\cal I}_t={1\over
1+\bar t t}\left(\matrix{-i(t-\bar t)J & -({\rm
Im}\,\tau)^{-1}((1-\bar t t)\omega_I^{-1}-(t+\bar
t)\omega_K^{-1})\cr {\rm Im}\,\tau ((1-\bar t t)\omega_I-(t+\bar
t)\omega_K)& i(t-\bar t)J^t\cr}\right).} (To get this formula, it
helps to know that on a hyper-Kahler manifold, if $a^2+b^2+c^2=1$,
then
$(a\omega_I+b\omega_J+c\omega_K)^{-1}=a\omega_I^{-1}+b\omega_J^{-1}+c\omega_K^{-1}$.)

The first important observation is that $I, K$, and $\omega_J$
have disappeared.  ${\cal I}_t$ depends only on $J$, $\omega_K$,
and $\omega_I$, or equivalently on $J$ and the holomorphic
two-form $\Omega_J=\omega_K+i\omega_I$ (and its complex
conjugate). But this data, as we noted in section \mhhyper, is
independent of the metric of $C$. This explains, from the
two-dimensional point of view, why the family of topological field
theories that we get by dimensional reduction from four dimensions
does not depend on the metric.

An analogous family of generalized complex structures can be
constructed on any hyper-Kahler manifold $X$ and is described in
section 4.6 of \gualtieri.  As is explained there (Proposition
4.34), the generic element in such a family is a $B$-field
transform of a generalized complex structure derived from a
symplectic structure. Indeed, a small calculation shows that
\eqn\huffy{{\cal I}_t={\cal M}(B_0)\left(\matrix{0 &
-\omega_0^{-1}\cr \omega_0 & 0\cr}\right){\cal M}(B_0)^{-1},}
where \eqn\nutty{\eqalign{\omega_0 & = {\rm Im}\,\tau{1-\bar
t{}^2t^2\over (1+t^2)(1+\bar t{}^2)}\left(\omega_I-{t+\bar t\over
1-\bar t t}\omega_K\right)\cr
                      B_0 &= -{\rm Im}\,\tau\left.{i(t^2-\bar t{}^2)\over(1+t^2)(1+\bar
                      t{}^2)}\right.
                      \left(\omega_I+{1-\bar t t\over t+\bar t}\omega_K\right).\cr}}
Therefore, for $t\neq \pm i$ the TQFT derived from the generalized
complex structure ${\cal I}_t$ is equivalent to an $A$-model with
symplectic structure $\omega_0$ and $B$-field $B_0$.  This fact
will enable us to understand from a two-dimensional point of view
the canonical parameter $\Psi$ introduced in section \canonpar.

In general, the $A$-model with symplectic form $\omega_0$ and
$B$-field $B_0$ depends only on the cohomology class
$[B_0+i\omega_0]$.  (On a Kahler manifold, this is called the
complexified Kahler class.)  In the present problem, since the
cohomology class of $\omega_K$ vanishes (as we explained in
section \mhhyper), we have \eqn\gutty{[B_0+i\omega_0]=-i\,{\rm
Im}\,\tau\left(t-t^{-1}\over t+t^{-1}\right) [\omega_I].} Thus,
when the four-dimensional $\theta$ angle vanishes (we took it to
vanish by starting with ${\cal I}_t$ rather than a $B$-field
transform thereof), the model depends on ${\Im}\tau$ and $t$ only
in the combination that appears in \gutty.

The four-dimensional $\theta$ angle induces, as we explained at
the end of section \mhhyper, an additional contribution
$B'=-(\theta/2\pi)\omega_I=-({\rm Re}\,\tau)\,\omega_I$ to the
two-dimensional $B$-field. This can be incorporated in the
generalized complex structure simply by conjugating with ${\cal
M}(B')$.  The resulting model depends on the cohomology class
$[B_0+B'+i\omega_0]$, which is $-[\omega_I]$ times
\eqn\loopy{\Psi={\rm Re}\,\tau +i\,{\rm Im}\,\tau
\left(t-t^{-1}\over t+t^{-1}\right).} This gives a two-dimensional
interpretation of why the model depends on $t$ and $\tau$ only in
the combination $\Psi$.

\bigskip\noindent{\it A Few Loose Ends}

Finally, let us wrap up a few loose ends.

First of all, if $t$ is real, the TQFT determined by complex
structures $J_+(t),J_-(t)$ is an $A$-model to begin with, and the
above argument showing that $\Psi$ is the only relevant parameter
did not really require generalized complex geometry.   We could have based our
reasoning on the real case together with holomorphy in $t$.     We chose not to do this since
 we think that generalized
complex geometry is a natural framework for understanding this problem, even though there are
some technical gaps in the current understanding of it.

Second, in the above analysis, we used the fact that the $A$-model
only depends on the cohomology class of $B+i\omega$.  This is
proved by writing the action as $\{Q,V\}+\int_\Sigma
\Phi^*(\omega-iB)$, where  $\Phi:\Sigma\to X$ is the sigma-model
map.

In the above derivation, the $B$-field is not just of type $(1,1)$
(in a complex structure in which the symplectic form is Kahler);
it also has components of type $(2,0)\oplus (0,2)$. In fact, the
$A$-model is sensitive to all components of the $B$-field,
including the part of type $(2,0)\oplus (0,2)$, but this point may
require some clarification.

If $\Phi:\Sigma\to X$ is a holomorphic map, and $B$ is a form on
$X$ of type $(2,0)\oplus (0,2)$, then $\Phi^*(B)=0$.  This makes
the $(2,0)\oplus (0,2)$ part of $B$ appear irrelevant, if one
interprets the $A$-model purely as a mechanism for computing
correlation functions by summing over holomorphic maps. But if one
considers branes (as we most definitely will do to understand the
geometric Langlands program), one immediately sees that all parts
of the Hodge decomposition of $B$ are relevant.  A Lagrangian
submanifold $N\subset X$ endowed with a Chan-Paton line bundle
${\cal L}$ of curvature $F$ gives an $A$-brane if $F+B|_N=0$. This
condition is certainly sensitive to the $(2,0)\oplus (0,2)$ part
of $B$.

Finally, our analysis showing that $\Psi$ is the only relevant
parameter is really not valid at $t=\pm i$, because of poles in
the formulas.  At these values of $t$, the model is actually not
the $B$-field transform of an $A$-model; it is a $B$-model in
complex structure $\pm J$. To complete our analysis for these
values of $t$, we will argue directly using the $B$-model.

At $t=\pm i$, we have $\Psi=\infty$, independent of $\tau$, so to
complete the demonstration that $\Psi$ is the only relevant
parameter, we must show that $\tau$ is completely irrelevant at
$t=\pm i$. ${\Im}\,\tau$ controls the Kahler class of $\MH$, and
this is certainly irrelevant in the $B$-model.  Varying ${\rm
Re}\,\tau$ adds to the $B$-field a multiple of $\omega_I$. To show
that this term is irrelevant in the $B$-model with complex
structure $J$, we write $\omega_I=-i\Omega_J+i\omega_K$. The
contribution from $\omega_K$ is irrelevant because the form
$\omega_K$ is exact. The contribution from $\Omega_J$ is
irrelevant because $\Omega_J$ is a form of type $(2,0)$.  But in
the $B$-model, the $B$-field contribution $-i\int_\Sigma\Phi^*(B)$
can be written as $\{Q,\dots\}$ if $B$ is of type $(1,1)$ or
$(2,0)$.  (By contrast, a $(0,2)$ component of the $B$-field does
affect the category of $B$-branes \kap.) Of course, in complex
structure $-J$, we make the same argument, starting with
$\omega_I=i\bar\Omega_J-i\omega_K$.

\subsec{Some Specializations}

\subseclab\somespec

In the rest of this paper except section \genotics, we focus
primarily on the distinguished values $\Psi=0$ and $\Psi=\infty$,
and the duality \eqn\rtu{S=\left(\matrix{0 & 1/\sqrt {n_{\frak
g}}\cr -\sqrt {n_{\frak g}}& 0\cr}\right)} that maps $\Psi\to
-1/n_{\frak g}\Psi$ and hence exchanges $\Psi=0$ and
$\Psi=\infty$. We recall that $n_{\frak g}=1$ for simply-laced
$G$. When we speak of $S$-duality, we will be a little imprecise
about whether we mean the full duality group $\Gamma$ generated by
$S$ and by \eqn\ntu{T=\left(\matrix{1 & 1\cr 0& 1\cr}\right),} or
just the subgroup generated by $S$. In fact,  a complete story
requires including also $T$, but $S$ is the main actor in the
geometric Langlands program.

To get $\Psi=\infty$, we take $t=i$ and thus $(w_+,w_-)=(-i,-i)$.  The resulting model
is the $B$-model in complex structure $J$.  This
statement is unaffected by the choice of $\tau$, but it is
convenient to take $\tau$ imaginary.

Applying the transformation $S$, we find that $\tau$ remains
imaginary, and $t$ is mapped to $t'=-t ({ \tau}/|\tau|)=-it=1$.
This leads to $(w_+,w_-)=(-1,1)$, and hence to the $A$-model in
complex structure $K$.  Thus $S$-duality implies that the
$B$-model for $\MH(G,C)$ in complex structure $J$ is equivalent to
the $A$-model for $\MH(\LG,C)$ in complex structure $K$. If we
combine $S$ with a transformation ${\EUI}:\varphi\to i\varphi$
that rotates complex structure $K$ back to complex structure $J$,
we get a mirror symmetry that maps the $B$-model in complex
structure $J$ to the $A$-model in complex structure $J$.  So the
complex manifolds $\MH(G,C)$ and $\MH(\LG,C)$ are a mirror pair in
complex structure $J$.

The automorphism ${\EUI}$ that we used here is not quite natural,
however, in the sense that it moves us out of the four-dimensional
family of TQFT's parametrized by $\Bbb{CP}^1_g$. It has to move us
out of this family, since the $A$-model in complex structure $J$
is not a member of this family!

Additionally, we already noted in discussing \yfog\ that $S$ (and
the whole duality group) leaves invariant the $B$-model in complex
structure $I$, and likewise the $A$-model in that complex
structure.  And finally, since $S$ exchanges the $B$-model in
complex structure $J$ with the $A$-model in complex structure $K$,
we may guess that it likewise exchanges the $A$-model in complex
structure $J$ with the $B$-model in complex structure $K$. In
fact, the automorphism ${\EUI}$ relates these two statements, but
instead of arguing this way, let us use eqn. \yfog.  The $A$-model
in complex structure $J$ corresponds to $(w_+,w_-)=(-i,i)$.  This
is mapped by $S$ to $(w_+,w_-)=(1,1)$, which indeed corresponds to
the $B$-model in complex structure $K$. We summarize these
statements in a table.

\vskip 1cm

\centerline{
\vbox{\hsize=2truein\offinterlineskip\halign{\tabskip=2em plus2em
minus2em\hfil#\hfil&\vrule#&\hfil#\hfil\tabskip=0pt\cr Model&depth
6pt&$S$-Dual\cr \noalign{\hrule} ${ I_B}$&height14pt&${ I_B}$\cr
${I_A}$&height14pt&${ I_A}$\cr ${ J_B}$&height14pt&${ K_A}$\cr ${
J_A}$&height14pt&${ K_B}$\cr ${ K_B}$&height14pt&${ J_A}$\cr ${
K_A}$&height14pt&${ J_B}$\cr \noalign{\medskip} }}} \nobreak
\centerline{ \vbox{\hsize=3truein \noindent Table 2.  A model and
its $S$-dual. ${ I_B}$, for example, represents the $B$-model in
complex structure $I$, which is its own $S$-dual.}}

\subsec{$S$-Duality Of the Hitchin Fibration}

\subseclab\sdualhitfib

Our aim here and in section \twodimint\ is to show that
$S$-duality acts classically on the base of the Hitchin fibration
$\pi:\MH\to {\EUBB}$, while acting by $T$-duality on the fibers.

We recall that the base of the Hitchin fibration is a complex
vector space ${\EUBB}$. As described in section \hitchfib, the
linear functions on ${\EUBB}$ are the commuting Hamiltonians of
the integrable system $\MH$. They are of the form
\eqn\utyr{H_{{\cal P},\alpha}=\int_C\alpha {\cal P}(\varphi),}
where ${\cal P}$ is one of the fundamental homogeneous invariant
polynomials on the Lie algebra $\g$ of $G$. If ${\cal P}$  of
degree $n$, then $\alpha$ is an element of $H^1(C,K_C^{1-n})$. The
group ${\cal U}\cong \Bbb{C}^*$, which acts on $\varphi$ by
$\varphi\to\lambda\varphi$, acts as ${\cal P}\to \lambda^n{\cal
P}$.  This group action endows the holomorphic polynomial
functions on ${\EUBB}$ with the structure of a graded ring.

Roughly speaking, we want to show that $S$-duality preserves not
only the Hitchin fibration but also this graded ring of functions
on the base.  This statement is trivial for $T$ (which acts
trivially on everything in sight) so it is really a statement
about the duality transformation $S$.  $S$  maps $G$ to
$^L\negthinspace G$, and the claim is that it maps the sigma-model
of $\MH(G,C)$ to that of $\MH(^L\negthinspace G,C)$, mapping one
Hitchin fibration to the other, and preserving the grading.

 However, we need to explain exactly how to interpret
these notions quantum-mechanically. After compactifying the
topological gauge theory on $\Sigma\times C$, we pick a point
$z\in \Sigma$ and then, if ${\hat P}$ is any polynomial in the
$H_{{\cal P},\alpha}$'s, we evaluate ${\hat P}$ at $z$ to get a
local operator ${\EUO}_{\hat P}(z)$ in the effective
two-dimensional sigma-model on $\Sigma$. The operators of this
type, for any $z$, form a graded ring ${{\eurm R}}$; holomorphy in
$\varphi$ ensures that there are no singularities in products of
these operators.   The claim we wish to justify is that $S$
establishes an isomorphism between the graded rings ${{\eurm
R}}(G)$ and ${{\eurm R}}(^L\negthinspace G)$ for the two dual
groups. (There are no local operators analogous to the
${\EUO}_{\hat P}$'s that can be similarly used to measure the
fibers of the Hitchin fibration, so there is no analogous way to
show that $S$-duality acts classically on the fibers. It hardly
can, given its relation to mirror symmetry!)

The claim can be established directly by considering the
sigma-model with target $\MH$.  We simply note that in the
$B$-model of complex structure $I$, the ring we have just defined
is the same as the subring of ghost number zero (or cohomological
degree zero) of what is customarily called the chiral ring
\ref\vlw{W. Lerche, C. Vafa, and N. Warner, ``Chiral Rings In
${\EUN}=2$ Superconformal Theories,'' Nucl. Phys. {\bf B324}
(1989) 427-474}. In general, in the $B$-model with target $X$, the
chiral ring is the bi-graded ring $H^q(X,\wedge^pTX)$ ($TX$ is the
tangent bundle of $X$). In particular, the subring of the chiral
ring with $p=q=0$  is just the ring of holomorphic functions on
$X$. But in complex structure $I$, a holomorphic function on $\MH$
must be constant on the fibers of the Hitchin fibration (which are
compact complex submanifolds) and hence must come from a
holomorphic function on the base.  So the holomorphic functions on
$\MH$, in complex structure $I$, are precisely the holomorphic
functions on ${\EUBB}$.  Now the duality transformation $S$
preserves the $B$-model in complex structure $I$, as we have
learned above, or more precisely it maps this $B$-model for
$\MH(G,C)$ to the corresponding model for $\MH(^L\negthinspace
G,C)$. So $S$ maps the chiral ring of $\MH(G,C)$ to that of
$\MH(^L\negthinspace G,C)$. Equivalently, the base ${\EUBB}$ of
the Hitchin fibration for $G$ maps to the analogous base
$^L\negthinspace{\EUBB}$ for $^L\negthinspace G$.  Moreover, $S$
commutes with the $R$-symmetry group ${\cal U}_1$ (the unitary
subgroup of ${\cal U}\cong \Bbb{C}^*$).  This plus holomorphy
ensures that the action of $S$ is compatible with the $\Bbb{C}^*$
action on the two sides.

What we have just seen is a typical example of exploiting the fact
that $\MH$ is a hyper-Kahler manifold.  We made the argument in
the $B$-model of complex structure $I$ even though we will apply
the results to the $B$-model in complex structure $J$ and the
$A$-model in complex structure $K$.

 One can argue the same result more explicitly
starting from four-dimensional gauge theory. The four-dimensional
argument gives more information; it will tell us precisely how the
duality acts on the base of the Hitchin fibration.

We begin with ${\EUN}=4$ super Yang-Mills theory
 on $\Bbb{R}^4$. For any positive integer $n$, let ${\EUD}_n$
(or $\EUD_n(G)$ if we wish to specify the gauge group) be the
space of local operators of the following form. An element of
${\EUD}_n$ is a $G$-invariant polynomial function ${\hat
Q}(\phi_0,\dots,\phi_5)$ in the $\phi_i$, homogeneous of degree
$n$, and moreover  constrained as follows: under $Spin(6)_{\cal
R}=SU(4)_{\cal R}$, $\hat Q$ transforms in the representation
${\rm Sym}_0^n{\bf 6}$ (the representation consisting of traceless
$n^{th}$ order polynomials in the ${\bf 6}$ or in other words the
irreducible representation whose highest weight is $n$ times the
highest weight of the ${\bf 6}$).

We interpret an element of ${\EUD}_n$ as a local operator that can
be evaluated at an arbitrary point $x\in \Bbb{R}^4$.  These
operators are precisely the  ``half-BPS'' operators of  dimension
$n$ in ${\EUN}=4$ super-Yang-Mills theory, that is, the operators
that are annihilated by one-half of the superconformal symmetries
that leave  fixed the point $x$.

Because the space ${\EUD}_n$ is defined by an intrinsic criterion
in terms of the action of supersymmetry, $S$-duality must
establish an isomorphism between $\EUD_n(G)$ and $\EUD_n(\LG)$. In
particular, if the Lie algebra of $G$ is selfdual (${\EUN}=4$
super Yang-Mills theory on $\RR^4$ depends only on the Lie algebra
of $G$), one could hope that $S$-duality acts trivially on
${\EUD}_n$. This is indeed true for simply-laced $G$, if the
scalars are normalized to have canonical kinetic energy, as can be
seen from string-theoretic derivations of $S$-duality. For $G_2$
and $F_4$, the duality transformation $S$ acts on ${\EUD}_n$ as a
nontrivial involution \AKS.

If the graded sum ${\EUD}=\oplus_n {\EUD}_n$ were  a graded ring
under operator products, we would use this ring to draw the
conclusions we want about the Hitchin fibration.  This actually is
not true, mainly because of the condition on how $\hat Q$
transforms under $Spin(6)_{\cal R}$.

We do, however, get a graded ring if we fix a Weyl chamber of
$Spin(6)_{\cal R}$  and restrict to those operators in ${\EUD}_n$
that transform as highest weight vectors. Indeed, consider the
subgroup $Spin(2)\times Spin(4)\subset Spin(6)_{{\cal R}}$, where,
in our construction of the twisted TQFT,  $Spin(2)$ rotates the
$\phi_2-\phi_3$ plane and $Spin(4)$ rotates the untwisted scalars.
(Thus, $Spin(2)$ is precisely the group ${\cal U}_1$ that acts by
phase rotations of $\varphi$.)   Then in the representation of
$Spin(6)$ spanned by the scalars $\phi_0,\dots,\phi_5$, the field
$\varphi$ can be interpreted as a highest weight vector, for some
choice of Weyl chamber. And the subspace ${\EUD}'_n\subset
{\EUD}_n$ of highest weight vectors consists of the
gauge-invariant polynomials $P(\varphi)$. Alternatively, these are
the $Spin(4)$-invariants with the largest possible eigenvalue of
$Spin(2)$.  Since $S$-duality commutes with $Spin(6)_{\cal R}$ and
hence with $Spin(2)\times Spin(4)$, these conditions are invariant
under $S$-duality.   Hence, $S$-duality maps ${\EUD}'_n$ to
itself.

Now ${\EUD}'=\oplus_{n\geq 0}{\EUD}'_n$ {\it does} form a graded
ring under operator products.  It is part of what is usually
called the chiral ring of the gauge theory (the operators
$P(\varphi)$ are chiral superfields with respect to one of the
supersymmetries, and their products are also chiral superfields).
The action of $S$ on ${\EUD}'$ preserves its structure as a graded
ring, since this structure is part of the operator product
expansion of the theory.

 The transformation of gauge-invariant polynomials in $\varphi$
under $S$-duality is, for dimensional and symmetry reasons, not
affected by twisting and compactification.  Moreover, a commuting
Hamiltonian $H_{{\cal P},\alpha}$ of degree $n$ in $\varphi$ is
obtained by integrating an element ${\cal P}(\varphi)\in\EUD'_n$
over $C$ (more exactly, over $z\times C\subset \Sigma\times C=M$,
for some point $z\in\Sigma$) with some weight $\alpha\in
H^1(C,K_C^{1-n})$. This process of integration over $C$ does not
affect the transformation under $S$-duality, so the action of
$S$-duality on the graded ring of holomorphic functions on the
base ${\EUBB}$ of the Hitchin fibration is simply determined by
the action on the four-dimensional graded ring $\EUD'$. In
particular, therefore, the transformation $S$ maps the base of the
Hitchin fibration of $G$ to the base of the Hitchin fibration for
$^L\negthinspace G$, intertwining between the two $\Bbb{C}^*$
actions.

\subsec{Two-Dimensional Interpretation Of $S$-Duality}

\subseclab\twodimint

Now we can obtain a useful characterization of how the operation
$S$ acts on the sigma-model of target $\MH$.  Because the
$S$-transformation reduces essentially to mirror symmetry, we will
follow the insight of Strominger, Yau, and Zaslow \ref\syz{A.
Strominger, S.-T. Yau, and E. Zaslow, ``Mirror Symmetry Is $T$
Duality,'' Mucl. Phys. {\bf B479} (1996) 243-259.}: the key point
is to understand the action of $S$ on zerobranes. In this
analysis, and in much of our later discussion of branes, it will
suffice to consider branes in the effective two-dimensional
sigma-model of target $\MH$.  How to interpret these branes in the
underlying gauge theory is briefly discussed in section \zapendix\
and will be further described in \witfur. The picture we describe
has been argued \refs{\vafa,\hm} previously and has been exploited
mathematically \thadhau.

We start with a zerobrane ${\cal B}_p$ supported at a point $p\in
\MH(G,C)$. Like any brane, it corresponds to a boundary condition
in the sigma-model.  The specific meaning of a zerobrane supported
at $p$ is as follows: if this boundary condition is imposed on a
component $\gamma$ of the boundary of $\Sigma$ (in which case we
say that $\gamma$ is {\it labeled} by ${\cal B}_p$), this means
that the sigma-model map $\Phi:\Sigma\to\MH(G,C)$ maps $\gamma$ to
the point $p$.

A point is a complex submanifold in every complex structure. So a
zerobrane is what we will call a brane of type $(B,B,B)$; that is,
it is a $B$-brane in each of the three complex structures $I,J,K$.

\ifig\trampo{\bigskip  A local operator ${\EUO}(z)$ inserted at a
point $z$ in the interior of a disc (shaded region) whose boundary
is labeled by a brane ${\cal B}$.  As $z$ approaches the boundary
of the disc, ${\EUO}(z)$ may approach a complex constant. If this
holds for suitable operators ${\EUO}_{H_\alpha}$, we say that
${\cal B}$ is supported on the fiber $\CMF$ of the Hitchin
fibration. } {\epsfxsize=2.5in\epsfbox{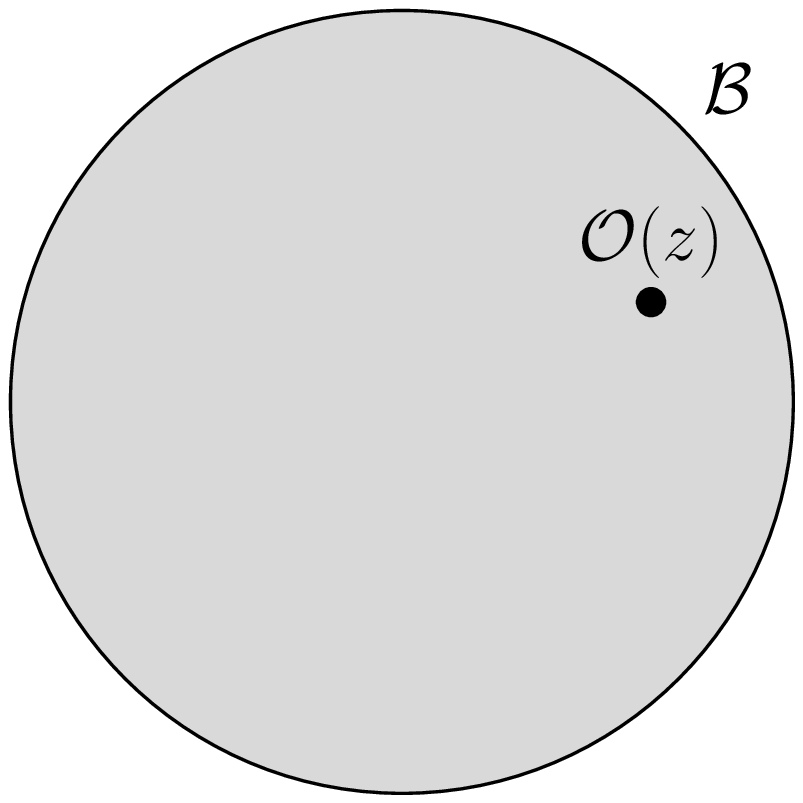}}

Because the zerobrane is supported at a single point, it lies on a
single fiber $\CMF_p$ of the Hitchin  fibration.  This fiber is
characterized by equations $H_\alpha = h_\alpha$, where $H_\alpha$
are the commuting Hamiltonians and $h_\alpha$ are complex
constants.  Quantum mechanically, we say that a brane ${\cal B}$
is supported on a fiber ${\CMF}_p$
 if  the
operators ${\EUO}_{H_\alpha}(z)$, as $z$ approaches a boundary
labeled by the brane ${\cal B}$, approach the $c$-numbers
$h_\alpha$ (\trampo).  The brane ${\cal B}_p$ certainly has this
property.

Now we consider the duality operation $S$.  It replaces $G$ with
the Langlands dual group $^L\negthinspace G$, and so turns the
zerobrane ${\cal B}_p$ into a dual brane $\tilde{\cal B}_p$ in the
sigma-model whose target is $\MH(^L\negthinspace G,C)$. Looking
back at Table 2, we see that, since ${\cal B}_p$ is a brane of
type $(B,B,B)$, $\tilde {\cal B}_p$ will be a brane of type
$(B,A,A)$, that is, a $B$-brane in complex structure $I$ and an
$A$-brane in complex structures $J$ and $K$.

The transformation $S$ gives a map $\Xi:{\EUBB}\to
{}^L\negthinspace{\EUBB}$, where ${\EUBB}$ is the base of the
Hitchin fibration for $G$ and $^L\negthinspace{\EUBB}$ is its
analog for $^L\negthinspace G$.  Concretely, if a point $v\in
{\EUBB}$ is defined by equations $H_\alpha-h_\alpha=0$, and $S$
maps $H_{\alpha}$ to $^LH_{\alpha}$, then $\Xi(v)$ is defined by
equations $^LH_\alpha-h_\alpha=0$. As explained above, the map
$\Xi$ is the identity map for simply-laced $G$.

Applying $S$ to the situation of \trampo, we observe that if for a
brane ${\cal B}$, the operator ${\EUO}_{H_\alpha}(z)$ approaches
$h_\alpha$ as $z$ approaches the boundary, then for the dual brane
$\tilde {\cal B}$, the dual operator ${\EUO}_{\tilde H_\alpha}(z)$
likewise approaches $h_\alpha$ in the same limit.
 Hence, if ${\cal
B}$ is supported on a fiber ${\CMF}$ of the the Hitchin fibration
for $G$, then its $S$-dual is a brane $\tilde {\cal B}$ supported
on the corresponding fiber $\tilde {\CMF}=\Xi({\CMF})$ of the
Hitchin fibration for $^L\negthinspace G$.

Therefore, $\tilde{\cal B}_p$ is a brane of type $(B,A,A)$ that is
supported on a fiber $\tilde {\CMF}$ of the dual Hitchin
fibration. Let us focus on complex structures $J$ and $K$ in which
$\tilde{\cal B}_p$  is an $A$-brane.  In general, the support of
an $A$-brane in a space $X$ is at least middle-dimensional. The
most familiar $A$-branes are supported on middle-dimensional
Lagrangian submanifolds; there are also more exotic coisotropic
$A$-branes, with support above the middle dimension \ref\kapor{A.
Kapustin and D. Orlov, ``Remarks On $A$-Branes, Mirror Symmetry,
and the Fukaya Category,'' J. Geom. Phys. {\bf 48} (2003) 84-99,
arXiv:hep-th/019098}. But the middle dimension is the lower bound
on the dimension of the support of an $A$-brane.

However, an $A$-brane such as $\tilde {\cal B}_p$ that is
supported on the fiber $\tilde {\CMF}$ has support that is {\it at
most} middle-dimensional. To reconcile these constraints,
$\tilde{\cal B}_p$ must have support that is precisely $\tilde
{\CMF}$. Indeed, as explained in section \hitchfib, $\tilde
{\CMF}$ is holomorphic in complex structure $I$ and Lagrangian
with respect to symplectic structures $\omega_J$ and $\omega_K$.
So a brane wrapped on $\tilde {\CMF}$ and endowed with a flat
unitary Chan-Paton bundle ${\cal L}$ is indeed a brane of type
$(B,A,A)$.

Let us next compare the moduli on the two sides, restricting
ourselves to branes supported on ${\CMF}$ on one side or on
$\tilde {\CMF}$ on the other. With this restriction, the moduli
space of ${\cal B}_p$ is just a copy of ${\CMF}$: $p$ can be any
point in ${\CMF}$.

Meanwhile, the moduli of the dual brane $\tilde{\cal B}_p$ are a
complex torus  ${\EUJ}_{\tilde {\CMF}}$ that parametrizes flat
Chan-Paton bundles on $\tilde {\CMF}$ of rank 1. (The rank is 1,
as otherwise the moduli space of $\tilde{\cal B}_p$ would have
dimension greater than that of ${\cal B}_p$, and $S$-duality could
not hold.) ${\EUJ}_{\tilde {\CMF}}$ has the same dimension as
${\CMF}$ or $\tilde {\CMF}$. Clearly, $S$-duality establishes an
isomorphism between ${\CMF}$ and ${\EUJ}_{\tilde {\CMF}}$.

Of course, we could run this backwards.  A zerobrane in
$\MH(^L\negthinspace G,C)$ at a point in $\tilde {\CMF}$ is
similarly transformed by $S$ to a brane in $\MH(G,C)$ that is
wrapped on ${\CMF}$, and endowed with a flat unitary Chan-Paton
bundle. So $S$-duality likewise establishes an isomorphism between
$\tilde {\CMF}$ and ${\EUJ}_{\CMF}$, the moduli space of flat
Chan-Paton bundles on ${\CMF}$ of rank 1.

This picture is  SYZ duality between torus fibrations.  ${\CMF}$
parametrizes flat Chan-Paton bundles of rank one on $\tilde
{\CMF}$, and vice-versa.  Informally, we can describe this by
saying that the operation $S$ acts by a map
$\Xi:{\EUBB}\to{}^L\negthinspace{\EUBB}$ from the base of one
Hitchin fibration to the base of the dual fibration, together with
a $T$-duality on the corresponding torus fibers. This is the usual
SYZ picture of mirror symmetry, except that the map $\Xi$ is
usually assumed to be trivial.

That the corresponding fibers of the Hitchin fibrations for $G$
and $^L\negthinspace G$ are dual complex tori has been shown for
unitary groups by Hausel and Thaddeus \thadhau. For the
exceptional Lie group $G_2$, this duality has been established
recently by Hitchin \ref\hitchinc{N. Hitchin, ``Langlands Duality And $G_2$ Spectral Curves,''
math.AG/0611524.}.  The
question has also been analyzed by Donagi and Pantev for any
semi-simple Lie group using an abstract approach to spectral
covers \ref\rdonagi{R. Donagi and T. Pantev,
``Langlands Duality For Hitchin Systems,'' math.AG/0604617.}.  See also the work of Faltings
on $G$-bundles
\ref\faltings{G. Faltings, ``Stable $G$-Bundles And Projective Connections,'' J. Alg. Geom.
{\bf 2} (1993) 507.}.

A number of important subtleties about this duality have been
omitted in this explanation. Some questions involve the center and
fundamental group of $G$ and are discussed briefly in section
\centertop. Other questions involve the role of a spin structure
on $C$ and will not be analyzed in this paper.  A discussion of
the role of spin structures and more detail on the role of the
center and fundamental group and the duality for unitary groups
will appear elsewhere \witfur.

\subsec{Branes On $\MH$}

\subseclab\mhbranes

We have  met several interesting examples of branes on $\MH$ --
the zerobrane and the brane wrapped on a fiber.  We will encounter
others.  As in our initial examples, most of the important branes
will  have supersymmetric properties with respect to all three
complex structures $I$, $J$, and $K$, and this is very useful in
understanding their behavior.  The branes of most interest
preserve half the supersymmetry of the effective two-dimensional
sigma-model, which in turn has half the supersymmetry of the
underlying gauge theory.

Examples include $(B,B,B)$-branes, which are $B$-branes in each
complex structure $I,J,K$, as well as branes of type $(B,A,A)$,
$(A,B,A)$, and $(A,A,B)$, which are of $ B$-type in one complex
structure and of $ A$-type in the two others. We have already seen
examples of $S$ exchanging a brane of type $(B,B,B)$ with one of
type $(B,A,A)$. {}From Table 2, we further see that $S$ maps a
brane of type $(A,B,A)$ to another brane of the same type, and
likewise for a brane of type $(A,A,B)$.

It is not difficult to give  examples of branes of each of these
four types; they will be the important branes in our study of the
geometric Langlands program.   Of course, we can more generally
have a brane that is of $B$ type for any linear combination of the
complex structures $I,J$, and $K$, and of $A$-type for the two
orthogonal complex structures.\foot{However, there are no branes
of type $(B,B,A)$.  A subvariety that is holomorphic in complex
structures $I$ and $J$ is automatically holomorphic in complex
structure $K$.  And there likewise are no branes of type
$(A,A,A)$.}

The zerobranes that we have already studied are the most obvious
branes of type $(B,B,B)$.   Other examples of $(B,B,B)$-branes are
space-filling branes, whose target is all of $\MH$, endowed with a
Chan-Paton vector bundle that is holomorphic in each of the three
complex structures. Bundles of that sort include the trivial
bundle as well as other examples whose role in the geometric
Langlands program will be discussed elsewhere \witfur.

A large class of examples  of $(B,A,A)$-branes is as follows. Let
$N$ be a complex submanifold of $\MH$ in complex structure $I$
which is a Lagrangian submanifold with respect to the holomorphic
symplectic structure $\Omega_I$, and hence with respect to
$\omega_J$ and $\omega_K$. Then  a brane supported on $N$ with
trivial Chan-Paton bundle is a $(B,A,A)$-brane. We have already
encountered one important example: $N$ might be a fiber of the
Hitchin fibration. For another example, we can take $N$ to be,
from the standpoint of complex structure $I$, the space of all
pairs $(E,\varphi)$  where $E$ is held fixed and $\varphi$ is
allowed to vary. {}From the point of view of gauge theory, this
means that $A_{\bar z}$ is held fixed up to a gauge
transformation, which ensures the vanishing of
$\Omega_I=(1/\pi)\int_C\,|d^2z|\,\Tr\,\delta
\phi_z\wedge\delta A_{\bar z}$.  Hence it gives us a brane of type
$(B,A,A)$.

Similarly, examples of $(A,B,A)$ and $(A,A,B)$-branes come from
complex Lagrangian manifolds with respect to the complex structure
$J$ or $K$. A simple example of a complex Lagrangian manifold in
complex structure $J$ is obtained simply by keeping ${\cal
A}_{\bar z}=A_{\bar z}+ i\phi_{\bar z}$ fixed and letting ${\cal
A}_z=A_z+i\phi_z$ vary. This gives us the family of all flat
$G_{{\Bbb{C}}}$ bundles that have a specified holomorphic
structure. Similarly, we get a complex Lagrangian manifold in
complex structure $K$ by specifying $A_{\bar z}-\phi_{\bar z}$ and
letting $A_z+\phi_z$ vary.  In section \abranes, we will construct
more sophisticated examples of branes of type $(A,B,A)$ or
$(A,A,B)$ as ``coisotropic branes.''

\newsec{Loop and Line Operators}

\seclab\loopop %

 What is really unusual about the four-dimensional TQFT's
introduced in section \twisting, compared to other theories with a
superficially similar origin, is that they admit operators that
are associated to oriented one-manifolds ${\cal S}\subset M$, and
depend on the homotopy class, not just the homology class, of
$\cal S$. The point is not that we want to study the fundamental
group of $M$, but that these one-manifold operators prove to have
very interesting properties in relation to branes.

If $M$ is compact and without boundary, then $\cal S$ is required
to be a closed loop. Reflecting this case, one-manifold operators
are commonly called loop operators.  More generally, however, on a
noncompact  four-manifold, we allow line operators that go off to
infinity. An important example is a static line operator, whose
trajectory spans all time at a given point in space.  Such a line
operator must be included in quantizing the theory to construct a
Hilbert space of physical states.  This will actually be the most
important case in our applications.\foot{Thus, other cases such as
line operators that end on branes, though they make sense in
general, will not be important in the present paper.}

\nref\blaut{M. Blau and G. Thompson, ``Aspects Of $N_T\geq 2$
Topological Gauge Theories And $D$-Branes,'' Nucl. Phys. {\bf
B492} (1997) 545-590, arXiv:hep-th/9612143.}%
 \nref\bln{L. Baulieu, A.
Losev, and N. Nekrasov, ``Chern-Simons And Twisted Supersymmetry
In Various Dimensions,'' Nucl. Phys. {\bf
B522} (1998) 82-104, arXiv:hep-th/9707174.}%
 \nref\malda{J. Maldacena,
``Wilson Loops In Large $N$ Field Theories,'' Phys. Rev. Lett.
{\bf 80} (1998) 4859-4862,
arXiv:hep-th/9803002.}%
\nref\rey{S.-J. Rey and J.-T. Yee, ``Macroscopic Strings As Heavy Quarks In Large $N$
Gauge Theory And Anti-de Sitter Supergravity,'' Eur. Phys. J. {\bf C22} (2001)
 379-394, hep-th/9803001.}%
In section \topwil, we discuss the most elementary loop or line
operators, usually called Wilson  operators. Wilson operators in
TQFT's similar to the ones we consider here, with the connection
modified by a scalar field to ensure the topological symmetry,
were first constructed by Blau and Thompson \blaut. Similar
gauge-invariant operators were also used in studying
five-dimensional supersymmetric Yang-Mills theory \bln. Such
operators are important in contemporary developments in string
theory (along with their magnetic duals, which we introduce in
section \tophooft). They were first introduced in the half-BPS
case in \refs{\malda,\rey}. The topological line operators that we
consider are most similar to the 1/16 BPS line operators defined
in \ref\zar{K. Zarembo, ``Supersymmetric Wilson Loops,'' Nucl.
Phys. {\bf B643} (2002) 157-171, arXiv:hep-th/0205160.} for any
loop ${\cal S}\subset \Bbb{R}^4$.

In section \tophooft, we describe the dual 't Hooft  operators. In
section \comptwo, we discuss the interpretation of such operators
after compactification to two dimensions. Finally, in section
\lineop, we study line operators as operators on branes.

\subsec{Topological Wilson Operators}

\subseclab\topwil

In gauge theory, the most elementary loop operator is the Wilson
loop operator.  We consider gauge theory with gauge group $G$ and
connection $A$ on a $G$-bundle $E\to M$. We let ${\cal S}$ be an
oriented loop in $M$, oriented say by the choice of a one-form
$ds$, where $s$ is a parameter along ${\cal S}$.  We pick an
irreducible representation $R$ of $G$, and let\foot{Since our
covariant derivative is $D=d+A$, the holonomy operator has a minus
sign in the exponent.  In terms of a hermitian gauge field $A'=iA$
(recall eqn. \boloko), the holonomy operator is ${\rm
Tr}_R\,P\,\exp\left(i\oint_{\cal S} A'\right)$.}
\eqn\ytro{W_0(R,{{\cal S}})={\rm Tr}_R
\,P\,\exp\left(-\oint_{{\cal S}} A\right).} In other words,
$W_0(R,{\cal S})$ is the trace, in the representation $R$, of the
holonomy of $A$ around ${\cal S}$.

Now let us look back to eqn. \bono\ from section \twisting, in
which we determined the transformation law of $A$ under the
topological symmetry: \eqn\likp{\delta_T
A_\mu=iu\psi_\mu+iv\tilde\psi_\mu.} Clearly, for any non-zero
$(u,v)$, the Wilson loop operator is not invariant.  However,
twisted ${\EUN}=4$ Yang-Mills theory also has an adjoint-valued
one-form $\phi=\sum_{\mu=0}^3\phi_\mu\,dx^\mu$, transforming as
\eqn\nikp{\delta_T\phi_\mu=iv\psi_\mu-iu\tilde\psi_\mu.} It is
possible to modify the connection $A$ by adding to it a multiple
of $\phi$.

Precisely if $t=v/u$ is equal to $\pm i$, a linear combination of
$A$ and $\phi$ is invariant.  For $t=i$, the invariant combination
is $\CA=A+i\phi$; at $t=-i$, it is $\overline\CA=A-i\phi$.  We
then define the supersymmetric Wilson loop operator as the
holonomy of $\CA$ or $\overline\CA$ around ${\cal S}$.  Thus
\eqn\ymop{ W(R,{\cal S})=\Tr_R\,P\exp\left(-\oint_{{\cal
S}}\CA\right)=\Tr_R\,P\exp\left(-\oint_{{\cal
S}}(A+i\phi)\right),~~ t=i} and \eqn\jymop{W(R,{{\cal S}})=
\Tr_R\,P\exp\left(-\oint_{{\cal S}}{\overline\CA}\right)
=\Tr_R\,P\exp\left(-\oint_{{\cal S}}(A-i\phi)\right), ~~t=-i.} The
central generator $\EUI$ of $SU(4)_{{\cal R}}$, which acts by
$t\to -t$ and $\phi\to -\phi$, transforms one formula into the
other.

The existence of Wilson loop operators possessing the topological
symmetry is extremely natural at $t=\pm i$.  At these values of
$t$, the topological equations \loopme\ assert the flatness of
$\CA$. So the holonomies  of $\CA$ or $\bar\CA$ around closed
loops are natural invariants. Equally well, such  holonomies are
not natural invariants if $t\not=\pm i$, since the topological
equations do not necessarily imply flatness.  (The vanishing
theorems of section \vanishing, which relate supersymmetric
configurations to complex flat connections for generic $t$,
require some global input and do not hold in all situations, for
instance involving branes, to which one might apply the TQFT.)

Though we formulated these definitions for ${{\cal S}}$ a closed
loop, they have a good analog for the open case. If ${{\cal S}}$
has endpoints $p$ and $q$, we define $W(R,{{\cal S}})$ not  as a
trace but as the matrix of parallel transport (of the connection
$\CA$ or $\bar\CA$) from the fiber of $E$ at $p$, taken in the
representation $R$, to the fiber at $q$. $W(R,{{\cal S}})$ is thus
a map from a vector space at $p$ to a vector space at $q$.  It is
then incorporated as part of a larger quantum construction
involving initial and final quantum states, as mentioned in our
introductory remarks. This extended version of the Wilson operator
will be essential when we get to the geometric Langlands program
in earnest.

  When the
context is clear, we abbreviate $W(R,{{\cal S}})$ as $W(R)$ or
simply $W$.

\subsec{Topological 't Hooft Operators}

\subseclab\tophooft

We have found Wilson operators in the topological field theory at
$t=\pm i$, that is at $\Psi=\infty$.  By  the $S$-duality
$S:\Psi\to -1/n_{\frak g}\Psi$, there must be dual operators at
$\Psi=0$. The topological Wilson operators were classified by the
choice of a representation $R$ of $G$.  So, as $S$ exchanges $G$
and $^L\negthinspace G$, the dual operators that appear in the
TQFT at $\Psi=0$ must be classified by a representation
$^L\negthinspace R$ of $^L\negthinspace G$.

The Wilson operator is an example of what is often called an
``order'' operator in statistical mechanics.  An order operator is
built from a classical expression (such as the holonomy of the
connection around a loop) that is then interpreted as a quantum
operator and included as a factor in a path integral.

The dual of an order operator is frequently a ``disorder''
operator.  A disorder operator is defined by specifying a
singularity that the classical fields are supposed to have, and
performing a path integral (or quantizing) in the presence of the
singularity.  Classically, the order and disorder operators seem
like completely different kinds of things.  But quantum
mechanically, they turn out to have, in suitable cases, the same
formal properties, and are frequently exchanged by duality. In our
problem, the dual of a Wilson operator is an 't Hooft  operator,
which is an example of a disorder operator.\foot{For an explicit
demonstration that the dual of a Wilson operator is an 't Hooft
operator in the case of $G=U(1)$, see \ref\witlec{E. Witten,
``Lectures On Quantum Field Theory,'' Lecture 10, in P. Deligne
et. al, eds., {\it Quantum Fields And Strings: A Course For
Mathematicians}, vol. 2 (American Mathematical Society, 1999).}.}

To define a Wilson loop operator associated with a loop ${\cal
S}\subset M$, we required an orientation of ${\cal S}$.  The 't
Hooft loop operator instead requires an orientation of the normal
bundle to ${\cal S}$.  In fact, it is convenient to identify a
small neighborhood of ${\cal S}$ with ${\cal S}\times \Bbb{R}^3$
and to write $\epsilon_3$ for a volume-form on $\Bbb{R}^3$.  The
definition of the 't Hooft operator will depend only on the
orientation of the normal bundle, not the details of these
choices. If $r$ denotes the distance from ${\cal S}$ (in some
metric), then we can also write \eqn\congo{\epsilon_3=dr\,d{\rm
Vol},} where $d{\rm Vol}$ is a volume-form on the two-sphere at
fixed $r$.

Throughout this paper, when we write explicit formulas for 't Hooft operators in $U(N)$
or $SU(N)$ gauge theory, we consider the gauge fields as connections on a rank $N$ vector
bundle.  This means that $A$ and $\phi$ are are one-forms valued in
$N\times N$ skew-hermitian matrices; for $N=1$, they are simply imaginary one-forms.

Let us first describe the most basic 't Hooft operator in $U(1)$
gauge theory. On $\Bbb{R}^4=\Bbb{R}\times \Bbb{R}^3$, we pick
coordinates $(s,\vec x)$.  The singular Dirac monopole is a
classical solution of Maxwell's equations that is invariant under
translations of $s$ and rotations of $\vec x$, and has a
singularity at the line $L$ defined by $\vec x=0$. If $F=dA$ is
the curvature of a $U(1)$ connection $A$, then Maxwell's equations
read $dF=d\star F=0$, and can be solved on $\Bbb{R}^3\backslash
\{0\}$ (that is, on the complement of the point $\vec x=0$ in
$\Bbb{R}^3$) by \eqn\izo{F={i\over 2}\star d\left({1\over |\vec
x|}\right).}  The use of the Hodge $\star$ operator shows
explicitly the dependence on the orientation of the normal bundle
to $L$.  To get a solution on $\Bbb{R}^4\backslash L$, we simply
use the projection $\Bbb{R}^4\backslash L\to \Bbb{R}^3\backslash
\{0\}$, and pull $F$ back to a solution of Maxwell's equations on
$\Bbb{R}^4\backslash L$.
 $F$ has been normalized so that
if $Y$ is a two-sphere enclosing the singularity at $\vec x=0$,
then $\int_Y \,iF/2\pi =1$.  Hence there exists a unitary line
bundle ${\cal L}\to \Bbb{R}^4\backslash L$  that admits a
connection of curvature $F$. This is the singular Dirac monopole.
It is characterized by $\int_Yc_1({\cal L})=1$.

The basic 't Hooft operator of abelian gauge theory is defined by
saying that the path integral is to be performed over connections
on a line bundle over $\Bbb{R}^4\backslash L$ that agree with the
singular Dirac monopole near $L$. Here there is no need to
consider only straight lines. $L$ can be replaced by any closed
curve  ${{\cal S}}$ in a general four-manifold $M$, or (if $M$ is
geodesically complete) any curve that goes off to infinity. The
path integral with an insertion of the 't Hooft loop operator
$T_{(0)}$ on the curve ${{\cal S}}$ is obtained by performing a
path integral for abelian gauge fields\foot{If the homology class
of ${\cal S}$ in $H_1(M,\Bbb{Z})$ is nonzero, there are no such
abelian gauge fields and we declare that this path integral
vanishes. The dual of this is that if the homology class of ${\cal
S}$ is nonzero, the Wilson operator $W(R,{\cal S})$ has a
vanishing expectation value after integrating over all
connections; this vanishing follows from the behavior in twisting
by a flat line bundle.} on $M\backslash {{\cal S}}$ such that the
curvature near ${{\cal S}}$ diverges as $F\sim {i\over 2}\star_3d
(1/r)$, where $r=|\vec x|$ is now the distance from ${{\cal S}}$
and $\star_3$ is an operation on forms that near ${{\cal S}}$
looks like the $\star$ operator on planes normal to ${{\cal S}}$.
More generally, picking an integer $m$ and requiring that the
singular part of $F$ is $F\sim{(im/ 2)}\star_3 d (1/r)$, we get
the 't Hooft loop operator $T_{(0)}(m)$ of charge $m$.

Now let us consider ${\EUN}=4$ super Yang-Mills theory with gauge
group $U(1)$.  We can define an 't Hooft operator by asking that
the gauge field have the Dirac monopole singularity while other
fields are non-singular.  But is such an operator compatible with
the topological symmetry at $\Psi=0$?

We can decide this question as follows. To get to $\Psi=0$, we
take $t=1$ and $\tau$ imaginary. In order for the 't Hooft
operator to preserve the topological symmetry, the singular
behavior of the fields near ${{\cal S}}$ must be compatible with
that symmetry.  Since the question concerns the local behavior
near ${{\cal S}}$, it is convenient to temporarily revert to the
case that ${{\cal S}}$ is a line $L\subset \Bbb{R}^4$.

 The conditions for a set of fields
$(A,\phi)$ to preserve the topological symmetry were found in
eqns. \loopme.  If those equations are compatible with the type of
singularity required by the 't Hooft operator, then the 't Hooft
operator preserves supersymmetry. Otherwise, it is impossible for
any field configuration in the presence of the 't Hooft operator
to be
 supersymmetric, and we say that the 't Hooft operator violates
supersymmetry.

At $t=1$ and for abelian gauge theory, the requisite conditions
read \eqn\nurot{(F+d\phi)^+=(F-d\phi)^-=0.} These equations are
not satisfied if $F$ has the Dirac monopole singularity and $\phi$
is non-singular.  To get an 't Hooft operator that preserves the
topological symmetry, just as in the Wilson case, we have to
include $\phi$ in a suitable fashion.

What sort of singularity of $\phi$ must accompany the Dirac
monopole singularity of $F$ in order for the equations \nurot\ to
be satisfied? This question has a simple answer: we should take
$\phi=(i/2|\vec x|)ds$, where $s$ is the ``time'' coordinate along
$L$.  Thus, the equations \nurot\ are obeyed if we take
\eqn\hildo{\eqalign{F& ={i\over 2}\star_3 d{1\over |\vec x|}\cr
                    \phi& = {i\over 2|\vec x|} ds,\cr}}
where $\star_3$ is the three-dimensional Hodge $\star$ operator in
the directions normal to $L$.

This tells us how to define, for a general closed one-manifold
${{\cal S}}\subset M$, the basic 't Hooft operator $T$ that
preserves the topological symmetry at $\Psi=0$: to evaluate a
matrix element of this operator, we perform a path integral over
fields that possess near ${{\cal S}}$ the singularity described in
\hildo. Similarly, for any integer $m$, to get a charge $m$
topological 't Hooft operator $T(m)$, we simply require that the
fields have a singularity along ${{\cal S}}$ that is $m$ times
what is described in \hildo.

\bigskip\noindent{\it Generalization To Any $G$ And Langlands
Duality}

The next step is to generalize this construction to arbitrary $G$.
This is done in a simple fashion \ref\otherkap{A. Kapustin,
``Wilson-'t Hooft Operators In Four-Dimensional Gauge Theories And
$S$-Duality,''
 arXiv:hep-th/0501015.}.  We
simply pick an arbitrary homomorphism $\rho:U(1)\to G$.   Then we
define a topological 't Hooft operator by asking that the fields
should have a singularity along ${{\cal S}}$ that looks like the
image under $\rho$ of the singularity in \hildo.

The operator defined this way depends on $\rho$ only up to
conjugacy.  But conjugacy classes of homomorphisms $\rho:U(1)\to
G$ are classified by highest weights of the dual group
$^L\negthinspace G$.  In other words, they are classified by
irreducible representations of $^L\negthinspace G$.

So for every irreducible representation $^L\negthinspace R$ of
$^L\negthinspace G$, and every one-manifold ${{\cal S}}$, we have
obtained a topological 't Hooft operator $T({}^L\negthinspace
R,{{\cal S}})$. As in the Wilson case, we refer to this operator
as just $T({}^L\negthinspace R)$ or $T$ if the context is clear.
The fact that 't Hooft operators of $G$ are classified by
representations of $^L\negthinspace G$ is the classic GNO duality
\gno, as recently reinterpreted in terms of operators rather than
states \otherkap.

The classification of 't Hooft operators is also intimately
related to Grothendieck's classification of $G$-bundles on ${\Bbb
{CP}}^1$.  According to this classification, a $G$-bundle on
${\Bbb {CP}}^1$ is associated to the fundamental line bundle
${\cal O}(1)\to\Bbb{CP}^1$ via a homomorphism $\rho:U(1)\to G$ (or
its complexification $\rho:\Bbb{C}^*\to G_{\Bbb{C}}$).  Let us
identify with $\Bbb{CP}^1$ a small two-sphere surrounding the
singularity that defines an 't Hooft operator. Every $G$-bundle on
a Riemann surface, even if unstable, admits a connection obeying
the Yang-Mills equations \ref\abott{M. F. Atiyah and R. Bott,
``Yang-Mills Equations Over Riemann Surfaces,'' Phil. Trans. R.
Soc. Lond. {\bf A308} (1983) 523-615.}, and on $\Bbb{CP}^1$ every
solution of the Yang-Mills equations is equivalent to one of the
abelian solutions used to construct 't Hooft operators.  So the
classification of 't Hooft operators is the same as the
classification of $G$-bundles on $\Bbb{CP}^1$.

\bigskip\noindent{\it Topologically Non-Trivial 't Hooft
Operators}

A holomorphic $G$-bundle over $S^2=\Bbb{CP}^1$ may be
topologically non-trivial, though the topological classification
of $G$-bundles is much less fine than the holomorphic
classification.

A $G$-bundle over $S^2$ is classified by a characteristic class
$\xi\in H^2(S^2,\pi_1(G))\cong \pi_1(G)$.  But $\pi_1(G)$ is
naturally isomorphic to the center of $^L\negthinspace G$.  The
bundle over $S^2$ that is used in constructing the 't Hooft
operator $T(^L\negthinspace R)$ is topologically non-trivial if
and only if the center of $^L\negthinspace G$ acts nontrivially on
$^L\negthinspace R$.

\bigskip\noindent{\it Combined Wilson-'t Hooft Operators: Abelian Case}

We have so far found Wilson line operators at $\Psi=\infty$ and 't
Hooft line operators at $\Psi=0$.  By applying
 duality transformations, we find other special values of
$\Psi$ at which line operators exist that preserve the topological
symmetry.  In fact, all rational values of $\Psi$, and only those,
can be obtained from $\Psi=0$ (or $\Psi=\infty$) by a duality
transformation.  This is true regardless of whether the duality
group $\Gamma$ is $SL(2,\Bbb{Z})$, which it is for $G$
simply-laced, or one of the Hecke groups that arise when $G$ is
not simply-laced.

To get to rational values of $\Psi$, we will take $t=1$ with
$\theta=2\pi \Psi$. Thus, to the action of the gauge field, we add
a term \eqn\into{{{\cmmib I}}_\theta=i{\Psi\over 4\pi}\int
\Tr\,F\wedge F.} At this stage, it is necessary to assume that $M$
is oriented, say by the choice of a volume-form $\epsilon_4$.
 ${\EUN}=4$ super Yang-Mills theory makes sense on an
unorientable four-manifold $M$.  But the interaction \into, whose
role we now wish to consider, can only be introduced if $M$ is
oriented, since this is required in order to define the integral
$\int_M\Tr\,F\wedge F$.  (More generally, $\Psi$ is odd under
orientation-reversal, as we noted at the end of section \canonpar,
so to get to nonzero rational $\Psi$, we require an orientation on
$M$.) So we assume that $M$ is oriented, and express the
orientation by the choice of a volume-form $\epsilon_4$.

The definition of a Wilson loop operator supported on ${\cal S}$
required an orientation of ${\cal S}$ by a one-form $ds$.  The
definition of an 't Hooft loop operator supported on ${\cal S}$
required an orientation $\epsilon_3$ of the normal bundle to
${\cal S}$.  Once $M$ is oriented, we can ask for the orientations
of ${\cal S}$ and its normal bundle to be compatible in the sense
that \eqn\delo{\epsilon_4=ds\wedge \epsilon_3} along ${\cal S}$.
Identifying a small neighborhood of ${\cal S}$ in $M$ with ${\cal
S}\times \Bbb{R}^3$, we can extend $ds$ over this neighborhood so
that \delo\ remains true throughout ${\cal S}\times \Bbb{R}^3$.
Then writing $\epsilon_3=dr\wedge d{\rm Vol}$ as in \congo, we
have \eqn\nelo{\epsilon_4=ds\wedge dr\wedge d{\rm Vol}=-dr\wedge
ds\wedge d{\rm Vol}.}

The line operators that preserve the topological symmetry at
rational values of $\Psi$ are called mixed Wilson-'t Hooft
operators.  Because these operators will combine the properties of
't Hooft and Wilson operators, to define them it is important to
choose compatible orientations on ${\cal S}$ and its normal
bundle, as we did in the last paragraph. To construct these
operators, we first consider the case $G=U(1)$.

At first sight it may appear that the 't Hooft operators as we
have defined them already possess the topological symmetry at any
$\Psi$.  This is in fact not true, because the singularity of the
't Hooft operator causes ${{\cmmib I}}_\theta$ not to be invariant
under the topological symmetry.

This requires some explanation.  We let $A_0$ be a fixed
connection whose curvature has the Dirac monopole singularity
along a curve ${{\cal S}}$.  A path integral in the presence of a
charge $m$ operator $T_m({{\cal S}})$ is evaluated as follows. One
takes the gauge field to be $A=mA_0+\hat A$, where $\hat A$ is
smooth near ${{\cal S}}$. The path integral is then evaluated by
integrating over $\hat A$, with $A_0$ fixed.  The precise choice
of $A_0$ does not matter (as long as it has the right
singularity), since we are going to integrate over all $\hat A$.

As a step in this direction, let us evaluate ${{\cmmib I}}_\theta$
as a function of $A$.  We write $F=mF_0+\hat F$, where $F_0$ and
$\hat F$ are, respectively, the curvatures of $A_0$ and $\hat A$.
Since we are going to encounter some singularities, we introduce a
cutoff. Letting $V_\epsilon$ be a tube of radius $\epsilon$
centered on ${{\cal S}}$, we will evaluate $\int_{M-V_\epsilon}
F\wedge F$ and consider the limit as $\epsilon\to 0$.

One term comes from $\int_{M-V_\epsilon}m^2 F_0\wedge F_0$.  This
term is not very interesting, because it is independent of the
integration variable $\hat A$.\foot{This contribution to ${\cmmib
I}_\theta$ actually converges as $\epsilon\to 0$, because the
singular part of $F_0$ is of rank two.} Another contribution that
is not interesting for our present purposes is
$\int_{M-V_\epsilon}\hat F\wedge \hat F$.  Since $\hat A$ is
smooth, this has a limit as $\epsilon\to 0$, namely $\int_M \hat
F\wedge \hat F$. The limiting functional is a topological
invariant, and so is certainly invariant under the change in $\hat
A$ generated by the topological symmetry.

The interesting contribution is the cross term $\tilde {\cmmib
I}_\theta =(im\Psi/2\pi)\int_{M-V_\epsilon}\,F_0\wedge \hat F$. We
want to determine if this expression is annihilated by the
generator $\delta_T$ of the topological symmetry. We define the
action of $\delta_T$ on $A_0$ and $\hat A$ by $\delta_T A_0=0 $
(since $A_0$ is a fixed gauge field) and $\delta_T\hat
A=\delta_TA$. We have $\delta_T \hat F=d (\delta_T\hat A)$, so
\eqn\juqu{\delta_T\tilde {{\cmmib I}}_\theta={im\Psi\over
2\pi}\int_{M-V_{\epsilon}}d\left(F_0\wedge \delta_T\hat
A\right)={im\Psi\over 2\pi}\int_{\partial V_\epsilon} F_0\wedge
\delta_T\hat A.} To justify the integration by parts here, we must
orient $\partial V_\epsilon$ by contracting the outward normal
vector to $M-V_\epsilon$ with the orientation form of $M$. As the
outward normal to $M-V_\epsilon$ is $-d/dr$, we see from \nelo\
that $\partial V_\epsilon$ must be oriented by $ds\wedge d{\rm
Vol}$.

Now the boundary of $V_\epsilon$ is a two-sphere bundle over the
curve ${{\cal S}}$.  The integral over the fibers of
$V_\epsilon\to {{\cal S}}$ can be performed using the fact that
the integral of $F_0$ over a fiber is $-2\pi i$.   So we get
\eqn\hugu{\delta_T {{\cmmib I}}_\theta=m\Psi \int_{{\cal
S}}\delta_T \hat A.}

So ${{\cmmib I}}_\theta$ breaks the topological symmetry in the
presence of the 't Hooft operator, or equivalently, the 't Hooft
operator spoils the topological symmetry when $\theta\not= 0$. How
can we restore the symmetry? Let us remember what one does with
the action ${{\cmmib I}}$ in quantum mechanics: one integrates
over all fields with a factor of $\exp(-{{\cmmib I}})$.\foot{In
Lorentz signature, this factor would be $\exp(i{{\cmmib I}})$, and
${{{\cmmib I}}}_\theta$ would have an additional factor of $i$,
leading to the same result.}  So it is useful to re-express \hugu\
as a formula for the variation of $\exp(-{{\cmmib I}}_\theta)$:
\eqn\jugu{\delta_T(\exp(-{{\cmmib
I}}_\theta))=-m\Psi\,\exp(-{{\cmmib I}}_\theta)\, \int_{{\cal S}}
\delta_T\hat A.}

{}From here, it is a short step to see that
\eqn\flugu{\delta_T\left(\exp\left(m\Psi\int_{{\cal S}} \hat
A\right)\exp\left(-{{\cmmib I}}_\theta\right)\right)=0.} So we can
restore the topological symmetry if we include a Wilson operator
$\exp(m\Psi\int_{{\cal S}} \hat A)$ as an additional factor in the
path integral.   (Notice that we have to use here a naive Wilson
operator, not a supersymmetric one with a contribution from
$\phi$.)

However, the expression $\exp(-n\int_{{\cal S}} \hat A)$ is not
gauge-invariant for generic real $n$.  It is gauge-invariant if
and only if $n$ is an integer. So $n=-m\Psi$ must be an integer if
we are to restore the topological symmetry. To put it differently,
$\Psi$ must be a rational number  $-n/m$. For rational
$\Psi=-n/m$, we have learned that the topological symmetry is
preserved by the product of a charge $n$ Wilson operator
$\exp(-n\int_S \hat A)$ and a charge $m$ 't Hooft operator.
 The product
is called a Wilson-'t Hooft operator. We will denote it as
$WT_{n,m}({{\cal S}})$.

For rational $\Psi$, we write $\Psi=-n/m$ with coprime integers
$n$ and $m$.  Under a duality transformation $\Psi\to
(a\Psi+b)/(c\Psi+d)$, $n$ and $m$ transform by
\eqn\yto{\left(\matrix{ m  & n \cr}\right)\to  \left(\matrix{ m &
n\cr}\right) \left(\matrix{ d & -b\cr -c & a \cr}\right) .} As in
eqn. \xilj, this is the standard transformation of the magnetic
and electric charges $\eurm m$ and $\eurm n$ under duality.  The
reason for this is that the state obtained by quantizing the
abelian gauge theory in the presence of a static Wilson-'t Hooft
operator $WT_{n,m}$ (and no other charges or sources) has $({\eurm
m}~{\eurm n})=(m~n)$.  For the magnetic charges, this is fairly
clear from the fact that $m$ and ${\eurm m}$ have both been
defined in terms of the first Chern class (measured either at
infinity or near the singularity). For the electric charges, it
can be demonstrated by solving Maxwell's equations  to determine
the asymptotic behavior of the electric field in the presence of
the Wilson operator.  The computation we have just performed shows
that $n=-\Psi m$ transforms under $\Psi\to\Psi+1$ as claimed in
\yto.

\bigskip\noindent{\it Combined Wilson-'t Hooft Operators:
Nonabelian Case}

Just as for 't Hooft operators, the generalization of Wilson-'t
Hooft operators to a general simple gauge group $G$ is made by
embedding the abelian construction in $G$, using a homomorphism
$\rho:U(1)\to G$.  This, however, requires some explanation.

We  write $A=\rho(A_0)+\hat A$, where $A_0$ is the singular $U(1)$
gauge field with the Dirac singularity along ${{\cal S}}$, and
$\hat A$ is smooth.  (For $G=U(1)$, the homomorphism $\rho$ is at
the Lie algebra level multiplication by $m$ for some integer $m$,
so we wrote $A=mA_0+\hat A$.)  Now, however, we must clarify what
it means for $\hat A$ to be smooth.  Because of the nonlinear term
in the curvature $F=dA+A\wedge A$, we must require
\eqn\milfo{[\rho(A_0),\hat A]=0} along ${{\cal S}}$, or else the
singular part of the curvature along ${{\cal S}}$ will depend on
$\hat A$ and will not coincide with the Dirac singularity.

For the condition \milfo\ to make sense, we must also restrict the
gauge transformations along ${{\cal S}}$ to the subgroup $H\subset
G$ that commutes with $\rho(U(1))$.  So, when restricted to
${{\cal S}}$, $\hat A$ has structure group $H$.

The Lie algebra of $H$ has an $H$-invariant projection $\pi_\rho$
to $\rho(\frak{u}(1))$, the Lie algebra of $\rho(U(1))$. Along
${{\cal S}}$, we can define the projection $\pi_\rho(\hat A)$, a
$\rho(\frak{u}(1))$-valued gauge field.

The same calculation as in the abelian case shows that the 't
Hooft operator that corresponds to $\rho$ is no longer invariant
under the topological symmetry  if $\Psi\not= 0$. Instead,
\eqn\yugo{\delta_T {{\cmmib I}}_\theta={\Psi}\int_{{\cal S}}
\Tr\,\rho(1)\,\delta_T\hat A.} Here $\rho(1)$ is simply the image
of $1\in \frak u(1)$ under the Lie algebra homomorphism
$\rho:\frak u(1)\to \frak g$.  The expression $\Tr\,\rho(1)\,
\delta_T \hat A$ is a multiple of $\pi_\rho(\delta_T\hat A)$.

Just as in the abelian case, topological invariance can be
restored if we multiply the 't Hooft operator by
 \eqn\yrot{\exp\left(\Psi\int_{{\cal S}} \Tr \,\rho(1)\hat A\right).} For this to be
gauge-invariant places a condition on $\Psi$, which informally is
that $\rho$ must be divisible by the denominator of $\Psi$.  For
any rational $\Psi$, this condition is obeyed for suitable $\rho$.

The operators obtained this way are called Wilson-'t Hooft
operators.  The underlying supersymmetric gauge theory has more
general Wilson-'t Hooft operators with linearly independent
electric and magnetic weights \otherkap.  The Wilson-'t Hooft
operators in the topological field theory all arise by duality
from Wilson operators at $\Psi=\infty$, so their electric and
magnetic weights are proportional.

See also \ref\mans{M. Henningson, ``Wilson-'t Hooft Operators And
The Theta Angle,'' arXiv:hep-th/0603188.} for a recent related
discussion of Wilson-'t Hooft operators.

\subsec{Compactification To Two Dimensions}

\subseclab\comptwo

We now want to consider the interpretation of loop or line
operators after compactification to two dimensions.  We set
$M=\Sigma\times C$, with $\Sigma$ and $C$ being Riemann surfaces.
Thinking of $\Sigma$ as being much bigger than $C$, we want to ask
what loop operators look like in an effective two-dimensional
theory on $\Sigma$. To begin with, we assume that $\Sigma$ has no
boundary.

\ifig\namp{A schematic depiction of the four-manifold
$\Sigma\times C$, with $\Sigma$ running vertically and $C$
horizontally.  (a) A Wilson line that propagates in the $C$
direction.  (b)  A Wilson line that propagates in the $\Sigma$
direction.  (c) A Wilson line that propagates in $\Sigma$, except
at two moments at which it loops around one-cycles in $C$.}
{\epsfxsize=4in\epsfbox{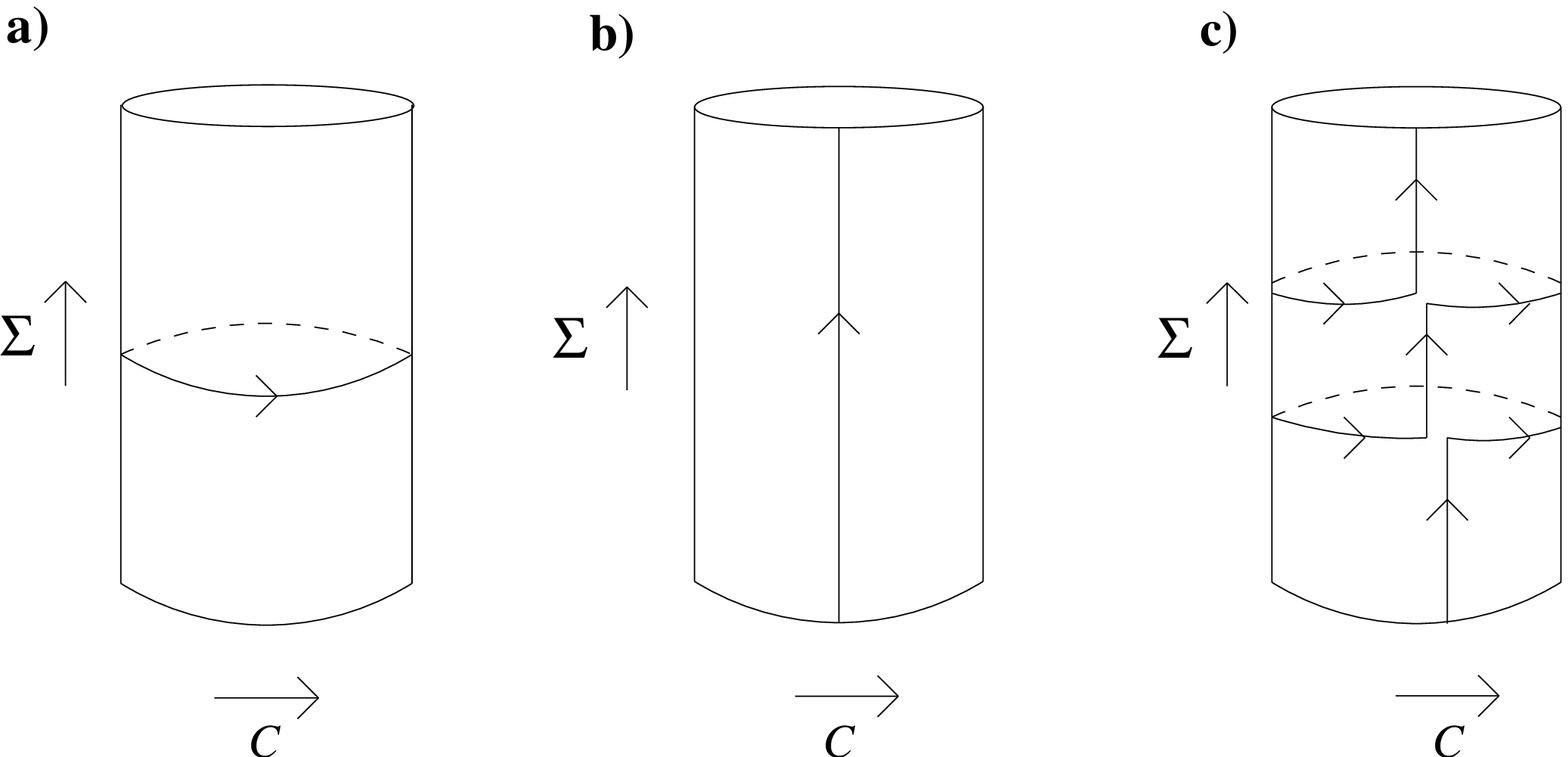}} Several cases can be
distinguished.  One case  (\namp(a)) is a loop of the form ${{\cal
S}}=p\times {{\cal S}}'$, where $p$ is a point in $\Sigma$ and
${{\cal S}}'$ is a loop in $C$. Clearly, an operator supported on
a loop of this kind simply looks like a pointlike local operator
in the effective theory on $\Sigma$.  All ordinary quantum field
theories and many TQFT's have such local operators, so in this
example, the loop operator, after compactification, turns into
something fairly ordinary in the effective theory on $\Sigma$.

The opposite case (\namp(b)) is ${{\cal S}}={{\cal S}}''\times q$,
with ${{\cal S}}''$ a loop in $\Sigma$ and $q$ a point in $C$.
Here the loop operator remains as a loop operator in the effective
two-dimensional theory.

The general case, of course, is a curve ${{\cal S}}$ that
propagates non-trivially in both $\Sigma$ and $C$.  Any such curve
is homotopic to a curve that propagates first on $\Sigma$, then on
$C$, then on $\Sigma$, and so on, as indicated in \namp(c). Let
$\bar {{\cal S}}$ be the projection of ${{\cal S}}$ to $\Sigma$. A
loop operator on ${{\cal S}}$ reduces in the effective theory on
$\Sigma$ to a loop operator on $\bar {{\cal S}}$ with local
operators inserted at distinguished points on $\bar {{\cal S}}$,
namely the points where ${{\cal S}}$ propagates around $C$. We
will momentarily describe a different interpretation of such a
loop operator.

\ifig\tonamp{(a) A line operator dividing the plane into two
regions, labeled by theories $Y$ and $Z$.  (b) A folded version of
the same picture, interpreted in terms of theory $Y\times \bar Z$
on a half-plane; the boundary is labeled by a $(Y,Z)$-brane.}
{\epsfxsize=4in\epsfbox{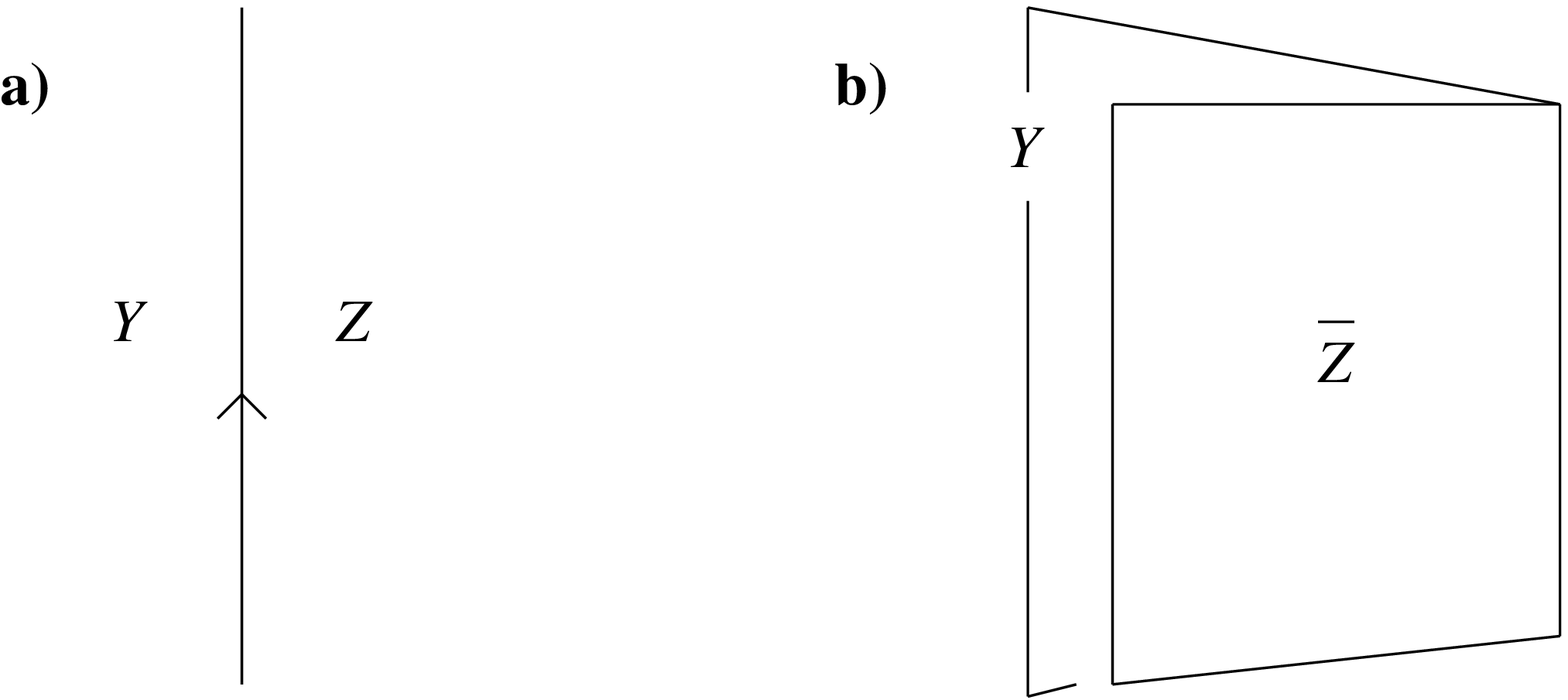}} The essential point is
clearly to understand the meaning of a loop or line operator in
two dimensions.  Here we should note that a two-manifold is
locally divided by a one-manifold into two disjoint regions.
Hence, a loop or line operator might  produce a long-range effect;
the couplings in the two-dimensional effective theory  might be
different on the two sides.  (How Wilson and 't Hooft line
operators can have such an effect is discussed in section
\wilbranes.) This possibility is incorporated in \tonamp(a), where
we sketch a two-dimensional line operator that divides the plane
into two regions labeled by two distinct theories, $Y$ and $Z$.

\nref\wong{E. Wong and I. Affleck, ``Tunneling In Quantum Wires: A Boundary Conformal Field
Theory Approach,'' Nucl. Phys. {\bf B417} (1994) 403-438.}%
\nref\aff{M. Oshikawa and I. Affleck, ``Boundary Conformal Field
Theory Approach To The Critical Two-Dimensional Ising Model With A
Defect Line,'' Nucl. Phys. {\bf B495} (1997) 533-582,
cond-mat/9612187.}%
\nref\dijk{C. Bachas, J. de Boer, R. Dijkgraaf, and H. Ooguri,
``Permeable Conformal Walls And Holography,'' JHEP {\bf 6} (2002) 1-32, arXiv:hep-th/0111210.}%
To put this situation in a more familiar framework, we can use a
``folding'' trick (see for example \refs{\wong-\dijk}).  At the cost of reversing the
orientation of region $Z$ to get what we will call theory $\bar
Z$, we can ``fold'' \tonamp(a) to get a similar figure in which
regions $Y$ and $Z$ are on the same side and end at the location
of the line operator (\tonamp(b)).

What we have now from an abstract point of view is a boundary
condition in the tensor product theory $Y\otimes \bar Z$.  For
purposes of this paper, by a ``brane,'' in general, we simply mean
a local boundary condition in a quantum field theory.  (In other
words, our branes are all $D$-branes.) We will refer to branes of
the product theory $Y\otimes \bar Z$ as $(Y,Z)$-branes. So the
line operator reduces in two dimensions to a $(Y,Z)$-brane.

Now we can also understand \namp(c) a little better. We previously
interpreted this configuration in terms of a line operator with
local operators inserted on it.  After folding, the effective
two-dimensional theory is formulated on a Riemann surface $\Sigma$
with boundary and with local operators inserted on the boundary.
Their existence is characteristic of brane physics; in general,
for any brane, there is a certain space of local operators that
can be inserted on a boundary component of $\Sigma$ that is
labeled by that brane.

\ifig\dunamp{(a) The space of $({\cal B}_1,{\cal B}_2)$ strings.
(b) Joining of strings to make a product ${\cal H}_{{\cal
B}_1,{\cal B}_2}\otimes {\cal H}_{{\cal B}_2,{\cal B}_3}\to {\cal
H}_{{\cal B}_1,{\cal B}_3}$. (c) The associativity relation that
comes by joining three strings.  (d) The space of local operators
${\EUO}$ that can be inserted on a boundary labeled by a brane
${\cal B}$ is the same as the space ${\cal H}_{{\cal B},{\cal
B}}$. To see this, we perform the path integral in a small region
around the insertion point of ${\EUO}$ (the unshaded region on the
left) to get a physical state $\Psi_{\EUO}$ that can be inserted
on the dotted line to reproduce the effects of ${\EUO}$.  The
resulting picture can be put in the form shown on the right. (e)
For every ${\cal B}$ and $\tilde{\cal B}$, the space ${\cal
H}_{{\cal B},\tilde{\cal B}}$ is a module for the algebra ${\cal
H}_{{\cal B},{\cal B}}$.} {\epsfxsize=4in\epsfbox{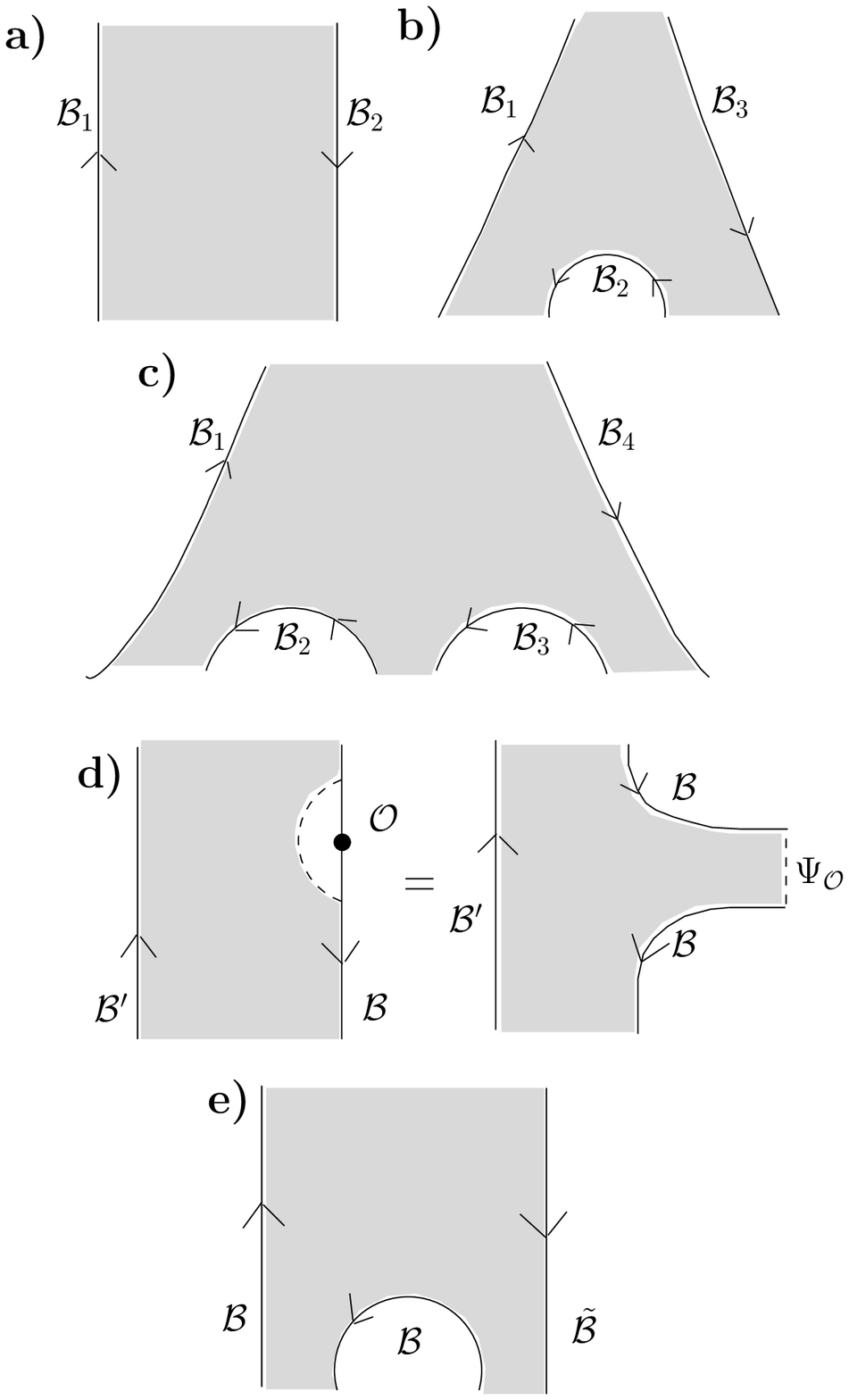}}

 Let us recall how to characterize such local operators. This will also help us
recall a few basic facts about branes. For every two-dimensional
quantum field theory $X$ and pair of branes ${\cal B}_1$ and
${\cal B}_2$, one defines a vector space ${\cal H}_{{\cal
B}_1,{\cal B}_2}$ of $({\cal B}_1,{\cal B}_2)$ strings. In unitary
quantum field theory, these spaces are Hilbert spaces; even
without unitarity, ${\cal H}_{{\cal B}_1,{\cal B}_2}$ is dual to
${\cal H}_{{\cal B}_2,{\cal B}_1}$. One defines ${\cal H}_{{\cal
B}_1,{\cal B}_2}$ by quantizing the theory $X$ on
$\Sigma=\Bbb{R}\times I$, with $I$ a unit interval whose ends are
labeled respectively by ${\cal B}_1$ and ${\cal B}_2$
(\dunamp(a)). ${\cal H}_{{\cal B}_1,{\cal B}_2}$ is also called
the space of physical states of the theory with the given boundary
conditions at the two ends of $I$.

Now in topological field theory, there are for any branes ${\cal
B}_1,{\cal B}_2,$ and $ {\cal B}_3$ natural maps ${\cal H}_{{\cal
B}_1,{\cal B}_2}\otimes {\cal H}_{{\cal B}_2,{\cal B}_3}\to {\cal
H}_{{\cal B}_1,{\cal B}_3}$ defined by joining Riemann surfaces
together (\dunamp(b)).  (In two-dimensional quantum field theory
without topological invariance, one must take into account the
metrics or conformal structures of the surfaces, as a result of
which an analogous discussion leads to the operator product
expansion.) These maps obey the obvious associativity relation,
which says that when three strings are joined  (\dunamp(c)), one
does not have to say which two joined first. In particular,
setting all the branes ${\cal B}_i$ equal to ${\cal B}$, we find
that ${\cal H}_{{\cal B},{\cal B}}$ always has the natural
structure of an associative algebra.

In fact, in TQFT, the space of local operators ${\EUO}$ that can
be inserted on a a boundary labeled by a brane ${\cal B}$ is the
same as ${\cal H}_{{\cal B},{\cal B}}$, since (\dunamp(d)) a
``cutting'' operation applied near such an ${\EUO}$ reveals an
element of ${\cal H}_{{\cal B},{\cal B}}$. By similar reasoning
(\dunamp(e)), we find that, for any ${\cal B}$ and  $ \tilde{\cal
B}$, ${\cal H}_{{\cal B}, \tilde{\cal B}}$ always has a natural
structure of left ${\cal H}_{{\cal B},{\cal B}}$-module.
Similarly, it has a natural structure of right ${\cal H}_{\tilde
{\cal B},\tilde{\cal B}}$ module. Indeed, ${\cal H}_{{\cal B},
\tilde{\cal B}}$ is a $({\cal H}_{{\cal B},{\cal B}},{\cal
H}_{\tilde{\cal B},\tilde{\cal B}})$ bimodule. This means simply
that one can act with ${\cal H}_{{\cal B},{\cal B}}$ by attaching
a string on the left, or by ${\cal H}_{\tilde{\cal B},\tilde{\cal
B}}$ by attaching a string on the right, and moreover these two
actions commute.

\subsec{Line Operator Near A Boundary}

\subseclab\lineop

The geometric Langlands program revolves around the case that
$\Sigma$ has a boundary and the line operator  runs parallel to
the boundary.

\ifig\trinamp{A Riemann surface with a boundary, labeled by a
brane ${\cal B}$, and a line operator $L$ parallel to the
boundary.}{\epsfxsize=3in\epsfbox{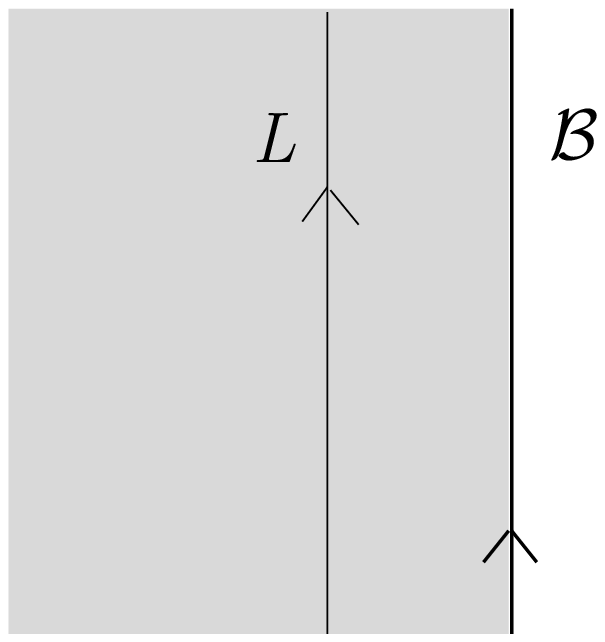}}
 Defining a quantum field theory on the Riemann surface $\Sigma$
with boundary requires a choice of boundary condition -- that is a
choice of a brane, which we will call ${\cal B}$. In \trinamp, we
sketch a Riemann surface $\Sigma $ with a boundary and a line $L$
parallel to the boundary. The boundary has been labeled by a brane
${\cal B}$, which defines the boundary conditions at that
boundary. A line operator $\EUX$ is supported on $L$. $\EUX$ may
be a Wilson, 't Hooft, or mixed Wilson-'t Hooft operator.

{}From a macroscopic point of view,  $\EUX$ and ${\cal B}$ simply
combine together into an effective boundary condition ${\cal B}'$.
One can characterize ${\cal B}'$ as the effective boundary
condition that one sees at a big distance due to the combined
effects of ${\cal B}$ and $\EUX$. Or one can simply take the limit
that $L$ approaches the boundary and let ${\cal B}'$ be the
limiting boundary condition that results.

Either way, we write ${\cal B}'=\EUX {\cal B}$ for the new brane
that is produced by acting with $\EUX$ on the old brane.   In this
way, we regard line operators as acting on boundary conditions to
produce new boundary conditions.

\ifig\quadramp{Action of a loop operator as an ordinary operator
on the space ${\cal H}_{\cal B,\tilde{\cal B}}$.}
{\epsfxsize=3in\epsfbox{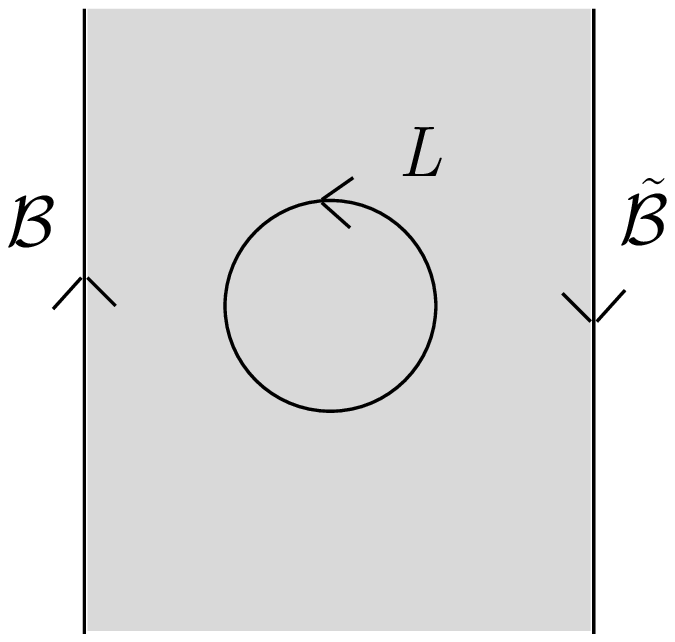}} This is a more abstract form
of action than the usual action of an operator on a vector space.
As we saw above, to get a  vector space ${\cal H}_{{\cal B},\tilde
{\cal B}}$ associated with ${\cal B}$ on which an operator  might
act, we need a pair of branes ${\cal B}$ and  $\tilde {\cal B}$.
Of course, we can just take $\tilde {\cal B}={\cal B}$. But even
once we construct a space ${\cal H}_{{\cal B},\tilde {\cal B}}$ of
physical states, it is really a {\it loop} operator, localized in
time, that acts on this space, by quantizing the picture of
\quadramp. An open line operator that runs in the time direction,
as in \trinamp, is instead part of the data that must be
incorporated in quantization to get the right space of physical
states.

If we {\it do} have a vector space ${\cal H}$ and an operator
${\EUO}:{\cal H}\to {\cal H}$, we can look for a vector $\psi\in
{\cal H}$ that is an eigenvector of ${\EUO}$:
\eqn\reot{{\EUO}\psi=\lambda\psi.} Here $\lambda$ is an ordinary
complex number.

In our problem, instead of an operator that acts on vectors in a
Hilbert space, we have a more exotic operator that acts on
theories, via ${\cal B}\to {\cal B}'=\EUX {\cal B}$.   An ordinary
complex eigenvalue does not make sense in this situation. Rather,
what plays the role of an ``eigenvalue'' is a vector space $V$.

For every brane ${\cal B}$ and vector space $V$, one defines a new
brane ${\cal B}\otimes V$.  An abstract way to characterize ${\cal
B}\otimes V$ is to say that for any brane $\tilde {\cal B}$, the
Hilbert space ${\cal H}_{\tilde {\cal B},{\cal B}\otimes V}$ is
equal to ${\cal H}_{\tilde {\cal B}, {\cal B}}\otimes V$.

We can be more concrete in the case of branes that are constructed
from geometry.  In the case that theory $X$ is a two-dimensional
sigma-model with a target space $\hat X$, a brane can be
constructed from a vector bundle ${\cal W}\to \hat X$ (or from a
more general sheaf, possibly suported on a submanifold). ${\cal
W}$ is known as the Chan-Paton bundle. Tensoring such a brane with
a fixed vector space $V$ just means replacing ${\cal W}$ by ${\cal
W}\otimes V$. This operation generalizes without any problem when
${\cal W}$ is replaced by a more general sheaf.

If we identify $V$ as $\Bbb{C}^k$ for some $k$, then tensoring the
Chan-Paton bundle with $V$ amounts to taking $k$ copies of the
original brane. So for ${\cal B}$ to be an eigenbrane for $\EUX$
means that $\EUX{\cal B}$ is isomorphic to the sum of $k$ copies
of ${\cal B}$.  However, it is much better to think of the
``eigenvalue'' as a vector space $V$, not just its integer
dimension $k$, since $V$ may in fact vary with some additional
parameters as the fiber of a vector bundle over some parameter
space.

Having established the notion of tensoring a brane by a vector
space, we now can define the  notion of eigenbrane.  A brane
${\cal B}$ is an eigenbrane for a line operator $\EUX$ if
\eqn\intico{\EUX {\cal B}={\cal B}\otimes V,} for some vector
space $V$.  This concept corresponds to the concept of an
eigensheaf as defined by Beilinson and Drinfeld in the geometric
Langlands program \bd.

\ifig\quintamp{Line operators $L_1$ and $L_2$ are said to commute
if they can be passed through each other without singularity.  In
topological field theory, this simply means that the two figures
pictured here can be considered equivalent.}
{\epsfxsize=4in\epsfbox{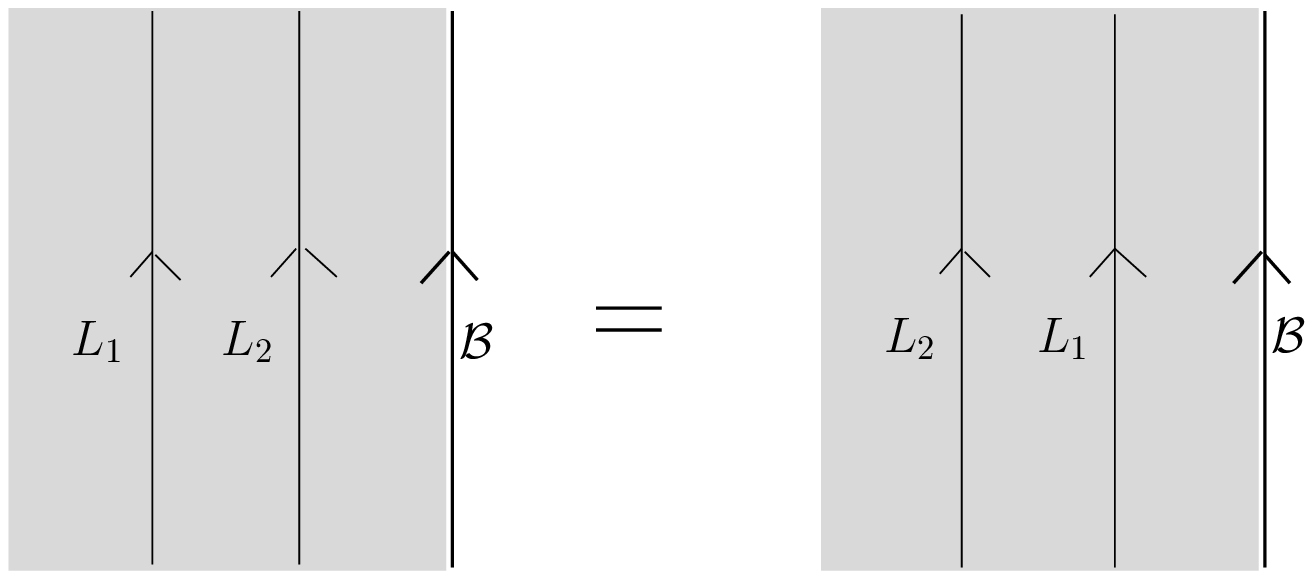}} If $\EUX_1$ and $\EUX_2$ are
two line operators, supported on parallel lines $L_1$ and $L_2$,
we say that they commute if for all ${\cal B}$ we have
\eqn\pintico{\EUX_1\EUX_2{\cal B}=\EUX_2\EUX_1{\cal B}.} This
statement can be illustrated in a convenient picture: it means
that the lines $L_1$ and $L_2$  can be passed through each other
without any discontinuity (\quintamp). Just as for ordinary
operators on a Hilbert space, if two line operators commute in
this sense, it is possible for them to have a simultaneous
eigenbrane, that is, a brane ${\cal B}$ obeying:
\eqn\pintico{\eqalign{\EUX_1{\cal B}& = {\cal B}\otimes V_1\cr
\EUX_2{\cal B}& = {\cal B}\otimes V_2.\cr}} More generally, any
collection of commuting line operators can have a simultaneous
eigenbrane.

\bigskip\noindent{\it Line Operators That Descend From Four
Dimensions}

Now in our problem, we are interested in a two-dimensional TQFT
that arises by reducing a four-dimensional TQFT on the Riemann
surface $C$. Moreover, our line operators all descend from line
operators in four dimensions.  This is enough to ensure some
special behavior.

First of all, each line operator $\EUX$  can be defined for any
point $p\in C$. To make this explicit, we write the line operator
as $\EUX_p$. If a brane ${\cal B}$ that respects the topological
symmetry is an eigenbrane of $\EUX_p$ for one $p$, it must be an
eigenbrane for every $p$, since $p$ is irrelevant anyway.  So
\eqn\hintico{\EUX_p{\cal B}={\cal B}\otimes V_p} for some $V_p$.
Here, locally, the vector space $V_p$ must be independent of $p$,
since $p$ is irrelevant. But topological invariance allows the
possibility that $V_p$ might have global monodromies as $p$ varies
on the Riemann surface $C$. So the $V_p$ are fibers of a flat
vector bundle over $C$.

To be more explicit about this, moving $p$ is equivalent to a
special case of changing the metric on $C$, and this changes the
underlying action by a BRST-trivial term $\{Q,\cdot\}$.  This fact
enables one to find a flat connection on the family of vector
spaces $V_p$ for $p\in C$ and thus to prove locally that $V_p$ is
independent of $p$. But this flat connection may have nontrivial
global monodromies.

Suppose that we are given several loop operators $\EUX_i$ that all
preserve the same topological symmetry.  For example, at
$\Psi=\infty$, they may be Wilson operators associated with
representations $R_i$ of $G$; at $\Psi=0$, they may be 't Hooft
operators associated with representations $^L\negthinspace R_i$ of
$^L\negthinspace G$. In any event, the $\EUX_i$ all depend on
points $p_i\in C$. As we are free to take the points $p_i$ to be
distinct, we can perform the interchange of \quintamp, which looks
potentially singular from a two-dimensional point of view, without
meeting any singularity. (The line operators do not meet in four
dimensions, so there is no possibility of a singularity. Moreover,
as $C$ is two-dimensional, there is enough room in moving the line
operators that we do not encounter monodromies that are local
along $C$.\foot{If $L$ and $L'$ are two line operators supported
at the same point $p\in C$, to move them past each other in
$\Sigma\times C$ without a singularity, we separate them to two
distinct points $p,p'\in C$, with $p'$ near $p$. This can be done
in a small open ball $U\subset C$. For $\Sigma=\Bbb{R}\times I$,
static line operators are supported at points in the
three-manifold $W=I\times C$ and the operation of moving $L$ past
$L'$ can be carried out in the subspace $W'=I\times U$. The space
of pairs of distinct points in $W'$ is simply-connected, so there
are no monodromies in this local model and no ambiguity about how
to move $L$ past $L'$. This argument would not work if $C$ were
one-dimensional.}) Hence, topological line operators with a
four-dimensional origin automatically commute.

\bigskip\noindent{\it Electric And Magnetic Eigenbranes}

As a special case, the topological Wilson operators at
$\Psi=\infty$ commute. It is therefore possible for them to have
simultaneous eigenbranes.  A joint eigenbrane of the Wilson line
operators will be called an electric eigenbrane.

Similarly, the topological 't Hooft operators at $\Psi=0$ commute
and can likewise have simultaneous eigenbranes.  A joint
eigenbrane of the 't Hooft operators will be called a magnetic
eigenbrane.

The $S$-transformation $S:\Psi\to -1/n_{\frak g}\Psi$ maps
topological Wilson operators to topological 't Hooft operators and
will hence map electric eigenbranes to magnetic eigenbranes.

Like eigenbranes of any family of line operators that descend from
four dimensions, electric and magnetic eigenbranes are
automatically associated with flat vector bundles over $C$.  This
is an important statement in the geometric Langlands program.  As
we have seen, it reflects the existence of an underlying TQFT
above two dimensions.

\bigskip\noindent{\it Algebras Of Commuting Line Operators}

The fact that the line operators that descend from four dimensions
commute raises another question.  What is the commutative algebra
that they generate?

 In the case of Wilson line operators, it is simply the tensor
algebra of representations of $G$.  In other words, let $R$ and
$\tilde R$ be two representations of $G$, and suppose that the
decomposition of $R\otimes \tilde R$ in irreducible
representations is \eqn\ico{R\otimes \tilde R=\oplus_\alpha n_\alpha R_\alpha.}
Here we take the $R_\alpha$ to be {\it distinct} irreducible
representations; the integer $n_\alpha$ is the multiplicity with which
$R_\alpha$ appears in the decomposition of $R\otimes \tilde R$.  (The relation \ico\ and
others below can be more precisely stated in terms of vector spaces $N_\alpha$ of dimension $n_\alpha$,
but we postpone this refinement to section \opprod.)

We consider parallel lines $L$ and $L'$ and consider the limit as
$L$ approaches $L'$. We claim that \eqn\tryo{\lim_{L\to
L'}W(R,L)\cdot W(\tilde R,L')=\sum_\alpha n_\alpha W(R_\alpha,L').} In fact,
because of supersymmetry, the limit can be evaluated classically.
Had we considered ordinary Wilson operators rather than
supersymmetric ones, there would be various quantum effects in the
limit $L\to L'$.  In the supersymmetric case, we can just set
$L=L'$ and calculate classically.

Given this, the calculation is clear sailing. In the case of a
closed loop, $W(R)$ and $W({\tilde R})$ are holonomies in
representations $R$ and $\tilde R$. Their product is a holonomy in
the representation $R\otimes \tilde R$, and in view of \ico, this
can be expanded as a sum of holonomies in the representations
$R_\alpha$, leading to \tryo. In the case of a line operator, $W(R)$ or
$W(\tilde R)$ acts by parallel transport on initial states in the
$R$ or $\tilde R$ representation, whose tensor product has a
direct sum decomposition as in \tryo.

More informally, the Wilson operators $W(R)$ and $W(\tilde R)$
describe external charges in representations $R$ and $\tilde R$.
Their product describes a pair of charges in the representation
$R\otimes\tilde R$, whose decomposition in irreducibles gives the
OPE of supersymmetric Wilson operators.

Dually, we expect that if $^L\negthinspace R$ and
$^L\negthinspace{\tilde R}$ are representations of
$^L\negthinspace G$, with \eqn\itro{^L\negthinspace R\otimes{}
^L\negthinspace\tilde R=\oplus_\alpha n_\alpha\, {}^L\negthinspace R_\alpha,}
then \eqn\hffo{\lim_{L\to L'}T({}^L\negthinspace
R,L)T({}^L\negthinspace\tilde R,L')=\sum_\alpha n_\alpha T({}^L\negthinspace
R_\alpha,L').}

\nref\lusztig{G. Lusztig, ``Singularities, Character Formula, And
A $q$-Analog Of Weight Multiplicities,'' in {\it Analyse et
Topologie Sur Les Espace Singuliers II-III}, Asterisque vol. 101-2
(1981) 208-229.}
\nref\ginzburg{V. Ginzburg, ``Perverse Sheaves On A Loop Group And
Langlands
Duality,'' alg-geom/9511007.}%
\nref\vilonen{I. Mirkovic and K. Vilonen, ``Perverse Sheaves On
Affine Grassmannians
And Langlands Duality,'' Math. Res. Lett. {\bf 7} (2000) 13-24, math.AG/9911050.}%

For some groups and representations, for example
in the cases studied in \refs{\malda,\rey}, duality between gauge
theory and strings makes this result clear.  But
we do not know of any attempts in the physics literature to
compute the operator product expansion of 't Hooft operators
directly in gauge theory.  In section \thooftheckeop, we will
establish a relation between 't Hooft operators and Hecke
transformations, and then we will see in section \opprod\ how the
statement \hffo\ is related to what is known in the mathematical literature \refs{\lusztig-\vilonen}.

We can get some information about the coefficients in \hffo\ by
studying the operator product expansion of 't Hooft operators on
the Coulomb branch, where the gauge group is broken down to an
abelian subgroup. Given two 't Hooft operators
$T({}^L\negthinspace R,L)$ and $T({}^L\negthinspace  R',L')$, let
$^L\negthinspace w$ and $^L\negthinspace w'$ be the highest
weights of the two representations $^L\negthinspace R $ and
$^L\negthinspace R'$.  On the Coulomb branch, these 't Hooft
operators act via singular monopole solutions with magnetic
charges that are weights $^L\tilde w$ and $^L\tilde w'$ of
$^L\negthinspace G$; here $^L\tilde w$ and $^L\tilde w'$ can be
any Weyl transforms of $^L\negthinspace w$ and $^L\negthinspace
w'$. The product of the two 't Hooft operators is represented by a
singular abelian monopole with magnetic charge $^L\tilde
w+{}^L\tilde w'$. In order for a representation $^L\negthinspace
R_i$ to appear in the operator product expansion \hffo, its
highest weight $^L\negthinspace w_i$ must equal $^L\tilde
w+{}^L\tilde w'$, for some choice of $^L\tilde w$ and ${}^L\tilde
w'$. Whichever choice we make for $^L\tilde w$ and $^L\tilde w'$,
their sum $^L\hat w={}^L\tilde w+{}^L\tilde w'$ is dominated by
$^L\bar w={}^L\negthinspace w+{}^L\negthinspace w'$ (meaning that
$^L\bar w -{}^L\hat w$ is a dominant weight, possibly zero). One
possibility for $^L\negthinspace R_i$ is that it may be the
representation $^L\negthinspace\bar R$ whose highest weight is
exactly $^L\bar w$.  This representation appears in the expansion
\hffo\ with a multiplicity that is precisely $n=1$, since there is
precisely one choice of $^L\tilde w$ and $^L\tilde w'$ that add up
to $^L\bar w$.  The other representations that can appear are the
representations whose highest weight  is dominated by $^L\bar w$.
We will call these the associated representations of
$^L\negthinspace\bar R$. There is no straightforward way to deduce
the coefficient in the operator product expansion for these
subdominant or associated representations. Ultimately, we see in
section \opprod\ that they arise from singularities of the space
of Hecke modifications, accounting for why it is subtle to
determine them.

\bigskip\noindent{\it Functors Acting On The Category Of Branes}

What we have said so far does not fully capture the formal
properties of line operators in two dimensions.

We have explained so far that a line operator $\EUX$ that respects
the appropriate topological symmetry has a natural action on
branes.  If ${\cal B}$ and ${\cal B}'$ are any two branes, we can
act on them with $\EUX$ to get new branes $\EUX{\cal B}$ and
$\EUX{\cal B}'$.  Associated with the pair of branes ${\cal B}$
and ${\cal B}'$ is a space of physical states ${\cal H}_{{\cal
B},{\cal B}'}$ obtained by quantizing the $({\cal B},{\cal B}')$
open strings.  Likewise, associated to the pair of branes
$\EUX{\cal B}$ and $\EUX{\cal B}'$ is a space ${\cal H}_{\EUX
{\cal B},\EUX{\cal B}'}$ of physical $(\EUX{\cal B},\EUX{\cal
B}')$ strings.

\ifig\zingop{(a) A line operator $\EUX$  supported on a line $L$
maps branes ${\cal B}$ and ${\cal B}'$ to branes $\EUX{\cal B}$
and $\EUX{\cal B}'$, and likewise maps $({\cal B},{\cal B}')$
strings to $(\EUX{\cal B},\EUX{\cal B}')$ strings. In the picture,
the operator ${\EUO}$ determines a $({\cal B},{\cal B}')$ string,
which is mapped by the line operator to an $(\EUX{\cal
B},\EUX{\cal B}')$ string. (b) These maps preserve the
associativity of open string multiplication. }
{\epsfxsize=4in\epsfbox{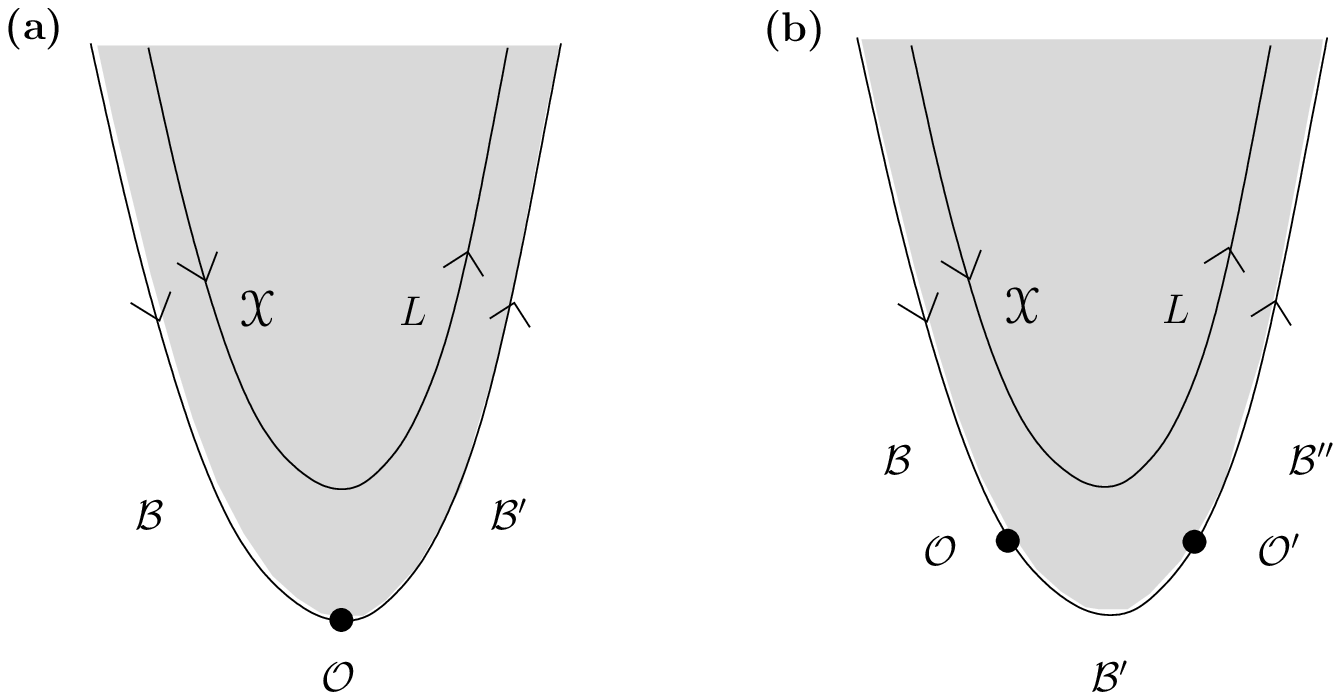}} Associated with the line
operator $\EUX$ is a natural map $\hat\EUX:{\cal H}_{{\cal
B},{\cal B}'}\to{\cal H}_{\EUX{\cal B},\EUX{\cal B}'}$. The
definition of $\hat \EUX$ should be clear from \zingop(a).  In
this picture, from a microscopic point of view, the operator
${\EUO}$ determines a state in ${\cal H}_{{\cal B},{\cal B}'}$,
but from a macroscopic point of view, as $L$ approaches the
boundary, what we see is a state in ${\cal H}_{\EUX{\cal
B},\EUX{\cal B}'}$.

Precisely if $\EUO$ is a multiple of the identity operator, there
is no distinguished point on the boundary of $\Sigma$ at which
$\EUO$ is inserted, and this is true before or after the line
operator $\EUX$ approaches the boundary. So $\hat\EUX$  maps the
identity in ${\cal H}_{{\cal B},{\cal B}}$ (that is, the open
string state that corresponds to the identity operator ${\EUO}=1$)
to the identity in ${\cal H}_{\EUX{\cal B},\EUX{\cal B}}$. It also
preserves the multiplication of open strings.  Indeed, given three
branes ${\cal B}$, ${\cal B}'$, and ${\cal B}''$, the associative
multiplication ${\cal H}_{{\cal B},{\cal B}'}\otimes{\cal
H}_{{\cal B}',{\cal B}''}\to {\cal H}_{{\cal B},{\cal B}''}$ is
defined by joining open strings as in \dunamp(b), or equivalently
by an operator product expansion of operators inserted on the
boundary, as in \zingop(b). This obviously commutes with bringing
a line operator up to the boundary.  The fancy way to summarize
these statements is to say that the line operator $\EUX$
determines a functor mapping the category of branes to itself.

Line operators as symmetries of the category of branes have been
discussed independently and from a different point of view in
\ref\fuchs{I. Runkel, J. Fuchs, and C. Schweigert,
``Categorification And Correlation Functions In Conformal Field
Theory,'' math.CT/0602079.}.  Their relation to duality
transformations will be discussed elsewhere \witfur.  For an example of the use
of Wilson-like operators to describe boundary perturbations in two-dimensional
conformal field theory, see section 5.5 of
\ref\mooreg{G. Moore, ``$K$-Theory From A Physical
Perspective,'' hep-th/0304018.}.

\bigskip\noindent{\it Line Operators and Enhanced Supersymmetry}

Special Wilson or 't Hooft operators can have more supersymmetry
than what we have claimed so far.

Let us go back to the definition \ymop\ of the supersymmetric
Wilson operator.  It is a function of ${\cal
A}_\mu=A_\mu+i\phi_\mu$, so it is invariant under any
supersymmetry that leaves $A_\mu+i\phi_\mu$ invariant. According
to the microscopic formula \hygo\ for the transformation of the
fields under supersymmetry, the variation of $A_\mu+i\phi_\mu$
under a supersymmetry generated by an infinitesimal parameter
$\epsilon$ is proportional to
$(\Gamma_\mu+i\Gamma_{\mu+4})\epsilon$.  The condition for
$\epsilon$ to generate a supersymmetry of an arbitrary Wilson line
operator is that this vanishes for $\mu=0,\dots,3$:
\eqn\pimbo{\Gamma_{\mu}\Gamma_{\mu+4}\epsilon=i\epsilon,~\mu=0,\dots,3.}
This condition leaves only one unbroken supersymmetry, which is
the one we have exploited so far.

Suppose, however, that $M=\Sigma\times C$ with a product metric.
In this case, the twisting preserves the four supersymmetries that
were characterized in \zonop.  Consider also a Wilson operator
defined on a one-manifold ${{\cal S}}$ of the form $\gamma\times
p$ where $\gamma$ is a curve in $\Sigma$ and $p$ is a point in
$C$. For an operator of this type, which we will call a special
Wilson operator, the conditions \pimbo\ collapse to
\eqn\relfo{\Gamma_{04}\epsilon=\Gamma_{15}\epsilon=i\epsilon.}
These imply the second condition in \zonop, and the combined set
of conditions leaves two unbroken supersymmetries.

We can easily see what the extra supersymmetry means.  Special
Wilson operators do not depend on $\varphi$, so they are invariant
under the ${\cal U}_1$ symmetry introduced in section \mhhyper,
which acts on the Higgs field by $\varphi\to\lambda\varphi$,
$|\lambda|=1$. This symmetry rotates complex structure $J$ to
complex structure $K$, so it maps the $B$-model in complex
structure $J$ to the $B$-model in complex structure $K$.  Hence
the two supersymmetries are the topological supercharge of the
$B$-models in complex structures $J$ and $K$.  The topological
supercharge of the $B$-model in any of the complex structures
$I_w$ is a linear combination of these.  So it follows that a
special Wilson operator is what we might call a line operator of
type $(B,B,B)$, that is, it preserves the topological
supersymmetry of the $B$-model in every complex structure of
$\MH$.

We can reason in the same way for special 't Hooft operators, that
is 't Hooft operators supported on a curve ${{\cal S}}$ of the
same type. Special 't Hooft operators are again independent of
$\varphi$ and so invariant under the group ${\cal U}_1$.  This
group rotates the $A$-model in complex structure $K$ into the
$A$-model in complex structure $J$.  General 't Hooft operators
preserve the $A$-type supersymmetry in complex structure $K$, so
special ones do so also in complex structure $J$.  The topological
supercharge of the $B$-model in complex structure $I$ is a linear
combination of those for the $A$-models in complex structures $J$
and $K$.  So actually, the special 't Hooft operators are of type
$(B,A,A)$, that is, they preserve the indicated topological
supersymmetries for complex structures $I$, $J$, and $K$.

This is completely in accord with expectations.  The
$S$-transformation maps Wilson operators to 't Hooft operators,
and as one can see from Table 2 of section \somespec, maps
supersymmetry of type $(B,B,B)$ to supersymmetry of type
$(B,A,A)$.

One can further specialize this situation by taking the metric on
$\Sigma$ to be flat and ${{\cal S}}$ to be a straight line.  In
this case, the line operator describes a static external electric
or magnetic charge.  A Wilson or 't Hooft operator of this form
actually preserves four supersymmetries. For example, for a Wilson
operator supported on a straight line that runs in the time
direction, the conditions for an unbroken supersymmetry reduce to
$\Gamma_{04}\epsilon=i\epsilon$ and
$\Gamma_{2367}\epsilon=\epsilon$.  This system is physically
sensible (if we interpret it with Lorentz signature); it is
described by a real, positive-definite Hilbert space with a
positive-definite, hermitian Hamiltonian $H$. The four unbroken
supersymmetries are the two topological supercharges $Q_i$,
$i=1,2$ described earlier, and their hermitian adjoints
$Q^\dagger_i$.  They obey a physical supersymmetry algebra
$\{Q_i,Q^\dagger_j\}=2\delta_{ij}H$,
$\{Q_i,Q_j\}=0=\{Q^\dagger_i,Q^\dagger_j\}$, for $i,j=1,2$.

\newsec{Fluxes and $S$-Duality}

\seclab\centertop

In this section, we will examine phenomena involving the center of
the gauge group that are relevant to understanding $S$-duality of
Hitchin moduli space and the geometric Langlands program.

We begin section \reviewthis\ with a review of discrete electric
and magnetic fluxes and the action of $S$-duality on them. More
detail can be found in \vafawitten , section 3, as well as in
\witfur. For discrete electric and magnetic flux in gauge theory,
the original reference is \ref\thooft{G. 't Hooft, ``On The Phase
Transition Towards Permanent Quark Confinement,'' Nucl. Phys. {\bf
B138} (1978) 1-25, ``A Property Of Electric And Magnetic Flux In
Nonabelian Gauge Theories,'' Nucl. Phys. {\bf B153} (1979)
141-160.}. We also explain some mathematical notions such as the
concept of the universal bundle. Then we develop our applications
in section \comptwop.

\subsec{Review}

\subseclab\reviewthis

\lref\Wittenindex{
  E.~Witten,
  ``Supersymmetric Index in Four-Dimensional Gauge Theories,''
  Adv.\ Theor.\ Math.\ Phys.\  {\bf 5}  (2002) 841-907, arXiv:hep-th/0006010.}

In this section, $G$ denotes a compact semi-simple Lie group;
$\bar G$ is its universal cover; and $G_{\rm {ad}}$ is the
corresponding adjoint group.  The center of $\bar G$, which is
also the fundamental group of $G_{\rm {ad}}$, is a finite abelian
group $\CZ$.  For simple $G$, $\CZ$ is a cyclic group in all cases
except for $\bar G=Spin(4n)$, as is summarized in the table.

One happy fact that we notice from the table is that some of the
subtleties with $S$-duality for non-simply-laced groups are
irrelevant, since $G_2$ and $F_4$ have trivial centers.

\vskip .5cm

\centerline{
\vbox{\hsize=2truein\offinterlineskip\halign{\tabskip=2em plus2em
minus2em\hfil#\hfil&\vrule#&\hfil#\hfil\tabskip=0pt\cr Group
&depth 6pt& Center\cr \noalign{\hrule} $
SU(N)$&height14pt&$\Bbb{Z}_N$\cr $ Spin(2n+1)$&height14pt&
$\Bbb{Z}_2$ \cr $Spin(4n+2)$&height14pt&$\Bbb{Z}_4$\cr
$Spin(4n)$&height14pt&$\Bbb{Z}_2\times\Bbb{Z}_2$\cr ${
Sp(n)}$&height14pt&$\Bbb{Z}_2$\cr ${ E_6}$&height14pt&${
\Bbb{Z}_3}$\cr ${ E_7}$&height14pt&${ \Bbb{Z}_2}$\cr
\noalign{\medskip} }}} \nobreak \centerline{ \vbox{\hsize=3truein
\noindent Table 3. Simple and simply-connected Lie groups with
non-trivial centers. $F_2$, $G_4$, and $E_8$, which are omitted,
have trivial center.}}

\vskip .5cm

An important property  is that the group ${\cal Z}$ is naturally
selfdual.  The dual of an abelian group $B$ is $B^\vee={\rm
Hom}(B,U(1))$.  A finite abelian group is always isomorphic to its
own dual, but not naturally.  The center $\CZ$ of  $\bar G$ always
has, however, a natural selfduality. This duality is best
expressed as a homomorphism \eqn\plook{\Upsilon:\CZ\times \CZ\to
U(1),} which is symmetric, $\Upsilon(a,b)=\Upsilon(b,a)$, and such
that any homomorphism of $\CZ$ to $U(1)$ is $a\to \Upsilon(a,b)$
for a unique $b$.  Such a $\Upsilon$ is called a ``perfect
pairing.'' $\Upsilon$ is constructed as follows.  If $G$ is
simply-laced, then $\CZ=\Lambda^\vee/\Lambda$, where $\Lambda$ is
the root lattice and $\Lambda^\vee$ is the weight lattice of $\bar
G$.  The pairing $\Upsilon$ is then defined as
\eqn\rofo{\Upsilon(a,b)=\exp(2\pi i\langle a,b\rangle),} where
$\langle~,~\rangle$ is the usual quadratic form on $\Lambda^\vee$
(which is even and integral when restricted to $\Lambda$ and takes
rational values on $\Lambda^\vee$). If $G$ is not simply-laced,
its center is trivial or $\Bbb{Z}_2$, and so admits precisely one
selfduality, which we call $\Upsilon$.

 A $\Gad$-bundle $E$ on $M$ has a characteristic class $\xi(E)$
that takes values in $H^2(M,\CZ)$. For example, if $G_{{\rm
ad}}=SO(3)$, $\xi(E)$ is the second Stieffel-Whitney class
$w_2(E)$.  One can define a partition function $Z_\xi$ for every
choice of $\xi$ by restricting the path integral to bundles with
given $\xi$.  The $Z_\xi$ transform in a unitary representation of
the $S$-duality group \refs{\vafawitten,\witfur}, but this is
something that we will not explain here.

Instead, we concentrate on the Hamiltonian description, and
specialize to $M=S^1\times W$ for a three-manifold $W$. In this
case, we have $H^2(M,\CZ)=H^1(W,\CZ)\oplus H^2(W,\CZ)$, so
$\xi(E)$ can be decomposed as $\xi={\bf a}\oplus {\bf m}$, with
${\bf a}\in H^1(W,\CZ)$, ${\bf m}\in H^2(W,\CZ)$.

Consider the pairing \eqn\junko{\hat \Upsilon:H^2(M,\CZ)\times
H^2(M,\CZ)\to H^4(M,U(1))=U(1)} obtained by composing the cup
product $H^2(M,\CZ)\times H^2(M,\CZ)\to H^4(M,\CZ\times\CZ)$ with
the map $\Upsilon:\CZ\times\CZ\to U(1)$. According to Poincar\'e
duality, $\hat\Upsilon$ is a perfect pairing, making $H^2(M,\CZ)$
a selfdual abelian group. More generally, for a closed oriented
manifold $Y$ of dimension $n$, Poincar\'e duality together with
selfduality of $\CZ$ gives perfect pairings
\eqn\unko{H^d(Y,\CZ)\times H^{n-d}(Y,\CZ)\to U(1), ~~0\leq d\leq
n.} In particular, we have the perfect pairing
\eqn\normo{\bar\Upsilon:H^1(W,\CZ)\times H^2(W,\CZ)\to U(1).} So
these are dual abelian groups, and in particular
\eqn\junok{H^2(W,\CZ)={\rm Hom}(H^1(W,\CZ),U(1)).}

Let us now recall how $S$-duality is implemented in a Hamiltonian
framework. The partition function of the gauge theory with ${\bf
m}$ specified has a natural Hilbert space interpretation, because
${\bf m}$ is part of the data at an initial time and is
independent of time. However, because ${\bf a}$ cannot be
expressed in terms of the data at a fixed time, the partition
function with fixed ${\bf a}$ cannot be given a Hamiltonian
interpretation.  Instead  one must introduce a character ${\bf e}$
of the finite abelian group $H^1(M,\CZ)$, that is, a homomorphism
\eqn\kilno{{\bf e}:H^1(W,\CZ)\to U(1).} For each choice of ${\bf
e}$ and ${\bf m}$, the sum $\sum_{{\bf a}}{\bf e}({\bf a})Z_{{\bf
a},{\bf m}}$ can be interpreted in terms of a trace in a Hilbert
space ${\cal H}_{{\bf e},{\bf m}}$. This sum has a Hilbert space
interpretation because summing over ${\bf a}$ with the weight
factor ${\bf e}({\bf a})$ is compatible with cutting and pasting
(or what  in quantum field theory is usually called cluster
decomposition) in the $S^1$ direction. ${\bf e}$ and ${\bf m}$ are
called respectively the discrete electric and magnetic flux
\thooft. Because of \junok, $\bf e$ can be alternatively viewed as
an element of $H^2(W,\CZ)$, just like $\bf m$.

So we get a Hilbert space ${\cal H}_{{\bf m},{\bf e}}$ for each
choice of \eqn\ytro{\eqalign{{\bf e}&\in {\bf E}={\rm
Hom}(H^1(W,\CZ), U(1))\cr
                             {\bf m}& \in {\bf M}=H^2(W,\CZ).\cr}}
Happily, the finite groups ${\bf E}$ and ${\bf M}$ are isomorphic
according to \junok. This makes it possible to have $S$-duality
symmetry. The duality transformation $S$ exchanges ${\bf E}$ and
${\bf M}$ and maps $({\bf e},{\bf m})\to (-{\bf m},{\bf e})$:
\eqn\coco{S:{\cal H}_{{\bf e},{\bf m}}\to {\cal H}_{-{\bf m},{\bf
e}}.} It also, of course, maps $\tau\to -1/\tau$.  (Eqn. \coco\ is
a discrete analog of \xilj, with $(\vec{\eurm m},\vec{\eurm n})$
replaced by their discrete counterparts $({\bf m},{\bf e})$.)

While it is possible to define all the Hilbert spaces ${\cal
H}_{{\bf e},{\bf m}}$, a given gauge theory construction may not
use all of them.  If we do gauge theory on $M=S^1\times W$ with
simply-connected gauge group $\bar G$, then we must set to zero
the characteristic classes ${\bf a}$ and ${\bf m}$ that enter in
the partition function $Z_{{\bf a},{\bf m}}(M)$. For the Hilbert
space, this means that ${\bf e}$, which arises by a Fourier
transform with respect to ${\bf a}$, is arbitrary, while ${\bf
m}=0$.  Hence \eqn\ytoto{{\cal H}(\bar G,W)=\oplus_{{\bf e}}{\cal
H}_{{\bf e},0}.} Taking into account the assumed transformation
law for ${\bf e},{\bf m}$ under $S$-duality, we see that the
Hilbert space of the dual theory must be \eqn\momjo{ \oplus_{{\bf
m}}{\cal H}_{0,{\bf m}}. } How is this Hilbert space related to a
path integral on $M=S^1\times W$? Setting $\bf e$ to zero means
summing over $\bf a$. So in terms of the variables ${\bf a},{\bf
m}$, \momjo\ means one must sum over all ${\bf a},{\bf m}$ with
equal weight. In other words, in the dual theory we sum over
$^L\Gad$-bundles with all possible $\xi$. Thus the dual theory has
gauge group $^L\Gad$, as expected.

\bigskip\noindent{\it The Universal Bundle}

We need another piece of background, which is the concept of the
universal bundle.

Consider first the moduli space $\M(G,C)$ of $G$-bundles over a
Riemann surface $C$.  For every point $p\in \M(G,C)$, there is a
corresponding $G$-bundle $E_p\to C$ parametrized by $p$; it is
determined up to isomorphism.  A universal bundle, if it exists,
is a $G$-bundle ${\EUE} \to \M\times C$ such that for any $p\in
\M$, ${\EUE}$ restricted to $p\times C$ is isomorphic to $E_p$.
Gauge theory gives \abott\ a very direct method to analyze the
universal bundle, though we will not explain this here.

Naively, we just define ${\EUE}$ so that its restriction to
$p\times C$ is ``the'' bundle over $C$ that is determined by $p$.
The reason that a problem arises is that, to begin with, $p$ only
determines an isomorphism class of $G$-bundle over $C$, not an
actual bundle. This causes no difficulty locally\foot{This statement holds
near a smooth point in $\M$.  A more precise
analysis than we will give shows that the class
$\zeta(\EUE_{\rm ad})$ that obstructs the universal bundle can have a local contribution
at singularities.};
we just pick a
bundle in the right isomorphism class at a particular $p$ and then
deform it in a small neighborhood. So for each small open set
$U_i\subset \M$, we can pick a universal bundle ${\EUE}_i\to
U_i\times C$, and moreover ${\EUE}_i$ is unique up to isomorphism.

The next step is to glue together the ${\EUE}_i$ over $ (U_i\cap
U_j)\times C$.  The bundles ${\EUE}_i$ and ${\EUE}_j$ are
isomorphic over $ (U_i\cap U_j)\times C$, so we pick an
isomorphism $\Theta_{ij}:{\EUE}_i\cong {\EUE}_j$ over this
intersection, with $\Theta_{ji}=\Theta_{ij}^{-1}$.

Now if on triple intersections $(U_i\cap U_j\cap U_k)\times C$,
the composition $\Theta_{ki}\Theta_{jk}\Theta_{ij}$ is equal to 1,
then we can consistently glue together the ${\EUE}_i$ to get the
desired universal bundle ${\EUE}$.

If the gauge group is the adjoint group $G_{{\rm ad}}$, there is
no problem in the gluing. A generic $G_{{\rm ad}}$ bundle has no
automorphisms (exceptions occur at singularities of $\M$). So
$\Theta_{ki}\Theta_{jk}\Theta_{ij}$, just because it is an
isomorphism of ${\EUE}_i$ (restricted to $(U_i\cap U_j\cap
U_k)\times C$) to itself, is the identity at a generic point of
the triple intersection and hence everywhere.

Thus, the universal   $G_{{\rm ad}}$-bundle ${\EUE}_{\rm {ad}}$
does exist. Now let us repeat this discussion, taking the gauge
group to be the simply-connected cover $\bar G$ (or more
generally, any cover of $G_{\rm{ad}}$ with a non-trivial center).
We start in the same way. For each small open set $U_i\in \M$, we
pick a universal bundle $\bar {\EUE}_i$ over $U_i\times C$, and
try to glue to make a universal $\bar G$-bundle $\bar {\EUE}$.

The reason that the result is different is that a generic $\bar
G$-bundle does have a non-trivial group of automorphisms; the
center $\CZ$ is a group of automorphisms of any $\bar G$-bundle
$E$.

Hence, in the above argument, the composite map
$\Theta_{ki}\Theta_{jk}\Theta_{ij}$ over a triple intersection is
not necessarily the identity; instead \eqn\yrf{
\Theta_{ki}\Theta_{jk}\Theta_{ij}=f_{ijk},   } with
$f_{ijk}\in\CZ$. The $f_{ijk}$ combine to a two-cycle defining an
element $\zeta\in H^2(\M,\CZ)$. The element $\zeta$ is known to be
non-trivial; it is the obstruction to the existence of the
universal $\bar G$-bundle $\bar{\EUE}$.

There is another way to explain the existence of $\zeta$.  Let
$\M(G_{\rm{ad}},C)_0$
  be the component of $\M(G_{\rm{ad}},C)$ that
parametrizes topologically trivial bundles -- the ones that when
restricted to $p\times C$, for $p\in \M(G_{\rm{ad}},C)$, can be
lifted to $\bar G$-bundles.    Consider the universal bundle
${\EUE}_{\rm{ad}}$ over   $\M(G_{\rm{ad}},C)_0\times C$.  If we
could lift ${\EUE}$ to a $\bar G$-bundle $\bar {\EUE}$, this
(after being pulled back to $\M(\bar G,C)\times C$, which is a
finite cover of $\M(G_{\rm {ad}},C)_0\times C$) would be the
universal $\bar G$-bundle. In general, the obstruction to lifting
a $G_{\rm ad}$-bundle ${\EUE}_{\rm{ad}}$ to a $\bar G$-bundle
$\bar {\EUE}$ is the characteristic class $\xi({\EUE}_{\rm{ad}})$.
So the obstruction to existence of the universal $\bar G$-bundle
is $\zeta=\xi({\EUE}_{\rm {ad}})$.

There is a similar story for Higgs bundles.  The universal Higgs
bundle is a pair $( {\EUE},\hat \varphi)$,
 with $ {\EUE} $ being a $G$-bundle over
$ \MH(G,C)\times C$, and $\hat\varphi\in H^0( \MH(G,C)\times
C,{\rm {ad}}({\EUE})\otimes K_C)$, obeying the following
condition. For each $p\in \MH(G,C)$, the restriction of $(
{\EUE},\hat\varphi)$ to $p\times C$ should be isomorphic to the
Hitchin pair $(E_p,\varphi_p)$ parametrized by $p$.

The arguments that we have already given can be carried over with
no essential change to show that for gauge group $G_{\rm{ad}}$,
the universal Higgs bundle $( {\EUE}_{\rm {ad}},\hat\varphi)$ does
exist. But for the simply-connected gauge group $\bar G$ (or any
nontrivial cover of $G_{\rm {ad}}$), the universal Higgs bundle
does not exist. It is obstructed by the fact that a generic
Hitchin pair $(E,\varphi)$, with $E$  a $\bar G$-bundle, has the
group $\CZ$ of automorphisms. The obstruction is just the
obstruction to lifting the ``bundle'' part of the universal
$G_{{\rm ad}}$ Higgs bundle $({\EUE}_{\rm{ad}},\hat\varphi)$ to a
$\bar G$-bundle. It is therefore
\eqn\plooo{\zeta=\xi({\EUE}_{\rm{ad}}).}

\bigskip\noindent{\it The Universal Bundle As A Twisted Vector
Bundle}

Although for $G$ not of adjoint type, the universal $G$-bundle
does not exist as a vector bundle or as a principal bundle, it
does exist as a twisted vector bundle.  We will describe this
concept in a very pedestrian way.

 One way to construct an ordinary vector
bundle $V$ of rank $N$ over a manifold $X$ is to cover $X$ with
small open sets $U_i$, on each of which we pick a rank $N$ trivial
bundle $V_i\cong U_i\times \Bbb{C}^N$. Then we glue $V_i$ to $V_j$
on the intersection $U_i\cap U_j$ via a gluing map $v_{ij}:V_i\cap
V_j\to U(N)$, with $v_{ji}=v_{ij}^{-1}$. If on triple
intersections we have \eqn\nofto{ v_{ki}v_{jk}v_{ij}=1,} then the
$V_i$ can be glued together consistently to make a rank $N$ vector
bundle $V\to X$.

Now suppose that we are given $b\in H^2(X,U(1))$.  Then $b$ can be
represented explicitly by a $U(1)$-valued cocycle $b_{ijk}$
defined on triple intersections $U_i\cap U_j\cap U_k$. A twisted
vector bundle is defined by the same sort of data $v_{ij}:V_i\cap
V_j\to U(N)$.  But now, instead of \nofto, we ask for
\eqn\hofto{v_{ki}v_{jk}v_{ij}=b_{ijk}.} Thus, a $b$-twisted vector
bundle $V$ is not a vector bundle in the usual sense.  However,
the associated adjoint bundle ${\rm ad}(V)$ is an ordinary vector
bundle, since the phase $b_{ijk}$ in \hofto\ disappears if we pass
to the adjoint representation.

Now let us return to the problem of constructing a universal
bundle.  In this paper, we are generally a little imprecise about
whether by a $G$-bundle, we mean a principal $G$-bundle, whose
fiber is a copy of $G$, or an associated vector bundle in some
faithful  representation of $G$.  Principal bundles make possible
a uniform analysis good for any $G$, but for a group like $U(N)$
that has a convenient faithful representation (the $N$-dimensional
representation) it is useful to think in terms of vector bundles.

In discussing the universal bundle, it is helpful to be more
precise.  We interpret the transition functions $\Theta_{ij}$ of
our discussion above as $\bar G$-valued functions, with no
particular choice of representation.  Their projection to $G_{ad}$
gives us transition functions for a universal principal
$G_{ad}$-bundle ${\EUE}_{\rm ad}$, but we cannot lift this to a
principal $\bar G$-bundle, because of the relation
\eqn\flutter{\Theta_{ki}\Theta_{jk}\Theta_{ij}=f_{ijk},} where the
$f_{ijk}$ define the class $\zeta=\xi({\EUE}_{ad})\in
H^2(\MH(G_{\rm ad},C),\CZ)$. Now let us pick an irreducible
representation $\varrho$ of $\bar G$.  We write
$\Theta_{ij}^\varrho$ for the transition functions $\Theta_{ij}$
evaluated in the representation $\varrho$. Likewise we write
$\varrho(f)$ for $f$ evaluated in the representation $\varrho$. In
the irreducible representation $\varrho,$ the center of $\bar G$
acts by scalar multiplication, so $\varrho(f)$ takes values in
$U(1)$.  We have
\eqn\glutter{\Theta^\varrho_{ki}\Theta^\varrho_{jk}\Theta^\varrho_{ij}=\varrho(f_{ijk}).}
The quantities $\varrho(f_{ijk})$ are a cocycle defining the
element $\varrho(\zeta)\in H^2(\MH(G_{\rm ad},C),U(1))$.

If $\varrho(f)=1$, the objects $\Theta^\varrho$ are transition
functions that define a vector bundle ${\EUE}_\varrho\to \MH(\bar
G,C)\times C$ that we may call the universal bundle in the representation
$\varrho$.  In general this is not the case.  However, comparing
eqns. \flutter\ and \hofto, we see that while ${\EUE}_\varrho$ may
not exist as an ordinary vector bundle, it does always exist as a
twisted vector bundle, twisted by $\varrho(\zeta)\in
H^2(\MH(G_{\rm ad},C),U(1))$.

\subsec{Compactification To Two Dimensions} \subseclab\comptwop

Now we compactify to two dimensions. Consider ${\EUN}=4$ super
Yang-Mills theory on $M=\Sigma\times C$ with gauge group
$G_{\rm{ad}}$. The data in this theory are a $G_{\rm{ad}}$-bundle
$E\to\Sigma\times C$, a connnection $A$ on $E$, and various other
fields.  Any choice of $E$ determines a class $\xi(E)\in
H^2(\Sigma\times C,\CZ)$.  Since $H^2(\Sigma\times
C,\CZ)=H^2(\Sigma,\CZ)\oplus H^1(\Sigma,H^1(C,\CZ))\oplus
H^2(C,\CZ),$ we have \eqn\ytro{\xi(E)\in H^2(\Sigma,\CZ)\oplus
H^1(\Sigma,H^1(C,\CZ))\oplus H^2(C,\CZ).} Relative to this
decomposition we write
$\xi(E)=\xi(E)^{2,0}+\xi(E)^{1,1}+\xi(E)^{0,2}$.

We can describe the same thing in the low energy effective
sigma-model.  Here we consider low energy data on $\Sigma\times
C$, which when restricted to $p\times C$ for a point $p\in \Sigma$
determine a point in $\MHAD$.  As $p$ varies, we get a map
$\Phi:\Sigma\to \MHAD$.   Composing $\Phi$ with the identity map
on $C$, we get a map $\Phi\times 1:\Sigma\times C\to \MHAD\times
C$. Now recall the universal bundle ${\EUE}_{\rm ad}$ and the
class $\zeta=\xi({\EUE}_{\rm {ad}})\in H^2( \MHAD\times C,\CZ)$.
Pulling it back to $\Sigma\times C$, we get a class $(\Phi\times
1)^*(\zeta)\in H^2(\Sigma\times C,\CZ)$. By chasing through the
definitions, one can see that \eqn\yrog{\xi(E)=(\Phi\times
1)^*(\zeta).} In fact, the universal bundle ${\EUE}_{ad}$ is
defined so that $(\Phi\times 1)^*({\EUE}_{{\rm ad}})=E$, so
$(\Phi\times 1)^*(\zeta)=(\Phi\times 1)^*(\xi(\EUE_{{\rm ad}}))
=\xi((\Phi\times 1)^*(\EUE_{{\rm ad}}))=\xi(E)$.

Let us further specialize to $\Sigma=S^1\times \tilde S^1$, where
the two factors are circles. We write $W=\tilde S^1\times C$. So
$\Sigma\times C=S^1\times W$. $\xi(E)$ has the familiar expansion
$\xi(E)={\bf a}+{\bf m}$ with ${\bf a}\in H^1(W,\CZ)$, ${\bf m}\in
H^2(W,\CZ)$. Now, however, we can further expand
$H^1(W,\CZ)=H^1(\tilde S^1,\CZ)\oplus H^1(C,\CZ)$, so we have
${\bf a}={\bf a}_0+{\bf a}_1$ with \eqn\mimot{\eqalign{{\bf
a}_0&\in H^1(\tilde S^1,\CZ) =\CZ,\cr {\bf a}_1& \in
H^1(C,\CZ).\cr}} And similarly, we can expand
$H^2(W,\CZ)=H^1(\tilde S^1,H^1(C,\CZ))\oplus H^2(C,\CZ)$, so we
have ${\bf m}={\bf m}_0+{\bf m}_1$ with \eqn\nimot{\eqalign{{\bf
m}_0&\in H^2(C,\CZ)=\CZ \cr
                     {\bf m}_1&\in
                     H^1(\tilde S^1,H^1(C,\CZ))=H^1(C,\CZ).\cr}}
Comparing back to the expansion \ytro, we see that
\eqn\zimot{\xi(E)^{2,0}={\bf a}_0,~\xi(E)^{0,2}={\bf m}_0.}

We can similarly decompose the character $\bf e$ that is dual to
$\bf a$. Clearly, we have ${\bf e}={\bf e}_0+{\bf e}_1$, where
${\bf e}_0$ is a character of ${\bf a}_0\in \CZ$ and ${\bf e}_1$
is a character of ${\bf a}_1\in H^1(C,\CZ)$.  Hence, using
 the selfduality of $\CZ$ and Poincar\'e duality, \eqn\tew{\eqalign{ {\bf e}_0 & \in
\CZ^\vee=\CZ\cr {\bf e}_1&\in H^1(C,\CZ)^\vee=H^1(C,\CZ).\cr}} In
view of \coco, the transformation under $S$-duality is
\eqn\niko{\eqalign{({\bf e}_0,{\bf m}_0)&\to (-{\bf m}_0,{\bf
e}_0)\cr ({\bf e}_1,{\bf m}_1)&\to (-{\bf m}_1,{\bf e}_1).\cr}}

Our next goal is to interpret ${\bf e}_0$, ${\bf e}_1$, ${\bf
m}_0$, and ${\bf m}_1$ in the effective two-dimensional
sigma-model with target $\MHAD$.  We will consider both the case
of a closed Riemann surface $\Sigma$, and the case when $\Sigma$
has a boundary. The latter case will enable us to understand the
implications of \niko\ for the geometric Langlands program.

\bigskip\noindent{\it Interpretation Of ${\bf m}_0$}

The easiest to interpret is ${\bf m}_0=\xi(E)^{0,2}$.  The target
space $\MH(G_{\rm ad},C)$ of our sigma-model is not connected. Its
components are labeled by the topological type of the $G_{\rm
{ad}}$-bundle $E\to C$.  But this is exactly what is measured by
${\bf m}_0$.

\bigskip\noindent{\it Interpretation Of ${\bf e}_0$}

Consider in general a sigma-model of maps $\Phi:\Sigma\to X$, for
some $X$.  A flat $B$-field is an element  $b\in H^2(X,U(1))$. A
flat $B$-field is incorporated in the sigma-model path integral as
follows, for the case that $\Sigma$ has no boundary.  Given a map
$\Phi:\Sigma\to X$, one pulls back $b$ to $\Phi^*(b)\in
H^2(\Sigma,U(1))=U(1)$, and then one includes in the path integral
a factor of $\Phi^*(b)$.  Thus, in this situation, incorporating
the flat $B$-field has the effect of weighting by phases the
different components of maps of $\Sigma$ to $X$.

Now ${\bf e}_0$ determines a flat $B$-field in the sigma-model of
maps $\Phi:\Sigma\to \MH(G_{\rm ad},C)$. Indeed, there is as we
have explained a natural class $\zeta\in H^2(\MH(G_{\rm
ad},C),\CZ)$, which expresses the  obstruction to lifting the
universal $G_{\rm{ad}}$ bundle  ${\EUE}_{\rm {ad}}\to\MH(G_{\rm
ad},C)\times C$ to a $\bar G$-bundle.  By composing $\zeta$ with ${\bf
e}_0:\CZ\to U(1)$, we get a flat $B$-field $b_{{\bf e}_0}={\bf
e}_0(\zeta)\in H^2(\MH(G_{\rm{ad}},C),U(1))$.  We claim that the
role of ${\bf e}_0$ in the effective sigma-model of maps
$\Sigma\to \MHAD$ is precisely to weight every map in the way that
one would expect for a sigma-model with the flat $B$-field
$b_{{\bf e}_0}$.

In fact, the definition of ${\bf e}_0$ is that the path integral
includes a phase factor ${\bf e}_0({\bf a}_0)={\bf
e}_0(\xi(E)^{2,0})$.   But $\xi(E)^{2,0}=\Phi^*(\zeta)$; this
follows upon restricting \yrog\ to a point in $C$.
So the phase factor induced in the path integral by
${\bf e}_0$ is ${\bf e}_0(\Phi^*(\zeta))=\Phi^*({\bf
e}_0(\zeta))=\Phi^*(b_{{\bf e}_0})$.  This justifies  our claim
that the effect of ${\bf e}_0$ in the sigma-model is to generate a
flat $B$-field $b_{{\bf e}_0}$.

\niko\ therefore means that $S$-duality exchanges the topological
class ${\bf m}_0$ of a flat $G_{\rm{ad}}$-bundle with the flat
$B$-field determined by ${\bf e}_0$.

\bigskip\noindent{\it Incorporation Of Branes}

For applications to the geometric Langlands program, the real
payoff is to understand the implications of all this for branes.

So we take $\Sigma$ to be a Riemann surface with boundary.
$M=\Sigma\times C$ is  therefore a four-manifold with boundary
$\partial M=\partial\Sigma\times C$.  On $\partial M$, we place
some supersymmetric boundary condition.

The effective two-dimensional description is a sigma-model of maps
$\Phi:\Sigma\to \MHAD$, with boundary condition corresponding to
some brane.  We would like to understand the role of ${\bf e}_0$
and ${\bf m}_0$ in this description.

There is little new to say about ${\bf m}_0$.  It labels the
components of $\MHAD$, whether $\Sigma$ has a boundary or not.

The role of ${\bf e}_0$ is more subtle.  As we have seen, the
sigma-model with target $\MHAD$ is endowed with a flat $B$-field
${\bf e}_0(\zeta)$. A flat $B$-field has a very interesting effect
on branes \ref\witkth{E. Witten, ``$D$-Branes And $K$-Theory,''
JHEP 9812:019,1998, arXiv:hep-th/9810188.}.  In the absence of a
$B$-field, a brane on $\MHAD$ has a Chan-Paton bundle, which is a
vector bundle over $\MHAD$  (or more generally a sheaf, perhaps
supported on a submanifold, that defines a $K$-theory class of
$\MHAD$). However, in the presence of a flat $B$-field, associated
with an element $b\in H^2(\MHAD,U(1))$, the Chan-Paton bundle
becomes a twisted vector bundle (or more generally a twisted sheaf
related to an element of the twisted $K$-theory of $\MHAD$),
twisted by $b$ in the sense of eqn. \hofto.  It is because of this
that we introduced the concept of a twisted vector bundle.

So in short, for ${\bf e}_0\not=0$,  the Chan-Paton bundle of a
brane is a twisted vector bundle, twisted by  $b_{{\bf e}_0}={\bf
e}_0(\zeta)$.  Luckily, from the analysis at the end of section
\reviewthis, we have a plentiful supply of such twisted vector
bundles.  If $\varrho$ is any irreducible representation of $\bar
G$ such that the character of the center of $\bar G$ defined by
$\varrho$ is equal to ${\bf e}_0$, then the universal bundle
${\EUE}_\varrho$ in the representation $\varrho$ is an example of
a twisted vector bundle for the flat $B$-field $b_{{\bf e}_0}$.

Other important examples of twisted branes can be constructed by
picking  a submanifold  $Y\subset \MHAD$ such that $b_{{\bf e}_0}$
is trivial when restricted to $Y$.  In this case, the definition
of a brane supported on $Y$ is independent of ${\bf e}_0$ (up to a
not quite canonical isomorphism).  For example, $Y$ could be a
point in $\MHAD$. Certainly $b_{{\bf e}_0}$ is trivial when
restricted to a point, so zerobranes exist for any ${\bf e}_0$.
{}From such zerobranes, we can form electric eigenbranes, as we
explain in section \electricbranes. A slightly more subtle example
is a magnetic eigenbrane, a brane of rank one supported on a fiber
${\CMF}$ of the Hitchin fibration. The Chan-Paton bundle of such a
brane should be a flat\foot{One can here interpret flatness to
mean, just as for ordinary line bundles, that the transition
functions are constants.} twisted line bundle. In \thadhau, it is
shown that the flat $b$-field $b_{{\bf e}_0}$ is trivial when
restricted to ${\CMF}$, as a result of which the space of flat
twisted line bundles is isomorphic, but not canonically
isomorphic, to the space of ordinary flat line bundles on
${\CMF}$.

We can now deduce from $S$-duality a statement about branes. The
duality transformation $S$ maps $({\bf e}_0,{\bf m}_0)\to (-{\bf
m}_0,{\bf e}_0)$, so it exchanges the topology of the component of
$\MH$ on which a brane is supported with the flat $B$-field
$b_{{\bf e}_0}$ by which its Chan-Paton bundle is twisted. It
also, of course, exchanges $G_{\rm ad}$ with\foot{By
$^L\negthinspace G_{\rm ad}$, we mean the adjoint form of the
group $^L\negthinspace G$. The statement we are describing here is
best expressed in terms of adjoint bundles on both sides to allow
all possible topologies. Momentarily we indicate explicitly
whether we want a given component of $\MH$ or its universal
cover.} $^L\negthinspace G_{{\rm ad}}$, and (as we explain in the
concluding remark of this section), exchanges $\MH(G_{{\rm ad}})$
with $\hat\M_H({}^L\negthinspace G_{{\rm ad}})$, which we define
to be the universal cover of $\MH({}^L\negthinspace G_{{\rm
ad}})$.

As a special case of this duality, a point on one side, contained
in a fiber $\CMF$ of the Hitchin fibration, is mapped on the other
side to a brane of rank one supported on the corresponding fiber
$^L\negthinspace\CMF=\Xi(\CMF)$ of the Hitchin fibration, and
endowed with a flat twisted line bundle. $\CMF$ is a union of
complex tori, labeled by the characteristic class ${\bf
m}_0=\xi(E)$ of the Higgs bundle. The choice of a component of
$\CMF$ on one side determines on the other side  the discrete
electric field ${\bf e}_0$ and hence the twist. The choice of a
point on $\CMF$ determines a flat twisted line bundle on
$^L\negthinspace\CMF$ (to which it maps under the duality
transformation $S$). Of course, this relationship between $\CMF$
and $^L\negthinspace\CMF$ is reciprocal.  This twisted duality
between $\CMF$ and $^L\negthinspace\CMF$ is in fact one of the
main results of Hausel and Thaddeus \thadhau.

\bigskip\noindent{\it Interpretation Of ${\bf e}_1$ And ${\bf
m}_1$}

\def\MHBARG{\MH(\bar G)}

Finally, let us discuss the interpretation of ${\bf e}_1$ and
${\bf m}_1$ in two-dimensional terms. It will be helpful to begin
by comparing the theories with gauge groups $\bar G$ and
$G_{\rm{ad}}$.

$\MHBARG$ is simply-connected.  It has a natural group of
symmetries ${\bf E}_C=H^1(C,\CZ)$.  Indeed, ${\bf E}_C$
parametrizes $\CZ$-bundles over $C$.  A $\bar G$ Higgs bundle
$(E,\varphi)$ can be tensored with a $\CZ$-bundle to make a new
$\bar G$ Higgs bundle.  (Concretely, this operation multiplies the
holonomies of $E$ around noncontractible loops in $C$ by elements
of the center of $\bar G$.)  So this gives a group ${\bf E}_C$ of
symmetries of the sigma-model with target $\MHBARG$. The Hilbert
space of this sigma-model can be decomposed in characters of ${\bf
E}_C$.

On the other hand, $\MHAD$ has no such geometrical symmetries. But
it has a fundamental group ${\bf M}_C=H^1(C,\CZ)$, which is
isomorphic to ${\bf E}_C$. To understand where this fundamental
group comes from, a shortcut is to note that one component
$\MHAD_0$ of $\MHAD$, namely the component that parametrizes Higgs
bundles that can be lifted to $\bar G$, is simply
$\MHAD_0=\MHBARG/{\bf E}_C$.  Dividing by ${\bf E}_C$ eliminates
the geometrical symmetries of $\MHBARG$, but of course it creates
a fundamental group.\foot{ ${\bf E}_C$ does not act freely, so $\M_H(G_{ad})$
 has orbifold singularities.  (It also has more severe singularities from reducible
 Higgs bundles.)
As is familiar in
sigma models, the fundamental group of $\M_H(G_{ad})$ must be understood in
an orbifold
sense.  Alternatively, one can rely on the four-dimensional gauge theory
instead of reducing
to the sigma model with its singularities.} So $\pi_1(\MHAD_0)\cong {\bf E}_C$.

Actually, the fundamental group is the same for any component of
$\MHAD$.  $\MHAD$ is defined by dividing the space of all
solutions of Hitchin's equations, for gauge group $G_{\rm {ad}}$,
by the group ${\cal G}_{\rm{ad}}(C)$ of all $G_{\rm{ad}}$-valued
gauge transformations on $C$. If one were to divide only by the
connected component of ${\cal G}_{\rm{ad}}(C)$, one would get the
universal cover $\hat\M_H(G_{\rm {ad}},C)$.  The fundamental group
of $\MHAD$ is therefore the group of components of ${\cal G}_{\rm
ad}(C)$, and this is ${\bf E}_C=H^1(C,\CZ)$, for any component of
$\MHAD$.

In sum, strings moving on $\MHAD$ have a discrete group of
conserved winding numbers ${\bf M}_C=\pi_1(\MHAD)=H^1(C,\CZ)$.
Likewise strings moving on $\MHBARG$ have a discrete group of
conserved momenta ${\bf E}_C=H^1(C,\CZ)$.

Let us compare the symmetries ${\bf E}_C$ and ${\bf M}_C$ to what
we can see in the underlying gauge theory. In $G_{\rm{ad}}$ gauge
theory on $M=\Sigma\times C=S^1\times \tilde S^1\times C$, the
bundle $E$ and other data determine a map $\Phi:S^1\times \tilde
S^1\to \MHAD$. In analyzing the topology of this situation, we
expanded the characteristic class $\xi(E)$ as $\xi(E)={\bf
a}\oplus {\bf m}$, where ${\bf m}$ is the restriction of $\xi(E)$
to $\tilde S^1\times C$ (more precisely, to $p\times \tilde
S^1\times C$, for a point $p\in S^1$). In the low energy
sigma-model, ${\bf a}$ and ${\bf m}$ are invariants of
$\Phi:\Sigma\to \MHAD$. In particular, ${\bf m}$ only depends on
the restriction $\Phi|_{\tilde S^1}$ of $\Phi$ to $\tilde S^1$.

In \nimot, we further expanded ${\bf m}={\bf m}_0\oplus {\bf m}_1$
with ${\bf m}_1\in H^1(\tilde S^1,H^1(C,\CZ))$.  ${\bf m}_1$ is a
topological invariant of $\Phi|_{\tilde S^1}$ that  vanishes for
constant maps of $\tilde S^1$ to $\MHAD$. So it measures the
homotopy class of the map $\Phi|_{\tilde S^1}$ in
$\pi_1(\MHAD)=H^1(C,\CZ)$.

Similarly, we could exchange the role of $S^1$ and $\tilde S^1$.
The restriction of $\xi(E)$ to $S^1 \times C$ (that is, to
$S^1\times q\times C$, for a point $q\in \tilde S^1$), is in the
above notation ${\bf m}_0\oplus {\bf a}_1$. In particular, ${\bf
a}_1\in H^1(S^1,H^1(C,\CZ))$ measures the winding of $\Phi$ in the
$S^1$ or ``time'' direction. The character ${\bf e}_1$ which is
dual to ${\bf a}_1$ therefore measures the conserved momentum of
the strings.

\bigskip\noindent{\it An Example}

For an important illustration of all this, consider the Langlands
dual pair $G_{\rm{ad}}$ and $^L\negthinspace\bar G$.  In
$G_{\rm{ad}}$ gauge theory, we set ${\bf e}=0$ and consider a
sigma-model with target $\MHAD$.  In this model, there are no
conserved momenta, but there is a symmetry ${\bf M}_C$ of string
windings.

In the dual picture, the gauge group is $^L\negthinspace\bar G$,
we set ${\bf m}=0$, and the sigma-model with target $\MHBARG$ has
no conserved string windings, but a symmetry group ${\bf E}_C$ of
discrete conserved momenta.

$S$-duality or Montonen-Olive or Langlands duality exchanges the
two pictures, exchanging ${\bf E}_C({}^L\negthinspace\bar G)$ with
${\bf M}_C(G_{\rm{ad}})$. The fact that the duality exchanges the
discrete conserved momenta and windings of strings is an aspect of
its relation to $T$-duality in two dimensions.

\bigskip\noindent{\it A Concluding Comment}

More generally, we can  simply specify ${\bf m}_0$ and ${\bf e}_0$
as we please. Then we consider branes on a component of $\MHAD$
labeled topologically by ${\bf m}_0$; the Chan-Paton bundles of
the branes are twisted by ${\bf e}_0$.

We still have two ways to proceed with the quantization.  If we
divide by all $G_{\rm {ad}}$-valued gauge transformations, then
the target space is the component of $\MH(G_{\rm ad},C)$ labeled
by ${\bf m}_0$.  In this case, there is a finite group ${\bf M}_1$
that classifies the string winding numbers, but there is no group
${\bf E}_1$ of geometrical symmetries.  If we divide by only the
connected gauge transformations, then the target is the universal
cover $\hat\M_H(G_{\rm ad},C)$.  In this case, there is a group
${\bf E}_1$ of geometrical symmetries, but no group ${\bf M}_1$ of
string winding numbers.  $S$-duality exchanges ${\bf E}_1$ and
${\bf M}_1$, so (in addition to exchanging $G$ and
$^L\negthinspace G$) it exchanges the two methods of quantization.

In particular, when we apply $S$-duality to branes of specified
${\bf e}_0$ and ${\bf m}_0$, we must exchange the two methods of
quantization in addition to exchanging the two adjoint groups
$G_{\rm ad}$ and $^L\negthinspace G_{\rm ad}$.

\newsec{Electric Eigenbranes}

\seclab\electricbranes

In section \loopop, we introduced the topological Wilson and 't
Hooft operators, their action on branes, and the concept of an
electric or magnetic eigenbrane. Our goal in section \wilbranes\
will be to understand explicitly how Wilson operators act on
branes.  Then in section \zerobelectric\ we will use this
information to identify zerobranes as electric eigenbranes.

\subsec{How Wilson Operators Act On Branes}

\subseclab\wilbranes

As always in this paper, the theory of interest is  twisted ${\cal
N}=4$ super Yang-Mills theory, with gauge group $G$, on a
four-manifold $M$. The fields include a connection $A$ on a
$G$-bundle $E\to M$. We assume $M$ to have a boundary that is
labeled by a brane ${\cal B}$ that respects the topological
symmetry, a condition that of course depends on the parameter
$\Psi$.

Topological Wilson line operators exist if $\Psi=\infty$, which we
achieve by setting $t=i$ with arbitrary gauge coupling $\tau$. Let
$R$ be a representation of $G$.
 A Wilson operator in the representation $R$ and supported
 on a curve ${{\cal S}}$ contributes in the
 path integral a factor \eqn\yrgy{W_{{\cal S}}(R)=P\exp\left(-\int_{{\cal S}}(A+i\phi)
 \right).}
 More explicitly, write
$E(R)$ for the bundle associated to $E$ in the representation $R$
of $G$.  The Wilson line operator is the matrix of parallel
transport  along ${{\cal S}}$, in the bundle $E(R)$, with the
connection $\CA=A+i\phi$.  The effect of including this operator
is to add an external charge, in the representation $R$, whose
trajectory in spacetime is ${{\cal S}}$.  If ${{\cal S}}$ is a
closed loop, we take the trace of $W_{{\cal S}}(R)$, and if
instead ${{\cal S}}$ goes off to infinity, we combine $W_{{\cal
S}}(R)$ with initial and final states at the ends of ${{\cal S}}$
so as to make a gauge-invariant expression.

Now we specialize to $M=\Sigma\times C$, so as to be able to
reduce the discussion to a two-dimensional sigma-model with target
$\MH(G,C)$. \nref\ouglas{M. R. Douglas, B. Fiol, and C.
Romelsberger,
``Stability And BPS Branes,'' JHEP {\bf 006} (2005) 1-14, arXiv:hep-th/0002037.}%
 \nref\douglas{M. R.
Douglas, ``$D$-Branes, Categories, and ${\EUN}=1$ Supersymmetry,''
J. Math. Phys. {\bf 42} (2001) 2818-2843,
arXiv:hep-th/0011017.}%
 In this  sigma-model, ${\cal
B}$ is represented by a Chan-Paton bundle $U$ (or possibly a more
general sheaf or complex of sheaves \refs{\ouglas,\douglas}) over
the target space $\MH$. $U$ is endowed with a connection $\alpha$.
Since the brane ${\cal B}$ is defined by its Chan-Paton bundle (or
sheaf), a Wilson line operator will have to act on branes by
acting on this bundle or sheaf in some way.

\def\QQ{\eurm Q}

 Let us recall how Chan-Paton bundles enter sigma-models.  We
consider a sigma-model of maps $\Phi:\Sigma\to \MH$, and a brane
${\cal B}$ that is endowed with a Chan-Paton bundle $U\to \MH$
with connection $\alpha$.   The quantum theory is defined by an
integral over the possible maps $\Phi$, along with certain
fermionic variables. One factor in the path integral comes from
the bulk action ${\cmmib I}$ and takes the form
$\exp\left(-\int_\Sigma {{\cmmib I}}\right)$. There also is a
boundary factor that involves parallel transport in the Chan-Paton
bundle. Let $\QQ$ be the part of the boundary of $\Sigma$ that is
labeled by the brane ${\cal B}$, and write $\Phi_\QQ$ for the
restriction of $\Phi$ to $\QQ$. The boundary factor in the path
integral involving $\QQ$ is given by the parallel transport or
holonomy along $\QQ$ of the bundle $\Phi_\QQ^*(U)$:
\eqn\tomfo{P\exp\left(-\int_\QQ \left(\Phi_\QQ^*(\alpha)
+\dots\right)\right).} If $\QQ$ is a closed circle, we take a
trace of this holonomy, and otherwise this factor combines at the
endpoints of $\QQ$ with other factors, depending on the precise
calculation that one chooses to perform, to make a gauge-invariant
expression. The ellipses in \tomfo\ are fermionic corrections to
the connection $\Phi_\QQ^*(\alpha)$ on $\Phi_\QQ^*(U)$.  They are
required by supersymmetry, rather as the shift $A\to {\cal
A}=A+i\phi$ was needed in section \topwil\ to define
supersymmetric Wilson operators.

There is an obvious analogy between the factor \tomfo\ by which
Chan-Paton bundles enter in sigma-models and the factor \yrgy\ by
which a Wilson line operator influences the underlying
four-dimensional gauge theory.  The analogy is even closer because
to define the action of the Wilson operator on the brane ${\cal
B}$, we must take the limit as ${{\cal S}}$ (or rather its
projection from $\Sigma\times C$ to $\Sigma$) approaches $\QQ$.

To get something precise from this analogy, we begin with the
following observation. When gauge theory on a $G$-bundle $E\to
M=\Sigma\times C$ is described in terms of a map $\Phi:\Sigma\to
\MH$,  the $G$-bundle $E$  can be identified as $(\Phi\times
1)^*({\EUE})$, where ${\EUE}$ is the ``bundle'' part of the
universal Higgs bundle $({\EUE},\hat\varphi)$ over $\MH$, and
$(\Phi\times 1)^*({\EUE})$ is its pullback via the map $\Phi\times
1:\Sigma\times C\to \MH\times C$. This statement just means that,
to the extent that the sigma-model is a good description, for each
point $q\in \Sigma$, the bosonic fields $A,\phi$ of the gauge
theory, when restricted to $q\times C$, are given by the solution
of Hitchin's equations corresponding to the point $\Phi(q)\in
\MH$. This solution is simply, up to a gauge transformation, the
restriction of the universal Higgs bundle $({\EUE},\hat\varphi)$
to $\Phi(q)\times C$.

To interpret the connection ${\cal A}=A+i\phi$ in \yrgy\ in terms
of the sigma-model, we note that in general, this connection
involves both $\cA_\Sigma$, the part of  the connection tangent to
$\Sigma$, and $\cA_C$, the part tangent to $C$. Here in the low
energy theory, we can assume that $\cA_C$ obeys Hitchin's
equations, and as long as we avoid singularities in $\MH$, the
fields $\cA_\Sigma$ are massive in the sigma-model. They can
therefore be integrated out in favor of the sigma-model fields
$\cA_C$.  For very large ${\rm Im}\,\tau$,  it is sufficient to
integrate out $\CA_\Sigma$ at the classical level. The part of the
gauge theory action which depends on only $A$ and $\phi$ was
written in \wingo. Assuming that $A_C$ and $\phi_C$ satisfy
Hitchin equations and dropping the terms which vanish as the
volume of $C$ goes to zero, we find a quadratic action for
$\cA_\Sigma$. The corresponding equations of motion read
\eqn\integrout{\cD_C{}^\dagger \cD_C \cA_\Sigma=\cD_C{}^\dagger
d_\Sigma \cA_C+\dots,} where $\cD_C$ is the covariant differential
with respect to the connection $\cA_C$.  The ellipses refer to
terms involving zero modes of the fermions $\psi$, $\tilde\psi$,
etc., of the four-dimensional gauge theory; we will not write
these terms explicitly.   A map $\Phi:\Sigma\to \MH$   determines
$\cA_C$ and hence also $d_\Sigma\cA_C$, and then, assuming we keep
away from singularities of $\MH$, the equation \integrout\ has a
unique solution for $\cA_\Sigma$.

So once $\Phi:\Sigma\to \MH$ is given (and assuming that we keep
away from singularities of $\MH$), the connection
$\CA=(\cA_\Sigma,\cA_C)$ is determined.  $\CA$ is, of course, a
connection on the bundle $E=(\Phi\times 1)^*({\EUE})$.
 The connection $\CA$ is actually the pullback by $\Phi\times 1$ of a
connection $\hat\CA$ on  ${\EUE}\to \MH\times C$. In fact, to
define $\hat\CA$, we must specify its components $\CA_{\MH},\CA_C$
tangent to $\MH$ and $C$.  $\CA_C$ is the appropriate solution of
Hitchin's equations, and $\CA_{\MH}$ is defined by generalizing
\integrout\ in an obvious way:\foot{The ellipses in \integrout\
involve fermionic zero modes, which represent tangent vectors to
$\MH$ and so have analogs in the case of $\MH\times C$.}
\eqn\pintegrout{\cD_C{}^\dagger \cD_C \cA_{\M_H}=\cD_C{}^\dagger
d_{{\eusm M}_H} \cA_C+\dots.}

Now let us specialize to the case that ${{\cal S}}=\gamma\times
p$, with $\gamma$ a curve in $\Sigma$ and $p$ a point in $C$.  We
write ${\EUE}_p(R)$ for the restriction of ${\EUE}(R)$ to
$\MH\times p$. We also write $\Phi_p$ for the restriction to
$\Sigma\times \{p\}\subset \Sigma\times C$ of the map $\Phi\times
1:\Sigma\times C\to \MH\times C$. We can replace the connection
$\CA=A+i\phi$ in \yrgy\ by $\Phi_p^*(\hat\CA)$.  Hence the factor
in the path integral that comes from the inclusion of a Wilson
operator on the contour ${{\cal S}}$ in the representation $R$ can
be written as \eqn\zico{W_R({{\cal S}})=
P\exp\left(-\int_\gamma\,\Phi_p^*(\hat\CA)\right).} In the limit that
$\gamma$ approaches the boundary $\QQ$ of $\Sigma$, this has the
same form as the term that comes anyway from the Chan-Paton bundle
$U$ of the original brane ${\cal B}$:\foot{The ellipses in \zomfo\
represent fermionic terms whose analog in \zico\ arises from the
ellipses in \integrout, which reflect fermionic contributions to
$\CA_\Sigma$. All these terms are uniquely determined by the
topological symmetry, so we do not need to worry about comparing
them.} \eqn\zomfo{P\exp\left(-\int_\QQ (\Phi_\QQ^*(\alpha)
+\dots)\right).}

So we learn how Wilson lines act on branes.  A Wilson line in the
representation $R$ and supported at a point $p\in C$ transforms
the Chan-Paton bundle $U$ of a brane ${\cal B}$ by \eqn\yuq{U\to
U\otimes {\EUE}_p(R).}

\bigskip\noindent{\it Transformation Of The $B$-Field}

As we have discussed in section \reviewthis,  ${\EUE}(R)$, and
hence also ${\EUE}_p(R)$, in general does not exist as a vector
bundle. But it always exists as a twisted vector bundle, twisted
by the flat $B$-field $\theta_{R}(\zeta)$, where $\theta_{R}$  is
the character of the center of the gauge group determined by $R$.

The category of branes depends on a choice of $B$-field, a fact
that we exploited in section \comptwop. For a given discrete
electric field ${\bf e}_0$, the background $B$-field in the
sigma-model on $\MH$ is $b_{{\bf e}_0}={\bf e}_0(\zeta)$, where
$\zeta=\xi({\EUE}_{\rm{ad}})$ is the obstruction to existence of a
universal $\bar G$ Higgs bundle. Tensoring with a twisted bundle
that is twisted by $\theta_R(\zeta)$ maps a brane that is twisted
by a flat $B$-field $b$ to a brane that is twisted by
$b+\theta_R(\zeta).$ Therefore, the action of a Wilson line
operator on branes changes the $B$-field, by $b\to
b+\theta_R(\zeta)$.  In other words, it changes the discrete
electric field studied in section \centertop\ by ${\bf e}_0\to
{\bf e}_0+\theta_R$.

We want to understand what this result means for $S$-duality.  So
we write it as a statement about the dual gauge theory, with gauge
group $^ LG$, and a Wilson line $W(^L\negthinspace R)$ determined
by a representation $^L\negthinspace R$.  This Wilson line
transforms the discrete electric field by \eqn\nuko{{\bf e}_0\to
{\bf e}_0+\theta(^L\negthinspace R),} where, for convenience, we
write $\theta(^L\negthinspace R)$ rather than
$\theta_{^L\negthinspace R}$.

\ifig\namp{Insertion of an 't Hooft operator changes the topology
of a $G$-bundle, as shown here.  Sketched is a three manifold with
boundary components consisting of two Riemann surfaces $C$ and
$C'$ and a small two-sphere $S$ enclosing a point at which an 't
Hooft operator is inserted.  Cobordism invariance of the
characteristic class implies that if $\bf{m}_0$ is the
characteristic class of the $G$-bundle $E\to C$, then the
characteristic class of $E\to C'$ must be ${\bf{m}}_0+\xi$, where
$\xi=\xi({}^L\negthinspace R)$ is the characteristic class
associated with the 't Hooft operator.}
{\epsfxsize=4in\epsfbox{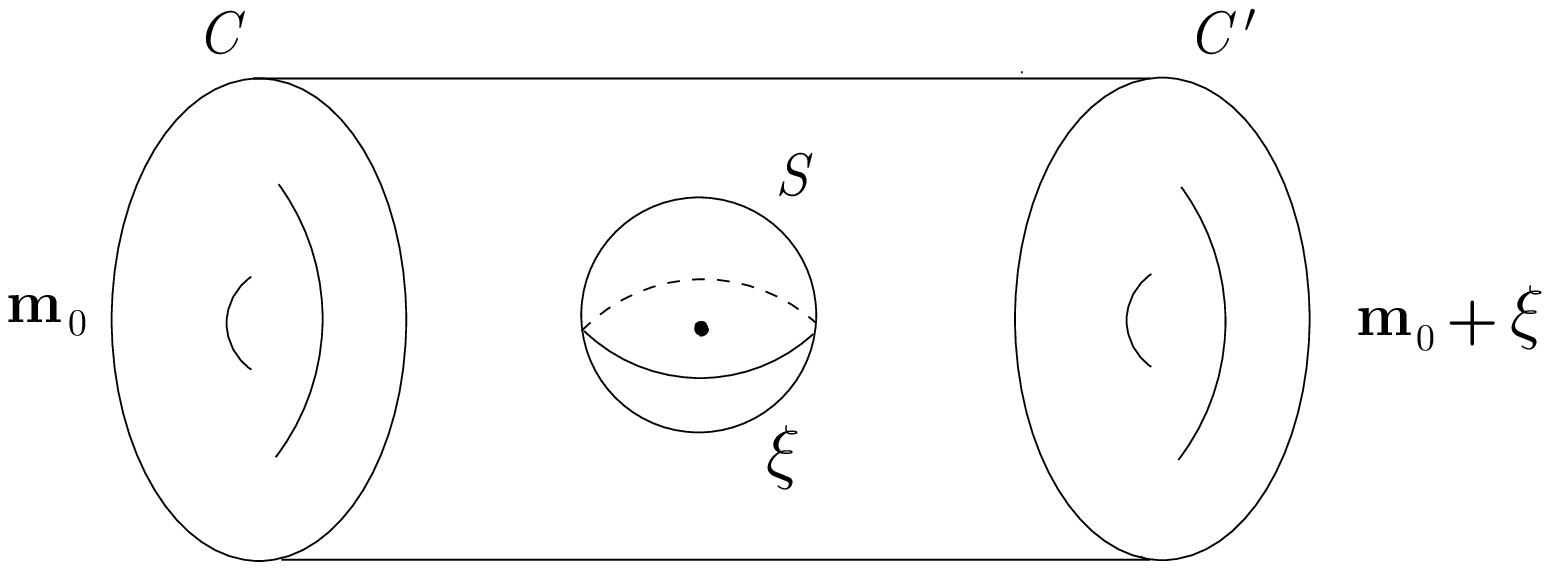}} Under $S$-duality, the fact
that a Wilson line operator can change the discrete electric field
${\bf e}_0$ maps to the fact that an 't Hooft line operator can
change the characteristic class ${\bf m}_0\in H^2(C,\pi_1(G))$
which classifies the topology of the $G$-bundle $E\to C$. Indeed,
as we explained in section \tophooft, an 't Hooft operator
$T(^L\negthinspace R)$ is constructed from a $G$-bundle, which we
may call $E(^L\negthinspace R)$, over $S^2\cong \Bbb{CP}^1$.  This
$G$-bundle has a characteristic class $\xi(^L\negthinspace
R)=\xi(E(^L\negthinspace R))$. The action of the 't Hooft operator
on ${\bf m}_0$ is \eqn\zuko{{\bf m}_0\to {\bf
m}_0+\xi(^L\negthinspace R).} This statement, which is the
$S$-dual of \nuko, just comes from the behavior of the
characteristic class under cobordism, as in \namp.

\bigskip\noindent{\it Wilson Operators And Supersymmetry}

Let us see what kind of supersymmetry the operation \yuq\
preserves.  For a generic choice of curve ${{\cal S}}\subset
\Sigma\times C$, the supersymmetric Wilson line $W(R,{{\cal S}})$
preserves $B$-type supersymmetry in the complex structure $J$ of
$\MH$ and nothing else. However, if we take ${{\cal
S}}=\gamma\times p$, for $\gamma$ a curve in $\Sigma$ and $p$ a
point in $C$, then the Wilson operator preserves more
supersymetry. In fact, as we showed at the end of section \lineop,
it preserves supersymmetry of type $(B,B,B)$, that is, it
preserves $B$-type supersymmetry in each complex structure.

We can now see explicitly how this is reflected in the action of
the Wilson operator on branes.  For a point $p\in C$, the bundle
${\EUE}_p(R)$ is holomorphic in each of the complex structures on
$\MH$.  (This can be naturally proved using the hyper-Kahler
quotient construction of $\MH$.) So the operation \yuq\ preserves
$B$-type supersymmetry for each complex structure.

\subsec{Zerobranes As Electric Eigenbranes}

\subseclab\zerobelectric

Now we can look for electric eigenbranes.  Consider a Wilson line
$W_q(R)$ in a representation $R$ and at a point $q\in C$.  It maps
a brane ${\cal B}$ with Chan-Paton sheaf $U$ to a brane with
Chan-Paton sheaf $U\otimes {\EUE}_q(R)$.  By definition, ${\cal
B}$ is an electric eigenbrane if $U\otimes {\EUE}_q(R)$ is
isomorphic to $U\otimes V$ for some fixed vector space $V$. The
condition, therefore, is that ${\EUE}_q(R)$ equipped with the
connection $\Phi_q^*(\hat \CA)$ must be holomorphically trivial --
isomorphic to a vector bundle with constant fiber $V$ -- when
restricted to the support of $U$.

Let $x$ be a point in $\MH$, and let ${\cal B}_x$ be a zerobrane
supported at $x$. In other words, ${\cal B}_x$ is a brane whose
Chan-Paton sheaf is a skyscraper sheaf $U_x$ supported at $x$. We
will suppose that $x$ is a smooth point in $\MH$, so that one can
integrate out the fields $A_\Sigma,\phi_\Sigma$ on the classical
level.

Any vector bundle is trivial when restricted to a point, and in
particular the bundle ${\EUE}_q(R)$ is trivial when restricted to
$x\in \MH$.  Its restriction to $x$, which we denote as
${\EUE}_q(R)|_x$, is just a fixed vector space.  So
\eqn\insto{U_x\otimes {\EUE}_q(R)=U_x\otimes {\EUE}_q(R)|_x.}
Therefore, ${\cal B}_x$ is an electric eigenbrane, the
``eigenvalue'' of the Wilson loop operator $W_q(R)$ being the
vector space ${\EUE}_q(R)|_x$.

Equivalently, the boundary conditions on the sigma-model fields
corresponding to the zerobrane say that $\cA_C$ is constant up to
a gauge transformation on the boundary of $\Sigma$. By virtue of
\integrout, this implies that ${\cal A}_\Sigma$ is trivial on the
boundary of $\Sigma$.

We argued in section \lineop, based on the underlying
four-dimensional topological field theory, that the ``eigenvalue''
of an electric or magnetic eigenbrane must vary with $q\in C$ as
the fiber of a flat bundle over $C$.  Indeed, as $q\in C$ varies,
the space ${\EUE}_q(R)|_x$ does vary as the fiber of a flat vector
bundle. In fact, taking complex structure $J$, the restriction
$({\EUE},\hat\varphi)|_x$ of the universal bundle to $x\times C$
gives a flat $G_{\Bbb{C}}$ bundle ${\EUE}|_x\to C$ (since a
solution of Hitchin's equations corresponds in complex structure
$J$ to a flat $G_{\Bbb{C}}$-bundle). By taking this bundle in the
representation $R$, we get the desired flat bundle whose fiber at
$q\in C$ is ${\EUE}_q(R)|_x$.  We picked here complex structure
$J$ because it is singled out by the underlying topological
supersymmetry.

In describing the zerobrane as an electric eigenbrane, we have
been slightly informal on one key point: the role of the discrete
$B$-field ${\bf e}_0(\zeta)$.  A $B$-field is trivial when
restricted to a point, so as long as one is considering
zerobranes, one can informally ignore the $B$-field, as we have
just done.  However, a more precise way to describe things is as
follows.  For every ${\bf e}_0$, the category of branes contains
an object that is a zerobrane supported at the (smooth) point
$x\in \MH$.  Because the action of a Wilson line operator can
change ${\bf e}_0$, the electric eigenbrane is really a sum of
twisted zerobranes with different ${\bf e}_0$ (all of which are
isomorphic, as the twisting is trivial when restricted to a
point). Dually, since an 't Hooft operator can change ${\bf m}_0$,
the magnetic eigenbranes will be sums of branes supported on
different components of $\MH$, for all possible ${\bf m}_0$.  In
standard approaches to the geometric Langlands program, one says
that the Hecke eigensheaves are supported on the union of all
topological components of $\M(G,C)$.

\bigskip\noindent{\it Relation To The Geometric Langlands Program}

Now we can get at least a glimmering of how all this is related to
the geometric Langlands program.

In the geometric Langlands program, one begins with the  group
$^L\negthinspace G$ and a homomorphism $\vartheta:\pi_1(C)\to{}
^L\negthinspace G_{\Bbb{C}}$. (One requires that this homomorphism
be semistable in a sense explained in section \complexstr.) The
space of such homomorphisms is the Hitchin moduli space
$\MH(^L\negthinspace G,C)$.  So $\vartheta$ defines a point
$x(\vartheta)$ in $\MH(^L\negthinspace G,C)$.

If $\vartheta$ is irreducible, then $x(\vartheta)$ is a smooth
point and a zerobrane ${\cal B}_{x(\vartheta)}$ supported at
$x(\vartheta)$ is an electric eigenbrane.  $S$-duality applied to
this electric eigenbrane will give a magnetic eigenbrane in the
sigma-model of target $\MH(G,C)$.  {}From sections \sdualhitfib\
and \twodimint, we know that the $S$-dual of a zerobrane of
$\MH(^L\negthinspace G,C)$ is a brane in $\MH(G,C)$ whose support
is a fiber ${\CMF}$ of the Hitchin fibration, endowed with a flat
Chan-Paton bundle of rank 1.  We refer to such a brane as a brane
of type ${\CMF}$.

The main claim of the geometric Langlands program is that a
homomorphism $\vartheta:\pi_1(C)\to {}^L\negthinspace G_{\Bbb{C}}$
is associated in a natural way to a sheaf on $\M(G,C)$ that is a
Hecke eigensheaf and also a holonomic ${\cal D}$-module.  Sections
\thooftheckeop -\abranes\ of this paper will be devoted to
explaining why a brane of type ${\CMF}$ on $\MH(G,C)$ has the
right properties. In sections \thooftheckeop\ and \bogeqheck, we
relate the 't Hooft operators of quantum gauge theory  to the
Hecke operators of the geometric Langlands program, showing how
our notion of a magnetic eigenbrane is related to the mathematical
concept of a Hecke eigensheaf.  In section \abranes, we argue that
by virtue of the existence of a certain coisotropic $A$-brane on
$\MH(G,C)$, the brane of type ${\CMF}$ is naturally associated to
a module for the differential operators on $\M(G,C)$.

\newsec{'t Hooft And Hecke Operators}
\seclab\thooftheckeop

 The goal of the present section is to begin the study of
the supersymmetric 't Hooft operators that appear at $\Psi=0$, a
value we reach by setting $t=1$ and $\theta=0$.  The main result
will be to show that 't Hooft operators correspond to the Hecke
operators of the geometric Langlands program.

We will mainly consider static 't Hooft operators. So our
four-manifold will be $M=\Bbb{R}\times W$, for some three-manifold
$W$, and our 't Hooft line operators will be supported on
one-manifolds of the form $\Bbb{R}\times p$, for some point $p\in
W$.  As we have discussed in section \loopop, line ``operators''
of this kind are not operators in the usual sense, acting on a
pre-existing vector space; rather, they must be incorporated in
the definition of the space of physical, supersymmetric states.

Our four-dimensional TQFT at $\Psi=0$ reduces in two dimensions to
the $A$-model with target $\MH$ in complex structure $K$.  So let
us first recall some facts about $A$-models.  In general, in a
two-dimensional $A$-model with target $X$, a supersymmetric
classical field configuration is a holomorphic map $\Phi:\Sigma\to
X$. Moreover, the first approximation to the space of
supersymmetric states is the cohomology of the space of
time-independent supersymmetric classical fields on
$\Sigma=\Bbb{R}\times I$, with $I$ being an interval.
Supersymmetric fields are holomorphic maps  $\Phi:\Sigma\to X$.
But a holomorphic map that is also time-independent is necessarily
a constant. So in the $A$-model, the classical approximation to
the space of physical states is simply the cohomology of the space
of constant maps to $X$ that obey the boundary conditions.   For
example, in Floer theory, one takes boundary conditions such that
the two boundary components of $\Bbb{R}\times I$ are mapped to two
specified Lagrangian submanifolds $N,N'\subset X$. A constant map
with these boundary conditions must map $\Sigma$ to the
intersection $N\cap N'$, so the classical approximation to the
space of supersymmetric states is simply the cohomology of $N\cap
N'$.  (One must shift the dimensions or ghost numbers of the
cohomology classes in order to account for the number of filled
fermion states in the vacuum. Mathematically, the shift involves
the Maslov index.)

In general, there are instanton corrections to this classical
intersection; their study is the main content of Floer theory. But
we will see that for the branes we consider, there are no
instanton corrections to the $A$-model at $\Psi=0$; the
supersymmetric configurations are all time-independent.

\bigskip\noindent{\it Reduction To Two Dimensions}

Aiming to reduce to a two-dimensional model on $ \Bbb{R}\times I$,
we specialize our four-manifold, which so far has been
$M=\Bbb{R}\times W$,  by taking $W=I\times C$, where $C$ is a
Riemann surface and $I$ an interval at whose ends we take boundary
conditions defined by choices of brane.  So altogether, we work on
$M=\Bbb{R}\times I\times C$. The conditions for supersymmetry are
familiar from section \familya:
\eqn\oops{\eqalign{(F-\phi\wedge\phi+tD\phi)^+ & = 0 \cr
                      (F-\phi\wedge\phi-t^{-1}D\phi)^-& = 0 \cr
                      D^* \phi & = 0.\cr}}
At $t=1$, which we generally assume in our study of 't Hooft
operators, the first two equations in \oops\ can be written
$F-\phi\wedge \phi+\star D\phi=0$.  In terms of the complex
connection $\CA=A+i\phi$ and curvature $\CF=d\CA+\CA\wedge\CA$,
this is equivalent to \eqn\nops{{\cal F}+i\star\bar\CF=0.}

In the absence of 't Hooft loops, and with suitable boundary
conditions, vanishing theorems similar to those of section
\vanishing\ show that solutions of these equations on
$M=\Bbb{R}\times I\times C$ are pulled back from $C$ and come from
a constant map of $\Sigma=\Bbb{R}\times I$ to $\MH$. This is part
of the reduction of the four-dimensional TQFT to a two-dimensional
sigma-model.

\ifig\botampo{A static 't Hooft line inserted at a point
$p=y\times p_0$ in $I\times C$.  Near this point, the solution of
the Bogomolny equation has a prescribed singularity.}
{\epsfxsize=4in\epsfbox{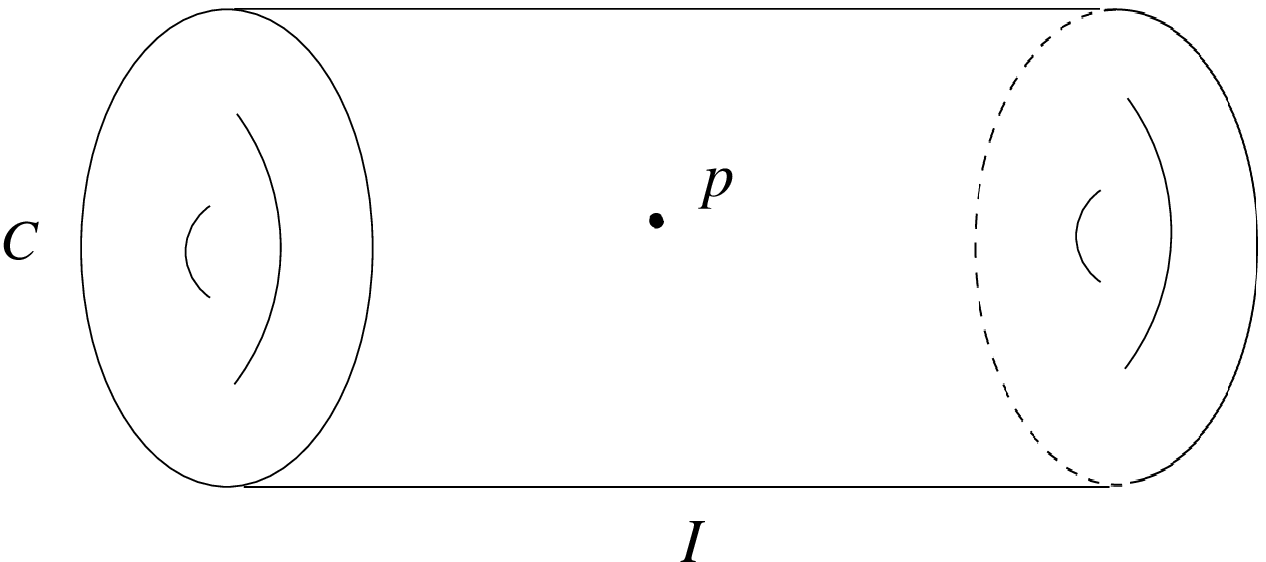}}

With 't Hooft lines included, things are more complicated
(\botampo). As we recall from section \tophooft, 't Hooft
operators create specified singularities in classical field
configurations. For an 't Hooft operator supported on the line
$\Bbb{R}\times p$, with some $p\in I\times C$, the singularity
will prevent us from  proving that the solution is a pullback from
$C$.  But for certain boundary conditions, a weaker vanishing
result still applies: the relevant solutions of \oops\ are
time-independent, that is, invariant under translations of
$\Bbb{R}$.

To explain this, recall that from the point of view of geometric
Langlands duality, the most important $A$-branes  are the branes
of type ${\CMF}$, supported on a fiber of the Hitchin fibration
with a flat Chan-Paton bundle of rank 1. They are $S$-duals of
zerobranes and hence are expected to be magnetic eigenbranes. As
explained in section \mhbranes, they are branes of type $(B,A,A)$,
i.e. the corresponding boundary conditions are compatible with two
linearly independent topological supercharges, suitable linear
combinations of which generate the topological supersymmetry of
the $B$-model in complex structure $I$ and the $A$-models in
complex structures $J$ and $K$.

When we consider BRST-invariant field configurations with
$(B,A,A)$ boundary conditions and an insertion of the 't Hooft
operator, we can distinguish between configurations which preserve
both fermionic symmetries and those which preserve only the
$A$-type  supersymmetry in complex structure $K$. In the latter
case, the broken supersymmetry generates an extra fermionic zero
mode that is not lifted by quantum corrections, since it is
generated by a symmetry, and will prevent such instantons from
contributing to any topological observables. (There may also be
other fermionic zero modes whose existence does not follow from
the fermionic symmetry.)  Hence, contributions of interest come
only from instantons that preserve both supersymmetries.

In fact, instantons with the requisite properties are
time-independent. One would guess this from the sigma-model, since
the $B$-model in complex structure $I$ has no non-trivial
instantons.  The result is also true in the context of the full
twisted supersymmetric gauge theory on $\Bbb{R}\times I\times C$,
even in the presence of static 't Hooft operators.  The most
general proof of this involves a moment map argument and will
appear elsewhere \witfur. Here we sketch another argument that is
adequate for the most important $(B,A,A)$-branes, such as branes
of type ${\CMF}$.

First we need a vanishing theorem for $\phi_1$, the component of
$\phi$ in the $I$ direction of $M=\Bbb{R}\times I\times C$. In any
classical instanton solution, $\phi_1$ will obey its second order
Euler-Lagrange equations, which can be deduced from the classical
action \nuseful\ and read \eqn\zungo{ -D_\mu
D^\mu\phi_1+[\phi_\mu,[\phi_1,\phi^\mu]]=0.} We suppose that
$\phi_1$ obeys Dirichlet or Neumann boundary conditions on each
brane; this is true for the most important branes.
 Multiplying  the equation by
$\phi_1$, taking the trace and integrating over space-time, we get
\eqn\wungo{\int d^4x\ \Tr\left( D_\mu\phi_1
D^\mu\phi_1+[\phi_1,\phi_\mu][\phi_1,\phi^\mu]\right)=0.}
Integration by parts produces no boundary terms because of the
assumed boundary conditions, and there is no problem with the
singularities associated with the 't Hooft operators because
$\phi_1$ is nonsingular near a static 't Hooft operator. Since the
integrand is negative-definite, the above identity implies that
$D_\mu\phi_1=[\phi_\mu,\phi_1]=0$, so that $\phi_1=0$ in the case
of an irreducible solution.

In this argument, we tacitly assumed that the integration by parts gives no
boundary term in the far past or future.  We do not know {\it a priori} that a supersymmetric
solution of the equations is time-independent.  However,
we at least assume that the desired solutions approach time-independent ``vacuum'' solutions in the
far past or the far future.  By first running the argument of the last paragraph for the
time-independent case, we learn that $\phi_1=0$ in each vacuum.  It then follows, since $W$ is
compact, that $\phi\to 0$ exponentially fast in the far past or the far future.  Hence, there is
no boundary term in the integration by parts used in the last paragraph to show that $\phi_1=0$
everywhere.

Now we want to show that a field that possesses the topological
supersymmetry of the $A$-model in both complex structures $J$ and
$K$ and also has $\phi_1=0$ is actually time-independent. The
supersymmetric equations of the $A$-model in complex structure $K$
are the equations \oops\ at $t=1$; those in the $A$-model in
complex structure $J$ are obtained from these by making a ${\cal
U}_1$ transformation $\varphi\to i\varphi$.  It is a short
exercise to write out the combined system of equations on
$M=\Bbb{R}\times I\times C$ and show that, if in addition
$\phi_1=0$, they imply that the fields are time-independent up to
a gauge transformation.

The time-independent supersymmetric fields can usefully be
analyzed in the more general context of $M=\Bbb{R}\times W$ with
static Wilson lines (without specializing to $W=I\times C$), even
though our argument that all relevant supersymmetric fields are
time-independent does not hold in this generality. When we
specialize to $M=\Bbb{R}\times W$, there is a further symmetry
${\EUT}$ of time-reversal invariance, acting as $s\to -s$ where
$s$ is a ``time'' coordinate on $\Bbb{R}$; ${\EUT}$ acts trivially
on $W$ and maps $(A,\phi)\to ({\EUT}^*A,-{\EUT}^*\phi)$. (This is
an example of the orientation-reversing symmetries discussed at
the end of section \canonpar.) The sign change of $\phi$ is needed
because ${\EUT}$ reverses the sign of the Hodge $\star$ operator
in \nops.

Therefore, we can consistently restrict ourselves to solutions
that are both time-independent and $\EUT$-invariant. In such a
solution, the $G$-bundle $E$ and connection $A$ are pullbacks from
$W$. Also, $\phi$ takes the form $\phi_0\,ds$, where $\phi_0$ is
an ${\rm ad}(E)$-valued zero-form on $W$.  The equations reduce to
\eqn\bogeq{F=\star D\phi_0.} Our orientation conventions are such
that in four dimensions, $\star(dx^0\wedge dx^1)=dx^2\wedge dx^3$,
and in three dimensions, $\star(dx^1)=dx^2\wedge dx^3$. The
equations \bogeq\ are the standard Bogomolny equations
\ref\bogeqns{E. B. Bogomolny, ``Stability Of Classical
Solutions,'' Sov. J. Nucl. Phys. {\bf 24} (1976) 449-454.} for
supersymmetric or BPS monopoles\foot{The name reflects the fact
that the basic supersymmetric monopole solution was also found in
\ref\prasom{ M. K. Prasad and C. Sommerfield, ``An Exact Classical
Solution For The 't Hooft Monopole And The Julia-Zee Dyon,'' Phys.
Rev. Lett. {\bf 35} (1975) 760-762.} as a solution of the second
order Yang-Mills-Higgs equations.} and have been widely studied,
for instance in \nref\bogeqv{E. J. Weinberg, ``Parameter Counting
For Multi-Monopole Solutions,'' Phys. Rev. {\bf D20} (1979)
936-944, ``Fundamental Monopoles And Multi-Monopole Solutions For
Arbitrary Simple Gauge
Groups,'' Nucl. Phys. {\bf B167} (1980) 500-524.}%
\nref\bogeqx{C. H. Taubes, ``Stability In Yang-Mills
Theories,'' Commun. Math. Phys. {\bf 91} (1983) 235-263.}%
\nref\bogeqa{M. F. Atiyah and N. Hitchin, {\it The Geometry And
Dynamics Of Magnetic
Monopoles} (Princeton University Press, 1988).}%
\refs{\bogeqv-\bogeqa}.  A more familiar way to obtain the
Bogomolny equations is to look for solutions of the Yang-Mills
instanton equation on $\Bbb{R}^4$ that are invariant under
translations in one direction.

In this section, we will first analyze static 't Hooft operators
in the context of the standard Bogomolny equations \bogeq.  We
will show that they implement Hecke transformations of
$G$-bundles, as usually defined in the geometric Langlands
program.   This will involve studying BPS monopole solutions with
point singularities due to the 't Hooft operators.  Such singular
solutions of the Bogomolny equations have been studied first in
unpublished work by Kronheimer \ref\kronheimermon{P. Kronheimer,
MSc. thesis (Oxford University, 1986), unpublished.}.  They can
arise as limits of smooth monopole solutions for larger gauge
groups \nref\polo{E. J. Weinberg, ``A Continuous Family Of
Magnetic Monopole Solutions,'' Phys. Lett. {\bf B119} (1982) 151-154.}%
\nref\golo{K.-M. Lee, E. J. Weinberg, and P. Yi, ``Massive And
Massless Monopoles With Nonabelian Magnetic Charges, Phys. Rev.
{\bf D54} (1996) 6351-6371;
 E. J. Weinberg, and P. Yi, ``Explicit Multimonopole Solutions In $SU(N)$ Gauge
Theory,'' Phys. Rev. {\bf D58}
(1998) 046001, arXiv:hep-th/9803164.}%
\nref\latwein{C. J. Houghton and E. J. Weinberg, ``Multicloud
Solutions With Massless And Massive Monopoles,'' Phys. Rev. {\bf
D66} (2002) 125002, arXiv:hep-th/0207141.}%
 \nref\bais{F. A. Bais and B. J. Schroers, ``Quantization Of
Monopoles With Nonabelian Magnetic Charge,'' Nucl. Phys. {\bf
B512} (1998) 250-294, arXiv:hep-th/9708004.}%
\refs{\polo-\bais}, and from certain string theory brane
configurations \ref\hanwit{A. Hanany and E. Witten, ``Type IIB
Superstrings, BPS Monopoles, And Three-Dimensional Gauge
Dynamics,'' Nucl. Phys. {\bf B492} (1997) 152-190,
arXiv:hep-th/9611230.}, which have motivated more recent study in
\nref\cherk{S. A. Cherkis and A. Kapustin, ``Singular Monopoles And Supersymmetric
Gauge Theories In Three-Dimensions,'' Nucl. Phys. {\bf B525} (1998) 215-234, hep-th/9711145.}%
\nref\cherktwo{S. A. Cherkis and A. Kapustin, ``$D(k)$ Gravitational Instantons And Nahm
Equations,'' Adv. Theor. Math. Phys. {\bf 2} (1999) 1287-1306, hep-th/9803112.}%
\nref\cherkis{S. A. Cherkis and A. Kapustin, ``Periodic Monopoles With
Singularities And $N=2$ Super QCD,'' Commun. Math. Phys. {\bf 234}
(2003) 1-35, arXiv:hep-th/0011081.}%
\refs{\cherk-\cherkis}. Then we will provide an
elementary introduction to Hecke modifications of $G$-bundles --
claiming neither novelty nor completeness from a mathematical
point of view!

In section \extbog, we will generalize the analysis to relax the
assumption of time-reversal symmetry. As we will see, in this case
the supersymmetric equations give a sort of complexified or
extended analog of the Bogomolny equations, which does not appear
to have been studied before. In the context of these extended
equations, static 't Hooft operators act by Hecke transformations
of Higgs bundles or in other words  of Hitchin pairs
$(E,\varphi)$. Hecke operators in this extended sense are natural
mathematically and have been considered before
\nref\donagi{R.
Donagi, unpublished.}%
\nref\olsh{A.Levin, M.Olshanetsky, and A.Zotov,
``Hitchin Systems -- Symplectic Hecke Correspondence and Two-dimensional
Version,'' Commun. Math. Phys. {\bf 236} (2003) 93-133, nlin.SI/0110045.}
\refs{\donagi,\olsh}.

Applications to the geometric Langlands program really depend upon
the extended Bogomolny equations and Hecke operators. We can
reduce to ordinary Bogomolny equations and Hecke operators only
when the boundary conditions (the choices of branes) ensure that
the Higgs field $\varphi$ vanishes. This is not the case for
typical applications.  However, we begin with ordinary Bogomolny
equations and Hecke operators and devote most of our attention to
them because this case is simpler and exhibits the main ideas.

\subsec{'t Hooft Operators And Hecke Modifications}

\subseclab\theck

To begin with, we consider the Bogomolny equations on
$W=\Bbb{R}^3$, which we regard  as $\Bbb{R}\times\Bbb{C}$, where
$\Bbb{C}$ is parametrized by $z=x^2+ix^3$, and $\Bbb{R}$ by
$y=x^1$.  We have \eqn\boffo{\eqalign{\star dy&={i\over 2}dz\wedge
d\bar z\cr
        \star dz&=-idz\wedge dy\cr
        \star d\bar z& = id\bar z\wedge dy.\cr}}
        Upon expanding $F=dz\wedge d\bar z \,F_{z\bar z}+dy\wedge dz
        F_{yz}+dy\wedge d\bar z F_{y\bar z}$, and similarly
        $D\phi_0=dy D_y\phi_0+dz D_z\phi_0+d\bar z D_{\bar z}\phi_0$,
the Bogomolny equations become \eqn\noffo{\eqalign{F_{z\bar
z}&={i\over 2}D_y\phi_0\cr
                    F_{yz}&=iD_z\phi_0\cr
                    F_{y\bar z}&=-iD_{\bar z}\phi_0.\cr}}

More generally, take $W= \Bbb{R}\times C$, with a Riemann surface
$C$. We write $z$ for a local complex coordinate on $C$, endow $C$
with a Kahler metric $h(z,\bar z)|dz|^2$, for some positive
function $h$, and take the metric on $W$ to be $h(z,\bar
z)|dz|^2+dy^2$. The first equations in \boffo\ and \noffo\ become
respectively \eqn\jungol{\eqalign{\star dy& ={ih\over 2}\,dz\wedge
d\bar z\cr F_{z\bar z}&={ih\over 2}D_y\phi_0.}} The other
equations are unchanged.

There is no integrability condition for $\bar\partial $ operators
in complex dimension 1.  So any connection $A$ on a $G$-bundle
$E\to C$ endows $E$ with a holomorphic structure.  We simply
define the $\bar \partial$ operator as $\bar D=d\bar z D_{\bar
z}=d\bar z(\partial_{\bar z}+A_{\bar z})$.

\ifig\otampo{In a solution of the Bogomolny equations on a
$G$-bundle $E$, the holomorphic type of $E_y$ -- the restriction
of $E$ to $ \{y\}\times C$ -- is constant except when one crosses
the position of an 't Hooft operator. Sketched are an 't Hooft
operator at $y=y_0$ and copies of $C_y=\{y\}\times C$ to the left
and right of $y_0$.  They are denoted $C_-$ and $C_+$.}
{\epsfxsize=4in\epsfbox{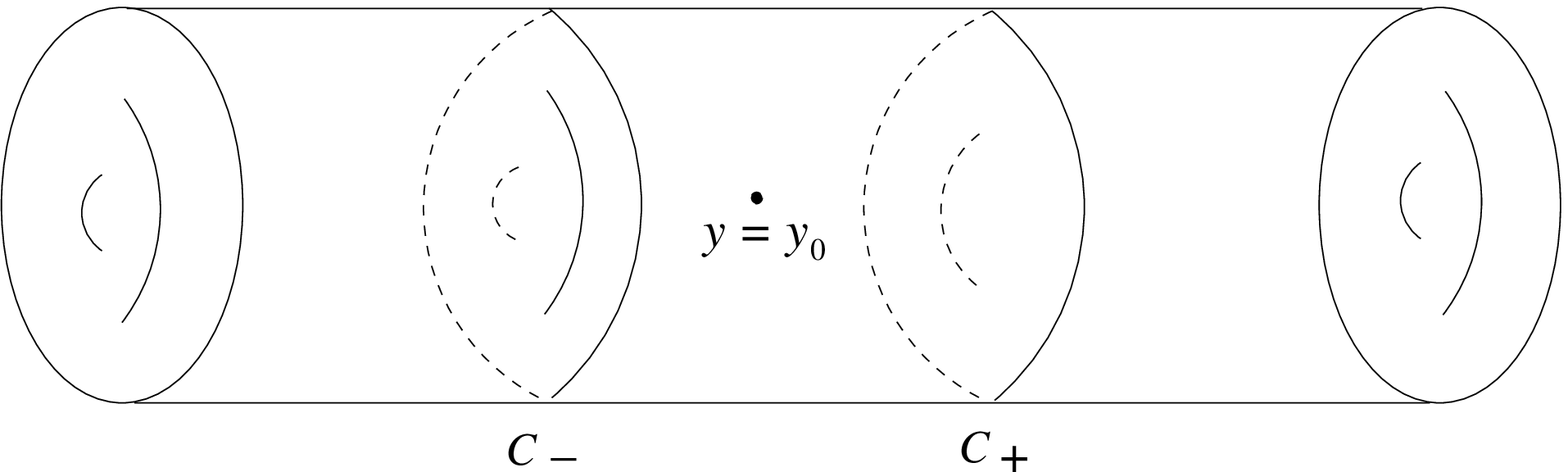}}

Thus, as $y$ varies in $ \Bbb{R}$, any connection $A$ on a
$G$-bundle $E\to  \Bbb{R}\times C$ defines a family of holomorphic
$G$-bundles $E_y\to C$. $E_y$ is simply the restriction of $E$ to
$C_y= \{y\}\times C$ (\otampo). Now suppose that $A$ obeys the
Bogomolny equations. Let us work in the gauge $A_y=0$. Then
$F_{y\bar z}=\partial_yA_{\bar z}$. Hence the last equation in
\noffo\ tells us that \eqn\troffo{{\partial\over
\partial y}A_{\bar z}=-iD_{\bar z}\phi_0.}
But the right hand side is the change in $A_{\bar z}$ under a
gauge transformation generated by $-i\phi_0$.  Hence the
holomorphic type of $E_y$ is independent of $y$.  (Another way to
state this argument is to note that $\partial_y\bar D=-i[\bar
D,\phi_0]$, showing that $\bar D$ is independent of $y$ up to
conjugation.)

Now let us see what happens if we incorporate an 't Hooft operator
at a point $p=y_0\times p_0\in \Bbb{R}\times C$. This means that
we consider a bundle $E$ and connection $A$ that are defined on
the complement of the point $p$, and which have a certain type of
singularity near $p$, as described in section \tophooft.

Away from $y=y_0$, the above argument applies and the holomorphic
type of $E_y$ is independent of $y$.  At $y=y_0$, because of the
singularity, the bundle $E_y$ is not defined.  In crossing
$y=y_0$, the holomorphic type of the bundle may jump.  However, it
can only jump in a very special way.  If we restrict $E$ to
$C\backslash p_0$ (that is, the complement of $p_0$ in $C$), then
the above argument applies and the restricted bundle $E_y$ has a
holomorphic type that is constant even at $y=y_0$.  So the
holomorphic type of $E_y$ may jump at $y=y_0$, but only in a way
that is trivial if we omit the point $p_0$ from $C$.

We will now try to describe precisely how $E_y$ can jump at
$y=y_0$.  We consider first the abelian case, and then the
nonabelian case.

\bigskip\noindent{\it The Abelian Case}

We begin with $G=U(1)$.  In this case, $E$ is a complex line
bundle ${\cal L}$. We write ${\cal L}_-$ for ${\cal L}_y$ with
$y<y_0$ and ${\cal L}_+$ for ${\cal L}_y$ with $y>y_0$.  The line
bundles ${\cal L}_-$ and ${\cal L}_+$ are isomorphic on the
complement of the point $p_0\in C$.

This means that the line bundle ${\cal L}_+\otimes {\cal
L}_-^{-1}$ is trivial away from $p_0$.  It hence has a section $s$
that has neither a zero nor a pole away from $p_0$; let $q$ be the
order of its zero at $p_0$.  Then ${\cal L}_+\otimes {\cal
L}_-^{-1}\cong {\cal O}(p_0)^q$.  Here as usual ${\cal O}(p_0)$ is
the line bundle whose holomorphic sections are  functions
holomorphic away from $p_0$ with a possible single pole at $p_0$.
So the relation between ${\cal L}_+$ and ${\cal L}_-$ is ${\cal
L}_+\cong {\cal L}_-\otimes {\cal O}(p_0)^q$, for some integer
$q$.

We recall that an 't Hooft operator $T(m)$ for the group $G=U(1)$
is classified by the choice of an integer $m$. (We write the
operator as $T(m;p_0)$ if we want to specify the point $p_0$.)
This suggests an obvious hypothesis -- that $m$ may coincide with
the integer $q$ of the last paragraph.

The operator $T(m)$ is defined by saying that near $p=p_0\times
y_0$, the curvature has the singular behavior \eqn\formigo{F\sim
\star d\left({im\over 2}{1\over |\vec x - p|}\right),} where
$|\vec x-p|$ is the distance from $p$ to a nearby point $\vec x\in
 \Bbb{R}\times C$.  This implies that if $S$ is a small sphere enclosing
the point $p$, as in the figure, then $\int_S c_1({\cal L})=m$.
The 't Hooft operators $T(m)$ are thus all topologically distinct
from one another.  This statement, which is far from being valid
if $G$ is nonabelian,   will enable us in the abelian case to
determine the action of $T(m)$ just on topological grounds.

\ifig\notampo{A cobordism between the two-cycles $C_-+S$ and
$C_+$, showing that the homology cycle $D=C_+-C_--S$ is a
boundary.} {\epsfxsize=3in\epsfbox{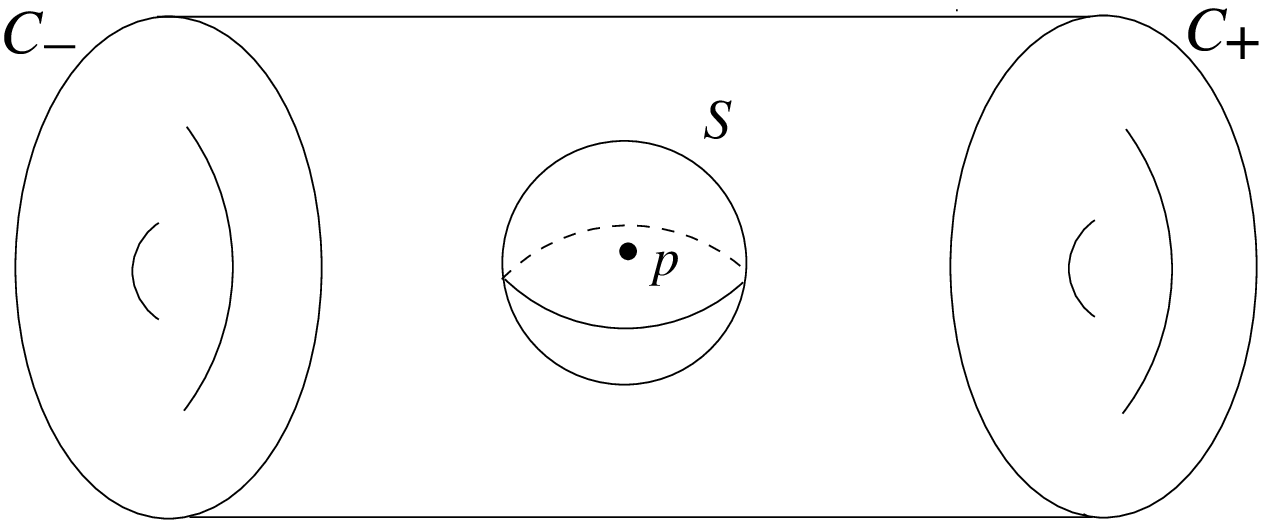}}

 Now pick $y_\pm$ with
$y_-<y_0<y_+$, and write $C_\pm$ for $ y_\pm\times C$, so that
${\cal L}_\pm $ is the restriction of ${\cal L}$ to $C_\pm$. Let
$q_\pm=\int_{C_\pm}c_1({\cal L})$. With an obvious choice of
orientations, as in  \notampo, the homology cycle $D=C_+-C_--S$ is
a boundary in $(\Bbb{R}\times C)\backslash p$. Hence
$0=\int_Dc_1({\cal L})=\int_{C_+}c_1({\cal L})-\int_{C_-}c_1({\cal
L})-\int_Sc_1({\cal L})=q_+-m-q_-$.

So $q_+=m+q_-$.  Consequently, ${\cal L}_+$ and ${\cal L}_-$,
which are isomorphic away from $p_0$, differ in first Chern class
by $m$. Hence ${\cal L}_+\cong {\cal L}_-\otimes {\cal O}(p_0)^m$.

Thus, we have determined the action of the 't Hooft operators for
$G=U(1)$.  The 't Hooft operator $T(m;p_0)$, that is the operator
$T(m)$ inserted at the point $p_0\in C$, acts by twisting with
${\cal O}(p_0)^m$.   This result agrees with the standard
definition of the Hecke operators for $U(1)$.

We see that, in accord with general arguments in section \lineop,
 't Hooft operators inserted at distinct points in $C$
 commute.  Moreover, the 't Hooft operators at
a given point $p_0\in C$ form a commutative group, with
$T(m;p_0)=T(1;p_0)^m$.

\bigskip\noindent{\it The Nonabelian Case: $U(N)$}

\def\VV{\eurm V}

Now let us discuss what 't Hooft operators do in the nonabelian
case.  We first consider the group $U(N)$.  There are two things
that make this group simple to analyze.  $U(N)$ has a convenient
representation, the $N$-dimensional representation that we will
call $\VV$. And $U(N)$ is its own Langlands dual.

The definition of an 't Hooft operator depends on a choice of
homomorphism $\rho:U(1)\to U(N)$, up to conjugation.  The most
general such homomorphism maps $\exp({i\alpha})\in U(1)$ to the
diagonal matrix ${\rm
diag}(\exp({im_1\alpha}),\exp({im_2\alpha}),\dots,
\exp({im_N\alpha}))$, with an $N$-plet of integers
$^L\negthinspace w=(m_1,m_2,\dots,m_N)$ that are unique up to
permutation. (We call this quantity $^L\negthinspace w$ because in
the general case, it will be a weight of the dual group
$^L\negthinspace G$.) We will usually order the $m_i$ so $m_1\geq
m_2\geq \dots\geq m_N$. At the Lie algebra level, $\rho$ maps
$1\in \frak{u}(1)$ to the diagonal matrix
\eqn\morphy{\left(\matrix{m_1& & & \dots & \cr
                                    &m_2& & \dots & \cr
                          & & m_3 &\dots & \cr
                & & \vdots & \dots& \cr
            & & & & m_N\cr}\right).}

The corresponding 't Hooft operator $T(^L\negthinspace w)$ is
defined by saying that near the point $p\in  \Bbb{R}\times C$, the
gauge field has a singularity obtained by embedding the basic
$U(1)$ singularity in $U(N)$, via this embedding.  For example,
the singular part of the curvature near $p$ is diagonal
\eqn\orphy{F\sim \star d\left({i\over 2}{1\over |\vec x -
p|}\right) \left(\matrix{m_1& & & \dots & \cr
                                    & m_2& & \dots & \cr
                          & & m_3 &\dots & \cr
                & & \vdots & \dots& \cr
            & & & & m_N\cr}\right).}
Near $p$, the Bogomolny equations reduce to equations in some
maximal torus ${\cal T}=U(1)^N$ of $U(N)$. (From a holomorphic
point of view, it is more natural to consider the corresponding
reduction to a complexified maximal torus in the complexification
of the gauge group, namely $U(N)_{\Bbb{C}}=GL(N,\Bbb{C})$.)
Corresponding to the reduction of the equations to a maximal
torus, the bundle $E_y$ splits up, near $p_0\times y \in C_y$, as
a sum ${\cal L}_1\oplus {\cal L}_2\oplus\dots \oplus {\cal L}_N$
of line bundles (each of which, of course, is trivial near $p$).
The effect of the 't Hooft operator on ${\cal L}_i$ is precisely
what it was in the abelian case, namely ${\cal L}_i\to {\cal
L}_i\otimes {\cal O}(p_0)^{m_i}$.

Thus, we have arrived at a description of how the 't Hooft
operator $T(^L\negthinspace w)$  acts on a bundle $E$.  Relative
to some decomposition of $E$ as $\oplus_{i=1}^N{\cal L}_i$ near
$p$, it acts by ${\cal L}_i\to {\cal L}_i\otimes {\cal
O}(p_0)^{m_i}$. This coincides with the usual mathematical
description of the action of the Hecke operators for $U(N)$.
This is the basic link between 't Hooft and Hecke operators.

In sections \spaceheck\ and \affgrass, we will address the
question of what sort of data is contained in the local
decomposition of $E$ as $\oplus_{i=1}^N{\cal L}_i$, and how much
of this data is relevant to the action of the 't Hooft operators.
The general answer to this question involves something called the
affine Grassmannian, which we briefly describe in section
\affgrass.

\bigskip\noindent{\it Some Examples}

In general, as we know from section \tophooft, representations of
the Langlands dual group $^L\negthinspace G$ correspond naturally
to 't Hooft operators of $G$. A representation $R(^L\negthinspace
w)$ with highest weight $^L\negthinspace w$ corresponds to an 't
Hooft operator that we write as $T(R(^L\negthinspace w))$ or
simply as $T(^L\negthinspace w)$. In the present discussion,
$^L\negthinspace G=G=U(N)$.

For example, the fundamental $N$-dimensional representation $\VV$
of $U(N)$ has, with our convention $m_1\geq m_2\geq \dots\geq
m_N$, the highest weight $^L\negthinspace w(1)=(1,0,0,\dots,0)$.
It corresponds to an 't Hooft operator $T(^L\negthinspace w(1))$
or simply $T_{(1)}$. The $k^{th}$ antisymmetric tensor
representation $\wedge^k\VV$, for $k=1,\dots,N$, has highest
weight \eqn\nyto{^L\negthinspace
w(k)=(\overbrace{1,1,\dots,1}^{\text {$k$ times}},0,\dots,0)} and
corresponds to an 't Hooft operator $T_{(k)}$.

The representation $\wedge^N\VV$ is one-dimensional; it is the
representation in which $g\in U(N)$ acts by multiplication by
$\det(g)$.  It can be raised to any integer power $u$, positive or
negative, to get a representation in which $g$ acts by
$(\det(g))^u$. This representation, which we denote as
$(\wedge^N\VV)^u$, corresponds to the weight $^L\negthinspace
w=(u,u,\dots,u)$. The 't Hooft/Hecke operator $T_{(N)}$
corresponding to $\wedge^N\VV$ simply acts by $E\to E\otimes {\cal
O}(p_0)$, as in the abelian case, and is obviously invertible. Its
$u^{th}$ power corresponds to the representation
$(\wedge^N\VV)^u$.  The $T_{(k)}$ with $k<N$ are not invertible.

The central element $\exp(i\alpha)$ of $U(N)$ acts on the
representation $\wedge^k\VV$ as multiplication by
$\exp(ik\alpha)$. More generally, it acts on any representation of
highest weight $^L\negthinspace w=(m_1,m_2,\dots,m_N)$ as
multiplication by $\exp(i\alpha\sum_{i=1}^Nm_i)$. A special case
of this statement is that the center of $U(N)$ acts trivially if
and only if \eqn\hoty{\sum_im_i=0.} $S$-duality, as we discussed
in section \electricbranes, maps the action of the center of
$^L\negthinspace G$ on a representation $^L\negthinspace R$ to the
action of the corresponding 't Hooft operator $T(^L\negthinspace
R)$ on the topology of a $G$-bundle.  Indeed, since
$T(^L\negthinspace w)$ locally maps $\oplus_i{\cal L}_i$ to
$\oplus_i\left({\cal L}_i\otimes {\cal O}(p_0)^{m_i}\right)$, it
changes $c_1(E)$ by \eqn\yorgo{c_1(E)\to c_1(E)+\sum_im_i,} and
thus the topology of $E$ is unchanged if and only if
$\sum_im_i=0$.

{}From $S$-duality, we expect 't Hooft operators of $G$ to form a
commutative algebra isomorphic to the representation ring of
$^L\negthinspace G$.  (In fact, as we discussed in section
\lineop, commutativity of 't Hooft operators follows from more
general considerations of four-dimensional topological field
theory, without resort to $S$-duality.)  We will examine this in
section \kahstr\ from the standpoint of the Bogomolny equations.
The representation ring of $U(N)$ is freely generated by the
representations $\wedge ^k\VV,\,k=1,\dots,N$ along with
$(\wedge^N\VV)^{-1}$. Accordingly the commutative algebra of 't
Hooft operators for $U(N)$ is freely generated by the $T_{(k)}$
together with $T_{(N)}^{-1}$. In the mathematical literature on
the geometric Langlands program, these generators are usually
taken as the basic Hecke operations of $U(N)$.

\bigskip\noindent{\it 't Hooft/Hecke Operators For Other Groups}

The generalization of this to any compact gauge group $G$ is
rather direct. To keep things relatively simple, we will pick a
representation $R$ of $G$ and and consider a $G$-bundle $E_R$
which we take in the representation $R$. (Thus, if $E$ denotes a
principal $G$-bundle, then $E_R=E\times_G R$.) Picking a maximal
torus ${\cal T}$  of $G$, let $w$ denote a weight of $G$, that is
a character $w:{\cal T}\to U(1)$. Let $R=\oplus_wR_w$ be the
decomposition of $R$ in weight spaces $R_w$ (all but finitely many
of which vanish).

An 't Hooft operator of $G$ is classified by a dominant coweight
$^L\negthinspace w:U(1)\to {\cal T}$, which is the highest weight
of a representation $^L\negthinspace R$ of $^L\negthinspace G$.
The composition  of $w:{\cal T}\to U(1)$ and $ ^L\negthinspace
w:U(1)\to {\cal T}$ is a homomorphism $w\circ{}^L\negthinspace
w:U(1)\to U(1)$ which takes the form $\exp(i\alpha)\to
\exp(i\langle ^L\negthinspace w,w\rangle\alpha)$ for some integer
$\langle ^L\negthinspace w,w\rangle$.

For any choice of a complexified maximal torus ${\cal
T}_{\Bbb{C}}$ of $G_{\Bbb{C}}$, the fiber of $E_R$ at $p_0\in C$
has a decomposition $E_R|_{p_0}=\oplus_w E_{R,w}|_{p_0}$ in weight
spaces. If we extend ${\cal T}$ to a holomorphically varying
family of maximal tori near $p_0$, then we get a corresponding
local decomposition of $E_R$ in subbundles of definite weight:
$E_R=\oplus_w {E}_{R,w}$. By solving the Bogomolny equations near
$p_0$, and taking account of the abelian nature of the
singularity, we find that relative to some choice of such a
decomposition, $T(^L\negthinspace w)$ acts by ${E}_{R,w}\to {
E}_{R,w}\otimes {\cal O}(p_0)^{\langle ^L\negthinspace
w,w\rangle}$.

We will make this more concrete for the case $G=SU(N)$,
$^L\negthinspace G=PSU(N)=SU(N)/\Bbb{Z}_N$, and the reciprocal
case $G=PSU(N)$, $^L\negthinspace G=SU(N)$.

A homomorphism $\rho:U(1)\to G=SU(N)$ takes the same form as for
$U(N)$, namely $\exp(i\alpha)\to{\rm
diag}(\exp({im_1\alpha}),\exp({im_2\alpha}),\dots,\exp(im_N\alpha))$,
except that now we must require $\sum_im_i=0$, as in \hoty.  The
action of the corresponding 't Hooft operator $T(^L\negthinspace
w)$ on an $SU(N)$ bundle $E$ is as before: relative to some
decomposition $E=\oplus_{i=1}^N{\cal L}_i$ near $p$, we have
${\cal L}_i\to {\cal L}_i\otimes {\cal O}(p_0)^{m_i}$.

With our usual ordering $m_1\geq m_2\geq\dots\geq m_N$, the
$N$-plet $^L\negthinspace w=(m_1,m_2,\dots,m_N)$ is a positive
integer combination of the simple roots $e_1=(1,-1,0,\dots,0)$,
$e_2=(0,1,-1,0,\dots,0),\dots ,e_{N-1}=(0,0,\dots,0,1,-1)$ of
$SU(N)$. Hence, it is the highest weight of a representation of
the adjoint group $PSU(N)$, which of course coincides with
$^L\negthinspace G$ in this case.

Now let us discuss the case that $G=PSU(N)$, and therefore
$^L\negthinspace G=SU(N)$.  A homomorphism $\rho:U(1)\to PSU(N)$
can be lifted to a homomorphism from an $N$-fold cover of $U(1)$
to $SU(N)$. Such a homomorphism takes the form $\exp(i\alpha)\to
{\rm
diag}(\exp(im_1\alpha),\exp(im_2\alpha),\dots,\exp(im_N\alpha))$,
but the $m_i$ are not integers; rather, they take values in
 ${1\over N}\Bbb{Z}$, with the differences $m_i-m_j$ being integers. Such an $N$-plet
$^L\negthinspace w=(m_1,m_2,\dots,m_N)$, ordered so that $m_1\geq
m_2\geq \dots\geq m_N$, is a dominant weight of $^L\negthinspace
G=SU(N)$. It corresponds to a representation $^L\negthinspace R$
of $SU(N)$ and an 't Hooft operator $T(^L\negthinspace R)$ of
$G=PSU(N)$. The representations $^L\negthinspace R$ on which the
center of $SU(N)$ acts nontrivially, and hence those for which
$T(^L\negthinspace R)$ changes the topology of a $PSU(N)$-bundle,
are precisely those for which the $m_i$ are non-integral.

Let $E$ be a principal $PSU(N)$ bundle.  As $PSU(N)$ does not have
an $N$-dimensional representation corresponding to the
$N$-dimensional representation $\VV$ of $SU(N)$, we cannot derive
from $E$ a rank $N$ complex vector bundle $E_\VV$.  However,
$E_\VV$ does exist as a twisted vector bundle, a notion that we
explained briefly  in section \comptwop.  The obstruction to
lifting $E_\VV$ to an ordinary vector bundle is the characteristic
class $\xi(E)$ introduced in section \reviewthis.  After picking a
family of maximal tori, $E_\VV$ has a local decomposition
$E_\VV=\oplus_{i=1}^N{\cal L}_i$ in terms of twisted line bundles
${\cal L}_i$ (all of which are twisted in the same way, since the
obstruction to lifting $E_\VV$ to a vector bundle is central).

The 't Hooft/Hecke operators act, relative to some such
decomposition, in the familiar way  ${\cal L}_i\to{\cal
L}_i\otimes {\cal O}(p_0)^{m_i}$.  Now, however, the $m_i$ may not
be integers.  We define ${\cal O}(p_0)^{1/N}$ as a twisted line
bundle, and we define ${\cal O}(p_0)^{m/N}$, for any integer $m$,
as the $m^{th}$ power of ${\cal O}(p_0)^{1/N}$.  In particular,
the $N^{th}$ power of ${\cal O}(p_0)^{1/N}$ is the ordinary line
bundle ${\cal O}(p_0)$.  Since the differences $m_i-m_j$ are
integers, the twisted line bundles ${\cal O}(p_o)^{m_i}$ are all
twisted in the same way, and the action of the 't Hooft/Hecke
operators by ${\cal L}_i\to {\cal L}_i\otimes {\cal O}(p_0)^{m_i}$
preserves the fact that the ${\cal L}_i$ all have the same twist.

Of course, we can avoid talking about twisted bundles by picking a
representation $R$ of $PSU(N)$, and applying the 't Hooft
operators to the ordinary vector bundle $E_R=E\times_{PSU(N)}R$.
For example, let $N=2$, so that $PSU(2)$ becomes $SO(3)$. The rank
two twisted bundle $E_\VV$ has the local decomposition ${\cal
L}\oplus {\cal L}^{-1}$ in terms of a twisted line bundle ${\cal
L}$. If we take $R$ to be the adjoint representation of $SO(3)$,
then the corresponding adjoint bundle $E_R$ is ${\rm
Sym}^2(E_\VV)$ (the symmetric part of $E_\VV\otimes E_\VV$) or
${\cal L}^2\oplus {\cal O}\oplus {\cal L}^{-2}$, where ${\cal
L}^2$ is an ordinary line bundle.  The transformation ${\cal L}\to
{\cal L}\otimes {\cal O}(p_0)^{1/2}$ acts on $E_R$  by $E_R\to
({\cal L}^2\otimes \CO(p_0))\oplus \CO\oplus ({\cal L}^{-2}\otimes
\CO(p_0)^{-1})$.  This transformation involves no fractional
exponents.

\subsec{The Space Of Hecke Modifications}

\subseclab\spaceheck

The rest of this section is devoted to a closer look at the
holomorphic data in this problem.

We will first give a direct description of the holomorphic data,
emphasizing simple examples, and then explain in section
\affgrass\ the more powerful framework in which the space of Hecke
modifications is usually placed in the mathematical literature.

The considerations will be local, so to keep things simple we
choose $C=\Bbb{CP}^1$, which we think of as the complex $z$-plane
plus a point at infinity.  We take the point $p_0$ to be $z=0$. To
start with, we take the gauge group to be $G=U(N)$.  We let $E_-$
be a trivial bundle of rank $n$. (Since the considerations will be
local, it does not matter what $E_-$ we start with.)  We want to
discuss how an 't Hooft/Hecke operator can modify $E_-$ to make a
new bundle $E_+$.
 We write a local holomorphic section of $E_-$
(that is, a section holomorphic near $p_0$), as
$s=(f_1,f_2,\dots,f_N)$, where the $f_i$ are local holomorphic
functions.

If $s$ is any holomorphic section of $E_-$ near $p_0$,  we can
define a line bundle ${\cal L}$ near $p_0$, as follows. ${\cal L}$
is a subbundle of $E_-$, defined, over a neighborhood of $p_0$, by
saying that sections of ${\cal L}$ are sections of $E_-$ of the
form $gs$ for an arbitrary holomorphic function $g$. We call
${\cal L}$ the line bundle generated by $s$. Now, suppose that
$s_1,\dots,s_N$ are $N$ local holomorphic sections of $E_-$ whose
restrictions to $z=0$ are linearly independent. (If we write
$s_i(z)=(f_{1i}(z),f_{2i}(z),\dots,f_{Ni}(z))$, the condition is
that the $N\times N$  matrix  $f_{ji}(0)$ has nonzero
determinant.) For each $i$, we let ${\cal L}_i$ be the line bundle
generated by $s_i$.  The linear independence ensures that near
$z=0$ we have a decomposition $E_-=\oplus_{i=1}^N{\cal L}_i$,
which is equivalent to saying that any local holomorphic section
$s$ of $E_-$ is of the form $s=\sum_i g_is_i$ with some local
holomorphic functions $g_i$.  In this situation, we say that the
sections $s_i$ generate $E_-$.

Conversely, any such local decomposition of $E_-$ takes this form.
If $E_-=\oplus_{i=1}^N{\cal L}_i$, we let $s_i$ be any section of
${\cal L}_i$ that is nonzero at $z=0$.  Then near $z=0$, ${\cal
L}_i$ is the line bundle generated by $s_i$, and so the
decomposition of $E_-$ takes the form described in the last
paragraph.

Now we consider the action of the 't Hooft/Hecke operator
$T(^L\negthinspace w)$ with $^L\negthinspace w=(m_1,m_2,\dots,
m_N)$.  With our chosen decomposition of $E$, it maps $E_-$ to the
bundle that near $z=0$ is $E_+=\oplus_i\left( {\cal L}_i\otimes
{\cal O}(p_0)^{m_i}\right)$. Away from $z=0$, $E_+$ is just the
same as $E_-$.

But ${\cal L}_i\otimes {\cal O}(p_0)^{m_i}$ is the line bundle
whose general section, near $z=0$, is $s_i/z^{m_i}$. Thus, near
$z=0$ a general section of $E_+$ takes the form $\sum_{i=1}^N
g_i(z) s_i(z)z^{-m_i}$, where the functions $g_i$ are holomorphic
near $z=0$.  Away from $z=0$, a holomorphic section of $E_+$ is
the same as a holomorphic section of $E_-$: it takes the form
$(f_1,\dots,f_N)$ with any holomorphic functions $f_1,\dots,f_N$.

Different decompositions may lead to the same $E_+$. This is
precisely what we want to investigate; we want to investigate the
space of all bundles $E_+$ to which $E_-$ can be mapped by a given
't Hooft operator.  In formulating this question, we must remember
that any Hecke modification $E_+$ of $E_-$ is naturally isomorphic
to $E_-$ away from $z=0$, so if $E_+$ and $\tilde E_+$ are two
Hecke modifications of $E_-$, they are naturally isomorphic to
each other away from $z=0$.  We consider $E_+$ and $\tilde E_+$
equivalent if and only if the natural isomorphism away from $z=0 $
can be extended over $z=0$.  We want to classify the possible
$E_+$'s up to this equivalence.  What this means in practice will
become clear as we consider examples.

We begin with a few simple examples with $N=2$. First we take
$^L\negthinspace w=(1,0)$. We suppose that
$s_1(z)=(e_1(z),f_1(z))$ and $s_2(z)=(e_2(z),f_2(z))$, with
$(e_1(0),f_1(0))$ and $(e_2(0),f_2(0))$ linearly independent.
$E_+$ is generated by $s_1'=z^{-1}s_1$ together with $s_2$.  In
particular, $s_1=zs_1'$ is a section of $E_+$, as of course is
$s_2$.  Since $s_1$ and $s_2$ generate $E_-$, this means that any
section of $E_-$ is a section of $E_+$.  But a section of $E_+$
may have a polar part proportional to $z^{-1}s_1(0)$.

In sum, in this example, any section of $E_+$ takes the form
\eqn\milok{s(z)={c\over z}s_1(0)+(h_1(z),h_2(z))} near $z=0$,
where $c$ is a complex constant and $h_1$, $h_2$ are local
holomorphic functions. Clearly, $E_+$ depends on the choice of
$s_1(0)=(e_1(0),f_1(0))$ only up to complex scaling.  So we should
think of $(e_1(0),f_1(0))$ as defining a point in $\Bbb{CP}^1$,
and this $\Bbb{CP}^1$ parametrizes all bundles $E_+$ that can be
constructed from $E_-$ by acting with $T(^L\negthinspace w)$ for
$^L\negthinspace w=(1,0)$.

We will say that $\Bbb{CP}^1$ is the space of all Hecke
modifications of type $^L\negthinspace w=(1,0)$. We denote this
space as ${{\EUY}}(^L\negthinspace w)$, or
${{\EUY}}(^L\negthinspace w,p_0)$ if we want to specify the point,
in this case $p_0$, at which the Hecke modification is made.  We
also denote the same space as ${{\EUY}}(^L\negthinspace R)$, where
$^L\negthinspace R$ is the representation of $^L\negthinspace G$
with highest weight $^L\negthinspace w$, and refer to it
alternatively as the space of Hecke modifications of type
$^L\negthinspace R$.

In this particular example, the space of Hecke modifications is
compact and smooth.  But that is not typical.  For another
example, take $^L\negthinspace w=(2,0)$.  The Hecke operator
$T(^L\negthinspace w)$, acting on $E_-$ with the same
decomposition as before, now produces a bundle $E_+$ that is
generated by $s_1'=z^{-2}s_1$ and $s_2$. Since $s_1=z^2s_1'$ is a
section of $E_+$, along with $s_2$, it is still true that any
section $(h_1,h_2)$ of $E_-$ defines a section of $E_+$.  In
addition, $E_+$ has a two-dimensional space of polar sections,
generated by $z^{-2}s_1$ and $z^{-1}s_1$.  Let us expand
$e_1(z)=u+zu'+{\cal O}(z^2)$, $f_1(z)=v+zv'+{\cal O}(z^2)$, with
$u,u',v,v'\in\Bbb{C}$. Then the general local holomorphic section
of $E_+$ takes the form \eqn\ilko{s(z)={c\over
z^2}(u+zu',v+zv')+{d\over z} (u,v)+(h_1(z),h_2(z)),} where $c$ and
$d$ are complex constants, and $h_1(z),h_2(z)$ are holomorphic
functions.

How can we describe the space of Hecke modifications in this case?
Obviously, an overall scaling of the four variables
\eqn\lorky{(u,u',v,v')\to\lambda (u,u',v,v')} is inessential; it
can be absorbed in scaling $c$ and $d$.  But there is another
equivalence relation that we should take into account. If we
transform \eqn\morky{(u',v')\to (u', v')+w(u,v),} for  $w\in
\Bbb{C}$, then this can be absorbed in $d\to d-wc$. So we should
regard this transformation as another equivalence relation on the
parameters $u,u',v,v'$.  The two symmetries both arise from the
fact that $E_+$ is invariant under $s_1\to gs_1$, where $g$ is a
holomorphic function that is invertible at $z=0$.

\def\k{b}
So the space of Hecke modifications is parametrized by
$(u,u',v,v')$, with $u$ and $v$ not both zero (since $s_1(0)$ must
be nonzero), and subject to the equivalences \lorky\ and \morky.
The invariants under the transformation \morky\ are $u$, $v$, and
$\k=vu'-uv'$.  So we can take this scaling into account by simply
using $u$, $v$, and $\k$ to parametrize the Hecke modifications.
$\k$ scales under \lorky\ with weight 2, $\k\to \lambda^2\k$. So
any triplet $(u,v,\k)$, not identically zero, defines, modulo the
scaling,  a point in a weighted projective space
$\Bbb{WCP}^2(1,1,2)$.  The space of Hecke modifications as we have
defined it so far is not all of the weighted projective space; we
are missing the single point $(0,0,1)$, since we do not allow
$u=v=0$.

Although it is possible to work with a noncompact space of Hecke
modifications, it is inconvenient.  The spaces of physical states
in the $A$-model are the cohomology groups of various moduli
spaces.  Compactness makes the definition of the appropriate
cohomology groups much more straightforward; without it, one needs
a detailed discussion of how the wavefunctions are supposed to
behave near infinity (i.e., near $(0,0,1)$ in the present case).

The space of Hecke modifications ${{\EUY}}(^L\negthinspace w)$ as
we have defined it so far does have a natural compactification,
coming from the fact that the bundle $E_+$ has a natural limit as
$u,v\to 0$. Introduce a small parameter $\epsilon$ and let
$u=\epsilon \bar u$, $v=\epsilon \bar v$, and $d=d'/\epsilon$. A
general section of $E_+$ takes the form \eqn\gilko{s(z)={c\over
z^2}(\epsilon \bar u+zu',\epsilon \bar v+zv')+{d'\over z} (\bar
u,\bar v)+(h_1(z),h_2(z)).} The limit of this as $\epsilon\to 0$,
keeping the other variables fixed, is just \eqn\pilko{
s(z)={c\over z}(u',v')+{d'\over z}(\bar u,\bar
v)+(h_1(z),h_2(z)).}

The bundle whose general section takes this form can be given a
more simple interpretation.  The point $(0,0,1)$ on the weighted
projective space has $\k\not=0$, which is equivalent to linear
independence of the vectors $(u,v)$ and $(u',v')$ or equivalently
$(\bar u,\bar v)$ and $(u',v')$. So when we approach $(0,0,1)$ by
taking a limit as $\epsilon\to 0$, we should assume this linear
independence. That being so, \pilko\ can be described more simply
by saying that $s(z)$ can have an arbitrary simple pole at $z=0$.
Consequently, the bundle in this limit is simply $E_+=E_-\otimes
{\cal O}(p_0)$.

But this is what we get if we make a Hecke modification of the
bundle $E_-$ with $^L\negthinspace w=(1,1)$.  The Hecke operation
$T(^L\negthinspace w)$ for this $^L\negthinspace w$ is simply the
operation of tensoring with ${\cal O}(p_0)$, as we have explained
in section \theck.  The conclusion, then, is that the space of
Hecke modifications with $^L\negthinspace w=(2,0)$ has a natural
compactification in which one also allows a Hecke modification
with $^L\negthinspace w=(1,1)$.  We will write $\bar
{{\EUY}}(^L\negthinspace w)$ for this sort of compactification of
the space of Hecke modifications of type ${{\EUY}}(^L\negthinspace
w)$ (we write more explicitly $\bar {{\EUY}}(^L\negthinspace
w,p_0)$ if we wish to specify the point at which the Hecke
modification is made).

We recall that in general an 't Hooft operator is specified by the
choice of a $G$-bundle over $\Bbb{CP}^1$; once such a choice is
given, one then solves the Bogomolny equations requiring a local
singularity determined by that choice of $G$-bundle. In
particular, for $G=U(2)$, the 't Hooft operator with
$^L\negthinspace w=(m_1,m_2)$ is defined using the $U(2)$ bundle
${\cal O}(m_1)\oplus {\cal O}(m_2)$ over $\Bbb{CP}^1$.  The only
topological invariant of such a bundle is its first Chern class,
which is $m_1+m_2$. The bundle ${\cal O}(2)\oplus {\cal O}$, which
corresponds to $^L\negthinspace w=(2,0)$, is an ``unstable
bundle''; it can be infinitesimally perturbed to change its
holomorphic type to ${\cal O}(1)\oplus {\cal O}(1)$, which
corresponds to $^L\negthinspace w =(1,1)$. The analysis that we
have just described shows that the Hecke modification
corresponding to $^L\negthinspace w=(1,1)$ must be included in
order to compactify the space of Hecke modifications with
$^L\negthinspace w=(2,0)$.

The other example that we considered is quite different.  The
bundle ${\cal O}(1)\oplus {\cal O}(0)$  cannot be perturbed to any
other holomorphic type. So the space of Hecke modifications with
$^L\negthinspace w=(1,0)$  is smooth and compact.

\bigskip\noindent{\it The Singularity of The Weighted Projective Space}

To understand more deeply the example with $^L\negthinspace
w=(2,0)$, let us reexamine the weighted projective space
$\Bbb{WCP}^2(1,1,2)$, which has homogeneous coordinates $(u,v,\k)$
of weights $(1,1,2)$. At the point $(0,0,1)$, this space has a
$\Bbb{Z}_2$ orbifold singularity.

The local structure is therefore that of the $A_1$ singularity
$\Bbb{C}^2/\Bbb{Z}_2$. $\Bbb{C}^2/\Bbb{Z}_2$ is a hyper-Kahler
orbifold.  Unlike a typical complex singularity, the $A_1$
singularity can be resolved and deformed in a compatible fashion.
This is most naturally understood  \ref\otherkron{P. Kronheimer,
``The Construction Of ALE Spaces As Hyper-Kahler Quotients,'' J.
Diff. Geom. {\bf 29} (1989) 665-683.} by exhibiting
$\Bbb{C}^2/\Bbb{Z}_2$ as a hyper-Kahler quotient of
$\Bbb{H}^2\cong \Bbb{C}^4$ by $U(1)$. Instead of following that
route, we will here just describe by hand the deformation and
resolution of the singularity of the weighted projective space
$\Bbb{WCP}^{2}(1,1,2)$.

To describe the deformation, we set $a_1=u^2,$ $a_2=uv$,
$a_3=v^2$. The variables $(a_1,a_2,a_3,\k)$ all have weight two,
and modulo scaling they parametrize an ordinary projective space
$\Bbb{CP}^3$.  However, they obey $a_1a_3-a_2^2=0$, so
$\Bbb{WCP}^2(1,1,2)$ is the subvariety of $\Bbb{CP}^3$ defined by
this equation.  The equation $a_1a_3-a_2^2=0$ can be deformed to
$a_1a_3-a_2^2=\epsilon \k^2$ with a small parameter $\epsilon$.
This last equation defines a smooth quadric in $\Bbb{CP}^3$, which
is isomorphic to\foot{By a linear change of variables, one can
arrange $a_1,a_2,a_3$ and $\k$ as the matrix elements of a
$2\times 2$ matrix $M$ such that the equation becomes $\det(M)=0$.
This equation says that $M$ is of rank 1, and so takes the form
$M_{ij}=x_iy_j$, where the $x_i$ and $y_j$ are homogeneous
coordinates for, respectively, the two factors in
$\Bbb{CP}^1\times \Bbb{CP}^1$.} $\Bbb{CP}^1\times \Bbb{CP}^1$, so
$\Bbb{WCP}^2(1,1,2)$ can be deformed to $\Bbb{CP}^1\times
\Bbb{CP}^1$.

To describe the resolution, we begin with $\Bbb{C}^4$ with
coordinates $u,v,\k$, and $\k'$.  We consider the action of
$U(1)\times U(1)$  by $(u,v,\k,\k')\to (\lambda u,\lambda
v,\lambda^2 \tilde \lambda \k,\tilde\lambda \k')$, where
$|\lambda|=|\tilde\lambda|=1$.  We define the moment maps
$\mu=|u|^2+|v|^2+2|\k|^2$, $\tilde\mu= |\k|^2+|\k'|^2$, and we
take a symplectic quotient of $\Bbb{C}^4$ by $U(1)^2$.  To do
this, impose the moment map equations
\eqn\ripo{\eqalign{|u|^2+|v|^2+2|\k|^2 & = 1\cr
                                        |\k|^2+|\k'|^2& = d\cr}}
for some constant $d$, and divide by $U(1)\times U(1)$. For small
positive  $d$, the equations do not permit $u=v=0$, so the pair
$(u,v)$ can be taken as homogeneous coordinates for a copy of
$B=\Bbb{CP}^1$.  The fiber over a given point in $\Bbb{CP}^1$ is,
for small $d$, another copy of $\Bbb{CP}^1$, parametrized by $\k$
and $\k'$.  Since $(\k,\k')$ transform to $(\lambda^2\k,\k')$
under $(u,v)\to (\lambda u,\lambda v)$, $\k$ and $\k'$ take values
in the line bundles ${\cal O}(2)$ and ${\cal O}$, respectively,
over $B$. So the symplectic quotient for small positive $d$ is
fibered over $B=\Bbb{CP}^1$ with the fiber being $Y=\Bbb{P}({\cal
O}(2)\oplus {\cal O})$, that is, the projectivization of the rank
two vector bundle $\CO(2)\oplus \CO$. In particular, the
symplectic quotient for small positive $d$ is a $\Bbb{CP}^1$
bundle over $\Bbb{CP}^1$, and thus a Hirzebruch surface.

If we increase $d$, nothing happens until we reach $d=1/2$. At
that point, the Hirzebruch surface develops a singularity and
reduces to the weighted projective space $\Bbb{WCP}^2(1,1,2)$. At
$d=1/2$, we simply write the second moment map equation as
$|\k'|^2=1/2-|\k|^2$, and observe that, by virtue of the first
equation, the right hand side is nonnegative. So a solution for
$\k'$ always exists and is unique up to the
action\foot{$\tilde\lambda$ acts freely on $\k'$ except at the
point $\k'=u=v=0$, where the action of $\tilde\lambda$ generates
the same orbit as $\lambda$.  So at $d=1/2$, upon fixing $\k'$ to
be a nonnegative real number, we can omit $\tilde\lambda$ from the
description.} of $\tilde\lambda$. The symplectic quotient at
$d=1/2$ is thus parametrized by $u,v$, and $\k$, subject to
$|u|^2+|v|^2+2|\k|^2=1$ and the equivalence $(u,v,\k)\cong
(\lambda u,\lambda v,\lambda^2 \k)$ for $|\lambda|=1$. The
combined operation gives the weighted projective space
$\Bbb{WCP}^2(1,1,2)$.

Running this in reverse, reducing $d$ from $1/2$ to a smaller
value resolves the singularity of $\Bbb{WCP}^2(1,1,2)$ to give the
Hirzebruch surface.  Moreover, this particular Hirzebruch surface
can be deformed to give a simple product $\Bbb{CP}^1\times
\Bbb{CP}^1$.  For this, we just observe that the bundle
$\CO(2)\oplus \CO$ is unstable and can be deformed to
$\CO(1)\oplus \CO(1)$.  But the projectivization of $\CO(1)\oplus
\CO(1)$ is the same as the projectivization of $\CO\oplus \CO$ and
so is a trivial $\Bbb{CP}^1$ bundle over $\Bbb{CP}^1$, that is a
product $\Bbb{CP}^1\times \Bbb{CP}^1$.

The conclusion is that, in contrast to a generic singular
algebraic variety, the singularity of $\Bbb{WCP}^2(1,1,2)$ can be
deformed or resolved to give the same result topologically, and
moreover the deformation and resolution can be carried out
simultaneously. Moreover, this is also related to the fact that
the local structure near the singularity is the hyper-Kahler $A_1$
singularity.  At the end of section \monbub\ and in section
\kahstr, we will explain this behavior from the point of view of
the Bogomolny equations.
 We also in section \opprod\
will use the structure near this singularity to illustrate the
computation of the operator product expansion of 't Hooft
operators.

\bigskip\noindent{\it General Case For $U(2)$}

The space of Hecke modifications of type $^L\negthinspace
w=(m_1,m_2)$ can be analyzed similarly.  First of all, the space
of Hecke modifications of type $(m_1+c,m_2+c)$ is the same as that
of type $(m_1,m_2)$, since adding $(c,c)$ just has the effect of
tensoring the output of the Hecke transformation with
${\CO}(p_0)^c$.  So it is enough to take $^L\negthinspace w=(a,0)$
with some integer $a$, which  we can assume nonnegative.
(Otherwise we replace $(a,0)$ by $(0,a)$ and then, adding $-a$ to
each weight, by $(-a,0)$.) In describing Hecke modifications of
type $(a,0)$, it is convenient, given a function $g$ with a pole
of order $n$ at $z=0$, to write $[g]_-$ for the polar part of $g$:
$[g]_-=g_{-n}z^{-n}+g_{-n+1}z^{-n+1}+\dots + g_{-1}z^{-1}$.

Applying a Hecke transform with $^L\negthinspace w=(a,0)$ relative
to the decomposition $E_-={\cal L}_1\oplus {\cal L}_2$ introduced
above, we get a bundle $E_+$ whose general section takes the form
\eqn\indon{s=[c_0s_1]_-+[c_1zs_1]_-+[c_2z^2s_1]_-+\dots
+[c_{a-1}z^{a-1}s_1]_-+(h_1(z),h_2(z)),} with complex coefficients
$c_0,c_1,\dots,c_{a-1}$.

The bundle $E_+$ therefore depends on $s_1$ modulo terms of order
$z^a$. This leaves a total of $2a$ complex parameters. However,
the bundle $E_+$ is unaffected if we replace $s_1$ by $gs_1$,
where $g$ is a holomorphic function that is invertible at $z=0$.
By choice of $g$, we can eliminate $a$ of the complex parameters
in $s_1$, generalizing what we did for $a=1,2$. So the space of
Hecke modifications of type $(a,0)$ has complex dimension $a$. To
compactify this space, one has to include Hecke modifications of
types $(a-1,1), \,(a-2,2),\dots, (a-[a/2],[a/2])$, which appear
when successive terms in the Taylor series expansion of $s_1$
vanish.

Some of what we have explained will be more transparent to most
physicists if we shift from $^L\negthinspace w=(a,0)$ to
$^L\negthinspace w=(a/2,-a/2)$. (The space of Hecke modifications
is of course unchanged in this shift.) As we have explained in
section \theck, weights $(a/2,-a/2)$ make sense for gauge group
$G=PSU(2)=SO(3)$, $^L\negthinspace G=SU(2)$, and moreover the 't
Hooft operator with these weights is $S$-dual to a Wilson loop of
$SU(2)$ in the representation of spin $a$.

In studying confinement in $SU(2)$ gauge theory, it is redundant
to consider Wilson loops based on an arbitrary representation $R$.
Two representations on which the center of $SU(2)$ act in the same
way are equivalent for that purpose, so the only non-trivial
representation of $SU(2)$ that one must consider is the
representation of spin $1/2$.   Dually, any 't Hooft operator of
$SO(3)$ is topologically equivalent to either the trivial one or
the operator with weights $(1/2,-1/2)$.  Precisely because the
higher 't Hooft operators are topologically unstable, they are
considered inessential in applications of 't Hooft operators (such
as 't Hooft's original work) that aim at understanding the phases
of gauge theories.

{}From our point of view, the unstable 't Hooft operators are
important, but can ``mix'' with lower operators of the same
topological type.  In section \monbub, we will explain what this
mixing means in terms of the Bogomolny equations.

The general nature of the mixing is most easily stated in case the
Lie groups $G$ and $^L\negthinspace G$ are simple.  Let
$^L\negthinspace R$ be a representation of $^L\negthinspace G$.
Let us say that a representation $^L\negthinspace R'$ of
$^L\negthinspace G$ is associated to $^L\negthinspace R$ if the
highest weight of $^L\negthinspace R'$ is a weight of
$^L\negthinspace R$ but not the highest weight. Writing
$^L\negthinspace w$ and $^L\negthinspace w'$ for the highest
weights of $^L\negthinspace R$ and $^L\negthinspace R'$, we also
say in this situation that $^L w'$ is associated to
$^L\negthinspace w$. The condition is that $^L\negthinspace
w-{}^L\negthinspace w'$ is a nonzero dominant weight of the
adjoint form of $^L\negthinspace G$.

Then in general, the natural compactification of the space
${{\EUY}}(^L\negthinspace R)$ of Hecke modifications of type
$^L\negthinspace R$ includes the Hecke modifications of type
$^L\negthinspace R'$ for all representations $^L\negthinspace R'$
associated to $^L\negthinspace R$.
 For $G=U(N)$ (or $SU(N)$ or $PSU(N)$), this
statement just means that for a weight $^L\negthinspace
w=(m_1,m_2,\dots,m_N)$, if $m_i\geq m_j+2$ for some $i,j$, the
natural compactification includes Hecke modifications with $m_i$
replaced by $m_i-1$ and $m_j$ replaced by $m_j+1$, as well as all
other weights that can be obtained by repeating this move. That
this move is needed for the compactification can be seen simply by
embedding in $U(N)$ a family of $U(2)$ Hecke modifications whereby
weights $(m_i,m_j)$ degenerate to $(m_i-1,m_j+1)$.   For more
general groups, one can argue similarly, using moves generated by
various $SU(2)$ subgroups of $^L\negthinspace G$.  The role of
these moves in the compactification can also be seen in terms of
the affine Grassmannian, a notion that we briefly introduce in
section \affgrass, or from the Bogomolny equations, as in section
\monbub.

\bigskip\noindent{\it Minuscule Representations}

Generalizing the weight $(1,0)$ of $U(2)$, we will describe a few
more examples where the space of Hecke modifications is smooth and
compact. For $G=U(N)$, let us consider as in \nyto\ the weight
$^L\negthinspace w(k)=(1,1,\dots,1,0,0,\dots,0)$, with the number
of 1's equal to $k$. Picking a decomposition
$E_-=\oplus_{i=1}^N{\cal L}_i$, where ${\cal L}_i$ is generated by
a section $s_i$, the Hecke operator $T({}^L\negthinspace w(k))$
maps $E_-$ to $E_+=\oplus_{i\leq k}\left({\cal L}_i\otimes
\CO(p_0)\right)\oplus\left(\oplus_{i>k}{\cal L}_i\right)$. The
general section of $E_+$ takes the form \eqn\nson{s={1\over
z}(a_1,\dots,a_N)+(h_1,\dots,h_N)} near $p_0$, where $\vec
a=(a_1,\dots,a_N)\in \Bbb{C}^N$ is a linear combination of
$s_1(0),\dots,s_k(0)$.  $E_+$ therefore depends on the choice of
$s_i$ precisely through the subspace $W\subset \Bbb{C}^N$
generated by $s_1(0),\dots,s_k(0)$.  $W$ can be any
$k$-dimensional subspace of $\Bbb{C}^N$, so the space
${{\EUY}}(^L\negthinspace w(k))$ of Hecke modifications of type
$^L\negthinspace w(k)$ is the Grassmannian ${\rm Gr}(k,N)$ of
complex $k$-planes in $\Bbb{C}^N$.

Clearly, then, the space of Hecke modifications of type
$^L\negthinspace w(k)$
 is smooth and compact.  This generalizes the $U(2)$
example with $^L\negthinspace w=(1,0)$.  As we noted in relation
to eqn. \nyto, $^L\negthinspace w(k)$ is the highest weight of the
representation $\wedge^k\VV$ of $^L\negthinspace G=U(N)$.  This
representation is ``minuscule,'' which means that its weights form
a single Weyl orbit.  In general, minuscule representations  are
associated to stable 't Hooft operators with smooth, compact
spaces of Hecke modifications.  Minuscule representations are
precisely those that have no associated representations, since the
highest weight of a minuscule representation is its only dominant
weight.

The highest weight $^L\negthinspace w'(k)$ of the analogous
representation $\wedge^k\VV$ of $SU(N)$ is obtained by subtracting
the constant $-k/N$ from each weight of $^L\negthinspace w(k)$.
The space of Hecke modifications is unchanged.  The
representations $\wedge^k\VV$ are minuscule and generate the
representation ring of $SU(N)$. However, $SU(N)$ is the only
simple and simply-connected Lie group whose representation ring is
generated by minuscule representations. For general simple $G$,
there is precisely one minuscule representation for each character
of the center of $G$; groups other than $SU(N)$ have relatively
small centers and few minuscule representations.

\bigskip\noindent{\it The Dimension Of The Space Of Hecke
Modifications}

For $U(N)$ (or $SU(N)$, or $PU(N)$), let us determine the
dimension of the space of Hecke modifications of an arbitrary
type.  For weight $^L\negthinspace w=(a_1,a_2,\dots,a_N)$, with
$a_1\geq a_2\geq \dots\geq a_N$, the dimension is
\eqn\ydro{\Delta_{^L\negthinspace w}=\sum_{i<j}(a_i-a_j).} To
justify this statement, we consider $N$ sections $s_i$ that
generate the trivial rank $N$ bundle $E_-=\CO\oplus
\CO\oplus\dots\oplus\CO$ near $z=0$.  We recall that this means
that the $s_i(0)$ are linearly independent.  Then we consider the
bundle $E_+$ whose general section is
\eqn\ytgo{q=\sum_{i=1}^Nz^{-a_i}g_is_i,} where the functions
$g_i(z)$ are holomorphic at $z=0$.  The bundle $E_+$ is invariant
under \eqn\gunjo{s_i\to s_i+\sum_{j\leq
i}h_js_j+\sum_{j>i}z^{a_i-a_j}h_js_j,} for generic functions $h_j$
that are holomorphic at $z=0$. Such a transformation of the $s_i$
can be absorbed in redefining the functions $g_i$ that appear in
\ytgo. Expanding $s_i$ in a Taylor series near $z=0$, the number
of coefficients that cannot be eliminated by the transformation
\gunjo\ is $\sum_{j=i+1}^N(a_i-a_j)$.  After summing over $i$, we
arrive at the formula \ydro\ for the dimension of the space of
Hecke modifications.

\subsec{The Affine Grassmannian}\subseclab\affgrass

\lref\PS{A.~Pressley and G.~Segal, {\it Loop Groups,} (Oxford
University Press, 1988).}

The space of Hecke modifications associated with an 't Hooft
operator can be described as a finite-dimensional subvariety in a
certain infinite-dimensional variety called the affine
Grassmannian. This description enables one to construct a natural
compactification of the space of Hecke modifications.  We will not
really use this framework in the present paper, but we explain it
here because it gives a powerful way of understanding some things
and is the framework in which the subject is usually placed
mathematically. The affine Grassmannian is familiar to physicists
primarily for its applications to two-dimensional conformal field
theory \PS.

For the sake of clarity, let us consider the case $G=U(N)$. In
this case, an 't Hooft operator is specified by a set of integers
$m_1,\ldots,m_N$ modulo permutations. Since the problem is local,
we can fix the curve $C$ to be $\Bbb{CP}^1$ and take $E_-$ to be
trivial.  A Hecke modification $E$ of $E_-$ at a point $p_0\in C$
is a bundle that is endowed with a chosen isomorphism
$\sigma:E\cong E_-$ outside of $p_0$; $\sigma$ may not extend over
$p_0$.

It is natural to regard the space of Hecke modifications for fixed
$m_1,\ldots,m_N$ as a subspace of the space of pairs $(E,\sigma)$,
where $E$ is a holomorphic vector bundle on $C$ and $\sigma$ is
its trivialization outside $p_0$. Let us recall an explicit
description of the space of such pairs $(E,\sigma)$, following
chapter 8 of \PS.

It is convenient to think of $C=\Bbb{CP}^1$ as a one-point
compactification of the complex $z$-plane $\CC$ and to set $p_0$
to be the point $z=0$. We also let \eqn\jmj{
U_\infty=\Bbb{CP}^1-\{0\},\quad U_0=\Bbb{CP}^1-\{\infty\} .} We
are given a trivialization $\sigma$ of $E$ over $U_\infty$.
Although $\sigma$ may not extend over $U_0$, we can pick a
trivialization $\sigma'$ of $E$ over $U_0$. Over $U_0\bigcap
U_\infty\simeq\CC^*$, $\sigma $ and $\sigma'$ are related by a
$GL(N,\CC)$ gauge transformation $g(z)$. This is a
$GL(N,\CC)$-valued function whose entries are holomorphic on
$\CC^*$ but may have poles at $0$ and $\infty$. ($\sigma'$ can
always be chosen so that the singularities of $g(z)$ are poles.)
Let us denote by ${\cal X}$ the ring of functions holomorphic on
$\CC^*$ and having a meromorphic extension to $\Bbb{CP}^1$; then
the group of holomorphic gauge transformations on $\CC^*$ can be
identified with $GL(N,{\cal X})$, and $g(z)$ is an element of this
group.

If we change $\sigma'$ by a gauge transformation $h_0(z)$ which is
holomorphic throughout $U_0$, then $g(z)$ is replaced by
$g(z)h_0^{-1}(z)$.  $h_0(z)$ takes values in $GL(N,\cO)$, the
group of invertible matrices whose entries are holomorphic
functions on $U_0\simeq \CC$. The set of holomorphic vector
bundles of rank $N$ over $\Bbb{CP}^1$ equipped with a fixed
holomorphic trivialization $\sigma$ outside $p_0$ is isomorphic to
the quotient $GL(N,{\cal X})/GL(N,\cO)$. The latter space is known
as the affine Grassmannian ${\rm Gr}_N$ for the group $GL(N)$.
Another name for it is the loop Grassmannian, because it is
isomorphic \PS\ to the space of based loops in $U(N)$.

The definition of the affine Grassmannian admits slight
variations. For example, since any function in ${\cal X}$ can be
expanded in a Laurent series, one can embed ${\cal X}$ into the
ring $\CC((z))$ of formal Laurent series. Similarly, one can embed
the ring of holomorphic functions $\cO$ into the ring $\CC[[z]]$
of formal power series. One can show that the quotient
$GL(N,\CC((z)))/GL(N,\CC[[z]])$ is isomorphic to the affine
Grassmannian ${\rm Gr}_N$. Intuitively, this happens because given
any two elements of $GL(N,{\cal X})$, one can tell whether they
are in the same $GL(N,\cO)$ orbit by studying a finite number of
terms in their Laurent expansions, and therefore it is immaterial
whether the Laurent series have a nonzero region of convergence.

Now we have to identify those points in ${\rm Gr}_N$ which can be
obtained from the trivial vector bundle by a Hecke modification of
weight $^L\negthinspace w=(m_1,\ldots,m_N)$. A Hecke modification
of the trivial vector bundle with this weight is obtained by
choosing a trivialization of the trivial vector bundle over $U_0$
by $N$ linearly independent holomorphic sections $f_1,f_2,\dots,
f_N$, and declaring that $E_+$ is generated over $U_0$ by sections
$s_1,\ldots,s_N$ which upon restriction to $U_\infty\bigcap U_0$
are given by \eqn\mcop{ z^{-m_j} f_j,\quad j=1,\ldots,N. } Here we
made use of the fact that on $U_\infty$ we are given an
isomorphism between $E_+$ and the trivial vector bundle. Thus if
$f_j=(f_j^1,f_j^2,\dots,f_j^N)$, then the matrix $g(z)$
corresponding to $E_+$ is given by \eqn\xcoxo{ g^i_j(z)=z^{-m_j}
f^i_j(z). } The simplest choice is $f^i_j(z)=\delta^i_j$; all
other choices can be obtained from this one by acting on $g(z)$
from the left by an element of $GL(N,\cO)$. We conclude that the
space of Hecke modifications ${{\EUY}}({}^L\negthinspace w;p_0)$
is the orbit of the point \eqn\gspec{ g^i_j(z)=z^{-m_j}
\delta^i_j} in the affine Grassmannian under the left action of
the group $GL(N,\cO)$. One can replace ${\cal X}$ with $\CC((z))$
and $\cO$ with $\CC[[z]]$ throughout and get an equivalent result.
Note that $g(z)$ given by \gspec\ is a homomorphism of $\CC^*$ to
$GL(N,\CC)$.  In fact, it is the complexification of the
homomorphism \eqn\xxconon{ \rho: U(1)\to U(N),\quad \rho:
e^{i\alpha}\to \left(\matrix{e^{i m_1\alpha} & & \ldots & \cr &
e^{i m_2\alpha} & \ldots &  \cr  & \vdots & \ldots & \cr & & &
e^{i m_N\alpha}\cr}\right) } which enters into the definition of
the 't Hooft operator $T(m_1,\ldots,m_N)$.

We have associated to each set of integers $m_1,\ldots,m_N$ an
orbit of $GL(N,\cO)$ in ${\rm Gr}_N$. It turns out that all
$GL(N,\cO)$ orbits in ${\rm Gr}_N$ are obtained in this way \PS .
Thus ${\rm Gr}_N$ is stratified by spaces
${{\EUY}}(m_1,\ldots,m_N)$, which are called Schubert cells.
Equivalently, by applying Hecke modifications to the trivial
vector bundle, one can obtain an arbitrary holomorphic vector
bundle on $\Bbb{CP}^1$ with an arbitrary trivialization on
$U_\infty$.

The examples of spaces of Hecke modifications discussed in section
\spaceheck\ suggest that these spaces are always
finite-dimensional. This can be shown in general as follows. Let
${\rm Gr}_N(i)$ be the subset of ${\rm Gr}_N$ defined by the
condition that $g(z)$ has at $z=0$ a pole of order not higher than
$i$. The variety ${\rm Gr}_N(i)$ is finite-dimensional,
$GL(N,\cO)$-invariant, and compact, and any point in ${\rm Gr}_N$
belongs to ${\rm Gr}_N(i)$ for some $i$. This implies that any
orbit of $GL(N,\cO)$ belongs to some ${\rm Gr}_N(i)$ and therefore
is finite-dimensional.

It may be helpful to mention at this point that the
infinite-dimensional space ${\rm Gr}_N$ has another stratification
with strata labelled by sets of integers $k_1,\ldots,k_N$. To
define this stratification, we recall that by a theorem of
Grothendieck any holomorphic vector bundle of rank $N$ on
$\Bbb{CP}^1$ is isomorphic to \eqn\Groth {\oplus_{i=1}^N
\cO(p_0)^{k_i}} for some integers $k_1,\ldots,k_N$. We define the
stratum in ${\rm Gr}_N$ corresponding to $k_1,\ldots,k_N$ as the
set of those pairs $(E,\sigma)$ where $E$ is isomorphic to \Groth.
In other words, a stratum is obtained by fixing $E$ and varying
the trivialization $\sigma$. Obviously, the strata are
infinite-dimensional in this case, in contrast with the
stratification given by ${{\EUY}}(m_1,\ldots,m_N)$. The relation
between the two stratifications of ${\rm Gr}_N$ is studied in \PS.

As we have seen, spaces of Hecke modifications or Schubert cells
are noncompact, in general. A natural way to compactify them is to
consider the closure of the corresponding orbits in ${\rm
Gr}_N(i)$ for sufficiently large $i$. For a given $\rho: U(1)\to
U(N)$, the closure of the orbit $C_\rho$ is in general a singular
variety which  is a union of $C_\rho$ and a finite number of
orbits of lower dimension. It is called a Schubert cycle.  As
discussed in section \spaceheck, the structure of the closure of
$C_\rho$ reflects the ``mixing'' between 't Hooft operators with
different $\rho$ but the same topological type.  We also will
describe this process in terms of monopole bubbling in section
\monbub.

The construction of ${\rm Gr}$ and the description of the spaces
of Hecke modifications as subvarieties in ${\rm Gr}$ can be
generalized to other gauge groups. For any simple compact Lie
group $G$, with complexification $G_{\Bbb{C}}$, one defines ${\rm
Gr}_G$ as the quotient \eqn\hxonoh{ {\rm Gr}_G=G_{\Bbb{C}}({\cal
X})/G_{\Bbb{C}}(\cO). } An 't Hooft operator is parametrized by
the conjugacy class of a homomorphism $\rho: U(1)\to G$, which can
be analytically continued to a homomorphism $\rho_\CC: \CC^*\to
G_{\Bbb{C}}$. Obviously, $\rho_\CC$ defines a point on ${\rm
Gr}_G$, and the orbit of this point under the left action of
$G_{\Bbb{C}}(\cO)$ depends only on the conjugacy class of $\rho$.
This orbit is the Schubert cell or space of Hecke modifications
${{\EUY}}(\rho)$.

\newsec{ The Bogomolny Equations And The Space Of Hecke
Modifications}

\seclab\bogeqheck

In this section, we return to the Bogomolny equations with
singularities and study them more closely. The first goal is to
exhibit the space of Hecke modifications of a given type as a
moduli space of solutions of the Bogomolny equations, with
suitable boundary conditions. This will enable us to get a clear
picture of many relatively subtle aspects of the spaces of Hecke
modifications. Then we relax the assumption of time-reversal
invariance and consider the complexified or extended Bogomolny
equations. These extended equations enable one to define the Hecke
transform of an arbitrary $A$-brane on $\MH(G,C)$ and therefore
are the important ones for geometric Langlands duality.

\subsec{Boundary Conditions}

\subseclab\boundcon

We will study the Bogomolny equations on  a $G$-bundle $E$ over a
three-manifold $W=I\times C$ (as usual, $I$ is an interval and $C$
a Riemann surface, and there are prescribed singularities at
points $p_i\in W$ where 't Hooft operators are inserted).  We
write $C_-$ and $C_+$ for the two ends of $W$,  $E_-$ and $E_+$
for the restriction of $E$ to the two ends, and $A_-$ and $A_+$
for the restrictions of the connection $A$.

\ifig\utampo{(a) On the three-manifold $W=I\times C$ with 't Hooft
operator insertions, a bundle $E_-$ is specified on the left, and
one wishes to determine possible Hecke modifications $E_+$ that
may appear on the right.  On the left, the connection $A_-$ obeys
Dirichlet boundary conditions and $\phi_0$  is undetermined; on
the right, their roles are reversed. (b) To describe a fiber of
the Hecke correspondence, one specifies both $E_-$ and $E_+$ and
leaves $\phi_0$ undetermined at each end.}
{\epsfxsize=4in\epsfbox{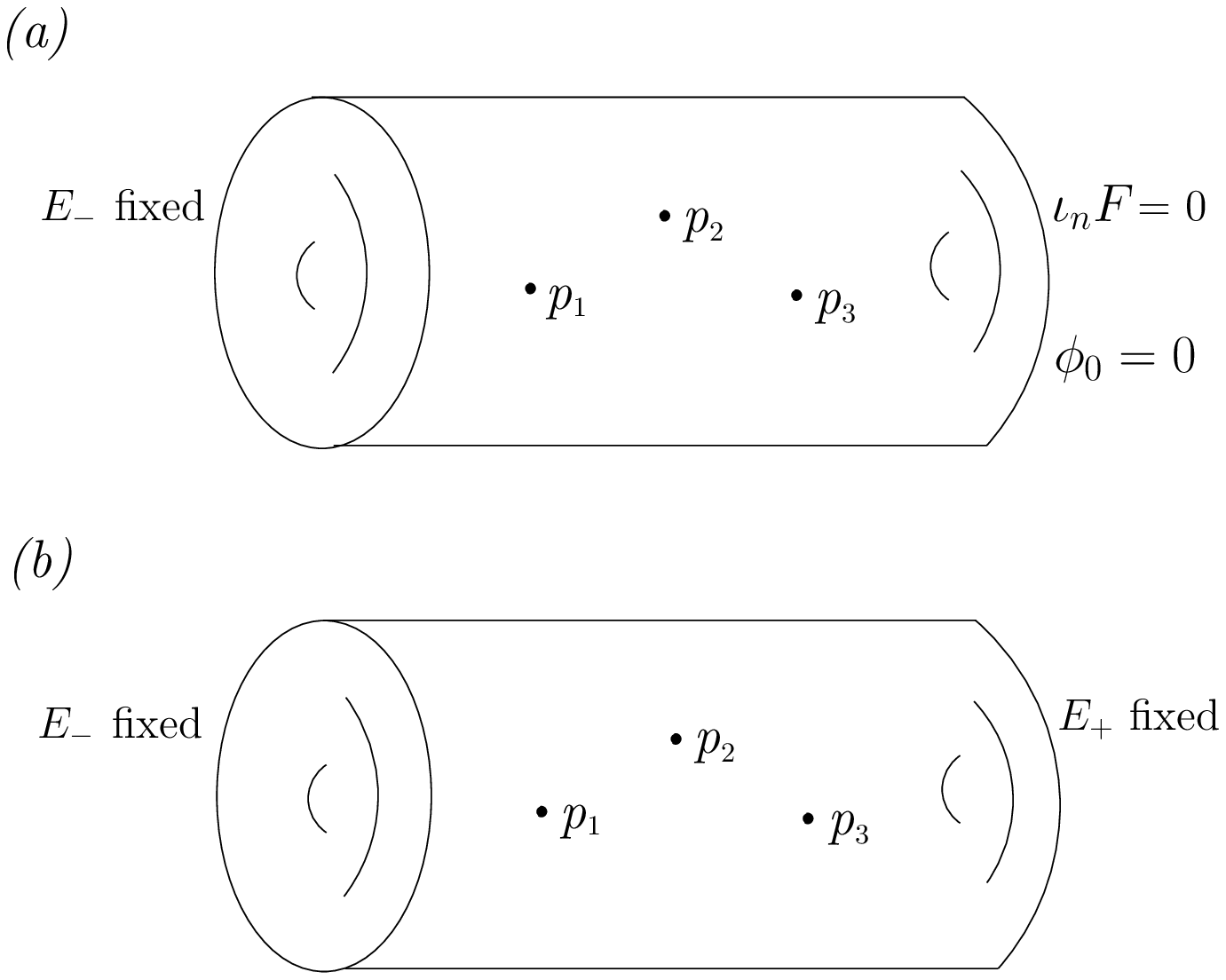}}

There are two closely related problems that we might consider. One
natural question  is to describe the action of the 't Hooft/Hecke
operators on bundles, and the second is to describe the Hecke
correspondence.

For the first problem (\utampo(a)), we specify a bundle $E_-$ at
the left end of $W$ and then ask what bundle $E_+$ can appear at
the other end as a result of solving the Bogomolny equations with
singularities. In this way we should recover the space of Hecke
modifications, as described in section \theck.  An alternative
problem (\utampo(b)) is to try to describe the Hecke
correspondence, which parametrizes pairs of bundles with a
suitable Hecke relation. We will describe boundary conditions
appropriate for either of these two problems.

To specify the bundle $E_-$, we specify the gauge field $A_-$ on
$C_-$. We want to specify an actual bundle $E_-$, not an
isomorphism class of bundles, so we specify an actual connection
$A_-$ on $C_-$. So we will only divide by gauge transformations
that are trivial on $C_-$.  It is convenient for some purposes to
assume that the bundle $E_-$ is stable, in which case we can
assume that the connection $A_-$ is flat.  We will not assume that
the output bundle $E_+$ that results from solving the Bogomolny
equations is stable.

We want the boundary conditions on $C_-$ to be elliptic (modulo
the gauge transformations). The simplest elliptic boundary
conditions in the Bogomolny equations are obtained by either
imposing Dirichlet boundary conditions on $A$ -- that is,
specifying its boundary values -- and leaving the boundary values
of $\phi_0$ unspecified, or vice-versa. If the boundary values of
$A$ are specified, then the Bogomolny equations determine the
normal derivative of $\phi_0$ at the boundary, and vice-versa. At
$E_-$, since we want to specify $A$, we leave $\phi_0$
unspecified.  If $A_-$ is flat, then the Bogomolny equations
require the normal derivative of $\phi_0$ to vanish at $C_-$.

At $C_+$, on the other hand, we want to allow all possible Hecke
modifications $E_+$ that may arise from $E_-$ by solving the
Bogomolny equations.  So the connection $A$ is unspecified on
$C_+$, and instead we set $\phi_0=0$ on $C_+$.   The Bogomolny
equations then require that $A$ obeys covariant Neumann boundary
conditions at $E_+$; that is, its  curvature $F(A)$ vanishes when
contracted with the normal vector to the boundary. We are only
interested in the output bundle $E_+$ up to gauge transformations,
so we allow the gauge transformations to be non-trivial on $C_+$.
These boundary conditions are summarized in \utampo(a). In section
\kahstr, we will understand more deeply the naturalness of these
boundary conditions.

\bigskip\noindent{\it The Hecke Correspondence}

What we have just described is the boundary condition that we will
generally use in employing the Bogomolny equations to study Hecke
operators. But we pause to discuss the differential geometric
analog of another formulation that is standard in algebraic
geometry. (The details are not needed for the rest of the paper.)
Instead of thinking of $E_-$ as input and $E_+$ as output, as we
have done above,  one can treat the two symmetrically by
describing the ``Hecke correspondence.'' In the usual formulation
in algebraic geometry, we let ${\rm Bun}_G$ be the ``stack'' of
all $G$-bundles over $C$, not necessarily stable. The Hecke
correspondence of type $^L\negthinspace w,z$ (for a weight
$^L\negthinspace w$ and a point $z\in C$) is then a variety ${\cal
Q}(^L\negthinspace w,z)$ that maps to ${\rm Bun}_G\times {\rm
Bun}_G$. The fiber of ${\cal Q}(^L\negthinspace w,z)$ over
$E_-\times E_+\in {\rm Bun}_G\times {\rm Bun}_G$ parametrizes ways
of obtaining $E_+$ from $E_-$ by a Hecke transformation of type
$^L\negthinspace w$ at $z$.

\def\EUPT{\eusm{PT}}
The relationship between $E_-$ and $E_+$ can be stated more
symmetrically, as $C_-$ and $C_+$ are exchanged by an
orientation-preserving automorphism $\EUPT$ that acts trivially on
$C$, exchanges the two ends of $I$, and also reverses the time
coordinate.   Such an automorphism maps $\phi$ to $\EUPT^*(\phi)$,
and maps an 't Hooft operator of weight $^L\negthinspace w$ to one
of weight $^L\negthinspace w'$, where $^L\negthinspace w'$ is a
dominant weight that is Weyl conjugate to $-{}^L\negthinspace w$.
So $E_+$ is produced from $E_-$ by a Hecke transformation of type
$^L\negthinspace w$ at $z$, or equivalently $E_-$ is produced from
$E_+$ by a Hecke transformation of type $^L\negthinspace w'$ at
$z$.

The fiber of the map ${\cal Q}({}^L\negthinspace w,z)$ to ${\rm
Bun}_G\times {\rm Bun}_G$ is generically empty if the weight
$^L\negthinspace w$ is small and the genus of $C$ is large. On the
other hand, for any $C$, this map is generically a fibration if
$^L\negthinspace w$ is large enough. One can extend the definition
of ${\cal Q}$ to allow several Hecke operators with specified
positions and weights, in which case the nature of the map ${\cal
Q}\to {\rm Bun}_G\times {\rm Bun}_G$ depends on the number and
weights of the 't Hooft operators.

In differential geometry, ${\rm Bun}_G$ corresponds to the space
$\EUA$ of all connections.  $\EUA$ is of course
infinite-dimensional, and ${\rm Bun}_G$ is an algebro-geometric
analog of an infinite-dimensional manifold (it includes bundles
that are arbitrarily unstable with no bound on the dimensions of
spaces of deformations and automorphism groups).  The Hecke
correspondence ${\cal Q}$ likewise is infinite-dimensional and
will not come from elliptic boundary conditions.  A
differential-geometric version of the Hecke correspondence is to
simply define ${\cal Q}$ as the space of all solutions of the
Bogomolny equations on $W=I\times C$, with specified
singularities, modulo gauge transformations that are trivial at
each end. The map of ${\cal Q}$ to $\EUA\times \EUA$ is obtained
by restricting a solution to the two ends of $W$.

${\cal Q}$ is infinite-dimensional, as in defining it we have
specified neither $A$ nor $\phi_0$ at either end; these boundary
conditions are not elliptic.  In differential geometry, it seems
more natural to define the fiber of the Hecke correspondence over
a pair of connections $A_-\times A_+\subset \EUA\times \EUA$. This
is the gauge theory analog of studying in algebraic geometry the
fiber of the map  ${\cal Q}\to {\rm Bun}_G\times {\rm Bun}_G$. For
this, we fix $A_-$ and $A_+$ at each end, and divide by gauge
transformations that are trivial at each end (\utampo(b)). The
space of all solutions of the Bogomolny equations with these
boundary conditions is the fiber of the Hecke correspondence over
the pair $A_-\times A_+$. The boundary conditions defining this
fiber are elliptic and correspond to simply using at both ends of
$W$ the boundary conditions that in \utampo(a) are imposed only on
$C_-$.

The virtual complex dimension of the fiber of the Hecke
correspondence can be computed via index theory, by modifying the
discussion presented below.  It equals $-(g-1){\rm dim}(G)$ plus
the dimension $\Delta$ (also calculated below) of the space of
Hecke modifications of type $^L\negthinspace w$. If the virtual
dimension is a negative number $-d$, it means that generically
${\cal Q}$ is a subvariety of $\EUA\times \EUA$ of codimension
$d$. If it is a positive integer $d'$, it means that generically
${\cal Q}$ fibers over $\EUA\times \EUA$ with fibers of dimension
$d'$.

For the rest of our analysis, however, we concentrate on
describing the space of possible Hecke modifications of a given
bundle $E_-$, and hence we use the boundary conditions of
\utampo(a).  Before investigating more detailed properties, first
we will see what can be learned from index theory.

\bigskip\noindent{\it Index Theory And The Virtual Dimension}

We let ${\EUZ}$ be the moduli space of solutions of the Bogomolny
equations with the elliptic boundary conditions of \utampo(a).
{}From our discussion of the relationship between 't Hooft and
Hecke operators in section \thooftheckeop, we anticipate that
${\EUZ}$ is a complex manifold. We demonstrate this in section
\kahstr. Linearization of the Bogomolny equations leads to an
elliptic complex studied in \refs{\bogeqv,\bogeqx}:
\eqn\tifor{0\to \Omega^0({\rm
ad}(E))\underarrow{~d_1~}\Omega^1({\rm ad}(E))\oplus \Omega^0({\rm
ad}(E))\underarrow{~d_2~}\Omega^2({\rm ad}(E))\to 0 .} (Here
$\Omega^i({\rm ad}(E))$ is the space of ${\rm ad}(E)$-valued
$i$-forms.  $d_1$ is the map from an element of $\Omega^0({\rm
ad}(E))$, the Lie algebra of gauge transformations, to the
linearization of $A$ and $\phi$, which take values in
$\Omega^1({\rm ad}(E))$ and $\Omega^0({\rm ad}(E))$, respectively.
$d_2$ is the linearization of the Bogomolny equations.) The
tangent space to ${\EUZ}$ is the first cohomology group of this
complex, namely $ H^1={\rm ker}(d_2)/{\rm im}(d_1)$. The index of
this elliptic complex is called the virtual dimension of ${\EUZ}$,
and coincides with the actual dimension if the other cohomology
groups $H^0={\rm ker}(d_1)$ and $H^2={\rm coker}(d_2)$ vanish.
$H^0$ consists of covariantly constant sections of $\Omega^0({\rm
ad}(E))$, which generate unbroken gauge symmetries. With our
boundary conditions, there are none, since we require the gauge
transformations to be trivial on one end of $W$. As we will see,
with the boundary conditions of \utampo(a), the virtual dimension
is nonnegative.  This being so, $H^2$ vanishes away from
singularities of ${\EUZ}$, and the smooth part of ${\EUZ}$ is a
manifold whose dimension equals the virtual dimension or the index
of the complex.

We first analyze the Bogomolny equations in the absence of any 't
Hooft operators. We assume for simplicity that $E_-$ is stable and
represented by a flat connection with $F=0$.  (We can always
deform to this situation without changing the index.) In this
case, a standard sort of argument shows that a solution of the
Bogomolny equations with any of the boundary conditions above is a
pullback from $C$, so in particular $E_+$ and $E_-$ are
isomorphic.  This is actually a special case of the vanishing
theorems of section \vanishing, but the argument is so simple that
we present it separately here.

If the Bogomolny equations $F-\star D\phi_0=0$ are obeyed, then
$0=\int_W\Tr\,(F-\star D\phi_0)\wedge \star (F-\star D\phi_0)$.
Expanding this out and integrating by parts, we have
\eqn\notty{-\int_W\Tr\,\left(F\wedge\star F+D\phi_0 \wedge \star
D\phi_0\right)=-2\int_{\partial W}\Tr\,\phi_0 F.} The boundary
term vanishes on a component of $\partial W$ on which either
$\phi_0$ or $F$ vanishes.  In \utampo(a), we have $F=0$ at one end
and $\phi_0=0$ at the other, so the boundary term vanishes at each
end. But the left hand side of \notty\ is positive semi-definite
and can only vanish if $F=D\phi_0=0$.  So, given that $E_-$ is
flat, any solution of the Bogomolny equations with these boundary
conditions and without 't Hooft operators is given by a flat
connection that is pulled back from $C$, with $\phi_0=0$.

In particular, $E_+$ is isomorphic to $E_-$, and ${\EUZ}$ is a
point, of dimension zero.   One can further verify that, in
expanding around such a trivial solution with the boundary
condition of \utampo(a), $H^2$ vanishes. Thus, in this case the
virtual dimension of ${\EUZ}$ is the same as the actual dimension,
namely zero.  Now, what happens if 't Hooft operators are
included? Each singularity associated with an 't Hooft operator
$T(^L\negthinspace R)$ shifts the virtual dimension of the moduli
space by an amount that only depends on the representation
$^L\negthinspace R$ of the dual group (and not the details of the
three-manifold in which the 't Hooft operator is inserted, or the
topology of the bundle $E$, or the possible presence of other 't
Hooft operators). This follows from general excision properties of
index theory \ref\atiyahsinger{M. F. Atiyah and I. M. Singer,
``The Index Of Elliptic Operators (I),'' Annals of Mathematics
{\bf 87} (1968) 484-530.}, as shown by Pauly \ref\pauly{M. Pauly,
``Monopole Moduli Spaces For Compact 3-Manifolds,'' Math. Ann.
{\bf 311} (1998) 125-146.}. Moreover, Pauly computed, in our
language, the contribution of an 't Hooft operator to the
dimension of moduli space
 for $G=PSU(2)=SO(3)$, $^L\negthinspace G=SU(2)$. The
answer\foot{This is Theorem I of \pauly, with $k$ corresponding to
what we call $a/2$.  The theorem is stated for even $a$, as the
gauge group is taken to be $G=SU(2)$ rather than $PSU(2)$, but the
proof also works for odd $a$ and in fact for all $G$, as we will
discuss.} is that an 't Hooft operator associated with a
representation of highest weight $^L\negthinspace w=(a/2,-a/2)$,
with  positive integer $a$, shifts the virtual (complex) dimension
of the moduli space ${\EUZ}$ of solutions of the Bogomolny
equations by  $a$.

The fact that the singularity increases the complex dimension by
$a$ agrees with the result we found in section \spaceheck\ by
considering the space of Hecke modifications. There we showed that
for $U(2)$ weight $(a,0)$, or equivalently $SU(2)$ weight
$(a/2,-a/2)$, the space of Hecke modifications has complex
dimension $a$.

Pauly's proof actually generalizes immediately to give the result
for any compact Lie group $G$.  To explain this, we must review
the technique behind Pauly's proof, which was introduced by
Kronheimer \kronheimermon. The basic idea is to consider
instantons on the four-manifold\foot{Kronheimer actually considers
a more general situation with $\Bbb{C}^2$ replaced by a more
general four-dimensional hyper-Kahler manifold with $U(1)$ action,
and gets a description of solutions of the Bogomolny solutions on
$\Bbb{R}^3$ with a more general set of singularities.}
 $\Bbb{C}^2\cong \Bbb{R}^4$ that are invariant under the action of ${\EUF}=U(1)$ on
$\Bbb{C}^2$ by $(z_1,z_2)\to (\exp(i\theta)z_1,\exp(i\theta)z_2)$.
The quotient $\Bbb{C}^2/{\EUF}$ is isomorphic to $\Bbb{R}^3$.
${\EUF}$ does not act  freely -- it has a fixed point at the
origin, $z_1=z_2=0$ -- but nonetheless the quotient
$\Bbb{C}^2/{\EUF}$ is a manifold $\Bbb{R}^3$.  The map from
$(z_1,z_2)\in \Bbb{C}^2$ to $\vec x\in \Bbb{R}^3$ is
\eqn\nigo{\vec x=\bar z\vec\sigma z,} where $\vec\sigma$ are the
$2\times 2$ traceless hermitian matrices (normalized to $\Tr
\,\sigma_i\sigma_j=2\delta_{ij}$ and known as the Pauli spin
matrices).  This suggests that ${\EUF}$-invariant instantons on
$\Bbb{C}^2$ might be related to interesting objects on
$\Bbb{R}^3$, but we should expect something special to happen at
the origin of $\Bbb{R}^3$, which corresponds to the fixed point at
the origin of $\Bbb{C}^2$.

The description of ${\EUF}$-invariant instantons on $\Bbb{C}^2$ is
somewhat subtle. If the action of ${\EUF}$ on $\Bbb{C}^2$ is
lifted to an action on a $G$-bundle $\hat E\to \Bbb{C}^2$, then in
particular ${\EUF}$ acts on the fiber $\hat E_{0}$ of $\hat E$ at
the fixed point. Such an action is characterized by a homomorphism
$\rho:{\EUF}\cong U(1)\to G$. We recall that such a homomorphism
can be interpreted as a weight of $^L\negthinspace G$ and
determines an 't Hooft operator $T(\rho)$. Kronheimer considers
${\EUF}$-invariant instantons on $\Bbb{C}^2$  with a given choice
of $\rho$, and shows that they are equivalent to solutions of the
Bogomolny equations on $\Bbb{R}^3$ with a singularity at the
origin which in our language represents the insertion of the
operator $T(\rho)$.

Pauly then shows that the contribution $\Delta_\rho$ of a
singularity of type $\rho$ to the virtual dimension of ${\EUZ}$
can be computed from the contribution of the fixed point at the
origin in $\Bbb{C}^2$ to the ${\EUF}$-equivariant index of the
linear operator that computes the deformations of instantons on
$\Bbb{C}^2$.  Let ${\rm ad}(\hat E)$ be the adjoint bundle derived
from $\hat E$, and let ${\rm ad}(\hat E)_0$ denote its fiber at
the origin. The fixed point contribution to the index involves a
trace in ${\rm ad}(\hat E)_0$. (The adjoint representation comes
in because deformation theory of instantons involves an elliptic
operator acting on the adjoint bundle.)  In view of Pauly's
computation in section 4.2, the result can be described as
follows. The action of ${\EUF}\cong U(1)$ on ${\rm ad}(\hat E)_0$,
which is obtained by composing $\rho:{\EUF}\to G$ with the adjoint
representation of $G$, decomposes as a sum of characters. Any
character of ${\EUF}$ takes the form $\exp(i\theta)\to
\exp(im\theta)$ for some integer $m$. As the adjoint
representation of $G$ is real, the nonzero integers appearing in
the decomposition of $\rho$ into characters come in pairs
$c_k,-c_k$, where we can take $c_k$ to be positive.
  Then the contribution of the
't Hooft operator $T(\rho)$ to the virtual dimension of $\EUZ$ is
\eqn\xinon{\Delta_\rho=\sum_kc_k.} This generalizes the case of
$G=PSU(2)$, in which only a single integer $c$ appears.  Pauly's
analysis immediately extends from $SU(2)$ to any $G$ because the
equivariant index theorem works for any $G$. The computation
involves a trace in the Lie algebra, which leads to \xinon.

With no 't Hooft operators at all, the virtual dimension with our
boundary conditions is zero. With a single 't Hooft operator
$T(\rho)$, the virtual dimension is therefore $\Delta_\rho$. But
with a single 't Hooft operator,  $\EUZ$ is just the space
${{\EUY}}(\rho)$ of Hecke modifications of type $\rho$. The
formula \xinon\ for the dimensions of spaces of Hecke
modifications agrees with the result \ydro\ for $U(N)$ (or $SU(N)$
or $PSU(N)$) and generalizes it to arbitrary compact $G$.

Since the contribution of an 't Hooft operator to the index theory
is local, we can immediately write the general result for the
dimension of the moduli space ${\EUZ}(\rho_1,\dots,\rho_n)$ of
solutions of the Bogomolny equations with an arbitrarily
prescribed set of 't Hooft operator insertions. If there are no 't
Hooft operators, the dimension of ${\EUZ}$ is zero. Each 't Hooft
operator $T(\rho_i)$ associated with a homomorphism
$\rho_i:U(1)\to G$ contributes $\Delta_{\rho_i}$ to the dimension,
so in general the dimension of ${\EUZ}(\rho_1,\dots,\rho_n)$ is
\eqn\migoty{\Delta=\sum_i\Delta_{\rho_i}.}

\subsec{Monopole Bubbling} \subseclab\monbub

In section \spaceheck, we found that the natural compactification
$\bar {{\EUY}}(^L\negthinspace w)$ of the space
${{\EUY}}(^L\negthinspace w)$ of Hecke modifications of some given
type $^L\negthinspace w$ is achieved by including Hecke
modifications of various associated types with smaller weights.

We can now see why this is true from the point of view of the
Bogomolny equations.  Consider a solution of the Bogomolny
equations on $\Bbb{R}^3$ with a specified singularity at the
origin.  It corresponds to an equivariant instanton on
$\Bbb{C}^2\cong \Bbb{R}^4$.  The instanton equations on
$\Bbb{R}^4$ are scale-invariant, so a phenomenon can occur that is
called ``bubbling'' in the mathematical literature: an instanton
can shrink to a point.  In the case of an ${\EUF}$-equivariant
instanton, this point will automatically be the origin in
$\Bbb{C}^2$.

When interpreted in three dimensions using the relation
\kronheimermon\ between equivariant instantons and solutions of
the Bogomolny equations in three dimensions with point
singularities (which we interpret in terms of 't Hooft operators)
this gives a phenomenon that we might call monopole bubbling.
Monopole bubbling (which was also discovered in unpublished work
by Kronheimer related to \kronheimermon) is relatively unfamiliar
because it does not occur for smooth solutions of the Bogomolny
equations on $\Bbb{R}^3$ (or other complete Riemannian manifolds)
with the usual boundary conditions. On the contrary, smooth
monopoles on $\Bbb{R}^3$ have a discrete topological
classification  and a characteristic size, which involves the
asymptotic behavior at infinity on $\Bbb{R}^3$.  Monopole bubbling
can occur only in the presence of an 't Hooft operator; it is a
process in which a quantized unit of charge becomes concentrated
at a point (the position of the 't Hooft operator) and disappears.

 We will now very briefly use the ADHM construction \ref\adhm{M.
F. Atiyah, V. Drinfeld, N. J. Hitchin, and Yu. I. Manin,
``Construction Of Instantons,'' Phys. Lett. {\bf 65A} (1978)
185-87.} of instantons on $\Bbb{R}^4\cong \Bbb{C}^2$ to determine
how bubbling of equivariant instantons -- or in other words
monopole bubbling -- affects the weight $^L\negthinspace w$ of an
't Hooft operator for the case of gauge group $G=SU(N)$. (A
similar analysis could be made for other groups for which there is
an ADHM construction, namely the classical groups $SO(N)$ and
$Sp(N)$. But for exceptional groups there is no ADHM
construction.) We will see that bubbling changes the weight in
precisely the expected fashion.

We will follow \ref\donkron{S. Donaldson and P. B. Kronheimer,
{\it The Geometry Of Four-Manifolds} (Oxford University Press,
1990).}, section 3.3. To describe $SU(N)$ instantons on
$\Bbb{R}^4$, let $U$ be a two-dimensional vector space (which
arises as one of the spin bundles of $\Bbb{R}^4$) on which
${\EUF}$ acts by the matrix \eqn\normalo{\left(\matrix{e^{i\theta}
& 0 \cr 0 & e^{-i\theta}\cr}\right).}  For any vector space $V$
with ${\EUF}$ action, we write $\chi(V)$ for the character of
${\EUF}$; for instance,
 $\chi(U)=e^{i\theta}+e^{-i\theta}$. To construct instantons of
instanton number $k$, let ${{\EUH}}$ be a $k$-dimensional vector
space.  And let $E_\infty$ be a vector space of rank $N$ (which
turns out to be the fiber at infinity of the instanton bundle).
The instanton is built from a special kind of complex
\eqn\onnos{0\,\underarrow{~~~~}\,{{\EUH}}\underarrow{\,\,\alpha(z)\,\,}{{\EUH}}\otimes
U\oplus E_\infty\underarrow{\,\,\beta(z)\,\,}
{{\EUH}}\,\underarrow{~~~~}\, 0,} where $\beta(z)\alpha(z)=0$. For
our limited purposes here, we do not need the details. $\alpha(z)$
and $\beta(z)$ depend linearly on the complex coordinates
$z_1,z_2$ of $\Bbb{C}^2$.  For a smooth instanton, the only
nonzero cohomology group of this complex, for any $z=(z_1,z_2)\in
\Bbb{C}^2$, is $H^1(z)={\rm ker}\,\beta(z)/{\rm im}\,\alpha(z)$.
The fiber at $z$ of the instanton bundle $E$ derived from the ADHM
data is $E_z=H^1(z)$.

To get an ${\EUF}$-equivariant instanton, we pick an action of
${\EUF}$ on $E_\infty$ and ${{\EUH}}$, and choose the complex
\onnos\ to be ${\EUF}$-equivariant. The fiber of the instanton
bundle at the origin, which we denote $E_0$, is in particular the
first cohomology group of the corresponding complex:
\eqn\otonos{0\longrightarrow{{\EUH}}\underarrow{\,\,\alpha(0)\,\,}{{\EUH}}\otimes
U\oplus E_\infty\underarrow{\,\,\beta(0)\,\,}
{{\EUH}}\longrightarrow 0,} For an equivariant complex
$0\underarrow{~~~} V_0\underarrow{\,\,\alpha\,\,}
V_1\underarrow{\,\,\beta\,\,} V_2\underarrow{~~~} 0$ whose only
cohomology is $H^1$, the character of $H^1$ can be expressed as
$\chi(H^1)=-\chi(V_0)+\chi(V_1)-\chi(V_2)$. So in particular
\eqn\onsno{\chi(E_0)=\chi(E_\infty)+\chi({{\EUH}})(e^{i\theta}+e^{-i\theta}-2).}
$\chi(E_0)$ is important because, according to Kronheimer's
construction, it determines the weights of the 't Hooft operator
that appears at the origin of $\Bbb{R}^3=\Bbb{C}^2/{\EUF}$ when
the equivariant instanton is interpreted in three dimensions. If
$\chi(E_0)=\sum_{i=1}^N\exp({im_i\theta})$, then  the weights are
$^L\negthinspace w=(m_1,m_2,\dots,m_N)$.

Instanton bubbling occurs when $\alpha$ and $\beta $ are varied in
such a way that, for some decomposition ${{\EUH}}={{\EUH}}_1\oplus
{{\EUH}}_2$, the complex \onnos\ decomposes as the direct sum of a
zero complex \eqn\teto{0\,\underarrow{~~~~}\,
{{\EUH}}_1\underarrow{\,\,\,0\,\,\,}{{\EUH}}_1\otimes
U\underarrow{\,\,\,0\,\,\,}{{\EUH}}_1\,\underarrow{~~~~}\, 0} and
a complex
\eqn\ponos{0\,\underarrow{~~~~}\,{{\EUH}}_2\underarrow{\,\,\alpha(z)\,\,}{{\EUH}}_2\otimes
U\oplus E_\infty\underarrow{\,\,\beta(z)\,\,}
{{\EUH}}_2\,\underarrow{~~~~}\, 0} with cohomology only in
dimension 1. The ``bubbled'' instanton bundle $E'$, whose
instanton number is reduced by the rank of ${{\EUH}}_1$, has for
its fiber the first cohomology group of this second complex, and
by analogy with \onsno, the character of its fiber at the origin
is
$\chi(E'_0)=\chi(E_\infty)+\chi({{\EUH}}_2)(e^{i\theta}+e^{-i\theta}-2)$.
Hence
\eqn\ponzo{\chi(E'_0)=\chi(E_0)+\chi({{\EUH}}_1)(2-e^{i\theta}-e^{-i\theta}).}

This formula determines the change in weights under instanton
bubbling.  The weights $m_i$ of $E_0$ and the weights $m'_i$ of
$E'_0$ are related by
\eqn\yhog{\sum_{i=1}^N\exp({im'_i\theta})=
\sum_{i=1}^N\exp({im_i\theta})+\chi({{\EUH}}_1)(2-e^{i\theta}-e^{-i\theta}).}
 For example, if ${{\EUH}}_1$ is
one-dimensional with character $\exp(im\theta)$ for some integer
$m$, then the list of weights of $E_0$ must include $m+1$ and
$m-1$ (since $\chi(E'_0)$ is a Fourier series with nonnegative
coefficients). \yhog\ implies that precisely two weights change in
this bubbling process, by $(m+1,m-1)\to (m,m)$. This corresponds
to the simplest degeneration of 't Hooft/Hecke operators that we
discussed in section \spaceheck. For another example, suppose
${{\EUH}}_1$ has rank two, with character $\exp({im
\theta})+\exp({im'\theta})$.  If $|m-m'|>1$, we just get a pair of
moves of the kind just described, with weights
$(m+1,m-1,m'+1,m'-1)$ replaced by $(m,m,m',m')$. But if $m'=m+1$,
then \yhog\ allows another move with weights $(m+2,m-1)$ replaced
by $(m+1,m)$. This corresponds to another expected degeneration of
't Hooft/Hecke operators. More generally, if ${{\EUH}}_1$ has rank
$s$ and weights $m,m+1,\dots,m+s-1$, then \yhog\ leads to a
move\foot{A complex of the appropriate ADHM form to realize this
move can be achieved with ${{\EUH}}={{\EUH}}_1$, ${{\EUH}}_2=0$
and $E'$ being the trivial bundle with fiber $E_\infty$. With the
characters of $E'_0$ and ${{\EUH}}_1$ given in the text, there is
an essentially unique ${\EUF}$-invariant choice of
$\tau_1,\tau_2,\sigma,$ and $\pi$ that obeys eqn. (3.3.11) of
\donkron\ and gives a smooth instanton bundle $E$, with instanton
number $s$, that can degenerate to $E'$ by bubbling.} in which
weights $(m+s,m-1)$ of $E_0$ are replaced by weights $(m+s-1,m)$
of $E'_0$. The expected degenerations of 't Hooft/Hecke operators
for $G=SU(N)$ are all compositions of moves of this type.  For a
general $G$, the expected degenerations can be generated by
bubbling of $SU(2)$ instantons embedded in various subgroups of
$G$.

Note in particular that while in the absence of
${\EUF}$-invariance, instantons can be assumed to bubble one at a
time, this is not so once one imposes ${\EUF}$-invariance. There
are irreducible processes that involve the simultaneous bubbling
of an arbitrarily large number of ${\EUF}$-invariant instantons.
See \refs{\golo,\latwein} for detailed description of some
Bogomolny moduli spaces on $\Bbb{R}^3$ in which monopole bubbling
phenomena can be seen, in a suitable limit.

\bigskip\noindent{\it Hyper-Kahler Structure At Singularities}

In section \spaceheck, we noted in an example that although the
compactified space $\bar {{\EUY}}(^L\negthinspace w)$ of Hecke
modifications of a given type $^L\negthinspace w$ is not a
hyper-Kahler manifold, its singularities do have a hyper-Kahler
structure.  Now we can explain why.  The singularities come from
monopole bubbling. Monopole bubbling is governed by the moduli
space of ${\EUF}$-equivariant instantons on $\Bbb{C}^2=\Bbb{R}^4$.
But instanton moduli spaces on $\Bbb{R}^4$ have a hyper-Kahler
structure inherited from the hyper-Kahler structure of $\Bbb{R}^4$
(and manifest in the ADHM construction).  Since the group ${\EUF}$
preserves the hyper-Kahler structure of $\Bbb{R}^4$, the space of
${\EUF}$-equivariant instantons is hyper-Kahler (and again, the
ADHM construction makes this manifest).

 We can also  argue in three-dimensional
terms, without using the relation to equivariant instantons in
four dimensions, that monopole bubbling, and hence the
singularities of $\bar {{\EUZ}}({}^L\negthinspace
w_1,\dots,{}^L\negthinspace w_n)$, is governed by a hyper-Kahler
moduli space. Indeed, the moduli space of solutions of the
Bogomolny equations on $\Bbb{R}^3$ is hyper-Kahler
\bogeqa.\foot{This can be shown using the same sort of arguments
we use momentarily to exhibit a Kahler structure for the moduli
space of solutions of the Bogomolny equations on $W=I\times C$.
Those arguments also work if $I$ is replaced by $\Bbb{R}$. If in
addition $C=\Bbb{R}^2$, then $W=\Bbb{R}^3$ can be written as
$\Bbb{R}^2\times \Bbb{R}$ in many different ways, leading to a
hyper-Kahler structure, not just the Kahler structure that we
obtain in section \kahstr. } The spaces of $\bar
{{\EUZ}}({}^L\negthinspace w_1,\dots,{}^L\negthinspace w_n)$ do
not come  from solutions of the Bogomolny equations on
$\Bbb{R}^3$, but on the more complicated three-manifold $W=I
\times C$ with rather particular boundary conditions. So $\bar
{{\EUZ}}({}^L\negthinspace w_1,\dots,{}^L\negthinspace w_n)$ is
not hyper-Kahler. However, monopole bubbling is a local
phenomenon. When a unit of monopole charge shrinks to a point, the
local structure does not depend on the details of the
three-manifold in which this is occurring.  So the structure near
the singularity is hyper-Kahler, just as if one is on $\Bbb{R}^3$.

We give further details presently on the deformation and
resolution of these hyper-Kahler singularities.

\subsec{Kahler Structure And The Moment Map}

\subseclab\kahstr

Our next goal is to exhibit ${\EUZ}({}^L\negthinspace
w_1,\dots,{}^L\negthinspace w_n)$ as a Kahler manifold, with a
complex structure that agrees with the appropriate moduli space of
Hecke modifications, as described in section \thooftheckeop. This
analysis will make many things clearer.  It will give a new
motivation for the choice of boundary conditions that we used in
defining ${\EUZ}$; it will give a more precise framework for
understanding the relation between 't Hooft and Hecke operators;
and it will make more transparent the properties of the
singularities of the space of Hecke modifications.

\ifig\xutampo{Comparing 't Hooft and Hecke operators on a
three-manifold $W=I\times C$.  On the left, a connection $A$ and
bundle $E_-$ are specified,  and the gauge transformations are
required to be trivial. On the right, one sets $\phi_0=0$ and
solves for $E_+$.} {\epsfxsize=4in\epsfbox{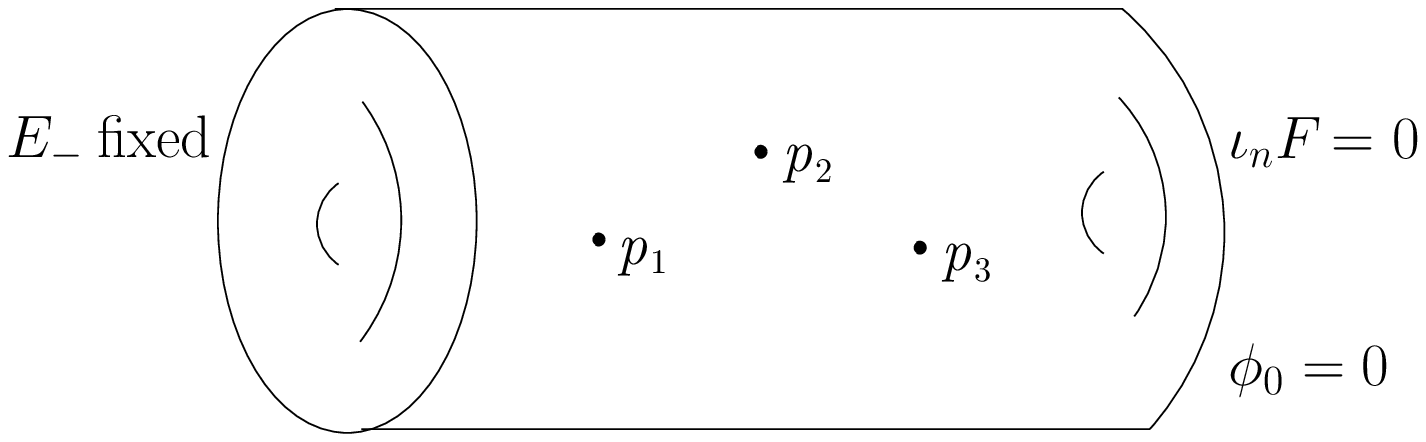}}

As in the introduction to section \thooftheckeop, we take the
metric of the three-manifold $W=I\times C$ to be of the form
$ds^2=h|dz|^2+dy^2$, where $z=x^2+ix^3$ is a local complex
coordinate on $C$ and $y$ parametrizes $I$. We consider the
Bogomolny equations $F=\star D\phi_0$, where $F$ is the curvature
of a connection $A$ on a $G$-bundle $E$.  At the left end of $W$,
we specify a flat $G$-bundle $E_-$, and divide only by gauge
transformations that equal 1 on the boundary; at the right end, we
require $\phi_0=0$ with and divide by all gauge transformations.

 We expand the
connection as $A=A_z dz+A_{\bar z}d\bar z+A_ydy=A_2dx^2+A_3
dx^3+A_ydy$. By rewriting \noffo\ and \jungol, the Bogomolny
equations in these coordinates can be written as a complex
equation \eqn\mirot{\left[D_y-i\phi_0, D_{\bar z}\right]=0 } and a
real equation \eqn\irot{F_{23}=hD_y\phi_0.}

On the space $\EUC$ of fields $(A,\phi_0)$, we introduce a complex
structure ${\cal I}$ in which the complex coordinates are $A_{\bar
z}$ and $B=A_y-i\phi_0$.  The rationale for this choice is that
the complex equation in \mirot\ is holomorphic in this complex
structure. Furthermore, we endow $\EUC$ with the symplectic
structure \eqn\uryt{\omega=\int_{W}\Tr\Bigl(\delta A_2\wedge
\delta A_3-h\,\delta A_1\wedge \delta\phi_0\Bigr) dx^2\,dx^3\,dy.}
The complex structure ${\cal I}$ and symplectic structure $\omega$
combine to a Kahler structure, the Kahler metric being
\eqn\ryt{-\int_W\Tr\Bigl( \delta A_2\otimes \delta A_2+\delta
A_3\otimes \delta A_3+h\,\delta A_y\otimes \delta A_y+h\,\delta
\phi_0\otimes \delta \phi_0\Bigr)dx^2\,dx^3\,dy.}

Now we want to find the moment map $\mu$ for the action of the
gauge group.  As usual, the moment map $\mu(v)$ for a symplectic
vector field $v$ is characterized by
\eqn\charby{d\mu(v)=\iota_v\omega,} where $\iota_v$ is the
operation of ``contracting with $v$.'' For the vector field
characterized by the infinitesimal gauge transformation that acts
by $\delta A=-D\epsilon$ and $\delta\phi_0 =[\epsilon,\phi_0]$,
with $\epsilon\in \Omega^0(C,{\rm ad}(E))$, we find the moment map
to be \eqn\prigo{\mu(\epsilon)=\int_W\Tr\left(\epsilon(
\,F_{23}-hD_y\phi_0)\right)dx^2\,dx^3\,dy+ \int_{\partial
W}\,\Tr\,(\epsilon\phi_0) \,h\,dx^2\,dx^3.} The calculation is
similar to that in deriving eqn. \joget, except that a boundary
term appears in integration by parts.

In the presence of 't Hooft operators at points $p_i\in W$, we
modify the definitions slightly.  $A$ and $\phi_0$ are required to
have prescribed singularities near  $p_i$.  And $\epsilon$ is
constrained near $p_i$ to generate a gauge transformation that
leaves fixed the chosen singularity. Our formulas for the moment
map remain valid in this situation.

We want to interpret the real Bogomolny equation \irot\ in terms
of the vanishing of the moment map.  In fact, for $\mu(\epsilon)$
to vanish does give us \irot\ together with another condition:
$\Tr\,\epsilon\phi_0$ must vanish on the boundary.  On any given
boundary component $C_\pm$ of $W$, there are two simple ways to
ensure that $\Tr\,\epsilon\phi_0=0$.  We can set $\epsilon=0$ on
the boundary, in other words we can allow only gauge
transformations that are trivial on this boundary component. Or we
can set $\phi_0=0$.  To define the moduli space ${\EUZ}$, we have
used (\xutampo) one of these boundary conditions on $C_-$, and the
other on $C_+$.  Thus we have a new rationale for our boundary
conditions: they are the simplest ones that allow us to interpret
the Bogomolny equations in terms of a holomorphic equation
together with the vanishing of the moment map.

Now we can invoke the usual correspondence between symplectic
quotients and complex quotients.  Let ${\EUC}_0$ be the subspace
of $\EUC$ characterized by vanishing of the holomorphic Bogomolny
equation \mirot.  The moduli space ${\EUZ}$ is defined by setting
$\mu $ to zero and dividing by the group $\cal G$ of $G$-valued
gauge transformations.  We can compare it to
${\EUZ}'={\EUC}_0/{\cal G}_{\Bbb{C}}$, the quotient of ${\EUC}_0$
by the complexified group ${\cal G}_{\Bbb{C}}$ of ${
G}_{\Bbb{C}}$-valued gauge transformations.

Of course, because of the boundary conditions, $\cal G$ is the
group of gauge transformations that are trivial at the left end of
$W$, and likewise ${\cal  G}_{\Bbb{C}}$ is the group of
$G_{\Bbb{C}}$-valued gauge transformations that are trivial at
that end.

 Usually, in comparing a symplectic quotient to the
corresponding complex quotient, one has to worry about possible
unstable and semistable orbits.  In the present case, however,
because the gauge transformations are restricted to be trivial at
one end of $W$, such issues are avoided.  ${\cal G}_{\Bbb{C}}$
acts freely and with closed orbits, so all orbits are stable and
${\EUZ}={\EUZ}'$.

It is straightforward to analyze ${\EUZ}'$.  In fact, we have
essentially done so in section \theck.  For any $y\in I$ away from
the location of an 't Hooft operator, let $E_y=E|_{ \{y\}\times
C}$. The complex Bogomolny equation says that the holomorphic type
of $E_y$ is independent of $y$, as long as one does not cross the
position of an 't Hooft operator.  Moreover, this holomorphic type
is the only ${\cal G}_{\Bbb{C}}$-invariant in an interval without
't Hooft operators.  In jumping across 't Hooft operators, $E_y$
undergoes Hecke modifications. So ${\EUZ}'$ is the space of Hecke
modifications of $E_-$ due to the prescribed singularities.

This comparison of complex and symplectic quotients thus gives a
precise framework for understanding the relationship between 't
Hooft and Hecke operators, which we compared more informally in
section \theck.  In addition, it will enable us to understand
better some properties of the spaces of Hecke modifications that
were described in section \spaceheck .

\bigskip\noindent{\it Changing the Complex And Symplectic
Structures}

Let us denote the points $p_i$ in $W=I\times C$ at which 't Hooft
operators are inserted as $y_i\times z_i$, with $y_i\in I$ and
$z_i\in C$.  We want to understand the dependence of the moduli
space ${\EUZ}$ on the points $p_i$.

It is convenient to introduce the modified connection
$A'=A-i\phi_0 dy$, the idea being that if we expand
$A'=dz\,A'_z+d\bar z \,A'_{\bar z} +dy\, A'_y$, then $A'{}_y$ and
$A'{}_{\bar z}$ are holomorphic in complex structure ${\cal I}$,
although $A'_z$ is not.  Since $A$ is real,  that is, it takes
values in the real Lie algebra of $G$, $A'_z$ (which equals $A_z$)
is the complex conjugate of $A'_{\bar z}$; of course, $A'_y$ obeys
no reality condition. To describe a holomorphic group action on
the space $\EUC$ of pairs $(A,\phi_0)$, which is the same as the
space of modified connections $A'$, it suffices to describe how
$A'_{\bar z}$ and $A'_y$ transform.  The transformation of $A'_z$
is then determined by the fact that it is the complex conjugate of
$A'_{\bar z}$.

 Consider
$W=I\times C$ as a fiber bundle over $C$, and let $D_I$ be the
group of  diffeomorphisms of $W$ that preserve this fibration.
Such a diffeomorphism, while keeping $z$ fixed, maps $y$ to
$\tilde y(y;z,\bar z)$, where $\tilde y$ agrees with $y$ on the
boundaries of $W$ and $\partial \tilde y/\partial y>0$.  We define
an action of $D_I$ on $\EUC$  by simply saying that $A'_{\bar z}$
and $A'_y$ transform in the natural way, by the ``pullback.''

Note that we cannot transform $A'$ by pullback; for $\beta\in
D_I$, the operation $A'\to \beta^*(A')$ would not preserve the
fact that $A'_z$ and $A'_{\bar z}$ are complex conjugates. Instead
we define $\beta(A')=\beta^*(A')+dz\,\varepsilon$, where
$\varepsilon$ is determined to retain the condition that $A'_z$
and $A'_{\bar z}$ are complex conjugates.

This action of $D_I$ on the space $\EUC$ of pairs $(A,\phi_0)$
preserves the complex structure ${\cal I}$.  It also preserves the
complex Bogomolny equation,  so it preserves the space ${\EUC}_0$
of solutions of that equation.  These statements depend on the
fact that diffeomorphisms in $D_I$ keep $z$ fixed. They therefore
preserve the foliation generated by $\partial/\partial y$ and
$\partial/\partial \bar z$ , and the assertion of flatness along
leaves of this foliation -- which is the content of the complex
Bogomolny equation -- is invariant. In addition, $D_I$ maps the
group ${\cal G}_{\Bbb{C}}$ of $G_{\Bbb{C}}$-valued gauge
transformations of $\EUC$  to itself. So ${\EUZ}$, defined as a
complex manifold by ${\EUZ}={\EUC}_0/{\cal G}_{\Bbb{C}}$, is
invariant under $D_I$.

However, the symplectic form $\omega$ on $\EUC$ is definitely not
$D_I$-invariant, and accordingly neither is the moment map for the
${\cal G}$ action, or the real Bogomolny equation.  So ${\EUZ}$,
understood as the symplectic quotient of ${\EUC}_0$ by the action
of ${\cal G}$, using the symplectic structure $\omega$, is not
invariant under $D_I$.

The conclusion is that the complex structure of ${\EUZ}$ is
$D_I$-invariant, but not its symplectic structure.

Now let us consider 't Hooft operators at points $p_i=y_i\times
z_i$.  The group $D_I$ can slide the $p_i$ up and down in the $y$
direction, in a fashion that is arbitrary except for one
restriction: if two $p_i$ are at the same value of $z$, then their
ordering in the $y$ direction is $D_I$-invariant. The conclusion
is that sliding the points $p_i$ in the $y$ direction, without
letting two points at the same $z$ cross each other, changes the
symplectic structure of $\EUZ$ but not its complex structure.

There is also a reverse version of this.  The definition of the
symplectic structure of $\EUC$ used the area form of $C$, but not
its complex structure.  The same is, therefore, true for the
moment map $\mu$.  So just given the area form of $C$, we can
define the symplectic quotient $\EUC/\negthinspace/{\cal
G}=\mu^{-1}(0)/{\cal G}$. This is an infinite-dimensional
symplectic manifold whose symplectic structure depends only on the
area form of $C$.  Now in general, in a symplectic manifold, a
family of symplectic submanifolds defined by a varying family of
equations has a fixed symplectic structure.  In our application,
the varying family of equations are the complex Bogomolny
equations, which depend on the complex structure of $C$. So as the
complex structure and the positions of the points vary, the
complex structure of ${\EUZ}$ may change, but the symplectic
structure does not.

If 't Hooft operators are inserted at point $p_i=y_i\times z_i\in
I\times C$, then changing the $z_i$ with the $y_i$ fixed, and
without changing whether or not $z_i=z_j$ for a given pair $i,j$,
is equivalent to a special case of changing the complex structure
of $C$.  So changing the $z_i$ in this way changes the complex
structure of ${\EUZ}$ but not its symplectic structure.

\bigskip\noindent{\it Comparison To Holomorphic Results}

We can use these observations to get a new perspective on some
facts that we described in section \spaceheck .

\ifig\xtampo{Insertion of two 't Hooft operators of weights
$^L\negthinspace w$ and $^L\negthinspace w'$ at points $p,p'\in
W$.} {\epsfxsize=4in\epsfbox{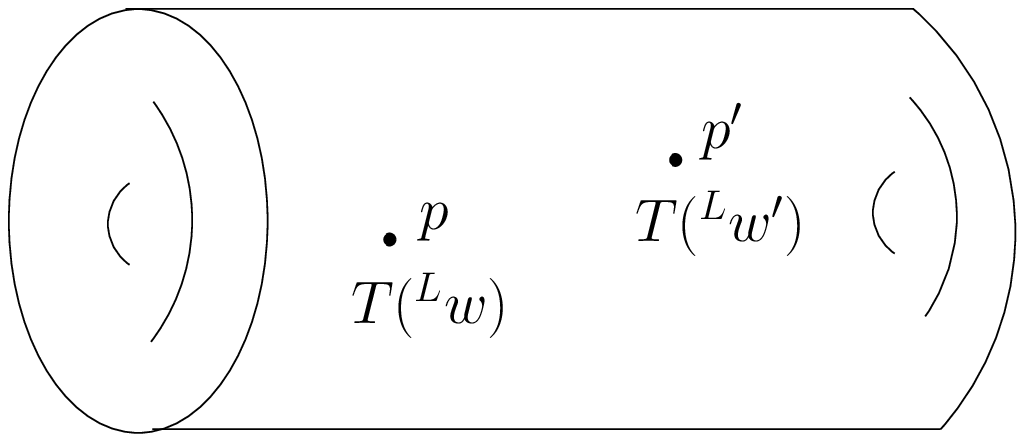}}

Consider $W=I\times C$ with insertion of two 't Hooft operators
$T(^L\negthinspace w,p)$ and $T(^L\negthinspace w',p')$ of the
indicated weights,  inserted
 at $p=y\times z$ and $p'=y'\times z'$ (\xtampo).
 We write $\bar{\EUZ}(^L\negthinspace w,p;{}^L\negthinspace w',p')$, or more briefly $\bar
 {\EUZ}(p;p')$, for
the corresponding compactified moduli space with some specified
bundle $E_-$ at the left end of $W$, and likewise we write
$\bar{\EUZ}(^L\negthinspace w,p)$ for a similar Bogomolny moduli
space with only one singularity.

We can vary the symplectic structure of $\bar\EUZ(p;p')$ by moving
the points in the $I$ direction keeping the projection to $C$
fixed. Or we can vary the complex structure by moving the points
in the $C$ direction keeping the projection to $I$ fixed.

Generically, $\bar\EUZ(p;p')$  is singular. Even the moduli space
of Hecke modifications associated with a single 't Hooft operator,
such as $T(^L\negthinspace w_1)$ or $T(^L\negthinspace w_2)$, is
generically singular, as we discussed in section \spaceheck.
Exceptions correspond to minuscule representations of
$^L\negthinspace G$.

So $\bar{\EUZ}(p;p')$ is typically singular even if $p\not= p'$.
But it becomes more singular (or develops a singularity if it was
previously smooth) when $p'\to p$.  Let us denote as
$\bar\EUZ{}^*$ the singular limit at $p'\to p$.  We can partially
resolve the singularities of $\bar\EUZ{}^*$ by displacing the
points in the $I$ direction, and we can partially deform the
singularities by displacing the points in the $C$ direction
(\xdtampo).  This partial deformation of singularities and partial
resolution of singularities are topologically equivalent, since we
can simultaneously move $p$ and $p'$ in both the $I$ and $C$
directions.

If $z\not=z'$, then $\bar\EUZ(p;p')$ is as a complex manifold a
simple product $\bar\EUZ({}^L\negthinspace w,p)\times
\bar\EUZ({}^L\negthinspace w',p')$, parameterizing independent
Hecke transformations at the two points $z$ and $z'$ in $C$.  If
$z=z'$ and $y<y'$, it is a fiber bundle over
$\bar\EUZ({}^L\negthinspace w,p)$ with fiber
$\bar\EUZ({}^L\negthinspace w',p')$, describing successive Hecke
modifications of $E_-$ at the point $z\in C$ with weights
$^L\negthinspace w$ and $^L\negthinspace w'$.  If $z=z'$ and
$y'<y$, then $\bar\EUZ(p;p')$ is again a fiber bundle, but the
roles of $^L\negthinspace w$ and $^L\negthinspace w'$ are
reversed.

\ifig\xdtampo{(a) Two partial resolutions of singularity. The
moduli space is a fiber bundle parameterizing successive Hecke
modifications at the same point $z\in C$. (b) A partial
deformation of singularity. The moduli space is a product,
parameterizing independent Hecke modifications at distinct points
$z,z'\in C$.} {\epsfxsize=4in\epsfbox{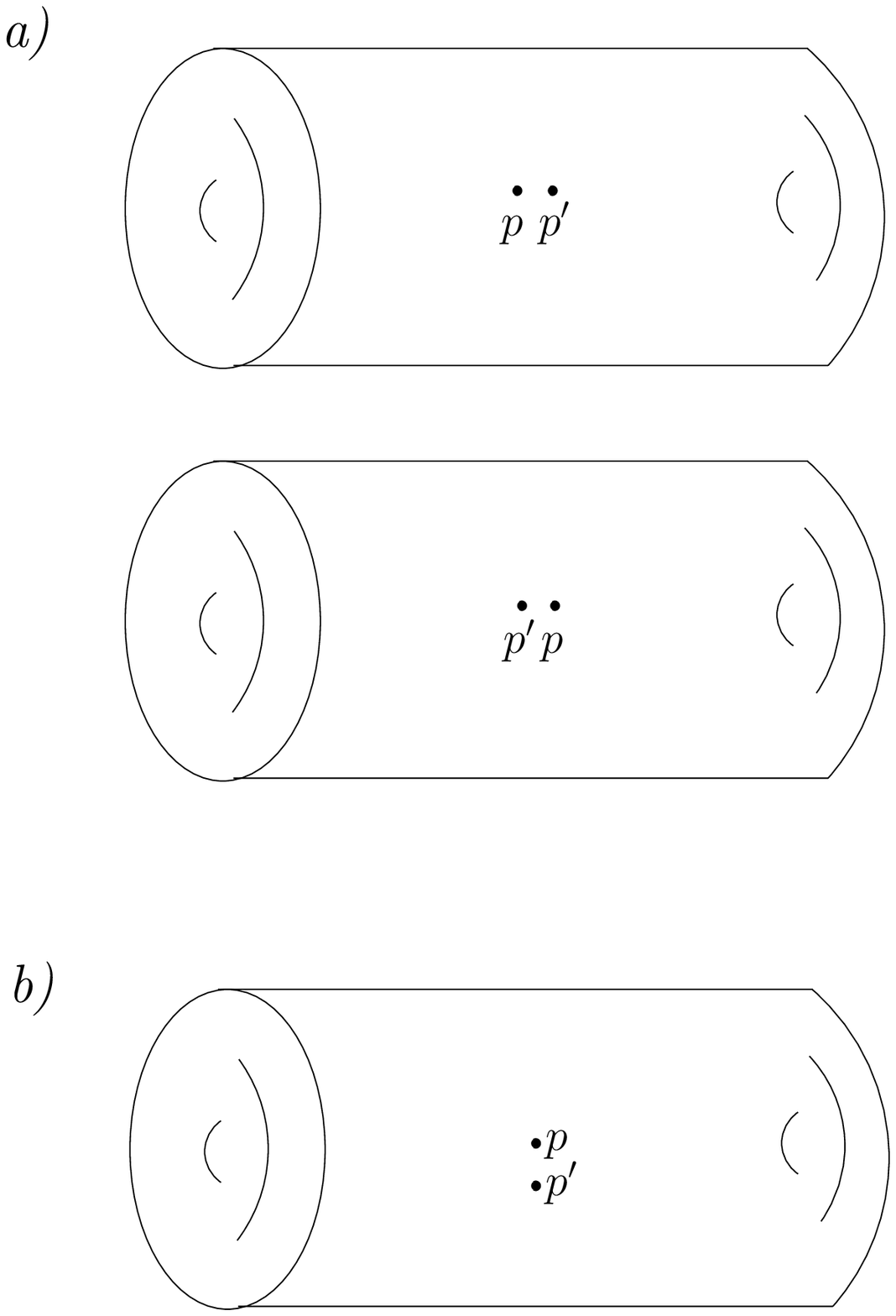}}

What is the space $\bar\EUZ{}^*$ that has the simultaneous
deformations and resolutions just described? It is in fact
$\bar{\EUZ}(^L\bar w,p')$, the compactified moduli space of
solutions of the Bogomolny equations in the presence of a single
't Hooft operator of weight $^L\bar w={}^L\negthinspace
w+{}^L\negthinspace w'$.

{}From a physical point of view, this statement is part of the
operator product expansion of the operators $T(^L\negthinspace
w,p)$ and $T(^L\negthinspace w',p')$. We discussed some general
properties of this expansion in section \lineop. Simple arguments
that we explained there suffice to show that the product
$T(^L\negthinspace w,p)\cdot T(^L\negthinspace w',p')$ has an
expansion as $\sum_{i\in \EUR}c_i T(^L\negthinspace w_i,p')$, for
some set $\EUR$ of weights $^L\negthinspace w_i$ and coefficients
$c_i$. $S$-duality predicts the detailed form of this expansion
(which should match the representation ring of $^L\negthinspace
G$), but it is difficult to verify this prediction directly.
However, the simple arguments of section \lineop\ do suffice to
show that the form of the expansion as $p\to p'$ is
\eqn\hittor{T({}^L\negthinspace w,p)\cdot T({}^L\negthinspace
w,p')\to T({}^L\bar w,p')+\sum_{i\in \EUR'} c_i
T({}^L\negthinspace w_i,p'),} where $^L\bar w={}^L\negthinspace
w+{}^L\negthinspace w'$, and the weights $^L\negthinspace
w_i,\,i\in \EUR'$ are the dominant weights such that $^L\bar
w-{}^L\negthinspace w_i$ is also dominant. This means that
$^L\negthinspace w_i$ is what we have called an associated weight
of $^L\bar w$.

The formula \xinon\ shows that the dimension of $\bar\EUZ(^L\bar
w,p')$ precisely equals the dimension of $\bar\EUZ(^L\negthinspace
w,p;{}^L\negthinspace w',p')$. This is consistent with the
statement that the former is the limit of the latter for $p\to
p'$.  Other spaces $\bar\EUZ(^L\negthinspace w_i,p')$, $i\in
\EUR'$ have strictly smaller dimension. In fact, in a certain
sense, the spaces $\bar\EUZ(^L\negthinspace w_i,p')$ all appear in
the limit of $\bar\EUZ(^L\negthinspace w,p;{}^L\negthinspace
w',p')$ for $p\to p'$. Since the $^L\negthinspace w_i$ are
associated weights of $^L\bar w_i$, the natural compactification
$\bar {{\EUY}}(^L\bar w)$ of the space of Hecke modifications of
type $^L\bar w$ includes the spaces $\bar {{\EUY}}(^L\negthinspace
w_i)$  of Hecke modifications of types $^L\negthinspace w_i$, for
all $i\in \EUR'$.  In terms of Bogomolny moduli spaces, this means
that the spaces $\bar\EUZ(^L\negthinspace w_i,p')$ are all
contained in the singular ``monopole bubbling'' locus of
$\bar\EUZ(^L\bar w,p')$.

This is why it is difficult to determine the operator product
expansion for 't Hooft operators; the subleading terms in \hittor\
all come entirely from singularities of $\bar\EUZ(^L\bar w,p')$.
In section \opprod, we will explain certain mathematical results
\refs{\lusztig - \vilonen} which from a
physical point of view are equivalent to determining the full
operator product expansion.

First, however, we want to recall an example analyzed in section
\spaceheck\ that provides a good illustration of these arguments.
Take $G=PSU(2)$, $\LG=SU(2)$, $^L\negthinspace w={}^L\negthinspace
w'=(1/2,-1/2),$ and $^L\bar w={}^L\negthinspace
w+{}^L\negthinspace w'=(1,-1)$. The weights $^L\negthinspace w$
and $^L\bar w$ correspond to the two-dimensional and the adjoint
representations of $\LG=SU(2)$, respectively. As we learned in
section \spaceheck, the  spaces $\bar {{\EUZ}}(^L\negthinspace
w,z)$ and $\bar {{\EUZ}}(^L\negthinspace w',z)$ are copies of
$\Bbb{CP}^1$. The space $\bar\EUZ{}^*=\bar {{\EUZ}}(^L\bar w,z)$
is the weighted projective plane $\Bbb{WCP}^2(1,1,2)$ and has an
orbifold singularity. We have found that this singular space can
be deformed into a product $\Bbb{CP}^1\times\Bbb{CP}^1$ and also
resolved into a Hirzebruch surface which is a $\Bbb{CP}^1$ bundle
over $\Bbb{CP}^1$. We also saw that the resolution and deformation
are compatible, in the sense that the Hirzebruch surface can be
deformed into $\Bbb{CP}^1\times\Bbb{CP}^1$. All this is in accord
with the above general arguments.  (However, in this particular
example, $^L\negthinspace w={}^L\negthinspace w'$, so the two
resolutions are equivalent.)

In this example, there is a unique way to write $^L\bar w$ as a
sum of nonzero positive weights.  In general, there will be
several ways to do this and thus several choices of partial
resolutions and desingularizations of $\bar {{\EUZ}}(^L\bar w,z)$
along these lines.  For a simple Lie group $^L\negthinspace G$ of
rank $r$, there are $r$ fundamental weights $^L\negthinspace
w_\gamma$, $\gamma=1,\dots,r$, such that any dominant weight
$^L\bar w$ can be uniquely expanded $^L\bar
w=\sum_{\gamma=1}^r t_\gamma {}^L\negthinspace w_\gamma$ with some
non-negative integers $t_\gamma$.  (For example, if
$^L\negthinspace G=SU(r+1)$, the $^L\negthinspace w_\gamma$ are
the highest weights of the minuscule representations
$\wedge^\gamma \VV$, $\gamma=1,\dots r$. These are called the
fundamental representations.) The maximum partial resolution and
desingularization of $\bar {{\EUY}}(^L\bar w,z)$ that we can make
using the Bogomolny equations is to relate it to a moduli space
with $t_\gamma$ insertions of 't Hooft operators of type
$^L\negthinspace w_\gamma$, for $\gamma=1,\dots,r$.  For
$SU(r+1)$, since the fundamental representations  are minuscule,
this gives a complete resolution or deformation of singularities
of the space of Hecke modifications of any given weight.

For other groups, the best we can do is to reduce to the
singularities of a product of 't Hooft operators
$T({}^L\negthinspace w_\gamma)$ with fundamental weights.  The
space of Hecke modifications for such an operator has
singularities, associated with monopole bubbling, that can be
locally modeled by a hyper-Kahler manifold, but there is no
obvious hyper-Kahler resolution of these singularities.

\subsec{Operator Product Expansion Of 't Hooft Operators}

\subseclab\opprod

Let us write the tensor product of representations of $^L\negthinspace G$ as
\eqn\yurgo{^L\negthinspace R\otimes {}^L\negthinspace
R'=\bigoplus_{\alpha\in \EUR}\left(N_\alpha \otimes{}^L\negthinspace R_\alpha\right).}
Here $N_\alpha$ is the vector space $N_\alpha={\rm Hom}_G({}^L\neg R\otimes {}^L\neg{R'},{}^L\neg
R_\alpha)$, where ${\rm Hom}_G$ is the space of $G$-invariant linear transformations. The set $\EUR$
consists of
 the representations for which $N_\alpha$ is nonzero.
It is
convenient to also write $n_\alpha$ for the integer which is the dimension of the vector space
$N_\alpha$.

As we discussed in section \lineop, the tensor product of representations is precisely mirrored
in the operator product expansion of Wilson operators.  In eqn.  \tryo, we wrote the operator
product expansion for Wilson {\it loop} operators in terms of $c$-number coefficients $n_\alpha$.
However, for a static Wilson {\it line} operator, which we regard as part of the problem
of quantization, rather than an operator acting on quantum states, we can write a closely related and
more precise statement:
\eqn\docus{W({}^L\neg R,p)\cdot W({}^L\neg\tilde R,p')\to
\bigoplus_{\alpha\in\EUR}\left(N_\alpha\otimes W({}^L\neg
R_\alpha,p')\right).}
We are here taking the gauge group to be $^L\neg G$, and $p$ and $p'$ are the positions of the static
Wilson operators.
\docus\ is a recipe for constructing the quantum Hilbert space in the presence
of two Wilson operators labeled by $^L\neg R$ and $^L\neg R'$, as a direct sum of contributions;
each contribution is the tensor product with a vector space $N_\alpha$ of the quantum Hilbert space
obtained by quantization in the presence of a single Wilson operator labeled by $^L\neg R_\alpha$.  The basis
for the statement \docus\ is that the effect of a static Wilson operator in the quantum theory is
to introduce an external charge in the appropriate representation.  If there are two Wilson static
Wilson operators, there are  two external charges; after decomposing the tensor product of the
appropriate representations in a direct sum of irreducibles, one arrives at \docus.

From $S$-duality, we expect the corresponding operator
product expansion of 't Hooft line operators to take the same form
\eqn\zurgo{T({}^L\negthinspace R,p)\cdot T({}^L\negthinspace
R',p')\to \bigoplus_{\alpha\in \EUR}\left(N_\alpha \otimes T({}^L\negthinspace
R_\alpha,p')\right).}

As  in section \lineop, even without $S$-duality,  since the 't
Hooft operators are the only line operators at $\Psi=0$, a
relation such as \zurgo\ must hold with some set $\EUR$ of
representations (which depends on $^L\negthinspace R$ and
$^L\negthinspace R'$) and some vector spaces $N_\alpha$. Moreover,
simple arguments suffice to determine the set $\EUR$, which
depends on $^L\negthinspace R$ and $^L\negthinspace R'$. So the
problem is to describe the  $N_\alpha$.

If we specialize to the case of ordinary 't Hooft loop operators which may act on quantum states (as
opposed to static line
operators that modify the definition of the quantum Hilbert space), the coefficients in the operator
product expansion are ordinary $c$-numbers $n_\alpha$ rather than vector spaces $N_\alpha$.
One way to make this specialization is to take the ``time'' direction to be a circle.  This leads
to a trace in all of the quantum Hilbert spaces, and the vector space $N_\alpha$ is replaced by its
integer dimension $n_\alpha$.  The static 't Hooft line operators become circular 't Hooft loop operators
(which can act on quantum states if we interpret a different direction as the ``time'' direction
for quantization), and we learn that the operator product expansion coefficients
of 't Hooft loop operators, or of any loop operators in a topological field theory, are integers.

To make the point in a slightly different way, let us go back to the case of Wilson operators.
A closed Wilson loop operator is defined via the trace of the holonomy.
Taking the trace replaces a vector
space by its dimension.  But an open Wilson operator (the important example for us is a static Wilson
line operator) is defined in terms of the holonomy itself,
regarded as an operator on a vector space.  In decomposing the tensor products of such operators,
the ``coefficients'' are vector spaces $N_\alpha$, not integers $n_\alpha$.  For some purposes,
the distinction is not important and one can replace $N_\alpha$ by its dimension.

\bigskip\noindent{\it Some Useful Branes}

Once the scene is properly set, the determination of the
$N_\alpha$ is a consequence of a mathematical theory that originated with
Lusztig's work \lusztig\ on the relation of multiplicities in tensor products
of irreducible representations of $G$ to the convolution
of intersection cohomology complexes of affine Schubert varieties on $^L\neg G$,
and was further refined by Ginzburg \ginzburg\ and Mirkovic and Vilonen \vilonen\
to an equivalence of tensor categories.  In physical terms, this refinement is related to
a description of the operator product expansion of 't Hooft operators in which the ``coefficients''
are vector spaces $N_\alpha$ (as in our equation \zurgo), as
opposed to a more conventional description in which only closed loop operators are considered and
the $N_\alpha$ are replaced by their integer dimensions $n_\alpha$ (as in our equation \tryo).
We will set the context for the
analysis, get as far as we can with purely physical reasoning, and
then attempt to elucidate the mathematical theory.

\ifig\xitampo{Insertion of several 't Hooft operators
$T({}^L\negthinspace R_i,p_i)$, with a brane ${\cal B}$ on the
left that specifies an ``initial'' bundle $E_-$, and a brane
${\cal B}'$ on the right that sets $\varphi=0$.}
{\epsfxsize=4in\epsfbox{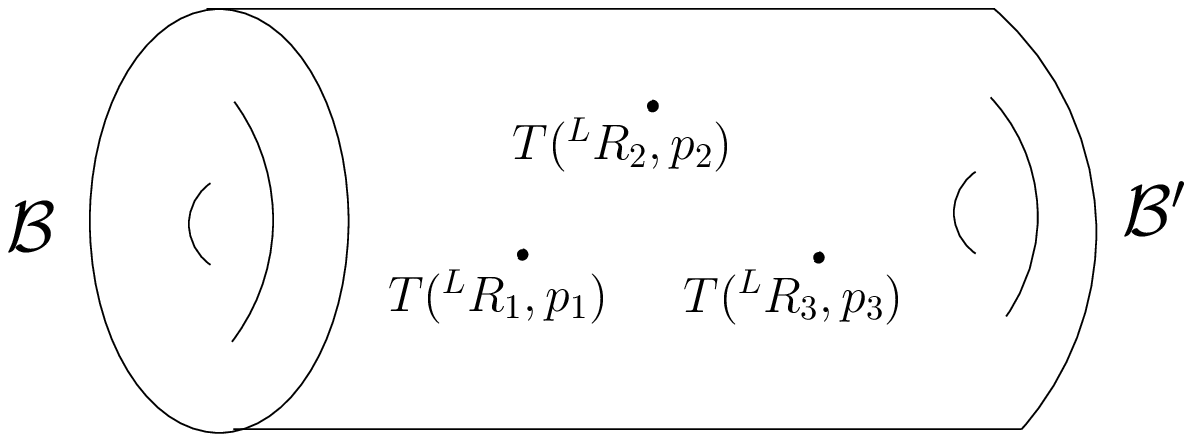}}

 To determine the $N_\alpha$, it suffices to consider any convenient and suitably rich
class of computations in which static 't Hooft operators play a role.
For this, we consider two or more 't Hooft operators in our
familiar three-manifold $W=I\times C$ and with suitable branes
${\cal B}$ and ${\cal B}'$ chosen at the two ends. We want to pick
branes that will simplify the computation.

't Hooft operators preserve supersymmetry of type $(B,A,A)$, so it
is most natural to choose ${\cal B}$ and ${\cal B}'$ to be branes
of that type.  For this, we can take ${\cal B}$ and ${\cal B}'$ to
be supported on subvarieties $U$ and $U'$ of Hitchin's moduli
space $\MH$ that are holomorphic in complex structure $I$ and
Lagrangian with respect to the symplectic form $\Omega_I$ that is
of type $(2,0)$ in this complex structure. (More briefly, $U$ and
$U'$ are complex Lagrangian submanifolds in complex structure
$I$.) We endow $U$ and $U'$ with  trivial Chan-Paton line
bundles.

We take $U$ to parametrize Hitchin pairs $(E,\varphi)$, where we
require $E$ to have a specified holomorphic type $E_-$ and
arbitrary $\varphi$. We require $E_-$ to be stable, but otherwise
the choice of $E_-$ does not matter.   And we take $U'$ to
parametrize pairs $(E,0)$ with any $E$ but with $\varphi=0$.

The first point is that $U$ and $U'$ intersect transversely at a
single point, corresponding to the Higgs bundle $(E_-,0)$. Moreover,
$U$ and $U'$ are complex symplectic manifolds in complex structure
$I$, so ${\cal B}$ and ${\cal B}'$ are branes of type $(B,A,A)$.

Finally, we have chosen ${\cal B}$ and ${\cal B}'$ so that the
space of supersymmetric configurations with these boundary
conditions and with some 't Hooft operators included is precisely
the familiar moduli space ${\EUZ}({}^L\negthinspace R_1,p_1;\dots;{}^L\negthinspace R_k,p_k)$ of
solutions of the Bogomolny equations.

 A few subtleties go into this statement.
First of all, supersymmetric configurations in general have
$\varphi\not= 0$ and must be described using the extended
Bogomolny equation that we described at the outset of section
\thooftheckeop, not the ordinary Bogomolny equations.  We have not
yet even analyzed the extended Bogomolny equations; this will be
the subject of section \extbog. When we do so, we will see that if
$\varphi$ vanishes at one end of $W$ -- and it does here, because
of the choice of $U'$ -- then $\varphi$ vanishes everywhere, and
the extended Bogomolny equations reduce to ordinary ones.

Second, in describing the branes ${\cal B}$ and ${\cal B}'$, we
only described the boundary conditions on the fields $(A,\varphi)$
of the sigma-model with target $\MH$.  The field $\phi_0$, which
is an important part of the Bogomolny equations, is present in the
gauge theory, but not in the sigma-model (except at singularities)
because in reduction to two dimensions it is ``massive.''  By
extending the analysis in section \zapendix, the branes ${\cal B}$
and ${\cal B}'$ can both be described in gauge theory \witfur, and
in that context it can be shown that the appropriate boundary
conditions for the field $\phi_0$ are the ones we use.

\bigskip\noindent{\it The Space Of Physical States}

In the $A$-model, in the absence of instanton corrections (they
are absent here for reasons we discussed at the beginning of
section \thooftheckeop), the space of physical states is the
cohomology of the space of time-independent supersymmetric
configurations.  In fact, the $A$-model as understood physically
leads most naturally to de Rham cohomology.

In other words, with branes ${\cal B}$ and ${\cal B}'$, and 't
Hooft operators $T(^L\negthinspace R_i,p_i)$, the space of
physical states is the cohomology of our friend, the moduli space
$\bar\EUZ(^L\negthinspace R_1,p_1;\dots;^L\neg R_k,p_k)$ of solutions of the Bogomolny
equations with the appropriate singularities. As $\bar\EUZ$ is
generically singular, we need to explain here what kind of
refinement of de Rham cohomology is meant. But first let us just
formally state what sort of result we expect.

For every finite set of representations $^L\negthinspace R_i$,
$i=1,\dots,k$, the cohomology of the moduli space
$\bar\EUZ=\bar\EUZ(^L\negthinspace R_1,p_1;\dots;{}^L\neg R_k,p_k)$
 gives us a space of physical
states ${\cal H}({}^L\negthinspace R_1,{}^L\negthinspace
R_2,\dots,{}^L\negthinspace R_k)=H^*(\bar\EUZ)$.
Notice that because of the underlying four-dimensional
topological quantum field theory (TQFT), the choices of the points
$p_i$ do not matter.  To be more exact, the TQFT says that locally
in $p_i$ the space of physical states is independent of the $p_i$
(there is a flat connection\foot{Moving the points is a special
case of changing the metric on $W$.  The change in metric is
trivial in the BRST cohomology, a fact which gives the flat
connection.} on the space of distinct points $p_i$). Since the
space of $n$-tuples in the three-manifold $W=I\times C$ is
simply-connected, there is also no room for monodromies, and thus
the $p_i$ are irrelevant. So we have labeled the space of physical
states ${\cal H}({}^L\negthinspace R_1,\dots,{}^L\negthinspace
R_k)$ just by the representations, not by the points.

Modulo questions about singularities, we do not really need
quantum field theory for this argument. In the case that the
representations $^L\negthinspace R_i$ are minuscule (and the $p_i$
are distinct), $\bar\EUZ(^L\negthinspace R_1,p_1;\dots;{}^L\neg R_k,p_k)$ is a family
of smooth manifolds. The cohomology of such a family is locally
constant, just as expected from the TQFT. In the general case with
singularities, we need to make sure to use a refinement of de Rham
cohomology in which this sort of argument is valid.  The property
of $\bar\EUZ(^L\negthinspace R_1,p_1;\dots;{}^L\neg R_k,p_k)$ that makes this possible
is that, although it is singular, its singular structure is
completely independent of the $p_i$.  The singularities come from
degenerations of individual Hecke transformations (or monopole
bubbling at positions of individual 't Hooft operators),
independent of the positions of the others.

Now the operator product expansion says that we can replace a
product of 't Hooft operators $T(^L\negthinspace R_1)\cdot
T(^L\negthinspace R_2)$ by a sum $\oplus_{\alpha\in \EUR}C_\alpha\otimes
T(^L\negthinspace R_\alpha)$, with some set $\EUR$ and some
vector spaces $C_\alpha$.  In the present context, for spaces of
physical states in the presence of the branes we have considered,
this means we must have a family of isomorphisms \eqn\mitrox{{\cal
H}({}^L\negthinspace R_1,{}^L\negthinspace
R_2,\dots,{}^L\negthinspace R_k)\cong\oplus_{\alpha\in
\EUR}\left(C_\alpha \otimes{\cal H}({}^L\negthinspace
R_\alpha,{}^L\negthinspace R_3,\dots,{}^L\negthinspace R_k)\right).}
Since the left hand side only depends on the unordered set of
representations $^L\negthinspace R_i$, the isomorphisms in
\mitrox\ are automatically compatible with associativity, that is
with taking repeated operator products to further reduce the
number of representations.

$S$-duality says that the set $\EUR$ and vector spaces $C_\alpha$
should be the ones that arise in the tensor product of
representations of $^L\negthinspace R$:
\eqn\omigron{^L\negthinspace R_1\otimes {}^L\negthinspace
R_2=\oplus_{\alpha\in \EUR}\left(N_\alpha\otimes{} ^L\negthinspace R_\alpha\right).}

We actually can be much more specific if we recall that\foot{These
assertions do not depend on the points $p_i\in W$, so we omit
them.} ${\cal H}({}^L\negthinspace R_1,^L\negthinspace
R_2,\dots,^L\negthinspace R_k)$ is the cohomology of
$\bar\EUZ({}^L\negthinspace R_1,{}^L\negthinspace
R_2,\dots,{}^L\negthinspace R_k)$  and
that topologically $\bar\EUZ({}^L\negthinspace R_1,{}^L\negthinspace
R_2,\dots,{}^L\negthinspace R_k)$ is a simple product
$\prod_{i=1}^k\bar\EUZ({}^L\negthinspace R_i)$.  (And
$\bar\EUZ({}^L\negthinspace R_i)$ is the same as the compactified
space $\bar {{\EUY}}({}^L\negthinspace R_i)$ of Hecke
modifications of type ${}^L\negthinspace R_i$.)  The de Rham
cohomology of a product is the product of the cohomologies of the
factors, so we get \eqn\kipro{{\cal H}({}^L\negthinspace
R_1,{}^L\negthinspace R_2,\dots,{}^L\negthinspace
R_k)=\otimes_{i=1}^k {\cal H}({}^L\negthinspace R_i).} This
implies that \mitrox\ must really come from a family of
isomorphisms \eqn\gitrox{{\cal H}({}^L\negthinspace R_1)\otimes
{\cal H}({}^L\negthinspace R_2)=\oplus_{\alpha\in \EUR}N_\alpha\otimes
{\cal H}({}^L\negthinspace R_\alpha).}

Comparing \omigron\ to \gitrox, we see that the ${\cal
H}({}^L\negthinspace R_i)$ obey the same algebra of tensor
products as the $^L\negthinspace R_i$.  This suggests that there
is an isomorphism between ${\cal H}({}^L\negthinspace R_i)$ and
$^L\negthinspace R_i$ that produces the automorphism between the
two algebras. Indeed, it can be shown \ref\deligne{P. Deligne and
J. Milne, ``Tannakian Categories,'' Lect. Notes in Math {\bf 900}
(1982) 101-228.} that this is the only way for the families of
finite-dimensional vector spaces $^L\negthinspace R_i$ and ${\cal
H}(^L\negthinspace R_i)$ to have the same algebra of tensor
products.  The isomorphism between ${\cal H}({}^L\negthinspace
R_i)$ and $^L\negthinspace R_i$ is unique up to conjugation by an
element  of $^L\negthinspace G$.

\bigskip\noindent{\it Grading By Ghost Number}

Since $^L\negthinspace G$ certainly acts on the $^L\negthinspace
R_i$, in a way compatible with the tensor products, this means
that there is a corresponding action of $^L\negthinspace G$ on
${\cal H}({}^L\negthinspace R_i)$ (defined up to conjugation).
What this means is mysterious from a physical point of view, since
$^L\negthinspace G$ is not a symmetry of the full $A$-model of
gauge group $G$ at $\Psi=0$, but only of the particular piece of
it that we have looked at to analyze the operator product
expansion. (Things might become clearer if one could understand
the $S$-duals of the branes ${\cal B}$ and ${\cal B}'$ that were
used in the construction.) However, the full $A$-model does have
one symmetry that is relevant.

This is the ``ghost number'' symmetry ${\EUK}$, whose origin in
topological twisting of ${\EUN}=4$ super Yang-Mills theory was
explained in section \twisting.  In the $A$-model, the ghost
number is, roughly speaking, the degree or dimension of a
cohomology class.\foot{For a Kahler target, the Hodge
decomposition $H^d(X)=\oplus_{p+q=d}\,H^{p,q}(X)$ gives a further
refinement  of the $A$-model, in general. However, this is not
relevant for our present discussion, as the cohomology of the
spaces  $\bar \EUZ({}^L\negthinspace R_1,\dots,{}^L\neg R_k)$ is all of type $(p,p)$.}
There is, however, one key subtlety.  For a complex manifold $X$
of complex dimension $n$, the degree of a differential form is
usually understood to run from $0$ to $2n$. But the space of
physical states of the $A$-model has a fermion conjugation
symmetry\foot{The symmetry in question is two-dimensional CPT
symmetry.  It is an exact symmetry of the physical supersymmetric
sigma-model with $(2,2)$ supersymmetry and target $X$. On a
non-flat Riemann surface $\Sigma$, this symmetry is spoiled by the
topological twisting that defines the $A$-model.  But the space of
physical states is obtained by quantizing the sigma-model on the
flat manifold $\Bbb{R}\times S^1$, which is unaffected by the
twisting.} that exchanges a class of dimension $d$ with one of
dimension $2n-d$ and maps ${\EUK}\to -{\EUK}$. The ghost number of
a state that is related to a cohomology class of degree $d$ is
actually\foot{ For completeness, let us note that if $X$ is a
Calabi-Yau manifold, then the $A$-model mapping from states to
operators involves a spectral flow that adds $\dim_{\Bbb{C}}(X)$
to ${\EUK}$.  So if $\psi$ is a class in the $d$-dimensional
cohomology of $X$, then the $A$-model state corresponding to
$\psi$ has ${\EUK}=d-{\rm dim}_{\Bbb{C}}(X)$, but the
corresponding operator ${\EUO}_\psi$ has ${\EUK}=d$. However, the
spaces $\bar \EUZ({}^L\negthinspace R_1,\dots,{}^L\neg R_k)$ are not Calabi-Yau,
and more importantly our discussion of branes and 't Hooft
operators refers to the physical states, not to corresponding
operators.} \eqn\insno{{\EUK}=d-{\rm dim}_{\Bbb{C}}(X).}

In the map from $^L\negthinspace R_i$ to ${\cal H}(^L\negthinspace
R_i)$, the symmetry ${\EUK}$ of the $A$-model  must correspond to
a derivation (that is, the generator of an automorphism) of the
tensor algebra of representations of $^L\negthinspace G$. Any such
derivation is actually an element of the complexified Lie algebra
$^L\negthinspace \frak{g}$. The element of $^L\negthinspace
\frak{g}$ corresponding to ${\EUK}$ is well-defined up to
conjugation.

For given $^L\negthinspace G$, one can find the right conjugacy
class by considering a specific example.  For example, for
$^L\negthinspace G=SU(N)$, consider the $N$-dimensional
representation $\VV$. The corresponding space of Hecke
modifications is $\bar {{\EUY}}(\VV)\cong \Bbb{CP}^{N-1}$, as we
learned in section \spaceheck. The de Rham cohomology of
$\Bbb{CP}^{N-1}$ is $N$-dimensional. The fact that this coincides
with the dimension of $\VV$ is an example of the correspondence
$^L\negthinspace R\leftrightarrow {\cal H}({}^L\negthinspace R)$.
$\Bbb{CP}^{N-1}$ has complex dimension $N-1$ and has cohomology in
dimensions $d=0,2,4,\dots,2N-2$. So the element of
$^L\negthinspace \frak{g}$ corresponding to ${\EUK}$  has
eigenvalues $N-1,N-3,N-5,\dots,-(N-1)$ in the representation
$\VV$: \eqn\unosn{\EUK=\left(\matrix{N-1& &&& \cr & N-3 &&&\cr &&
N-5 &&\cr &&&\ddots & \cr &&&& -(N-1)\cr}\right).} For
$^L\negthinspace G=SU(2)$, the conjugacy class of ${\EUK}$ is
simply \eqn\mitlgo{{\EUK}\cong\left(\matrix{1 & 0 \cr 0 & -1
\cr}\right).}

\bigskip\noindent{\it Intersection Cohomology}

 A mathematical theory that was initiated by Lusztig \lusztig, and
 later refined by   Ginzburg \ginzburg\ and Mirkovic and
Vilonen \vilonen, in effect determines the ``coefficients'' in
the operator product expansion of 't Hooft operators. Describing
these results in detail is a task better left to others. Instead,
here we will try to express a few of the ideas in a language that
physicists might find illuminating.

\nref\gormac{M. Goresky and R. MacPherson, ``Intersection Homology
Theory,'' Topology {\bf 19} (1980) 135-162.}%
\nref\kirwan{F. Kirwan,  {\it An Introduction To Intersection
Homology Theory} (Longman, Harlow, 1988).}%

 First of all, the cohomology theory used in this theorem is intersection cohomology
\refs{\gormac,\kirwan}.
 Roughly speaking,
and modulo standard conjectures, the intersection cohomology of an
algebraic variety $X$ is the ${\Bbb{L}}^2$ cohomology of the
smooth part of $X$.  One considers square-integrable differential
forms on the smooth part of $X$. The intersection cohomology
$IH^*(X)$ is the cohomology of the exterior derivative operator
$d$ in the space of such square-integrable forms.  (It cannot
necessarily be expressed in terms of harmonic forms.)  To match
intersection cohomology with quantum field theory, it is useful to
make the same shift in dimensions that arises naturally in the
$A$-model: we define a grading of $IH^*(X)$ such that an ${\Bbb{
L}}^2$ cohomology class that is a $d$-form is considered to have
degree $\EUK=d-{\rm dim}_{{\Bbb C}}(X)$.

To be more precise, the ${\Bbb{L}}^2$ cohomology of the smooth
part of $X$ is the intersection cohomology of $X$ relative to its
standard intersection cohomology complex.  We might denote this
$IH^*(X,IC_X)$.  One also defines other intersection cohomology
complexes on $X$.  We will really only need the simplest
construction. If $Y\subset X$ is a subvariety, one defines an
intersection cohomology complex $IC_Y$ which is supported on $Y$
and such that $H^*(X,IC_Y)$ is the ${\Bbb{L}}^2$ cohomology of the
smooth part of $Y$, shifted in dimension by $-{\rm
dim}_{\Bbb{C}}(Y)$. Thus, $IC_Y$ is the standard intersection
cohomology complex of $Y$, but embedded in $X$ (and ``extended by
zero'' away from $Y\subset X$).

In the case of the spaces $\bar {{\EUY}}({}^L\bar w)$ of Hecke
modifications, the subvarieties of interest are the spaces $\bar
{{\EUY}}({}^L\negthinspace w_\alpha)$, where $^L\negthinspace
w_\alpha$ is a weight associated to $^L\negthinspace\,\bar w$. So
for each associated weight, one defines an intersection cohomology
complex $IC({}^L\negthinspace w_\alpha)$, supported on the
subspace of $\bar {{\EUY}}({}^L\bar w)$ on which monopole bubbling
has reduced the weight from $^L\bar w$ down to $^L\negthinspace
w_\alpha$. We also denote the standard intersection cohomology
complex of $\bar {{\EUY}}({}^L\negthinspace\,\bar w)$ as
$IC({}^L\negthinspace\,\bar w)$.

It will help at this point to recall our knowledge of the spaces
of Hecke modifications and consider a simple example (whose
relevant properties were described in detail in section
\spaceheck). Let $^L\negthinspace G=SU(2)$ and let $^L \bar
w=(1,-1)$ be the weight associated to the adjoint representation.
The compactified space of Hecke modifications is then a weighted
projective space $\bar {{\EUY}}({}^L\negthinspace\bar
w)=\Bbb{WCP}^2(1,1,2)$. This space is an orbifold, so its de Rham
cohomology is well-defined (with no need to resort to
${\Bbb{L}}^2$ cohomology). The cohomology is three-dimensional,
with generators in degrees $d=0,2,4$.  Hence the values of
${\EUK}=d-{\rm dim}_{\Bbb{C}}(\Bbb{WCP}^2(1,1,2))=d-2$ are
${\EUK}=2,0,-2$. This is in accord with expectations; the adjoint
representation of $SU(2)$ is three-dimensional, and the element
${\EUK}$ of its Lie algebra that is defined in the two-dimensional
representation in \mitlgo\ indeed has eigenvalues $2,0,-2$ in the
adjoint representation.

\ifig\xytampo{A pair of Hecke modifications made at the same value
of $z$ but at distinct points in $I$.}
{\epsfxsize=4in\epsfbox{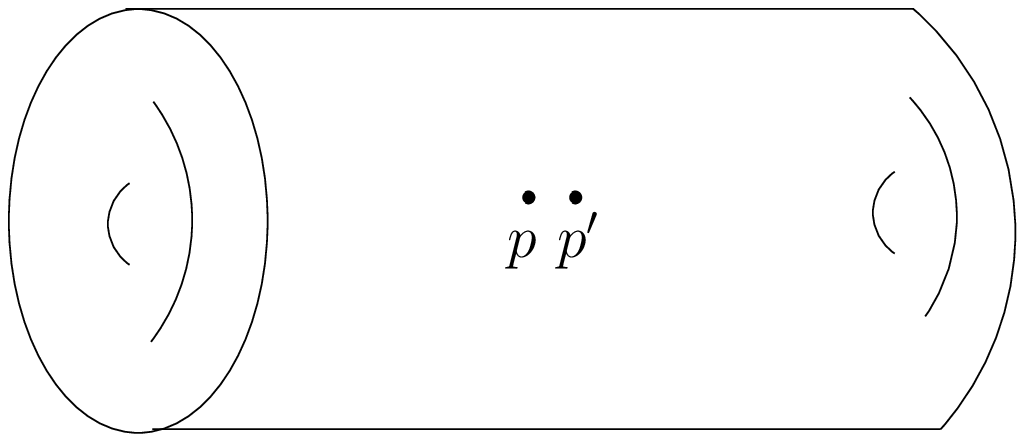}}

But so far we have not said anything that depends on the
singularity of the weighted projective space. To see its role, let
$^L\negthinspace w=(1/2,-1/2)$ be the weight corresponding to the
two-dimensional representation.  The operator product expansion of
the corresponding Wilson operator with itself should read
\eqn\murtog{T({}^L\negthinspace w)\cdot T({}^L\negthinspace
w)=T({}^L\negthinspace\bar w)\oplus T(0),} where $0$ is the zero weight,
and $T(0)=1$ is the corresponding trivial 't Hooft operator. (The coefficients are
really one-dimensional vector spaces, but we have omitted them.)  Let
us see how this comes about.  The space of Hecke modifications of
type $^L\negthinspace w$ is isomorphic to $\Bbb{CP}^1$. If
(\xytampo) we make repeated Hecke modifications of type
$^L\negthinspace w$ at the same point $z\in C$ but different
points in $I$, we get a moduli space $\bar\EUZ({}^L\negthinspace
w,z,y;{}^L\negthinspace w,z,y')$ that is a $\Bbb{CP}^1$ bundle
over $\Bbb{CP}^1$.  Its cohomology is four-dimensional, with
classes in dimensions $d=0,2,2,4$, so the eigenvalues of $\EUK$
are $-2,0,0,2$.

Now let us try to take the limit as $y\to y'$. For any $y<y'$, we
can use Hodge theory; the space of harmonic forms on $\bar\EUZ$ is
four-dimensional.  At $y=y'$, an $A_1$ orbifold singularity
develops; the local structure looks like $\Bbb{R}^4/\Bbb{Z}_2$.
For $y$ slightly less than $y'$ (compared to the length of the
interval $I$), this singularity is resolved.  The resolution can
be locally modeled by  a  hyper-Kahler manifold $\EUWW$
(discovered in \ref\eguchi{T. Eguchi and A. J. Hansen, ``Selfdual
Solutions To Euclidean Gravity,'' Ann. Phys. {\bf 120} (1979)
82-106.}) that in one of its complex structures can be identified
with $T^*\Bbb{CP}^1$. The metric on $\EUWW$ is complete, and
moreover, is asymptotic at infinity to the flat metric on
$\Bbb{R}^4/\Bbb{Z}_2$.

Crucially for us, the space of ${\Bbb{L}}^2$ harmonic forms on
$\EUWW$ is one-dimensional, generated  \eguchi\ by a two-form
$\omega$. (Note that here the ${\Bbb{L}}^2$ condition has to do
with the behavior at infinity, while in the definition of
intersection cohomology it has to do with the behavior near
singularities.) Being in ${\Bbb{L}}^2$, this form has support near
the exceptional divisor, and as $y\to y'$ (and the exceptional
divisor shrinks), it converges to a form with delta function
support at the singularity. Having $d=2$, it corresponds to a
cohomology class with $\EUK=0$, the value corresponding to the
trivial representation of $^L\negthinspace G=SU(2)$. The
${\Bbb{L}}^2$ harmonic two-form on $\EUWW$ exists for topological
reasons that can be understood from arguments in \ref\gsegal{G.
Segal and A. Selby, ``The Cohomology Of The Space Of Magnetic
Monopoles,'' Commun. Math. Phys. {\bf 177} (1996) 775-787; G.
Segal, ``Topology Of The Space Of $SU(2)$-Monopoles In
$\Bbb{R}^3$,'' in {\it Geometry And Physics}, ed. J. E. Anderson
et. al. (Marcel Dekker, New York, 1997) 141-147.}.

Now we can understand the meaning of the operator product relation
\murtog.  The left hand side is associated with the space
$\bar\EUZ$ of repeated Hecke modifications.  For $y<y'$, it is a
$\Bbb{CP}^1$ bundle over $\Bbb{CP}^1$, and has a four-dimensional
space of harmonic forms, which is the physical Hilbert space
${\cal H}$.  A three-dimensional subspace of this space consists
of harmonic forms that, as $y\to y'$, converge to ${\Bbb{L}}^2$
harmonic forms on the smooth part of
$\bar\EUZ{}^*=\Bbb{WCP}^2(1,1,2)$ (which is the limit of
$\bar\EUZ$ for $y\to y'$). They reflect the contribution of
$T({}^L\negthinspace\bar w)$ to ${\cal H}$. The fourth harmonic
form can be approximated for $y$ near $y'$ by the $\Bbb{L}^2$
harmonic form on $\EUWW$ and is supported near the exceptional
divisor. The support of this fourth harmonic form converges as
$y\to y'$ to the singular point of $\bar\EUZ{}^*$, which is the
space of Hecke modifications of weight $^L\negthinspace w=0$. It
is the contribution of $T(0)$ to ${\cal H}$. The fact that the
coefficient of $T(0)$ is 1 expresses the fact that the space of
harmonic two-forms that converge to the singularity as $y\to y'$
is precisely one-dimensional.  To be more exact, the coefficient of $T(0)$ is
this one-dimensional space of harmonic forms.

More generally, suppose we are given any pair of weights
$^L\negthinspace w$, $^L\negthinspace w'$. We want to describe the
operator product $T({}^L\negthinspace w)\cdot T({}^L\negthinspace
w')$, which is expected to take the form
\eqn\mitory{T({}^L\negthinspace w)\cdot T({}^L\negthinspace
w')=T({}^L\negthinspace\,\bar w)\oplus\left(\oplus_{\alpha\in \EUR'}\,N_\alpha
\otimes T({}^L\negthinspace w_\alpha)\right),} where $^L\negthinspace\,\bar
w={}^L\negthinspace w+{}^L\negthinspace w'$, and the sum runs over
weights associated to $^L\negthinspace\,\bar w$.   As $y\to y'$,
$\bar\EUZ({}^L\negthinspace w,y,z;{}^L\negthinspace w,y',z)$
converges as a space to $\bar\EUZ({}^L\negthinspace\,\bar w,y,z)$
(whose dependence on $y$ and $z$ is inessential). Some of the
${\Bbb{L}}^2$ harmonic forms on the moduli space converge to such
forms on the smooth part of $\bar\EUZ({}^L\negthinspace\,\bar w)$.
These are the contribution of $T({}^L\negthinspace\,\bar w)$ in
the operator product expansion. Others converge to have delta
function support on $\bar\EUZ({}^L\negthinspace w_\alpha)$ for
some associated weight $^L\negthinspace w_\alpha$ of
$^L\negthinspace\bar w$.  The forms that for $y\to y'$ have
support on $\bar\EUZ({}^L\negthinspace w_\alpha)$ and not on
$\bar\EUZ({}^L\negthinspace w_\beta)$ for any weight
$^L\negthinspace w_\beta$ associated to $^L\negthinspace w_\alpha$
are the contribution of $T({}^L\negthinspace w_\alpha)$ in the
operator product.

Let us simply conclude by describing how this idea has been
expressed mathematically \refs{\lusztig -\vilonen}.  The  moduli
space $\bar\EUZ({}^L\negthinspace w,z,y;{}^L\negthinspace
w',z,y')$ is a fibration with fiber $\bar\EUZ ({}^L\negthinspace
w',z,y')$ and base $\bar\EUZ({}^L\negthinspace w,z,y)$, as we
explained in section \kahstr. As $y\to y'$, this moduli space is
``blown down'' to make the more singular space
$\bar\EUZ({}^L\negthinspace\,\bar w,z,y')$, whose dependence on
$z$ and $y'$ is inessential. The blowdown is a rational map
\eqn\normo{\pi:\bar\EUZ({}^L\negthinspace w,z,y;{}^L\negthinspace
w',z,y')\to \bar\EUZ({}^L\negthinspace\,\bar w).}

We write $IC({}^L\negthinspace w)\star IC({}^L\negthinspace w')$
for the standard intersection cohomology complex on
$\bar\EUZ({}^L\negthinspace w,z,y;{}^L\negthinspace w',z,y')$.
(The notation is motivated by the fiber bundle structure of this
space.)  There is a pushdown map $\pi_*$ in intersection
cohomology.  As the sheaf $IC({}^L\negthinspace w)\star
IC({}^L\negthinspace w')$ is invariant under the group called
$GL(N,{\cal O})$ in section \affgrass\ or its analog $G({\cal O})$
for other $G$, and $\pi_*$ commutes with this action, the pushdown
$\pi_*(IC({}^L\negthinspace w)\star IC({}^L\negthinspace w'))$ is
likewise invariant under $G({\cal O})$.  It therefore is possible
to expand it as a sum of the standard intersection cohomology
complex $IC({}^L\negthinspace\bar w)$ of
$\bar\EUZ({}^L\negthinspace\bar w)$, and the complexes
$IC({}^L\negthinspace w_\alpha)$ that are supported on the
singular orbits with weights $^L\negthinspace w_\alpha$ that are
associated to $^L\negthinspace\bar w$.  (These are the only
$G({\cal O})$-invariant intersection cohomology complexes on
$\bar\EUZ({}^L\negthinspace\bar w)$.)

So we must have\foot{We have shifted all complexes in dimension to
be symmetric between positive and negative dimensions.  The map
$\pi_*$ preserves this symmetry.  So we do not need to be
concerned about further shifts.}
\eqn\yrox{\pi_*\left(IC({}^L\negthinspace w)\star
IC({}^L\negthinspace w')\right)=IC({}^L\negthinspace\bar
w)\oplus \left(\bigoplus_\alpha\, \tilde N_\alpha \otimes \,IC({}^L\negthinspace w_\alpha)\right).} The
``coefficient'' of the leading term $IC({}^L\negthinspace\bar w)$ is
a trivial one-dimensional vector space (which we can omit),
since the map $\pi$ is an isomorphism over the smooth part of
the moduli space. The nature of the leading term has an elementary
explanation explained in the discussion of eqn. \hffo. The other
``coefficients'' are vector spaces $\tilde N_\alpha$. This formulation is analogous to the
general structure \hffo\ or \zurgo\ of the product of 't Hooft
operators, with the intersection complex $IC({}^L\negthinspace
w_\alpha)$ representing the contribution of the 't Hooft operator
$T({}^L\negthinspace w_\alpha)$.  Of course, the hard part of the
proofs \refs{\lusztig - \vilonen} is to describe the spaces $\tilde N_\alpha$ and
establish their isomorphism with the analogous spaces $N_\alpha$ that appear in \omigron.

\subsec{The Extended Bogomolny Equations}

\subseclab\extbog

So far, in our study of the action of 't Hooft operators on
branes,  we have considered only supersymmetric solutions that are
time-independent and time-reversal invariant.  With this
restriction, the BPS equations \oops\ reduce to the ordinary
Bogomolny equations, and the 't Hooft operators reduce to the
usual geometric Hecke operators.

The assumption of time-independence can actually be justified in
general, but requiring time-reversal invariance puts a severe
restriction on the allowed boundary conditions.  It is only
adequate if branes are chosen (as we did in section \opprod\ in
discussing the product of 't Hooft operators) to ensure that the
relevant solutions have $\varphi=0$. The most important $A$-branes
for the geometric Langlands duality (the generic fibers of the
Hitchin fibrations and the c.c. brane described in section
\abranes) do not have their support limited to $\varphi=0$, so in
order to study the action of 't Hooft operators on them, we need
to drop the condition of time-reversal invariance.  This is the
purpose of the following discussion.  Considerably more detail
will appear elsewhere \witfur.

We consider again the BPS equations on a four-manifold
$M=\RR\times W$ and write the Higgs field $\phi$ as $\phi_0\, dx^0
+\pi^*\tilde\phi$, where $\pi$ is the projection to $W$, and
$\tilde\phi$ is a one-form Higgs field on $W$. Likewise, we write
the gauge field as $A=A_0\,dx^0+\tilde A$, where $\tilde A$ is a
three-dimensional connection with curvature $\tilde F$. The
time-independent BPS equations for $t=1$ read
\eqn\BPSstaticgen{\eqalign{\tilde F-\tilde\phi\wedge\tilde\phi&=
\star\left(D\phi_0-[A_0,\tilde\phi]\right),\cr \star
D\tilde\phi&=[\phi_0,\tilde\phi]+D A_0,\cr
D^*\tilde\phi+[A_0,\phi_0]&=0.}} The exterior derivative $D$, the
Hodge star operator $\star$, and the operator $D^*=\star D\star$
are all understood in the three-dimensional sense.

These equations can be greatly simplified using suitable vanishing
theorems.  In fact, one can show that $\phi_1=A_0=0$ using the
same arguments, and the same type of assumptions, as in the
time-reversal invariant case (recall eqn. \wungo).

So we set $\phi_1=A_0=0$, choose a local complex coordinate
$z=x^2+ix^3$ on $C$, denote $y=x^1$,  and work in the gauge
$\tilde A_y=0$. We also now drop the tildes from our notation.
And, as in section \theck, we take the metric on $W$ to be $ds^2=h
|dz|^2+ dy^2$, for some function $h$. Then the extended Bogomolny
equations \BPSstaticgen\ reduce to
\eqn\BPSextsimple{\eqalign{D_\bz\phi_z&=0,\cr
\partial_y A_\bz&=-i D_\bz \phi_0,\cr
\partial_y\phi_z&=-i[\phi_z,\phi_0],\cr
F_{z\bz}-[\phi_z,\phi_\bz]&={ih\over 2}\partial_y\phi_0.}}

The first of the equations \BPSextsimple\ says that
$\varphi=\phi_z\,dz$, restricted to $C_y=\{y\}\times C$, is a
holomorphic section of $\End(E)\otimes K_C$ for any $y\in I$. So
denoting as $E$ the holomorphic bundle over $C_y$ defined by the
$\bar\partial$ operator $\bar D=d\bar z(\partial_{\bar z}+A_{\bar
z})$, the pair $(E,\varphi)$ is a Higgs bundle for any $y$. The
next two equations have an interpretation that should be easy to
anticipate based on our experience with the ordinary Bogomolny
equations in section \theck: they say that evolution in the $y$
direction changes the Higgs pair $(E,\varphi)$ only by a gauge
transformation.  Thus until we encounter the singularity
associated with an 't Hooft operator, the point in $\MH$
determined by a solution of the extended Bogomolny equations is
independent of $y$.

Now suppose that there is an 't Hooft operator of weight
$^L\negthinspace w$ at, say, $y=y_0$, $z=z_0$.  When $y$ is
increased past $y=y_0$, the bundle $E$ undergoes an ordinary Hecke
modification, as described in section \theck.  In fact, the
evolution equation for $A_{\bar z}$ is the same as it was in
section \theck, and so is the singularity due to the 't Hooft
operator. The pair $(E,\varphi)$ likewise undergoes what we will
call a Hecke modification.  The role of $\varphi$ is, however,
quite different from the role of $E$.

A holomorphic bundle can be modified at a single point, while
preserving holomorphy, and this is what the Hecke or 't Hooft
operators do.  There is a moduli space of possible such
modifications, the familiar space $\bar {\EUY}({}^L\negthinspace
w;z_0)$ parameterizing singular solutions of the Bogomolny
equations (and isomorphic to a Schubert variety). But a
holomorphic field, in this case a holomorphic section $\varphi$ of
${\rm ad}(E)\otimes K_C$, cannot be modified at a point while
preserving holomorphy. Hence, if $\varphi$ is given away from
$z=0$, it has at most a unique holomorphic extension over $z=0$.

So including $\varphi$ does not cause the Hecke modification due
to the 't Hooft operator to depend on additional parameters.  On
the contrary, including $\varphi$ generically eliminates some
parameters.  The reason is that even if $\varphi$ is holomorphic
at $z=z_0$ for $y<y_0$, it will generically have a pole at $z=z_0$
for $y>y_0$.  To see how this happens, let us take $G=U(N)$ and
consider an 't Hooft operator with weight $^L\negthinspace
w=(m_1,\ldots,m_N)$. The holomorphic type of the bundle $E$ jumps
in crossing $y=y_0$; we write $E_-$ and $E_+$ for the bundles
``before'' and ``after.'' Relative to some local decomposition
$E_-\cong \oplus_{i=1}^N \cL_i,$ the bundle $E_+$ has a local
decomposition as $\oplus_{i=1}^N\cL_i\otimes \CO(z_0)^{m_i}$. That
is, if $\cL_i$ is locally generated by a holomorphic section
$s_i$, $i=1,\ldots,N$, then $E_+$ is generated by holomorphic
sections \eqn\zilfog{ t_i=(z-z_0)^{-m_i} s_i. } Here we used a
natural isomorphism between $E_+$ and $E_-$ for $z\neq 0$.

Suppose that $\varphi$ is holomorphic at $z_0$ for $y<y_0$. The
products $\varphi \,s_i$ are holomorphic sections of $E_- \otimes
K_C$ (where $K_C$ will play no essential role as we can trivialize
it near $z=z_0$), so they can be expanded in terms of the $s_i$:
\eqn\bongo{\varphi\cdot s_i=\sum_j f_i{}^j\,s_j.}  Holomorphy of
$\varphi$ at $z=z_0$ for $y<y_0$ says that the $f_i{}^j$ (which
are sections of $K_C$) are holomorphic at $z=z_0$.  They are
subject to no other constraint.

Similarly, for $y>y_0$, we can expand $\varphi\,t_i$ as a linear
combination of the $t_j$.  Given the relation \zilfog\ between the
$s_i$ and the $t_i$, we can immediately write the form of the
expansion: \eqn\bzongo{\varphi\cdot t_i=\sum_j
(z-z_0)^{m_j-m_i}f_i{}^j\,t_j.} For $\varphi$ to be holomorphic at
$y>y_0$, the functions $(z-z_0)^{m_j-m_i}f_i{}^j$ must all be
regular at $z=z_0$.  If we order the $m_i$ in the usual way
$m_1\geq m_2\geq\dots\geq m_N$, then the number of conditions on
the Taylor series coefficients of the $f_i{}^j$ near $z=z_0$ is
$\sum_{i<j}(m_i-m_j)$.

This is the same as the dimension of the space of Hecke
modifications of bundles as determined in eqn. \ydro.  This
strongly suggests that, in sharp contrast to the case of Hecke
modifications of bundles, the Hecke modifications of a Higgs
bundle $(E,\varphi)$ are generically a finite set.  In fact, this
is the case if $\varphi(z_0)$ is regular and semisimple (which
means for $G=U(N)$ that it is diagonalizable with distinct
eigenvalues). The most systematic description of the result is to
say that $\varphi$ determines a vector field on the variety $\bar
{\EUY}({}^L\negthinspace w,z_0)$ that parametrizes Hecke
modifications of $E$ of type $^L\negthinspace w$ at the point
$z_0$; the zeroes of this vector field are the possible Hecke
modifications of the pair $(E,\varphi)$. (The vector field is not
quite canonically determined, but its zeroes are.)

Deferring a detailed explanation of this to \witfur, we will
briefly describe the simple example with $G=U(2)$,
$^L\negthinspace w=(1,0)$. In this example, as we learned in
section \spaceheck, if we take the bundle $E_-$ to be trivial,
then $E_+$ can be characterized by saying that a section of $E_+$
holomorphic at $z=0$ has the form \eqn\hxonxno{ s_+=(z-z_0)^{-1}
u+(h_1(z),h_2(z)), }  where $u$ is some nonzero element of
$\Bbb{C}^2$ and $h_1$ and $h_2$ are holomorphic near $z=z_0$.  Up
to scaling by $\Bbb{C}^*$, $u$ parametrizes the space of possible
Hecke modifications of $E$, which is $\Bbb{CP}^1$ in this example.
$\varphi$ is holomorphic at $z=z_0$  for $y>y_0$ if and only if
the product $\varphi\cdot s_+$ is of the same form, or in other
words if and only if $\varphi u$ is a multiple of $u$. In the
regular semisimple case, this means that $u$ must be one of the
two eigenvectors of $\varphi$.

Going back to the extended Bogomolny equations \BPSextsimple, an
immediate consequence of the relation
$\partial_y\phi_z=-i[\phi_z,\phi_0]$ is that $\phi_z$ is
everywhere zero if it zero for some $y$.  We assumed this in our
analysis of the product of 't Hooft operators in section \opprod.
A further consequence of the same equation is that the
characteristic polynomial of $\varphi$ is independent of $y$, even
in crossing an 't Hooft operator.  As a result, Hecke
transformations of Higgs bundles map each fiber of the Hitchin
fibration to itself.

Now consider a brane wrapped on a fiber of the Hitchin fibration,
with a flat Chan-Paton bundle. These are the branes that according
to $S$-duality must be magnetic eigenbranes. From what we have
said in the last paragraph, the support of such a brane is
invariant under Hecke transformations.  This observation is a step
toward proving directly, without relying on $S$-duality, that such
branes are magnetic eigenbranes, as will be shown in detail
elsewhere \witfur.

\bigskip\noindent{\it Hecke Correspondence For Higgs Bundles}

Just as in our discussion of the ordinary Bogomolny equations, we
can consider the possible Hecke modifications of a given Higgs
bundle $(E,\varphi)$ -- which has been our point of view so far --
or we can describe the initial and final states more symmetrically
and describe a Hecke correspondence ${\EUQ}$ on the space of Higgs
bundles. We define ${\EUQ}$ by solving the extended Bogomolny
equations on $W=I\times C$ with suitable boundary conditions at
the two ends and in the presence of specified 't Hooft operators.
We specify both $A$ and $\varphi$ (but not $\phi_0$) at both ends
of $W$, and divide only by gauge transformations that are trivial
at each end.

Let $\EUW_0$ be the space of all pairs $(A,\varphi)$ consisting of
a gauge field and Higgs field on $C$ satisfying $\bar D\varphi=0$.
Then $\EUW_0$ is a gauge theory version of the ``stack'' ${\rm
Higgs}_G$ of all Higgs bundles, just as the space of all
connections is a gauge theory version of ${\rm Bun}_G$, the stack
of all $G$-bundles.  By restricting a solution of the extended
Bogomolny equations to its boundary values, ${\EUQ}$ has a pair of
maps to $\EUW_0$, or in other words, a map to $\EUW_0\times
\EUW_0$. This is the Hecke correspondence for Higgs bundles.

${\EUQ}$ is very complicated, in general, with a variety of
components of different dimension.  However, if we limit ourselves
to a good region of $\EUW_0$ (where each Higgs bundle has only
finitely many possible Hecke modifications of the chosen type up
to a gauge transformation), then ${\EUQ}$ can be regarded as a
submanifold of $\EUW_0\times \EUW_0$. In this good region,
moreover, we can divide by gauge transformations and  ${\EUQ}$
becomes a middle-dimensional submanifold of $\MH\times\MH$.

't Hooft operators preserve supersymmetry of type $(B,A,A)$, as we
explained at the end of section \lineop.  This means that it must
be possible to interpret ${\EUQ}$ as a brane of type $(B,A,A)$ in
$\MH\times \MH$.  In particular, it must be holomorphic in complex
structure $I$ and Lagrangian\foot{Above the middle dimension,
there are $A$-branes that are not Lagrangian \kapor. But in the
good region, ${\EUQ}$ is middle-dimensional and must be
Lagrangian.} with respect to complex structures $J$ and $K$ or
equivalently with respect to the holomorphic symplectic form
$\Omega_I=\omega_J+i\omega_K$ defined in eqn. \ninto.  More
exactly, if $\pi_i:{\EUQ}\to\MH\times\MH$, $i=1,2$, are the two
projections, then ${\EUQ}$ must be Lagrangian with respect to the
holomorphic symplectic form
$\Omega=\pi_1^*(\Omega_I)-\pi_2^*(\Omega_I)$ on $\MH\times \MH$.
The minus sign here ensures that the diagonal in $\MH\times \MH$,
which is the trivial Hecke correspondence with no 't Hooft
operators, is Lagrangian.

Letting $C_+$ and $C_-$ denote the two ends of $W=I\times C$, we
can write the restriction of $\Omega$ to ${\EUQ}$ as
\eqn\symplevolume{{1\over\pi}\left(\int_{C_+}
-\int_{C_-}\right)|d^2 z|\,\Tr\left(\delta\phi_z\wedge \delta
A_\bz\right)={1\over \pi}\int_Idy\int_C|d^2z|\,
\partial_y \Tr\left(\delta\phi_z\wedge\delta A_\bz\right).}
To compute the restriction of this form to $\EUQ$, we can assume
that the variations $\delta\phi_z$ and $\delta A_\bz$ satisfy the
linearization of \BPSextsimple . The linearizations of the first
three equations in \BPSextsimple\ read
\eqn\BPSlin{\eqalign{D_\bz\delta\phi_z&=-[\delta
A_\bz,\phi_z],\cr
\partial_y\delta A_\bz&=-iD_\bz\delta\phi_0 - i [\delta
A_\bz,\phi_0]\cr
\partial_y\delta\phi_z&=-i[\delta\phi_z,\phi_0]-i[\phi_z,\delta\phi_0]
.\cr}} Using these equations, the Jacobi identity, and integration
by parts, one can verify that \symplevolume\ vanishes, so that
${\EUQ}$ is complex-Lagrangian, as expected.

\newsec{$A$-Branes and ${\cal D}$-Modules}
\seclab\abranes

\nref\kligcbrane{A.~Kapustin and Y.~Li, ``Open String BRST
Cohomology for Generalized Complex Branes,''
arXiv:hep-th/0501071.}
\nref\wittenhalftw{E.~Witten,
``Two-Dimensional Models With (0,2) Supersymmetry: Perturbative
Aspects,'' arXiv:hep-th/0504078.}
 \nref\kaphalftw{A.~Kapustin,
``Chiral de Rham Complex and the Half-Twisted Sigma-Model,''
  arXiv:hep-th/0504074.}
  \nref\nek{N. Nekrasov, ``Lectures On Curved Beta-Gamma Systems,
  Pure Spinors, And Anomalies,'' arXiv:hep-th/0511008.}

  \lref\laumon{G.\ Laumon, ``Transformation de
  Fourier g\'{e}n\'{e}ralis\'{e}e,'' arXiv:alg-geom/9603004.}
  \lref\rothstein{M.~Rothstein, ``Sheaves with Connections on Abelian Varieties,''
  Duke\ Math.\ J.\ {\bf 84}, 565 (1996), arXiv:alg-geom/9602023. }

  \lref\hitchinflat{N.~J.~Hitchin, ``Flat Connections And Geometric Quantization,''
  Commun.\ Math.\ Phys.\  {\bf 131} (1990) 347-380.}

  \lref\jackiwrebbi{R.~Jackiw and C.~Rebbi, ``Solitons With Fermion Number 1/2,''
  Phys.\ Rev.\ D {\bf 13} (1976)  3398-3409. }

A crucial ingredient is missing from what we have presented so
far. According to the geometric Langlands program, there is a
natural correspondence between homomorphisms of $\pi_1(C)$ into
$^L\negthinspace G$ and Hecke eigensheaves on ${\EUM}(G,C)$ which
are also ${\cal D}$-modules.  A ${\cal D}$-module is a module for
the sheaf of differential operators; the concept is further
elucidated below.

Concerning the left hand side of the correspondence, we learned in
section \electricbranes\ that homomorphisms of $\pi_1(C)$ to
$^L\negthinspace G$ correspond to electric eigenbranes.  Their
duals, which are branes of type ${\CMF}$ -- that is, branes
wrapped on a fiber of the Hitchin fibration, with a flat
Chan-Paton bundle -- must therefore be magnetic eigenbranes.
This statement follows from $S$-duality and can also be verified
directly \witfur, using  the abelianization that is provided by
the spectral cover construction.

However, the magnetic eigenbranes that we get this way are branes
of type $(A,B,A)$ on $\MH(G,C)$, while in the conventional
statement of the geometric Langlands correspondence, the right
hand side is supposed to involve instead a ${\cal D}$-module on,
roughly speaking, $\M(G,C)$, the moduli space of $G$-bundles.

This section is devoted to supplying the missing link. We will
show that every $A$-brane in complex structure $K$ on $\MH(G,C)$
automatically gives rise to, roughly speaking, a ${\cal D}$-module
on $\M(G,C)$.

One can regard ${\cal D}$-modules on $\M(G,C)$ as $B$-branes on a
noncommutative deformation of the cotangent bundle of $\M(G,C)$.
It has been argued previously \okap\ that for a certain class of
holomorphic symplectic manifolds, the category of $A$-branes is
equivalent to the category of $B$-branes on the noncommutative
deformation of the same manifold. The connection between
$A$-branes on $\M_H(G,C)$ and $\cal D$-modules on $\M(G,C)$ can be
viewed as a special case of this.

Actually, in the geometric Langlands program, it is not sufficient
(nor is it possible, because of the way the Hecke operators act)
to work only on $\M(G,C)$, the moduli space of stable bundles. One
instead constructs a ${\cal D}$-module on ${\rm Bun}_G(C)$, the
``stack'' of all $G$-bundles on $C$.  The gauge theory analog of
the stack of all $G$-bundles is simply the space of all
connections. The analog in our construction of the usual statement
that the ${\cal D}$-modules are defined on ${\rm Bun}_G(C)$ is
that the branes of interest can be defined in the underlying gauge
theory, and not only in the sigma-model with target space
$\MH(G,C)$. Showing this for the important branes is in fact the
goal of section \zapendix.

\subsec{The Canonical Coisotropic $A$-Brane}

Our starting point is the observation made in \kapor\ that for a
general symplectic manifold $X$ there may exist $A$-branes which
are supported not on a Lagrangian submanifold but on a more
general coisotropic submanifold. A coisotropic submanifold $Y$ of
a symplectic manifold $X$ is a submanifold  defined locally by
first-class constraints, in other words it is defined locally by
the vanishing of functions $f_i$, $i=1,\dots,d$ that are
Poisson-commuting.  The maximal number of independent
Poisson-commuting functions on $X$ is one half the dimension of
$X$, so the dimension of $Y$ is at least half the dimension of
$X$.  If the dimension of $Y$ is precisely half the dimension of
$X$, then $Y$ is Lagrangian, and the $A$-branes supported on $Y$
are the usual Lagrangian $A$-branes.  Coisotropic $A$-branes with
support above the middle dimension also exist, in general, though
the conditions for their existence are rather special.

We will need only  the case that $Y=X$ and the $A$-brane has rank
one. Thus its Chan-Paton bundle is a $U(1)$ bundle ${\cal L}\to
X$. The Chan-Paton gauge field takes values in the Lie algebra of
$U(1)$, which is a one-dimensional real vector space. So this
gauge field is locally an ordinary real one-form $A$ with
covariant derivative $D=d+A$; its curvature is defined as $F=dA$.
In a unitary representation of $U(1)$, the element $1$ of the Lie
algebra $\frak{u}(1)$ acts by the $1\times 1$ anti-hermitian
matrix $-in$ for some integer $n$, and parallel transport is
$\exp(in\int A)$.  We interpret the Chan-Paton bundle over $X$ as a
line bundle associated to the representation $n=1$.  For a
sigma-model map $\Phi:\Sigma\to X$, the factor in the worldsheet
path integral associated with the Chan-Paton bundle is the
holonomy of $\Phi^*(A)$ around $\partial\Sigma$ in the
representation with $n=1$, or
$\exp(i\oint_{\partial\Sigma}\Phi^*(A))$.

In order for a brane with $Y=X$ and Chan-Paton line bundle ${\cal
L}$ to be an $A$-brane, its curvature $F$ must satisfy
\eqn\coisobrane{ (\omega^{-1}F)^2=-1.} That is, the $(1,1)$ tensor
$\omega^{-1}F$ is an almost complex structure. The forms $F$ and
$\omega$ are closed, and together with \coisobrane\ this implies
that the almost complex structure $\hat N=\omega^{-1}F$ is
integrable \kapor. It is easy to see that both $\omega$ and $F$
are of type $(2,0)\oplus(0,2)$ with respect to $\hat N$.
Furthermore, the complex form $\omega-iF$ is of type $(2,0)$ and
so is a holomorphic symplectic form on $X$. Thus a coisotropic
$A$-brane of this kind may exist only if $\omega$ is the real part
of a holomorphic symplectic form.  See also \gualtieri\ for a
construction of branes of this type in generalized complex
geometry.

We are interested in the $A$-model on $\MH(G,C)$ with the
symplectic structure $\omega=(\Im\tau)\, \omega_K$. Then there is
an obvious solution to eqn. \coisobrane : \eqn\yetro{ F=
(\Im\tau)\,\omega_J,\quad \hat N=\omega_K^{-1}\omega_J= I. }
(There is also a second solution with the sign of $F$ and $\hat N$
reversed.) The two-form $\omega_J$ is exact, as we explained in
section \mhhyper, so $F$ is the curvature two-form of a connection
on a trivial line bundle ${\cal N}\to\MH(G,C)$. The connection
form is \eqn\exactcc{ \alpha={\Im\tau\over {2\pi}}\int_C |d^2z|\,
\Tr \left(\phi_z\delta A_\bz +\phi_\bz \delta A_z\right)} and
$F=d\alpha$. Although a more general two-form \eqn\lombo{
(\Im\tau)\left(a\omega_J+b\omega_I\right), \quad a^2+b^2=1, }
solves eqn. \coisobrane, it is not exact for $b\neq 0$, and for
general values of $\tau$ does not have periods which are integral
multiples of $2\pi$. So in general, this form is not the curvature
two-form of any connection, and cannot be used to construct an
$A$-brane.

We will call the $A$-brane on $\MH$ with $F=(\Im\tau)\,\omega_J$
the canonical coisotropic brane, or c.c. brane, for short. We also
sometimes denote it as ${\cal B}_{c.c.}$. If $G$ is
simply-connected, then so is $\MH(G,C)$, and the c.c. brane is
unique.  Otherwise, it is unique up to twisting its Chan-Paton
bundle by a flat line bundle.

Our first observation about the c.c. brane is that it is an
$(A,B,A)$-brane, i.e. it is an $A$-brane in complex structures $I$
and $K$ and a $B$-brane in complex structure $J$. To check that it
is an $A$-brane in complex structure $I$, we only need the
identity \eqn\hombo{ (\omega_I^{-1}\omega_J)^2=-1.} This means
that the c.c. brane obeys the conditions of a coisotropic brane in
complex structure $I$, just as it does in complex structure $K$.
To check that it is a $B$-brane in complex structure $J$, it is
sufficient to note that the curvature of the gauge field on the
brane is a multiple of  $\omega_J$, which has type $(1,1)$ in
complex structure $J$.

These formulas really only define the c.c. brane on the smooth
part of $\MH$.  However, the c.c. brane can be defined by a
boundary condition in the underlying gauge theory; see section
\xgoto\ for details.  This statement is the quantum field theory
equivalent of saying that the construction works on the whole
``stack'' of $G$-bundles over $C$.

Next, we would like to understand topological open strings with
both endpoints on the c.c. brane. They are represented by
BRST-invariant boundary operator insertions.  In general, as we
sketched in \dunamp\ of section \comptwo, for any brane ${\cal
B}$, the $({\cal B},{\cal B})$ strings form an associative, but
not necessarily commutative, algebra. In mathematical terms, this
is the endomorphism algebra of the brane ${\cal B}$ regarded as an
object of the category of branes on $\MH(G,C)$. We would like to
compute this endomorphism algebra in the case of the c.c. brane.

As a first step, we will argue that worldsheet instantons do not
contribute to the computation. The boundary condition defining
coisotropic branes in the sigma-model \kapor\ says that strings
ending on a space-filling coisotropic brane such as we consider
here must obey \eqn\melbo{ G_{IJ}\partial_1\Phi^J=-i
F_{IJ}\partial_0\Phi^J,} where $\partial_1$ and $\partial_0$ are
normal and tangential derivatives at the worldsheet boundary.
Imposing also the instanton equation (that is, the map $\Phi$
should be holomorphic in complex structure $K$), to relate
$\partial_1\Phi$ to $\partial_0 \Phi$, we get \eqn\zimbolo{
(\omega-iF)\partial_0\Phi=0.} Since both $\omega$ and $F$ are
nondegenerate, the instanton must be constant on the boundary. But
then, by analyticity, it must be constant everywhere, hence
trivial.

Since there are no instanton corrections, anything interesting
will have to come from worldsheet perturbation theory.  But
worldsheet perturbation theory, to any order in $\hbar$, generates
corrections to sigma-model operators and couplings that are local
in the target space. Hence, at least to all orders in world-sheet
perturbation theory, it is very natural to consider boundary
observables that may be defined only locally in target space. Such
observables form a sheaf over the target space, so instead of an
algebra of observables, we can get, to all orders in perturbation
theory, something more powerful: a sheaf of algebras. (In the
closed-string case, such localization in target space has been
discussed in \refs{\wittenhalftw -\nek}, interpreting earlier
mathematical work \ref\malikov{F. Malikov, V. Schechtman, and A.
Vaintrob, ``Chiral de Rham Complex,'' Commun. Math. Phys. {\bf
204} (1999) 439-473, math.AG/9803041.}.) Along with analyzing this
sheaf in perturbation theory, we will discuss what happens in the
exact theory.

The  space of BRST-invariant boundary operators for an arbitrary
coisotropic $A$-brane $Y$ has been determined in the classical
limit in \refs{\kapor, \kligcbrane}. For our purposes, it will be
sufficient to consider boundary operators of ghost number zero.
Then, in the case $Y=X$, such observables are simply holomorphic
functions on $X$, where $X$ is regarded as a complex manifold with
complex structure $\hat N$. Specializing to this case and
localizing in target space, we conclude that additively (ignoring
the ring structure) the sheaf of topological boundary operators
for the c.c. brane is the sheaf of functions on $\MH(G,C)$ which
are holomorphic in complex structure $I$. Note that while we are
working in the $A$-model in complex structure $K$, the complex
structure which is relevant here is $\hat N=I$. Locally the sheaf
of boundary operators has no higher cohomology and cannot be
deformed, so it cannot have quantum corrections. But globally it
has quantum corrections, which we will analyze.

\nref\seiwit{N. Seiberg and E. Witten, ``String Theory And
Noncommutative Geometry,'' JHEP 9909:032,1999 arXiv:hep-th/9908142.}%
The sheaf of holomorphic functions on $\MH$ has an obvious
commutative algebra structure, which is the correct one in the
classical limit. In the context of ``physical'' string theory, in
the presence of the gauge field strength $F$, the algebra of open
strings becomes noncommutative \seiwit.

Let us first discuss informally how
this would work in the present context.  The usual noncommutative parameter, called $\theta$ in
\seiwit, is the antisymmetric
part of $(g+F)^{-1}$.  In the present context, this becomes a multiple of $\omega_J^{-1}$.  $\omega_J$
is the real part of $\Omega_I$, the holomorphic $(2,0)$-form in complex structure $I$.  When we multiply
observables derived from functions
that are holomorphic in complex structure $I$ (which are the topological
observables), $\omega_J^{-1}$ can be replaced by $\Omega_I^{-1}$.  So the multiplication of such
observables  is deformed via the Poisson brackets derived from
$\Omega_I$, as we will find shortly.

In physical string theory, for any $F\not=0$, the multiplication of open strings contains noncommutative
phases coming from $(g+F)^{-1}$, but generally there are many other stringy effects.  A precise
description in terms of a noncommutative deformation of the algebra of functions on the target space
usually arises \seiwit\ only when $|F|>>|g|$.
We are not in that limit in the present problem; rather, we have $|F|=|g|$.  However,
once we restrict to topological observables, we will get a simplification
similar to what usually occurs in physical string theory for $|F|>>|g|$.
In the context of the $B$-type topological
strings, this was first discussed in \kap.  In essence, the elimination of stringy excitations that
in physical string theory occurs only if $|F|>>|g|$ arises here because of restricting to topological
observables.

Now we turn to a more precise analysis for the topological theory.
The action for the
$A$-model on a disk can be written in the form \eqn\mortop{
S=\int_\Sigma \Phi^*(\omega-iF)+\int_\Sigma d^2\sigma \{Q,V\}, }
where $Q$ is the BRST operator or topological supercharge. In our
case, the form $\omega-iF=-i \Omega_I \Im\tau$ is exact (and of
course this is always true locally). If we introduce the potential
$\varpi$ for $\Omega_I$, so that $\Omega_I=d\varpi$, then the
action takes the form \eqn\ortop{
S=-i\,\Im\tau\int_{\partial\Sigma}\Phi^*(\varpi) +\int_\Sigma
d^2\sigma \{Q,V\}.} Further,  $\varpi$ is proportional to the
canonical holomorphic one-form on $\M_H(G,C)$: \eqn\honop{
\varpi={1\over \pi}\int |d^2 z|\, \Tr \left(\phi_z\delta
A_\bz\right).}  If we restrict to $T^*\M(G,C)$, we can be more
explicit.  If $q^\alpha$ are local holomorphic coordinates on
$\M(G,C)$ and $p_\alpha$ are linear functions on the fibers of
$T^*\M(G,C)$ that are canonically conjugate to the $\phi_\alpha$,
then $\varpi=\sum_\alpha \,p_\alpha\,dq^\alpha$ and $S=-i\,{\rm
Im}\,\tau\int p_\alpha\,dq^\alpha+\{Q,\dots\}$.  Thus, up to
$Q$-exact terms, the $A$-model action is the holomorphic version
of the quantum mechanical action for a particle on $\cM(G,C)$
(with zero Hamiltonian).

We undoubtedly lose some important structure by specializing to
$T^*\M\subset \MH$, since the c.c. brane is defined on all of
$\MH$, and even in the four-dimensional gauge theory, not just on
$T^*\M$.  But let us see what structure remains after we make this
specialization.

The form of the action in \ortop\ is standard, and the result is
familiar: to first order in ${1/ {\rm Im}\,\tau}$, the effect of
the perturbation is to deform the commutative algebra of
locally-defined holomorphic functions by the Poisson brackets
derived from this action.  So in perturbation theory in ${1/{\rm
Im}\,\tau}$, the c.c. strings (or more fastidiously, the $({\cal
B}_{c.c.},{\cal B}_{c.c.})$ strings) form a sheaf of
noncommutative algebras. But precisely what sheaf do we get?

A powerful tool is the $\CC^*$ group ${\cal U}$ that was
introduced at the end of section \complexstr\ and used at various
points in this paper. It acts by \eqn\utyp{ \phi_z\to \lambda
\phi_z} and the associated grading on the space of observables was
used in section \sdualhitfib\ to analyze the action of $S$-duality
on zerobranes. $\varpi$ is of degree 1 under this scaling;
therefore,  ${\cal U}$ can be promoted to a symmetry of the
quantum theory if we assign degree $1$ to ${1/{\rm Im}\,\tau}$,
that is to Planck's constant $\hbar$ in the analog
quantum-mechanical problem.

At the classical level, the multiplication of observables
preserves the grading by ${\cal U}$; the product of observables of
degrees $m$ and $n$ is an observable of degree $m+n$. At the
quantum level, there can also be terms of lower degree in the
product, if they are accompanied by positive powers of  Planck's
constant. Thus the algebra of quantum observables is only
filtered, not graded.

In order to effectively use the $\Bbb{C}^*$ symmetry to analyze
the algebra, we need to consider observables that are defined in a
$\Bbb{C}^*$-invariant open set.  For this, we pick an open set
${{\cal V}}\subset {\EUM}$ and consider observables that are
defined on $T^*{{\cal V}}\subset \MH$.  Thus, we are only going to
partly sheafify the space of strings ending on the c.c. brane,
allowing only open sets in $\MH$ of this form. The advantage of
this partial sheafification is that it works nonperturbatively, as
we will see shortly.  A complete sheafification in which one
defines an associative algebra of observables for an arbitrary
open set ${\cal W}\subset \MH$ works fine to all orders in $1/{\rm
Im}\,\tau$, but it is not realistic to expect it to work in the
exact theory. For a general discussion of such matters, see
\ref\konts{M. Kontsevich, ``Deformation Quantization Of Algebraic
Varieties,'' Lett. Math. Phys. {\bf 56} (2001) 271-294,
math.AG/0106006.}.

To understand why in general there is an obstruction to
sheafification of deformation quantization beyond perturbation
theory, consider the simple example of deformation quantization of
$\Bbb{C}^{2}$ with coordinates $x,y$ and symplectic form $dx\wedge
dy$. Deformation quantization in this case can be carried out
exactly using the Moyal-Wigner product\foot{For reprints of their
papers and many other original papers, along with a survey, see
\ref\zfc{C. K. Zachos, D. B. Fairlie, and T. L. Curtwright, {\it
Quantum Mechanics In Phase Space: An Overview With Selected
Papers} (World-Scientific, 2005).}.} if one restricts oneself to
polynomial functions on $\Bbb{C}^{2}$. To sheafify this situation,
one picks an open set ${\cal W}\subset\Bbb{C}^{2}$, and tries to
define a Moyal-Wigner product of functions holomorphic in ${\cal
W}$. There is no problem to any finite order in $\hbar$, but the
fact that the Moyal-Wigner product involves an infinite sum causes
a problem in getting an exact result.  To see the problem, try to
evaluate at the origin the Moyal-Wigner product of the functions
$1/(x-a)$ and $1/(y-b)$, each of which are holomorphic in suitable
open sets in $\Bbb{C}^2$.

So we limit ourselves to partial sheafification in which we
consider an open set in $\MH$ of the form $T^*{{\cal V}}$, ${{\cal
V}}\subset\M$.  Moreover, we restrict ourselves to functions on
$T^*{{\cal V}}$ that have polynomial growth along the fibers of
the projection $T^*{{\cal V}}\to{{\cal V}}$.  The algebra of such
functions is generated by elements of degree less than or equal to
$1$, that is, by functions that are at most linear along the
fibers.

Elements of degree $0$ are simply holomorphic functions on ${{\cal
V}}$.  There is no possibility to deform the algebra of
multiplication of such functions, because there are no observables
in $T^*{{\cal V}}$ of negative degree.  As $\hbar$ has degree 1, a
correction to the multiplication of functions of degree 0 would
have to involve an observable of negative degree.

Elements of degree $1$ are linear functions on the fibers of the
cotangent bundle.  So if $q^\alpha$ are local coordinates on
${{\cal V}}$ and $p_\alpha$ are the conjugate linear functions on
the fibers, then an element of degree 1 takes the form
$\kappa=\sum_\alpha p_\alpha\, V^\alpha(q)$ and is in natural
correspondence with a holomorphic vector field $\sum_\alpha
V^\alpha{\partial\over\partial q^\alpha}$ on $\cM$. In multiplying
an object $f$ of degree $0$ and an object $\kappa$ of degree 1,
there is only ``room'' for a first order deformation -- since
terms of order $\hbar^2$ in the product would again multiply
operators of negative degree. The first order deformation is
characterized by the Poisson bracket \eqn\hembo{ [\kappa,
f]=-i(\Im\tau)^{-1}\kappa(f).}

To proceed, we write $\hat w$ for the open string state
corresponding to a holomorphic function $w$ on $T^*{{\cal V}}$.
The formula \hembo\ determines the $\star$ product (that is, the
algebra of open string multiplication) of the ring elements $\hat
f$, $\hat\kappa$ corresponding to $f$ and $\kappa$, but not quite
uniquely, since (while preserving the way the algebra is filtered)
we could add to $\hat\kappa$ an observable of degree zero. If for
$\kappa=V^\alpha(q) p_\alpha$, we define $\hat\kappa =\hat
V^\alpha\star \hat p_\alpha$, then the algebra is uniquely
determined and is \eqn\yembo{\eqalign{\hat f\star\hat\kappa&
=\hat{f\kappa}\cr \hat\kappa\star \hat f&
 =\hat{f\kappa}-i(\Im\tau)^{-1}\hat{\kappa(f)}.\cr}}
 Of course, $\hat{ f\kappa}$ and $\hat{\kappa(f)}$ are the ring
 elements that correspond to the functions
 $f\kappa$ and $\kappa(f)=V^\alpha\,\partial f/\partial
q^\alpha$. We still have not completely used the freedom to
redefine $\hat\kappa$; the remaining freedom is that for the
linear functions $p_\alpha$ on the cotangent bundle, we could
replace \eqn\remob{\hat p_\alpha\to \hat p_\alpha+\hat A_\alpha}
for some functions $A_\alpha(q)$.

When we multiply two functions $\kappa$ and $\kappa'$ that are
each of degree 1, something new happens. There is a first order
deformation, which is given again by the Poisson bracket, but
there now is also room for a second order deformation.  The
Poisson bracket of $p_\alpha $ and $p_\beta$ vanishes, but the
commutator $[\hat p_\alpha,\hat p_\beta]$ in the $\star$ algebra
might be $\hbar^2 F_{\alpha\beta}(q)$ for some functions
$F_{\alpha\beta}$. One can show using associativity of the algebra
that the two-form $F_{\alpha\beta}dq^\alpha\,dq^\beta$ is closed,
and hence can be set to zero (for a small open set ${{\cal V}}$)
by the transformation \remob.

The algebra associated with $T^*{{\cal V}}$ for a small open set
${\cal V}\subset \MH$ is therefore isomorphic simply to the
algebra of holomorphic differential operators on ${{\cal V}}$,
with functions on $\M$ acting by multiplication and
\eqn\mebolo{\hat p_\alpha=-{i\over {\rm
Im}\,\tau}{\partial\over\partial q^\alpha}.}

Now suppose that we cover $\M$ with small open sets ${{\cal
V}}^{(i)}$.  In overlap regions ${{\cal V}}^{(i)}\cap {{\cal
V}}^{(j)}$, the degree zero ring elements $\hat f$ must agree, but
the degree 1 elements $\hat p_\alpha^{(i)}$ and $\hat
p_\alpha^{(j)}$ may differ by a transformation \remob.  Since we
require $[\hat p_\alpha,\hat p_\beta]=0$ in each open set, they
must be related by \eqn\hecox{\hat p_\alpha^{(i)}=\hat
p_\alpha^{(j)}+{1\over {\rm Im}\,\tau}\hat{\partial s^{(ij)}\over
\partial q^\alpha}} for some local holomorphic functions $s^{(ij)}$ defined on
${{\cal V}}^{(i)}\cap {{\cal V}}^{(j)}$.   Consistency of this
relation implies that in triple overlaps the quantities
$c^{(ijk)}=s^{(ij)}+s^{(jk)}+s^{(ki)}$ are complex constants.
Moreover, they automatically obey a cocycle condition and define
an element of $H^2({\EUM},\Bbb{C})$.

If those constants take values in $2\pi\Bbb{Z}$, then the
quantities $\exp(-is^{(ij)})$ are the transition functions of a
complex line bundle ${\cal L}\to \MH$.  In this case, the sheaf of
algebras of open strings is the sheaf of differential operators
acting on sections of the line bundle ${\cal L}$.  We call this
sheaf of algebras ${\cal D}_{\cal L}$.

${\cal L}$ is not quite uniquely determined.  In general, the
sheaf of algebras ${\cal D}_{\cal L}$ is invariant under twisting
${\cal L}$ by a flat line bundle, since a flat line bundle has
constant transition functions that commute with the momenta $\hat
p_\alpha$.

There is no reason that the constants $c^{(ijk)}$ must be properly
quantized, that is, must take values in $2\pi\Bbb{Z}$.  For
example, beginning with a properly quantized case leading to the
sheaf ${\cal D}_{\cal L}$ of differential operators on a line
bundle ${\cal L}$, we can define a new sheaf ${\cal D}_{\cal
L^\gamma}$ by multiplying the $s^{(ij)}$ by a complex number
$\gamma$. If ${\cal L}$ is topologically nontrivial, a line bundle
${\cal L}^\gamma$ does not exist (except for integer or perhaps
rational values of $\gamma$) but the corresponding sheaf of
algebras ${\cal D}_{{\cal L}^\gamma}$ does exist.

Since $H^2({\EUM}(G,C),\Bbb{C})$ is one-dimensional, in this
specific example the most general sheaf of algebras that could
arise from this sort of construction is of the form ${\cal
D}_{{{\frak L}}^\gamma}$ where $\gamma$ is a complex number and
${{\frak L}}$ is the fundamental line bundle over $\EUM$, which we
loosely call the determinant line bundle. If ${\EUM}$ is replaced
by a more general space $X$, a slight generalization is possible
involving a product of complex powers of line bundles.

At any rate, as we will see, the basic case of the geometric
Langlands program involves the differential operators on a certain
ordinary line bundle.  But a generalization that we consider in
section \genotics\ does involve the complex powers of a line
bundle.

Notice that in the above analysis, there is no question of whether
perturbation theory converges, since everything is determined by
what happens in second order.  Had we tried to completely
sheafify, associating an algebra with every open set in $\MH$, we
would have had trouble with convergence of perturbation theory.

\bigskip\noindent{\it Time Reversal Symmetry}

These somewhat abstract considerations do not quite determine the
sheaf of algebras of c.c. strings. It is the sheaf of differential
operators on some line bundle (or complex power of a line bundle),
but which one? To determine it, we will use time-reversal
symmetry.  It is natural to get an extra constraint from
time-reversal symmetry, because as we see in eqn. \zongor, this is
a symmetry of the twisted topological field theory precisely at
$t=1$, the value at which we are working in the present
discussion.

In gauge theory, time-reversal, which we  call ${\EUT}$, reverses
the time coordinate of the four-manifold $M$ while acting
trivially on the space coordinates.  If $M=\Bbb{R}\times I\times
C$, then ${\EUT}$ just acts as $-1$ on $\Bbb{R}$. In the effective
two-dimensional theory, ${\EUT}$ is an orientation-reversing
symmetry of the string worldsheet $\Sigma$, which maps the
boundary of $\Sigma$ to itself while reversing the orientation of
the boundary.  Since our considerations are local, we can assume
that $\Sigma$ or $M$ has such a symmetry.

Under ${\EUT}$, the bosons $(A,\phi)$ transform to
$({\EUT}^*A,-{\EUT}^*\phi)$, as we saw in eqn. \yoro.  The terms
in \mortop\ and \ortop\ that involve $\omega$, $F$, and $\varpi$
would be odd under reversal of worldsheet orientation if ${\EUT}$
were to map $\phi$ to ${\EUT}^*\phi$. But with the extra minus
sign, these terms are invariant.

Time-reversal reverses the order of operator insertions on the
worldsheet boundary, so it maps the algebra of c.c. strings to its
``opposite'' algebra, in which the elements are the same but
multiplication goes in the opposite order. Thus, to every element
$x$ of an algebra ${\eurm C}$, the opposite algebra ${\eurm
C}^{op}$ has an element $x^{op}$, with $(x\star
y)^{op}=y^{op}\star^{op} x^{op}$.  Here $\star $ is the
multiplication in ${\eurm C}$ and $\star^{op}$ is the
multiplication in ${\eurm C}^{op}$.

  \lref\beibern{A. Beilinson and J. Bernstein, ``A Proof of Jantzen's Conjectures,''
  I. M. Gelfand Seminar, 1-50,
  Adv. Sov. Math. 16, Part 1, AMS, 1993.}
The opposite of the algebra of differential operators on a line
bundle $\cL$ over a complex  manifold $X$ is \beibern\ isomorphic
to the algebra of differential operators on $\cL^{-1}\otimes K_X$,
where $K_X$ is the canonical line bundle of $X$ (that is, the
bundle of holomorphic forms of top degree). To see this, let $s$
and $u$ be compactly supported (not holomorphic) sections,
respectively, of $\cL$ and $\cL^{-1}\otimes K_X$, and let ${\eurm
F}$ be a holomorphic differential operator acting on $\cL$.  Then
$s\, {{\eurm F}}u$ is a section of $K_X$. A section of $K_X$ is a
differential form of middle dimension, and can be integrated over
any real slice $Z$ of $X$. Moreover, by integrating by parts, we
can form a ``transpose'' operator ${{\eurm F}}^t$, acting on
sections of $\cL^{-1}\otimes K_X$, and obeying $\int_Z u\,{{\eurm
F}} s=\int_Z ({\eurm F}^t u)s$ for all $u$ and $s$. ${{\eurm
F}}^t$ is a differential operator with holomorphic coefficients,
just like ${{\eurm F}}$, and does not depend on the choice of $Z$.
The map ${{\eurm F}}\to {{\eurm F}}^t$ reverses the order of
multiplication, showing that the opposite sheaf of algebras to
${\cal D}_{\cal L}$ is ${\cal D}_{{\cal L}^{-1}\otimes K_X}$.

Time-reversal symmetry of the c.c. brane means that the sheaf of
algebras obtained from open strings ending on the c.c. brane must
be isomorphic to its opposite algebra.  This, plus the fact that
it is the sheaf of differential operators acting on some line
bundle, implies that this algebra must be ${\cal
D}_{K_{\EUM}^{1/2}}$, the sheaf of differential operators acting
on $K_{\EUM}^{1/2}$, the square root of the canonical bundle of
$\M$. $\M$ actually is a spin manifold, so $K_\M^{1/2}$ is an
ordinary line bundle.  The spin structure of $\M$ is not
necessarily unique, but since the sheaf of differential operators
acting on a line bundle is invariant under twisting by a flat line
bundle, the sheaf of algebras ${\cal D}_{K_\EUM^{1/2}}$ does not
depend on the choice of $K_{\EUM}^{1/2}$.

\bigskip\noindent {\it Global Algebra of Topological Strings}

Now we want to discuss the {\it global} c.c. strings.

How do we get the algebra of global observables from the sheaf of
observables?  We simply take the global sections of the sheaf of
observables, or more generally, the cohomology of $\M$ with values
in this sheaf. To keep things simple, we will just consider global
sections, that is observables of ghost number or cohomological
degree zero (but any scaling degree under the $\Bbb{C}^*$ symmetry
${\cal U}$).

Classically, the sheaf of observables is just the sheaf of
holomorphic functions on $\MH$, and its global sections are simply
the global holomorphic functions on $\MH$.  They are the commuting
Hamiltonians of the integrable system, which we described in
section \hitchfib.

Likewise quantum mechanically, to get the global observables, we
must take the global sections of the sheaf of local observables.
But there is a quantum correction to the gluing law that defines
this sheaf, relative to what we have classically.  The quantum
correction entered our analysis explicitly in eqn. \hecox. So
there is potentially a quantum correction to the description of
the global observables.

Having understood that the sheaf of c.c. strings is the sheaf of
differential operators on $\MH$, we can understand concretely what
this correction means.  Suppose that we are given a holomorphic
function on $\MH$, for example a function of degree three that
once we pick local complex coordinates on $\M$ looks like
$w=\sum_{\alpha,\beta,\gamma} f^{\alpha\beta\gamma}(q)p_\alpha
p_\beta p_\gamma$.  To promote it to a global $({\cal
B}_{c.c.},{\cal B}_{c.c.})$ string, we need to find a third order
differential operator $\hat w$ on $\M$ whose ``symbol'' is $w$. In
other words, we want $\hat w=(-i/{\rm
Im}\,\tau)^3\,f^{\alpha\beta\gamma}\partial^3 /\partial
q^\alpha\partial q^\beta\partial q^\gamma+\dots$, where the
ellipses refer to a differential operator of order two or less.
Because of the absence of a canonical system of local complex
coordinates $q^\alpha$ on $\M$, such an operator is not unique
locally, and may not exist globally.

If a holomorphic differential operator $\hat w$ with symbol $w$
does exist, we say that the function $w$ can be ``quantized.'' The
commuting Hamiltonians on $\MH$ can all be quantized.  This was
first proved for functions of degree two by Hitchin \hitchinflat\
and for functions of any degree by Beilinson and Drinfeld \bd;
this result is part of the rationale for the title of their
paper.\foot{Their argument is based on two-dimensional conformal
field theory, more precisely on current algebra of the group $G$
at level $k=-h$, with $h$ the dual Coxeter number of $G$.} We will
give an argument that involves further use of the time-reversal
symmetry $\EUT$. This argument also shows that the resulting
differential operators commute, as first proved in the above-cited
references.

We use the fact that $({\cal B}_{c.c},{\cal B}_{c.c.})$ strings
correspond to boundary operators that can be inserted at a
boundary labeled by the c.c. brane.  Moreover, these boundary
observables can be derived from operators of the four-dimensional
gauge theory, which is a powerful tool because it has properties
of locality that are lost upon reduction to two dimensions.  This
will enable us to give a simple argument.  In fact, as we recall
from section \hitchfib, the commuting Hamiltonians take the form
$\int_C \alpha\,{\cal P}(\varphi)$, with ${\cal P}$ a homogeneous
gauge-invariant polynomial of some degree $k$, and $\alpha\in
H^1(C,K_C^{1-k})$.  To show that such an operator is
$Q$-invariant, it suffices to show that the local operator ${\cal
P}(\varphi)$ is $Q$-invariant.  We do this in section \xgoto,
simply by classifying the possible operators and using
time-reversal symmetry.  From this point of view, it is clear that
the differential operators obtained by quantizing two classical
expressions $\int_C \alpha \,{\cal P}(\varphi)$ and $\int_C\tilde
\alpha \,\tilde {\cal P}(\varphi)$ commute, since one can take
$\alpha$ and $\tilde\alpha$ to have disjoint support.

To illustrate the nontrivial nature of the fact that the classical
holomorphic functions on $\M$ can be quantized to get differential
operators on $K_\M^{1/2}$, let us note \refs{\hitchinflat,\bd}
that they cannot be quantized to get differential operators on any
line bundle other than $K_\cM^{1/2}$.   The fact that they can be
quantized to get {\it commuting} differential operators means that
Hitchin's classical integrable system on $\MH$ can be
``quantized'' to make a quantum integrable system on $\M$.

\subsec{${\cal D}$-modules Corresponding to $A$-Branes}
\subseclab\dmab

\lref\lazaroiu{C.~I.~Lazaroiu,
 ``On the Structure of Open-Closed Topological Field Theory in Two
 Dimensions,''
 Nucl.\ Phys.\ B {\bf 603}, 497-530 (2001), arXiv:hep-th/0010269. }
 \lref\mooresegal{G. Moore and G. Segal, ``$D$-Branes And $K$-Theory In 2D
 Topological Field Theory,'' hep-th/0609042;
see also lectures by G. Moore,  at
http://online.itp.ucsb.edu/online/mp01. }

The existence of the c.c. brane means that every $A$-brane in
complex structure $K$ is a module for the sheaf of differential
operators on $K_\M^{1/2}$.  In fact, in general, if ${\cal B}$ and
${\cal B}'$ are any two branes, then the $({\cal B},{\cal
B}')$-strings form a module for the algebra of $({\cal B},{\cal
B})$ strings.  This idea is illustrated in \dunamp\ of section
\comptwo, and is part of the usual axioms of open-closed
topological field theory \refs{\lazaroiu,\mooresegal}. To the
extent that sheafification is possible, the $({\cal B},{\cal B})$
strings form a sheaf of algebras, not just an  algebra, and the
$({\cal B},{\cal B}')$ strings form a sheaf of modules for this
sheaf of algebras. This statement just means that the ring and
module structures can be defined for open strings that are regular
only in a suitable open set in the target space.

We want to apply this construction for the case that ${\cal
B}={\cal B}_{c.c.}$ is the canonical coisotropic brane.  In this
case, we claim that for any brane ${\cal B}$ that is of $A$-type
in complex structure $K$, there are no instanton corrections to
$({\cal B}_{c.c},{\cal B}')$ strings.  The argument also applies
to higher order topological couplings (such as cubic Yukawa
couplings) involving the c.c. brane. The absence of instantons can
be argued in much the same way as for strings ending entirely on
the c.c. brane. The relevant disk instantons have a part of their
boundary on the c.c. brane, and on this part of the boundary the
instanton must be constant. But then, by analyticity, the
instanton must be constant everywhere.

We also know, from our previous investigation,  what sort of
sheafification is possible.  We can associate an algebra and a
module to an open set in $\MH$ of the form $T^*{{\cal V}}$,
${{\cal V}}\subset \M$, but not necessarily to more general open
sets. So an $A$-brane on $\MH(G,C)$ in complex structure $K$ gives
a sheaf of modules for the sheaf of algebras ${\cal D}_{K_{
\EUM}^{1/2}}$ over $\M$.

Now we will give a few examples of this. (For some examples worked
out in detail of open string quantization involving coisotropic
branes, see \ref\zasloww{M. Aldi and E. Zaslow, ``Coisotropic
Branes, Noncommutativity, and the Mirror Correspondence,''  JHEP
{\bf{0506}} (2005) 019, arXiv:hep-th/0501247.}.)

Our first example is an $A$-brane ${\cal B}'$ defined by the
condition $\varphi=0$, with trivial Chan-Paton bundle. This
submanifold is a copy of $\cM(G,C)$ and is Lagrangian in complex
structures $K$ and $J$ and complex in complex structure $I$. Thus
${\cal B}'$ is an example of a $(B,A,A)$-brane. (We used this
brane in section \opprod\ in analyzing the operator product
expansion of 't Hooft operators.) We take the flat Chan-Paton
connection on ${\cal B}'$ to be trivial, for simplicity. As
described in \exactcc, the c.c. brane has a Chan-Paton line bundle
${\cal N}$ that is topologically trivial, but endowed with a
non-trivial connection.

To compute the spectrum of open strings on the classical level,
one can reduce the supersymmetric sigma-model to supersymmetric
quantum mechanics, i.e. retain only the zero modes. In this
approximation, open-string states with $({\cal B}_{c.c.},{\cal
B}')$ boundary conditions are sections of the tensor product with
${\cal N}^{-1}$ of a vector bundle obtained by quantizing the
space of fermion zero modes. The connection on ${\cal N}$ is zero
when restricted to $\varphi=0$, so we can omit the factor of
${\cal N}^{-1}$. The space of fermion zero modes is the tangent
space to $\M\subset \MH$ (fermions associated with normal
directions to $\M$ obey opposite boundary conditions at the two
ends of a string and have no zero modes). When we quantize the
space of fermion zero modes, we get the spinors on $\M$. Viewing
$\M$ as a complex manifold, its spin bundle is the same as
$K_\M^{1/2}\otimes(\oplus_{i=0}^{{\rm dim}\,\M}\Omega^{0,j}(\M))$.
Here $\Omega^{0,j}$ is the sheaf of $(0,j)$-forms on $\M$.

The BRST operator or topological supersymmetry $Q$ is in this
situation simply the $\bar\partial$ operator acting on the
$(0,j)$-forms with values in $K_\M^{1/2}$.  The sheaf of physical
 $({\cal B}_{c.c.},{\cal B}')$ strings is therefore the sheaf of
 holomorphic
sections of\foot{The $\Omega^j$-forms with $j>0$ do not contribute
to the cohomology over a small open set ${{\cal V}}$, so they can
be omitted here.} $K_\M^{1/2}$. This is of course a sheaf of
modules for the sheaf of rings ${\cal D}_{K_{ \EUM}^{1/2}}$;
indeed, it is the sheaf of modules by which this sheaf of rings
was defined. This construction gives, possibly, a more direct
explanation of why the sheaf of rings derived from the c.c. brane
is precisely ${\cal D}_{K_{ \EUM}^{1/2}}$, rather than ${\cal
D}_{\cal L}$ for some other ${\cal L}$.

It may appear that we have implicitly used again the time-reversal
symmetry $\EUT$ in claiming that quantization of the space of
fermion zero modes gives precisely the spinors on $\M$, rather
than spinors with values in some line bundle ${\cal L}$. Actually,
we can give an alternative argument for this point, though it is
an argument that uses another discrete symmetry of the theory. The
branes ${\cal B}_{c.c.}$ and ${\cal B}'$ are physically sensible,
unitary branes in the four-dimensional supersymmetric gauge theory
defined on $M=\Bbb{R}\times I\times C$; here $\Bbb{R}$ is the time
direction, and $I$ is an interval with boundary conditions at the
two ends defined respectively by ${\cal B}_{c.c.}$ and ${\cal
B}'$.    To define the theory on $M$, there is some twisting to
preserve supersymmetry in the compactification on $C$, but no
twisting that involves the time direction.  So along with the
topological supersymmetry $Q$, the theory on $M$ is also invariant
under the adjoint supersymmetry $Q^\dagger$ (they obey
$\{Q,Q^\dagger\}=H$, where $H$ is the Hamiltonian).  The physical
theory has a  ``charge conjugation'' symmetry that exchanges $Q$
and $Q^\dagger$.  Invariance under this symmetry implies that the
Chan-Paton bundle obtained by quantizing the fermion zero modes is
precisely the spin bundle of $\CM$, not the tensor product of the
spin bundle with an additional line bundle.\foot{If $\pi_1(G)$ is
nontrivial, we can modify the construction by taking the
Chan-Paton bundle on the brane ${\cal B}'$ to be a flat line
bundle.  Then we get spinors on $\M$ with values in a flat line
bundle; as a special case of this, we get all the possible spin
structures on $\M$. Because of the relation of $D$-branes to
$K$-theory \witkth, the choice of spin structure of $\M$ is really
part of a careful description of the Chan-Paton bundle of the
brane ${\cal B}'$.} This type of argument was first made in
\jackiwrebbi\ in quantizing solitons.

Note that in this computation it was important that the support of
${\cal B}'$ is not only a Lagrangian submanifold with respect to
$\omega_K$, but is also a complex submanifold with respect to the
complex structure $\hat N=I$ determined by the c.c. brane. Otherwise,
the topological supercharge or BRST operator would not reduce to the $\bar\partial$ operator
in complex structure $I$.

Our second example is the case that is important for the geometric
Langlands program: a brane of type ${\CMF}$, that is a brane
${\cal B}_{\CMF}$ supported on a fiber ${\CMF}$ of the Hitchin
fibration with a flat Chan-Paton line bundle.  This is a brane of
type $(A,B,A)$, so in particular it is an $A$-brane in complex
structure $K$. Therefore, it gives rise to a sheaf of modules for
${\cal D}_{K_{ \EUM}^{1/2}}$. We will explain at the end of
section \genotics, by a further elementary argument, how to
convert this to a sheaf of modules for ${\cal D}$, the sheaf of
ordinary differential operators on $\M$. So the brane of type
${\CMF}$ has the two key properties: it is a magnetic eigenbrane
because it is the $S$-dual of a zerobrane; and it gives rise to a
sheaf of ${\cal D}$-modules over $\M$.  These are the basic claims
of the geometric Langlands program.

It is difficult to explicitly describe the sheaf of ${\cal D}$ or
${\cal D}_{K_{ \EUM}^{1/2}}$-modules that comes from a brane of
type $\CMF$. But we can do this in the the abelian case $G=U(1)$.
In this special case, the sigma-model is a free field theory, and
the boundary conditions are linear, so the computation of the
spectrum and module structure of the open strings is
straightforward.

For $G=U(1)$, we can think of $A$ and $\phi$ as real one-forms
($A$ is only defined up to gauge transformations, of course, while
$\phi$ is gauge-invariant).  The Hitchin equations for $A$ and
$\phi$ decouple, and in complex structure $I$, the Hitchin moduli
space is a product of the Jacobian ${\rm Jac}(C)$ of $C$ (which is
the moduli space of topologically trivial holomorphic line bundles
on $C$) and the vector space ${\EUBB}=H^0(C,\Omega^1)$. ${\rm
Jac}(C)$ is a complex torus of dimension $g_C$, the genus of $C$,
and the Hitchin moduli space ${\rm Jac}(C)\times {\EUBB}$ can be
identified with its cotangent bundle. The Hitchin fibration is the
projection to ${\EUBB}$.

Given $p\in {\EUBB}$ and the corresponding fiber ${\CMF}_p\simeq
{\rm Jac}(C)$ of the Hitchin fibration, we can compute the space
of $({\cal B}_{c.c.},{\cal B}_{{\CMF}_p})$ strings much as in the
previous example. One difference is that although the Chan-Paton
line bundle $\cL\to{\CMF}_p$ of the brane ${\cal B}_{{\CMF}_p}$ is
still topologically trivial, we now allow an arbitrary unitary
flat connection on it. Another difference is that the restriction
of the Chan-Paton line bundle ${\cal N}$ of the c.c. brane to
${\CMF}_p$ is holomorphically nontrivial, in general. On the other
hand, the canonical bundle $K_{{\CMF}_p}$ is trivial as ${\CMF}_p$
is a torus. Consequently, BRST-invariant open string states with
ghost number zero are holomorphic sections of $T_p=\cL\otimes
{\cal N}^{-1}\vert_{{\CMF}_p}$.

\def\bz{{\bar z}}
\def\Im{{\rm Im}}
\def\bpartial{{\bar\partial}}
But how does the  sheaf of  $({\cal B}_{c.c.},{\cal
B}_{c.c.})$ strings acts on sections of $T_p$?  We will answer this question
next. We work on $\RR\times I$ with the left
boundary on the brane ${\cal B}_{c.c.}$ and the right boundary on
${\cal B}_{{\CMF}_p}$.  The fiber $p$ of the Hitchin fibration
is defined by $\phi=\upsilon$, where $\upsilon=\upsilon_z\,dz+\upsilon_\bz\,d\bz$ is a real
harmonic one-form. The Chan-Paton
line bundle ${\cal L}\to{\CMF}_p$ of the brane ${\cal
B}_{{\CMF}_p}$ has a flat unitary connection that can be
conveniently represented as \eqn\zoncob{ \beta={1\over
{2\pi}}\int_C |d^2 z|\left(a_z \delta A_\bz+a_\bz \delta
A_z\right)} where $a=a_z\,dz+a_\bz\,d\bz$ is a real harmonic
one-form on $C$.   Similarly, the connection on the line bundle
${\cal N}$ on the c.c. brane is\foot{To get this formula and the next one,
we use eqn. \tomog\ for $\omega_J$ and $\omega_K$.  In eqn. \tomog, ${\rm Tr}$ represents
a trace in the $N$-dimensional representation of $U(N)$, and $A$ and $\phi$ are understood
to be skew-hermitian.  Since here we consider $A$ and $\phi$ as real one-forms,
we must include a minus sign.} \eqn\homifly{ \alpha=-{\Im\,\tau\over
2\pi}\int_C |d^2 z|\left(\phi_z \delta A_\bz+\phi_\bz \delta
A_z\right).} Finally, the symplectic form $\omega=({\Im}\,\tau)\omega_K$ is the
exterior derivative of a one-form \eqn\nomigly{ \zeta=-{i\,
\Im\,\tau\over 2\pi}\int_C |d^2 z|\left(\phi_\bz \delta A_z-\phi_z
\delta A_\bz\right).}
The action of the $A$-model, up to $Q$-exact
terms, is an integral over the boundary of $\Sigma$. The contribution to the action from
the right
boundary, which we call $\partial_R\Sigma$, is  \eqn\omitro{ \int_{\partial_R\Sigma}
\Phi^*(\zeta -i\beta)={i\,\Im\,\tau\over 2\pi}\int_{\partial_R\Sigma}
\left(\left(\upsilon_z-(\Im\,\tau)^{-1}a_z\right) \dot{A}_\bz
-\left(\upsilon_\bz+(\Im\,\tau)^{-1}a_\bz\right) \dot{A}_z\right)ds.}
Here we used that on the right boundary, $\phi=\upsilon$.
Also, $s$ is a ``time'' coordinate on $\Sigma=\Bbb{R}\times I$, and
$\dot A=\partial A/\partial s$.
 The left boundary $\partial_L\Sigma$, having the opposite orientation,
contributes
\eqn\bofloc{
-\int_{\partial_L\Sigma} \Phi^*(\zeta-i\alpha)=-{i\,\Im\,\tau\over \pi}\int_{\partial_L\Sigma}
 \phi_z
\dot{A}_\bz\,ds.}
The supersymmetric string states come from zero modes along the string, so to determine
them, the distinction
between $\partial_L\Sigma$ and $\partial_R\Sigma$ is unimportant and we can just add
the two contributions to the action.
The total action is accordingly
\eqn\nofloc{
{i\,\Im\,\tau\over \pi}\int \left(-\phi_z+{1\over
2}(\upsilon_z-(\Im\,\tau)^{-1}a_z)\right)\dot{A}_\bz \,ds-{i\,\Im\,\tau\over
2\pi}\int(\upsilon_\bz+(\Im\,\tau)^{-1}a_\bz) \dot{A}_z\,ds.}
We see that the action depends on the parameters
$\upsilon_z,\upsilon_\bz,a_z,a_\bz$ in the combinations
\eqn\numiko{\upsilon_z-(\Im\,\tau)^{-1}a_z,\ \upsilon_\bz+(\Im\,\tau)^{-1}a_\bz.}

The first term in \nofloc, upon quantization, tells us how $\phi_z$ acts:
\eqn\xococox{\phi_z\to -i\pi(\Im\,\tau)^{-1}\left({\delta\over\delta
A_\bz}+{i\,\Im\,\tau\over 2\pi}\upsilon_z-{i\over 2\pi} a_z\right).}
This is a covariant $\partial$ operator on a topologically trivial line bundle
over ${\rm Jac}(C)$. The second term indicates that the
wavefunctions are sections of a topologically trivial but holomorphically nontrivial
 line bundle
 $T_p\to{\rm Jac}(C)$, which is, in fact, the same line bundle $T_p=\cL\otimes
{\cal N}^{-1}\vert_{{\CMF}_p}$ that we identified
 before. Indeed, this term is a total derivative, so if
we absorb it into the initial and final state wavefunctions, we
find that the wavefunctions are no longer independent of $A_z$ but
are annihilated by the operator
\eqn\bilbox{
{\delta\over\delta A_z}-{i\,\Im\,\tau\over 2\pi}\upsilon_\bz-{i\over
2\pi} a_\bz .}
This is a covariant $\bpartial$ operator on a trivial line bundle
over ${\rm Jac}(C)$. Obviously, these two operators define a flat
connection on ${\rm Jac}(C)$. It is unitary if and only if
$\upsilon=0$.

So in short, quantization of the
$({\cal B}_{c.c.},{\cal B}_{{\CMF}_p})$ strings
has given us a
${\cal D}$-module associated to a choice of complex flat connection
on a trivial line bundle over ${\rm Jac}(C)$.  On the other hand,
$S$-duality identifies this family of $A$-branes with the set of
all zerobranes on $\MH(U(1),C)$, which in complex structure $J$ is
the moduli space of flat $\Bbb{C}^*$-bundles over $C$.  In other
words, in the abelian case $S$-duality establishes a natural
correspondence between gauge-equivalence classes of flat $\CC^*$
connections on $C$ and those on ${\rm Jac}(C)$.

This correspondence can be seen directly.  A flat $\CC^*$
connection on a manifold $M$ is the same as a homomorphism from
$H_1(M,\Bbb{Z})$ to $\CC^*$. Thus it is sufficient to show that
$H_1(C,\Bbb{Z})$ is isomorphic to $H_1({\rm Jac}(C),\Bbb{Z})$. But
since ${\rm Jac}(C)\simeq H^1(C,\RR)/H^1(C,\Bbb{Z})$, this is the
same as proving that $H_1(C,\Bbb{Z})\simeq H^1(C,\Bbb{Z})$, which
is Poincar\'e duality.

\subsec{Generalizations of the c.c. Brane and Twisted Differential
Operators}\subseclab\genotics

We have seen that geometric Langlands duality can be understood as
a result of the $S$-transformation exchanging $\Psi=0$ and
$\Psi=\infty$. More generally, the transformation $S$ maps a gauge
theory with gauge group $G$ and canonical parameter $\Psi$ to the
gauge theory with gauge group $\LG$ and the canonical parameter
$-1/n_{\frak g}\Psi$ (where, as we explained in section \reviewsd,
$n_{\frak g}=1$ if $G$ is simply-laced, and otherwise $n_{\frak
g}=2$ or 3). Our goal here is to describe what sort of
generalization of the geometric Langlands program arises when we
depart from $\Psi=0,\infty$. (See \frenkel, sections 6.3 and 8.6,
for a discussion of this generalization via conformal field
theory, and \ref\polro{A. Polishchuk and M. Rothstein, ``Fourier
Transform For ${\cal D}$-Algebras,''  Duke Math. J. {\bf 109}
(2001) 123-146.} for the abelian case.)

The generalization of the geometric Langlands program that we will
explore is perhaps most interesting for the case of rational
$\Psi$, since, as we saw in section \tophooft, this is the case in
which there are interesting line operators. Moreover, the full
$S$-duality group, which as we showed in section \canonpar, is
generated by $T:\Psi\to\Psi+1$ along with $S:\Psi\to -1/n_{\frak
g}\Psi$, maps rational values of $\Psi$ to rational values. In
fact, all rational values of $\Psi$ are related by this group, a
fact which also ensures that at a rational value of $\Psi$ there
are plenty of branes, just as there are at $\Psi=0$ and
$\Psi=\infty$.

The duality nonetheless works for all complex-valued $\Psi$. For
simplicity, however, here we will take $\Psi$ to be real. One can
obtain an arbitrary real value of $\Psi$ by letting $t=1$ and
keeping the $\theta$-angle arbitrary: \eqn\melo{ \Psi={\theta\over
{2\pi}}=\Re\tau.} $S$-duality in general transforms ${\rm
Re}\,\tau=\theta$, ${\rm Im}\,\tau=4\pi/e^2$, and $t$. But $\Psi$
is the only important parameter, so as long as $\Psi$ is real, we
can think of $S$-duality as acting  on $\theta$ only.  In this
set-up, the ``electric'' and ``magnetic'' theories look much more
symmetric: they both have $t=1$ but their $\theta$-angles are
inversely related. Upon compactification on the Riemann surface
$C$, one obtains an $A$-model in complex structure $K$ with target
$\MH(G,C)$ and $B$-field \eqn\pelbo{ B=-\omega_I \Re\tau,} as
explained in eqn. \palooka. The $S$-transformation is implemented
by replacing $\Re\tau$ by $-1/n_{\frak g}\Re\tau$ and $G$ by
$^L\negthinspace G$.

The $B$-field given by \pelbo\ is not flat along the fibers of the
Hitchin fibration; indeed, it is a multiple of the Kahler form in
complex structure $I$ and hence is nondegenerate when restricted
to a fiber. An important consequence is that the fibers of the
Hitchin fibration are no longer valid $A$-branes for general
$\Psi$. To explain this, recall that the sigma-model with target
$X$ in the presence of a $D$-brane wrapped on a submanifold $Y$ is
invariant under a gauge transformation \eqn\zebol{B\to
B-d\lambda,\quad A\to A+\lambda\vert_Y,\quad \lambda\in
\Omega^1(X).} The gauge-invariant combination is $F+B\vert_Y$. Now
suppose $X$ is symplectic and $Y$ is Lagrangian. If $B=0$, then a
single $D$-brane wrapped on $Y$ can be an $A$-brane only if $F=0$.
The  gauge-invariant generalization of this condition, which is
also valid for $B\neq 0$, is \eqn\subol{ F+B\vert_Y=0.} If the
cohomology class of $B|_Y/2\pi$ is not integral, then this
equation cannot be satisfied, for any $U(1)$ gauge field, and
therefore a single $A$-brane cannot be wrapped on $Y$. In our
case, this means that for $\theta\neq 2\pi n$, the fibers of the
Hitchin fibration are not valid $A$-branes, regardless of the
choice of Chan-Paton bundle.

For non-integer $\Psi$, the objects of study will therefore not be
branes of rank one supported on a fiber of the Hitchin fibration.
 As a consistency check, note that their naive $T$-duals, i.e.
points on $\MH(\LG,C)$, are also not valid topological branes at
$t=1$ and $\theta\neq 0$.  In fact, a point is never an $A$-brane,
since the support of an $A$-brane is always at least
middle-dimensional.  For rational $\Psi$, there are $A$-branes of
finite rank supported on a fiber of the Hitchin fibration; we can
find what they are by following the duality starting at $\Psi=0$.
In any case, $S$-duality still maps  $A$-branes for $\MH(G,C)$ to
$A$-branes for $\MH(\LG,C)$, even though examples of $A$-branes
may be scarce.

But what can we say about $A$-branes when $\Psi\not=0$ that
generalizes what we have found at $\Psi=0$? In the remainder of
this subsection we will answer this question.
 We will perform all computations near the classical limit $e^2\to
0,\ \Im\,\tau\to \infty$, but since the theory depends only on
$\Psi=\Re\tau$, the results will be valid for all values of $e^2$.

We will see that an analogue of the c.c. brane exists for all
$\Psi$.  We will call this brane ${\cal B}_{c.c.}^\Psi$. Its
algebra of open strings will turn out to be ${\cal
D}_{K_{\EUM}^{1/2}\otimes {{{\frak L}}}^\Psi}$, that is, the sheaf
of holomorphic differential operators on $K_\M^{1/2}\otimes{\frak
L}^\Psi$, where ${\frak L}$ is the determinant line bundle on
$\cM(G,C)$ (see the discussion of eqn. \frox). Given any other
$A$-brane ${\cal B}'$ on $\MH(G,C)$, the sheaf of $({\cal
B}_{c.c.}^\Psi,{\cal B}')$ strings is a module for ${\cal D}_{K_{
\EUM}^{1/2}\otimes {{{\frak L}}}^\Psi}$.  So an $A$-brane at
general $\Psi$ gives a sheaf of modules for this algebra. Just as
at $\Psi=0$, the construction really comes from four-dimensional
gauge theory, which in algebraic geometry means that these twisted
${\cal D}$-modules can be extended to the stack of all
$G$-bundles. $S$-duality then maps  twisted ${\cal D}$-modules for
$G$ to those of  $^L\negthinspace G$, with the exchange $\Psi\to
-1/n_{\frak g}\Psi$.

The gauge field on a space-filling  $A$-brane must solve the
equation \eqn\juko{ (\omega^{-1}(F+B))^2=-1,} where
$\omega=(\Im\tau)\, \omega_K$. An obvious solution is \eqn\uko{ F=
\Im\tau\, (\cos{\eurm q})\, \omega_J,\quad \sin{\eurm q}=-{{\Re\,
\tau}\over {\Im\, \tau}}.}  We take this solution to define the
brane ${\cal B}_{c.c.}^\Psi$.  Note that, as ${\Im\,\tau}$ can be
arbitrarily large, ${\eurm q}$ is extremely small, and ultimately
the interesting effects will be linear in $\Psi=\Re\,\tau$, which
is of order ${\eurm q}$.

If we ignore the algebra structure, then the sheaf of open strings
is locally isomorphic to the sheaf of holomorphic functions on
$\MH(G,C)$ in complex structure\foot{Comparing to eqn. \wyre, we
see that $I(\Psi)$ is the same as $I_w$ with $w=i\tan({\eurm
q}/2)$.} \eqn\muko{ I(\Psi)=\omega^{-1}(F+B)=I \cos{\eurm q} -
J\sin{\eurm q}.} The corresponding holomorphic coordinates on
$\MH(G,C)$ are \eqn\suko{ A'_\bz=A_\bz - i \tan{{\eurm q}\over
2}\, \phi_\bz,\quad \phi'_z=\phi_z+i\tan{{\eurm q}\over 2}\, A_z.}
As usual, $A$ is a connection on a $G$-bundle $E\to C$, and $\phi$
is a one-form valued in ${\rm ad}(E)$. What we have in section
\hitchfib\ called Hitchin's second fibration is the forgetful map
\eqn\xnonko{\pi_0: (A_\bz,\phi_z)\to A_\bz} from $\MH$ to $\M$. It
is plain that $\pi_0$ is not holomorphic with respect to $I(\Psi)$
except for $\Psi=0$. To rectify the situation, let us define a new
$\bpartial$-connection on $E$ by \eqn\hoxo{ D'_\bz=D_\bz-i
\tan{{\eurm q}\over 2}\,\phi_\bz=\partial_\bz+A_\bz - i
\tan{{\eurm q}\over 2}\,\phi_\bz, } and map the pair $(A,\phi)$
into the holomorphic bundle on $C$ defined by $D'_\bz$. Let us
denote this map $\pi_\Psi$; it is obviously holomorphic with
respect to $I(\Psi)$ and for $\Psi=0$ reduces to $\pi_0$. As
usual, we will restrict this fibration to the subset of Higgs
bundles for which $E$, endowed with the modified $\bar\partial$
operator, is stable.

What does the fiber of the map $\pi_\Psi$ look like? To understand
this, we note that the obvious candidate for the fiber coordinate,
namely $\phi_z+i\tan{{\eurm q}\over 2}\, A_z$, transforms
inhomogeneously under gauge transformations. The object which
transforms homogeneously is \eqn\gyko{ D'_z=\phi_z+i\tan{{\eurm
q}\over 2}\, D_z,} where $D_z$ is the holomorphic covariant
derivative on $E$. Up to a factor $i\tan{{\eurm q}\over 2}$,
$D'_z$ is a $\partial$-connection over the bundle $E$. We choose
not to divide by $i\tan{{\eurm q}\over 2}$, because we would like
the formulas to make sense even for $\Psi=0$. Then $D'_z$ is what
is called a holomorphic $\lambda$-connection on $E$, where
$\lambda=i\tan{{\eurm q}\over 2}$. For any complex manifold $X$, a
holomorphic $\lambda$-connection on a holomorphic bundle $E\to X$
is a linear map $\nabla: \Gamma(E)\ra \Gamma(E\ot T^*X)$ that
commutes with the $\bar\partial$ operator of $E$ and such that
\eqn\hyfo{ \nabla (f\cdot s)=f\nabla s + \lambda
\partial f\wedge s,} for any function $f$ and any  section $s$ of
$E$. For $\lambda=1$, a $\lambda$-connection is an ordinary
$\partial$ operator on $E$, while for $\lambda=0$ it is a section
of $\End(E)\ot T^*X$.

By virtue of Hitchin's equations, $D'_z$ commutes with $D'_\bz$
(this is closely related to the discussion of complex structure
$I_w$ at the end of section \complexstr) and therefore defines a
holomorphic $\lambda$-connection on the holomorphic bundle
$E'=\pi_\Psi(E,\phi)$. By the usual logic, this implies that the
fiber over $E'$ is the moduli space of holomorphic
$\lambda$-connections on $E'$.

The space of holomorphic $\lambda$-connections on a fixed vector
bundle $E'\to C$ is naturally an affine space modeled on the space
of holomorphic sections of $\End(E')\ot K_C$, where $K_C$ is the
canonical line bundle of $C$. (This statement means that to a
holomorphic $\lambda$-connection, we can add a holomorphic section
of $\End(E')\otimes K_C$, and any two holomorphic
$\lambda$-connections are related in this way.)  The space
$H^0(C,\End(E')\ot K_C)$ can be identified with the cotangent
space to $\cM(G,C)$ at the point $E'$. Thus in complex structure
$I(\Psi)$, the moduli space $\MH(G,C)$ looks like an affine
deformation of $T^*\cM(G,C)$; in other words, it is a bundle of
affine spaces, with the underlying bundle of vector spaces being
$T^*\CM(G,C)$. To summarize, if we ignore the algebra structure,
the sheaf of $({\cal B}_{c.c.}^\Psi,{\cal B}_{c.c.}^\Psi)$ strings
is the sheaf of holomorphic functions on a ``twisted'' cotangent
bundle of $\cM(G,C)$.

It remains to determine the algebra structure of the sheaf of
$({\cal B}_{c.c}^\Psi,{\cal B}_{c.c.}^\Psi)$ strings. We restrict
ourselves to functions of polynomial growth along the fibers of
the affine bundle.\foot{This restriction amounts to picking a
particular algebraic structure on  $\MH$ endowed with the complex
structure $I(\Psi)$. In this algebraic structure, it is called the
moduli space of $G$-bundles with $\lambda$-connection. There is a
second algebraic structure that is compatible with the same
complex structure, namely one in which $\MH$ is equivalent to the
moduli space of stable homomorphisms $\pi_1(C)\to G_{\Bbb{C}}$. In
the second algebraic structure, the traces of holonomies are
regarded as the algebraic functions. What we have just encountered
is the one point in the present paper in which we have to choose
between the different algebraic structures compatible with a fixed
complex structure on $\MH$.  This choice plays a more prominent
role in other approaches to the geometric Langlands program.} We
can follow the same logic as at $\Psi=0$.
 Locally, the fact that we are quantizing a twisted cotangent
bundle rather than an ordinary one is irrelevant. Globally, it
might (and as we will see, it does) modify the transition
functions in \hecox. So the sheaf of c.c. strings at any $\Psi$ is
the sheaf of holomorphic differential operators acting on a power
of some holomorphic line bundle over $\cM(G,C)$.

We can be more precise if $\Psi$ is an integer. Shifting $\Psi$ to
$\Psi+n$ shifts the $B$-field, or more precisely $B/2\pi$, by the
first Chern class of the line bundle ${{{\frak L}}}^{-n}$. (We can
see this from \palooka; the shift $\Psi\to \Psi+n$, which is
$\theta\to\theta+2\pi n$, shifts $B$ by $-n\omega_I$, where
$\omega_I/2\pi$ is the first Chern class of ${\frak L}$.) This
observation enables us to define a convenient $A$-brane   in
complex structure $K$. At $\Psi=0$, one of the important branes
was the brane ${\cal B}_0$ supported at $\phi=0$ with trivial
Chan-Paton bundle. At $\Psi=n$, we likewise can consider a brane
${\cal B}_n$ supported at $\phi=0$, but now with Chan-Paton bundle
${{{\frak L}}}^n$, endowed with its natural connection.  This
obeys \subol, and so is an $A$-brane in complex structure $K$.

Since we have set $\Psi=n$, we write ${\cal B}_{c.c.}^n$ instead
of ${\cal B}_{c.c.}^\Psi$. Now we quantize the $({\cal
B}_{c.c.}^n,{\cal B}_n)$ strings. Everything is as before, except
that we must include the Chan-Paton bundle of the brane ${\cal
B}_n$.  So the physical states are now sections of
$K_{\EUM}^{1/2}\otimes {\frak
L}^n\otimes\left(\oplus_{j=0}^n\Omega^{0,j}(\M)\right)$.  This is
a module for ${\cal D}_{K_{\EUM}^{1/2}\otimes {{{\frak L}}}^n}$,
and not for ${\cal D}_{K_{\EUM}^{1/2}\otimes {{{\frak L}}}^m}$ for
any other $m$. So the sheaf of algebras of $({\cal
B}_{c.c.}^n,{\cal B}_{c.c.}^n)$ strings is precisely ${\cal
D}_{K_{\EUM}^{1/2}\otimes {{{\frak L}}}^n}$.

What happens if $\Psi$ is not an integer?  We cannot hope to find
for general $\Psi$ a brane that leads to a module consisting of
sections of $K_{\EUM}^{1/2}\otimes {{{\frak L}}}^\Psi$, since
there is no such module unless $\Psi$ is an integer. However, we
know that the algebra of c.c. strings at any $\Psi$ is the sheaf
of differential operators acting on ${K_{\EUM}^{1/2}}\otimes
{\frak L}^{f(\Psi)}$ for some function $f(\Psi)$.  So if we can
prove that $f(\Psi)$ is a linear function, this, together with the
fact that $f(\Psi)=\Psi$ for integer $\Psi$, implies that
$f(\Psi)=\Psi$ for all $\Psi$.

To show that $f(\Psi)$ is linear, consider the Euclidean action of
the sigma-model: \eqn\cumber{ S=\int_\Sigma\Phi^*
(\omega-iF-iB)=-i\,\Im\tau \int_\Sigma \Phi^*\left(\cos{\eurm q}\,
\omega_J+\sin{\eurm q}\, \omega_I + i\omega_K\right).} (We have
omitted terms of the form $\{Q,V\}$.)
 For $\Psi=0$, we have ${\eurm q}=0$, and the two-form in
the integrand becomes an exact two-form on $T^*\cM(G,C)$. The
corresponding one-form potential $\varpi$ is, up to a numerical
factor, the canonical one-form $p_\alpha\, dq^\alpha$.  For
$\Psi\neq 0$, the best we can do is to write this two-form as an
exact form on $T^*\cM(G,C)$ plus the pull-back of a closed
two-form from $\cM(G,C)$: \eqn\hexo{ \Omega'={{\Im\tau}\over
\pi}\delta\int_C \,\Tr\left(\phi_z \delta A'_\bz\right)-
i\,\Re\tau \int_C\, \Tr\left(\delta A'_z\delta A'_\bz\right), }
where we defined $A'_z$ as minus the Hermitian conjugate of
$A'_\bz$. The second term in this formula is $\Re\tau$ times a
$(1,1)$-form  on $\M(G,C)$ which is closed but not exact.  It is a
multiple of the curvature of the natural line bundle ${\frak
L}\to\M$. To compute the transition functions \hecox, we do
perturbation theory using a propagator that is the inverse of the
form $\Omega'$. \hecox\ was deduced from the commutator $[\hat
p_\alpha,\hat p_\beta]$, which is proportional to $\hbar^2$ or
${1/ ({\rm Im}\,\tau)^2}$. So we need to do perturbation theory up
to order $1/({\rm Im}\,\tau)^2$. In that order, perturbation
theory is linear in $\Psi={\rm Re}\,\tau$, since the inverse of
$\Omega'$ can be expanded as a series in ${\rm Re}\,\tau$ and
${\rm Im}\,\tau$ that schematically is of the form $\sum_{k\geq
0}a_k({\rm Re}\,\tau)^k/({\rm Im}\,\tau)^{k+1}$. This completes
the argument.

The canonical line bundle over $\cM(G,C)$ is isomorphic to ${\frak
L}^{-2h}$, where $h$ is the dual Coxeter number of $G$
\hitchinflat . Thus we can also say that the algebra of c.c.
strings at any given value of $\Psi$ is the sheaf of holomorphic
differential operators on the line bundle ${\frak L}^{\Psi-h}$. In
particular, if we let $\Psi =h$, the algebra of c.c. strings is
the sheaf of ordinary differential operators, that is,
differential operators acting on functions. It follows that by
combining $S$-duality with a transformation by $T^h$, which shifts
$\Psi$ by $h$, we can map a zerobrane on $\MH(\LG,C)$ to an
ordinary (untwisted) ${\cal D}$-module on $\cM(G,C)$. This is a
more standard formulation of the geometric Langlands duality. For
certain $G$, it is also a more precise statement, as one sees if
one considers carefully \refs{\bd,\witfur} the dependence on the
spin structure of $C$.

\newsec{Branes From Gauge Theory}
\seclab\zapendix

So far we have always considered branes in the effective
two-dimensional sigma-model with target $\MH$.  Our goal in this
concluding section is to show that the important branes actually
can be defined in four-dimensional gauge theory.

Branes of any interest always preserve at least one supersymmetry.
For example, they are $B$-branes in complex structure $J$ or
$A$-branes in complex structure $K$.  But the branes important for
geometric Langlands duality have special properties with respect
to all three complex structures. For example, the zerobrane on
$\MH(\LG,C)$ is a $(B,B,B)$-brane: it is a $B$-brane with respect
to complex structures $I,J,K$. Its mirror is a fiber of the
Hitchin fibration, which is complex with respect to $I$ and
Lagrangian with respect to $J$ and $K$. Thus it is a
$(B,A,A)$-brane. The canonical coisotropic brane is an
$(A,B,A)$-brane, and so is its mirror.  Branes of any of these
types preserve two topological supercharges (linear combinations
of which give, for example, for a brane of type $(B,B,B)$, the
requisite $B$-type supersymmetry in complex structures $I$, $J$,
$K$).

We will focus on these four kinds of brane and describe the
corresponding boundary conditions in the gauge theory.  The lift
to gauge theory is not necessarily unique; in fact, in some cases
we describe several distinct boundary conditions in the gauge
theory which upon reduction to two dimensions apparently become
equivalent, at least away from singularities.

To discuss the twisted theory as a topological field theory, one
replaces four-dimensional Minkowski space by a general
four-manifold $M$.  In topological field theory, it  is most
convenient to take $M$ to have Euclidean signature, for the
following reason. In going to a general four-manifold, in order to
preserve some supersymmetry, one ``twists'' by interpreting the
first four scalar fields $\phi_0,\dots,\phi_3$ as a section of the
tangent bundle of $M$. Since the scalar fields naturally have a
positive signature, this twisting is much more natural with
Euclidean signature on $M$. (Alternatively, one could possibly
make a Wick rotation in $\phi$ space, but this seems
unfelicitous.)  Additionally, any compact four-manifold admits a
positive signature metric, while admitting a metric of Lorentz
signature is a severe topological restriction.

For the present discussion, we will consider branes in a more
restricted situation.  As usual, we compactify to two dimensions
on a Riemann surface $C$.  This gives an effective two-dimensional
theory, and we want to understand how branes in this theory can be
interpreted in the underlying four-dimensional gauge theory.  Near
the boundary, there is a natural ``time'' direction (normal to $C$
and tangent to the boundary) and no major advantage in a Wick
rotation. On the contrary, we wish to avoid the Wick rotation in
order to make clear that all branes we consider are physically
sensible and unitary.

This being so, whenever a topological supercharge $Q$ is preserved
by our boundary conditions, its adjoint $Q^\dagger$ is also
preserved.  They obey a physical supersymmetry algebra
$\{Q,Q^\dagger\}=H$, with $H$ the Hamiltonian.  Since our boundary
conditions preserve two topological supercharges, they preserve a
total of four supersymmetries.  This is one-fourth of the 16
global supersymmetries of ${\EUN}=4$ super Yang-Mills theory, so
such branes are called 1/4 BPS branes.  From the point of view of
the sigma model, which has only one-half as much supersymmetry,
they are called 1/2 BPS branes.  Going back to gauge theory,
however, in some instances our boundary conditions in the gauge
theory preserve eight supersymmetries locally along $C$, and the
reduction to four supersymmetries comes from the twisting and
curvature of $C$.

To describe a brane in the two-dimensional effective theory, we
formulate this theory on (say) $\Bbb{R}^2_+$, with coordinates
$x^0$, $x^1$, where $-\infty< x^0< \infty$, $0\leq x^1 <\infty$. A
supersymmetric  brane is obtained by specifying a supersymmetric
boundary condition on $x^1=0$.  To obtain such a construction from
the underlying four-dimensional gauge theory, we simply consider
this  theory on $M=\Bbb{R}^2_+\times C$,  and again describe
supersymmetric boundary conditions at $x^1=0$.

We write $x^2,x^3$ for local coordinates on $C$.   As we take
$\Bbb{R}^2_+$ to be flat, the twisting only affects the two scalar
fields $\phi_2,\phi_3$ (or $A_6$, $A_7$; the notation was
explained in section \reviewsym). The supersymmetry left unbroken
by the compactification and twisting is generated by those spinors
$\epsilon$ that are covariantly constant on $C$ in the appropriate
sense.  The condition, in the twisted theory, is simply that
$\epsilon$ should obey \eqn\ucuv{\Gamma_{2367}\epsilon=\epsilon,}
as in eqn. \yty.

\subsec{General Properties Of Boundary Conditions}

First let us recall a few generalities about what a boundary
condition is supposed to be.

A boundary condition constrains the values of the fields (and
their normal derivatives, in the case of bosons that obey second
order equations of motion) in such a way that the boundary terms
in the equations of motion vanish.  For example, for gauge fields
on an $n$-manifold $M$ with metric $g$ of signature $-++\dots +$,
the minimal action is \eqn\upco{{{\cmmib I}}_B={1\over 2e^2}\int_M
d^nx \sqrt g\,\Tr \,F_{IJ}F^{IJ}.} $M$ is an $n$-manifold with
$x^0$ as the ``time'' direction and with a boundary defined by
$x^1=0$. The boundary terms in the variation $\delta {{\cmmib
I}}_B$ of the action are \eqn\ofof{(\delta {\cmmib
I}_B)_{bdry}={2\over e^2}\int_{\partial M}d^{n-1}x \sqrt
g\,\Tr\left(g^{IJ}\sum_{I,J\not=1}\delta A_I F_{1J}\right).}  A
boundary condition must set to zero a linear combination of the
boundary values of $F_{I1}$ and $\delta A_I$, for $I\not= 1$, such
that $(\delta {{\cmmib I}}_B)_{bdry}=0$.  What it means to set
$\delta A_I=0$ for some $I$ is to specify the boundary values of
$A_I$ or in other words to impose Dirichlet boundary conditions on
$A_I$, with some prescribed boundary values. Conversely, in a
gauge with $A_1=0$ near the boundary of $M$, the condition that
$F_{I1}=0$ means that the normal derivative of $A_I$ vanishes.  So
$F_{I1}=0$ amounts to a gauge-invariant version of Neumann
boundary conditions on $A_I$. For quantization, we do not want to
overconstrain the boundary values, so we set to zero a minimal
linear combination of $\delta A_I$ and $F_{I1}$ such that $(\delta
{{\cmmib I}}_B)_{bdry}=0$.  (The boundary conditions therefore
define, in a suitable sense, a Lagrangian submanifold of the
boundary data.) It is also possible to add boundary terms to
${{\cmmib I}}_B$, thus modifying $(\delta {{\cmmib I}}_B)_{bdry}$;
the boundary conditions always define a maximal subspace of
allowed boundary values (of $A_I$ and $F_{I1}$, $I\not= 1$), such
that $(\delta {{\cmmib I}}_B)_{bdry}=0$. We call such a boundary
condition a hyperbolic boundary condition.

There is, of course, a similar story for fermions. The minimal
action is \eqn\olpo{{{\cmmib I}}_F=-{i\over e^2}\int_M d^nx \sqrt
g\,\Tr\,{\bar\lambda} \Gamma^I D_I\lambda,} to which one may add
additional boundary terms. Using the minimal action, the boundary
term in the variation of ${{\cmmib I}}_F$ is \eqn\rolpo{\delta
{{\cmmib I}}_F=-{i\over e^2}\int_{\partial M}d^{n-1}x\sqrt
g\,\Tr\,{\bar \lambda} \Gamma_1\delta\lambda.}  For fermions, a
suitable boundary condition sets to zero the boundary values of
one-half the components of $\lambda$ (and therefore of
$\delta\lambda$) such that ${\bar \lambda}
\Gamma_1\delta\lambda=0$.

In the bulk theory, supersymmetry means (for $M$ flat or after
suitable twisting for $M$ curved) that there is a conserved
supercurrent $J_I$. For supersymmetric Yang-Mills theory,
\eqn\noggin{J_I={1\over 2}\G^{KL}F_{KL}\G_I\lambda.} In the
presence of a boundary, the condition that the boundary condition
is compatible with supersymmetry is that the normal component of
$J$ vanishes at the boundary, so that the flux of the supercurrent
does not disappear at the boundary.  More precisely, the condition
is \eqn\hobbo{ \bar \epsilon J_1=0} for suitable supersymmetry
generators $\epsilon$.

This automatically ensures that the boundary conditions are
consistent with the supersymmetric variations of both bosons and
fermions.  If, therefore, one knows what boundary conditions one
wants on bosons and fermions, the condition $J_1=0$ is the basic
one.  In practice, since we do not know to begin with what
boundary conditions we want, we will start by postulating a simple
boundary condition on the fermions (chosen to ensure that ${\bar
\lambda}\Gamma_1\delta\lambda=0$), deduce the corresponding
boundary conditions on bosons to ensure that the boundary
conditions on the fermions are preserved by supersymmetry, and
then verify that $J_1=0$.

If $J_1=0$, we have found boundary conditions that will enable us
to define conserved supercharges (if $M$ is flat or in a suitably
twisted theory if $M$ is curved).  However, we want more.  A
boundary condition that forces supersymmetry to be spontaneously
broken will not be of much interest if our goal is to study
supersymmetric states or topological field theory (in which only
the supersymmetric states are of interest). Dirichlet boundary
conditions for bosons are generically incompatible with unbroken
supersymmetry. Dirichlet boundary conditions mean, as we have
discussed, that the boundary values of some components $A_I$ of
the gauge fields, and hence (possibly) of some components $F_{IJ}$
of the field strength, are specified.  We are mainly interested in
boundary conditions in which the boundary values of $F_{IJ}$ are
fixed in a way that is compatible with unbroken supersymmetry,
that is with the vanishing of the supersymmetry variation
\eqn\undop{\delta\lambda =\half\G^{IJ}F_{IJ}\epsilon} for suitable
supersymmetry generators $\epsilon$.

\subsec{Branes of Type $(B,B,B)$}

Our branes will always be constructed to preserve those
supersymmetries whose generators obey a suitable condition
$T\epsilon=\epsilon$ as well as\foot{Replacing the condition
$T\epsilon=\epsilon$ with $T\epsilon=-\epsilon$ would lead to
nothing essentially new, as there always will be symmetries of the
theory that reverse the sign of $T$.  For example, in our first
example, $T=\Gamma_{04}$, such a symmetry would be  a reflection
in the $49$ plane.} $\Gamma_{2367}\epsilon=\epsilon$.
 These combined
conditions will allow four supersymmetries. $T$ will commute with
$\Gamma_{2367}$. In some examples, $T$ will be chosen so that
$T^2=1$.  Then the condition $T\epsilon=\epsilon$ would, by
itself, allow eight supersymmetries, and the boundary condition
will preserve those eight supersymmetries.  Even when that is so,
the compactification to two dimensions and the ``twisting'' will
impose the second condition $\Gamma_{2367}\epsilon=\epsilon$ on
unbroken supersymmetries, leaving only four supersymmetries in the
low energy theory.

We will explain illustrative choices of $T$ that lead to branes of
the various types.  We certainly do not claim to describe all
interesting constructions of branes in the four-dimensional gauge
theory, only some simple ones that are sufficient to exhibit in
the gauge theory the branes that will be most important in the
present paper.

To obtain branes of type $(B,B,B)$, we consider first $T=\G_{04}$
with the boundary condition \eqn\unvu{T\lambda\vert
=\lambda\vert.} For any field $\Phi$, we write $\Phi|$ for the
restriction of $\Phi$ to $x^1=0$. (However, we omit this symbol
when confusion seems unlikely.)  This boundary condition on
$\lambda$ is only compatible with supersymmetries such that the
supersymmetric transformation $\lambda\to \lambda+\delta\lambda$
preserves the condition $T\lambda\vert=\lambda\vert$.  The
requirement is that $(T-1)\delta\lambda=0$. Since
$\delta\lambda=\half\G^{IJ}F_{IJ}\epsilon$ and
$T\epsilon=\epsilon$, we have
 \eqn\bogus{(T-1)\delta\lambda =\half [T,\G^{IJ}F_{IJ}]\epsilon.}
Upon evaluating the commutator, we find that we require
\eqn\nogus{\sum_{I\not=0,4}(\Gamma_{0I}F_{4I}-\Gamma_{4I}F_{0I})\epsilon=0.}
Since $\epsilon$ is constrained precisely by
$\Gamma_0\Gamma_4\epsilon=\epsilon$, this condition is equivalent
to \eqn\hogus{F_{0I}+F_{4I}=0~~{\rm for}~I\not=0,4.}

To pick a hyperbolic boundary condition compatible with what has
just been described, we set (in a suitable gauge) the boundary
values of $A_0+A_4$ to zero, and for $A_I$, $I\not= 0,1,4$, we
impose Dirichlet boundary conditions with some specified,
time-independent, boundary values.  Finally, \hogus\ tells us that
$F_{01}+F_{41}=0$, corresponding to gauge-invariant Neumann
boundary conditions on $A_0+A_4$. (Since $A_0+A_4$ corresponds to
a ``null'' direction, it is possible in a hyperbolic boundary
condition to set to zero both $A_0+A_4$ and its normal derivative.
This depends on the minus sign in the signature, the fact that
$g^{IJ}={\rm diag}(-1,1,1,\dots,1)$ in \ofof.)

So far, starting with the boundary conditions on the fermions, we
have guessed what the boundary conditions on the bosons must be.
At this stage, we can verify the basic condition that the normal
component of the supercurrent vanishes on the boundary:
\eqn\jolygo{{\bar\epsilon}\G^{MN}F_{MN}\G_1\lambda=0.} Because
$\G_{04}\lambda=\lambda$ and $\G_{04}\epsilon=\eps$, the only
components of $F_{MN}$ that contribute are those with precisely
one of $M$ and $N$ equal to 0 or 4; using this, one can verify
that \jolygo\ is obeyed if $F_{0I}+F_{4I}=0$, $I\not= 0,4$.

This implies that the boundary conditions on both fermions and
bosons are compatible with supersymmetry. We have already seen
this for the fermions, and we can readily verify it for the
bosons.  The supersymmetry variation of the bosons
\eqn\guffo{\delta A_I=i\bar\epsilon\Gamma_I\lambda} implies, with
$(\Gamma_{04}-1)\lambda=(\G_{04}-1)\epsilon=0$, that $\delta
A_I=0$ for $I\not= 0,4$, and $\delta A_0+\delta A_4=0$, whence
$\delta(F_{0I}+F_{4I})=0$ for $I\not=0,4$.  Hence our condition
$F_{0I}+F_{4I}=0$, $I\not= 0,4$, is consistent with supersymmetry.

So far, we have obtained a boundary condition that ensures the
existence of conserved supercharges.  We also would like the
boundary condition to allow for existence of a supersymmetric
state.  For this, the Dirichlet boundary conditions on $A_I$,
$I\not=0,1,4$, are highly constrained.  Given the supersymmetric
variation $\delta \lambda= \half
\sum_{IJ}\Gamma^{IJ}F_{IJ}\epsilon$, we want to pick the boundary
values such that \eqn\humbo{\sum_{I,J\not=
0,1,4}\G^{IJ}F_{IJ}\epsilon=0.} If this were supposed to be so for
all $\epsilon$, we would have to set $F_{IJ}=0$, $I,J\not= 0,1,4$.
Since $\epsilon$ is constrained to
$\Gamma_{2367}\epsilon=T\epsilon=\epsilon$, the constraint on $F$
is less severe.  It suffices to take $F$ to be selfdual in the
subspace generated by $x^2,x^3,x^6$, and $x^7$. Thus, we must
impose the Hitchin equations, which in this notation read
\eqn\ibob{\eqalign{F_{23}-F_{67}& =0\cr F_{27}-F_{36}& = 0 \cr
 F_{26}-F_{73}& = 0 . \cr}}  Thus, the boundary values of the gauge fields
define a point $p$ in $\MH$, the moduli space of Higgs bundles.
Once we impose \ibob, \humbo\ is obeyed if in addition $F_{IJ}|=0$
unless both $I$ and $J$ are in the set ${\EUR}=\{2,3,6,7\}$. In
particular, vanishing of $F_{IJ}|$ for $I\in {\EUR}$, $J\notin
{\EUR}$ implies that $A_J|$, for $J=5,8,9$ is covariantly constant
(on the Riemann surface $C$ with local coordinates $x^2,x^3$) and
commutes with the Higgs field, whose components are $\phi_2=A_6 $
and $\phi_3=A_7$.   If $p$ is a smooth point in $\MH$
corresponding to an irreducible Higgs bundle, these conditions
imply that $A_J|=0$ for $J=5,8,9$.  This furthermore implies the
vanishing of $F_{IJ}|$ for $I,J\notin {\EUR}$. So for an
irreducible Higgs bundle, the boundary condition (of this type) is
uniquely determined.  If $p$ corresponds to a reducible Higgs
bundle, we only learn in general that $A_5,$ $A_8$, and $A_9$
generate symmetries of the Higgs bundle and commute with each
other.

The boundary conditions we have obtained so far correspond to a
zerobrane on $\MH$ supported at the point $p$.  This is a basic
type of $(B,B,B)$-brane.  As expected, such a brane is unique if
$p$ corresponds to a smooth point in $\MH$.

\bigskip\noindent{\it Another Construction Of The Zerobrane}

Somewhat surprisingly, there is another gauge theory construction
of zerobranes on $\MH$.  It seems to lead to the same brane in the
low energy theory as long as one is at a smooth point in $\MH$,
but may lead to something new in the case of a zerobrane supported
at a singular point in $\MH$.

Let us replace $T$ by $\tilde T = T
\Gamma_{2367}=\Gamma_{023467}$. The supersymmetries preserved by
the boundary condition will now be those whose generators obey
$\tilde T\epsilon=\epsilon$.  So the boundary condition in the
four-dimensional gauge theory will preserve a different set of
eight supercharges than what we had before. However, the unbroken
supersymmetries in the low energy theory are those that obey the
combined conditions $T\epsilon=\Gamma_{2367}\epsilon=\epsilon$.
These conditions are invariant under replacing $T$ by $\tilde T$,
so the change in boundary conditions will not change the unbroken
supersymmetry in the low energy theory and the construction will
give a new $(B,B,B)$-brane.

We will choose the boundary condition on the fermions to be
$\tilde T\lambda\vert = -\lambda\vert$. Asking for this condition
to be invariant under supersymmetry, we require that
$\{\Gamma^{IJ}F_{IJ},\tilde T\}\epsilon=0$. Let us assume in
addition that $\Gamma_{2367}\eps=\eps$. Then the supersymmetry
condition is satisfied if \eqn\kipo{\sum_{I,J\in
\{0,2,3,4,6,7\}}\Gamma^{IJ}F_{IJ}\vert\epsilon=0} and
\eqn\kipok{\sum_{I,J\in
\{1,5,8,9\}}\Gamma^{IJ}F_{IJ}\vert\epsilon=0.} \kipo\ is obeyed if
the $A_I$ obey Dirichlet boundary conditions with $A_4\vert=0,$
and with the boundary values of $A_I$, $I\in \{2,3,6,7\}$
constrained to obey the Hitchin equations \ibob.

Note that if $\eps$ does not obey $\Gamma_{2367}\eps=\eps$, the
supersymmetry constraint is not satisfied, in general. This means
that even in flat spacetime, this $(B,B,B)$-brane  preserves only
four supercharges.

To obey \kipok,  we {\it cannot} ask for $F_{IJ}|=0$, $I,J\in
\{1,5,8,9\}$ as these conditions are overdetermined. Indeed,
$F_{15}|=F_{18}|=0$ means that $A_5$ and $A_8$ obey Neumann
boundary conditions and are free to fluctuate on the boundary;
this being so, we cannot also ask for $F_{58}\vert=0$. However,
the condition $\tilde T\epsilon=\epsilon$, together with the
chirality condition $\hat\Gamma\epsilon=\epsilon$ of
ten-dimensional super Yang-Mills theory (here
$\hat\Gamma=\Gamma_0\Gamma_1\cdots \Gamma_9$), implies
$\Gamma_{1589}\epsilon=-\epsilon$.  With this constraint on
$\epsilon$, \kipok\ is equivalent to an anti-selfduality condition
on the boundary values in the $1589$ plane:
\eqn\ubb{\eqalign{F_{15}\vert+F_{89}\vert & = 0 \cr
                  F_{18}\vert+F_{95}\vert & = 0 \cr
                  F_{19}\vert+F_{58}\vert & = 0. \cr}}
This is a modified version of Neumann boundary conditions on
$A_5,$ $A_8$, and $A_9$, and is {\it not} overdetermined.

Since $A_2,A_3, A_6$, and $A_7$ obey Dirichlet boundary conditions
determined by a specific solution of the Hitchin equations, what
we obtain in this way is a zerobrane supported at a point $p$ in
$\MH$.  What distinguishes it from the previous construction of
the zerobrane is the boundary condition on $A_4,A_5,A_8$, and
$A_9$, as well as the fact that in flat spacetime the former
construction preserves half as many supersymmetries as the latter.
If $p$ is a smooth point in $\MH$ corresponding to an irreducible
Higgs bundle, the fields $A_4,A_5,A_8$, and $A_9$ have no zero
modes and vanish in the low energy theory, regardless of the
microscopic boundary conditions. Hence the new construction of the
zerobrane should be equivalent to the old one as long as $p$ is a
smooth point in $\MH$.  If $p$ corresponds to a reducible Higgs
bundle whose stabilizer is a subgroup of $G$ of positive rank,
then $A_4$, $A_5$, $A_8$, and $A_9$ do have zero modes -- which
generate symmetries of the Higgs bundle -- and there is no reason
to expect the two zerobranes to be equivalent.

One can get yet another $(B,B,B)$-brane by taking the boundary
condition for the fermions to be $\tilde T\lambda\vert =
\lambda\vert$. It turns out that it corresponds to a space-filling
brane with a flat gauge field. Since this brane does not play a
major role in the present paper, we omit the details.

\subsec{Branes Of Type $(B,A,A)$}

We can get branes of type $(B,A,A)$ by applying $S$-duality to
branes of type $(B,B,B)$.  Though it may be in general unclear how
to transform a particular brane under $S$-duality, there is no
problem in transforming the unbroken supersymmetries under
$S$-duality.  In view of eqn. \tokio, under the transformation
$S$, for ${\rm Re}\,\tau=0$, the generators $\epsilon_{\ell,r}$ of
supersymmetry transform as \eqn\bilj{ \epsilon_\ell\to
\epsilon_\ell\cdot\exp\left(-{i\pi\over 4}\right),\quad
\epsilon_r\to \epsilon_r\cdot\exp\left({i\pi\over 4}\right), } or
equivalently \eqn\yogo{\epsilon\to\epsilon'={1+\HG\over \sqrt
2}\epsilon} where $\HG=\Gamma_0\Gamma_1\Gamma_2\Gamma_3$ measures
the four-dimensional chirality.  If $T\epsilon=\epsilon$, then
$\epsilon'={1+\HG\over \sqrt 2}\epsilon={1+\HG\over \sqrt
2}T\epsilon={1+\HG\over \sqrt 2}T{1-\HG\over \sqrt 2}\epsilon'$,
so $T'\epsilon'=\epsilon'$ where \eqn\uvuf{T'={1+\HG\over \sqrt
2}T{1-\HG\over \sqrt 2}.} If $T=\Gamma_0\Gamma_4$, then
\eqn\polyo{T'=\Gamma_1\Gamma_2\Gamma_3\Gamma_4.}

We impose the boundary condition $T'\lambda|=-\lambda|$ on the
fermions. Then supersymmetry requires that
$\{T',\Gamma^{IJ}F_{IJ}\}\epsilon=0$, or
\eqn\bngo{\left(\sum_{I,J\in
\{1,2,3,4\}}\Gamma^{IJ}F_{IJ}+\sum_{I,J\notin
\{1,2,3,4\}}\G^{IJ}F_{IJ}\right)\epsilon=0.}

This condition implies that $F_{0I}|=0$ for $I>4$, so we impose
Dirichlet boundary conditions on the $A_I$ of $I>4$.  We simply
set $A_I\vert=0$ for $I=5,8,9$, and we look for a boundary
condition that gives a nonzero value for the ``Higgs fields''
$\phi_2=A_6,$ $\phi_3=A_7$.  Assuming that \bngo\ is supposed to
be true for all $\epsilon$ obeying
$\Gamma_{1234}\epsilon=\epsilon$ (without the further condition
$\G_{2367}\epsilon=\epsilon$), it implies that
$[\phi_2,\phi_3]=F_{67}=0$.  Given this, the gauge-invariant
content of $\phi_2$ and $\phi_3$ is in the characteristic
polynomial of $\phi_z={1\over 2}(\phi_2-i\phi_3)$.   We pick
Dirichlet boundary conditions in which this characteristic
polynomial is specified.

On the other hand, $A_{2}$, $A_{3}$, and $A_{4}$ obey modified
Neumann boundary conditions \eqn\yimgo{\eqalign{ F_{12}\vert -
F_{34}\vert
 &=0\cr
 F_{13}\vert - F_{42}\vert
 &=0\cr
 F_{14}\vert - F_{23}\vert
 &=0,\cr}}
 which follow from imposing \bngo\ with
 $\Gamma_{1234}\epsilon=\epsilon$.

We also want to require the supersymmetries which satisfy
$\Gamma_{2367}\eps=\eps$ to be unbroken. This puts constraints on
the boundary values of the fields which satisfy Dirichlet
conditions. Requiring the vanishing of the supersymmetry
variations of the fermions on the boundary, we find
\eqn\dirbaa{\eqalign{F_{26}|-F_{73}|&=0,\cr F_{27}|-F_{36}|&=0.}}
These equations are equivalent to $D_\bz\phi_z=0$, which is the
``complex'' Hitchin equation.

These boundary conditions mean that in the Higgs bundle $E$
determined by a pair $(A_{\bar z},\phi_z)$, the field $A_{\bar z}$
which determines the holomorphic structure of $E$ is allowed to
fluctuate, while keeping fixed the characteristic polynomial of
$\phi_z$.  Presumably, the corresponding brane is supported on a
fiber of the Hitchin map $\pi:\MH\to {\EUBB}$.  This gives us an
archetypical $(B,A,A)$-brane.

The condition $[\phi_2,\phi_3]=0$ that arose in this classical
analysis is not one of the Hitchin equations defining $\MH$.
Instead, we should have $[\phi_2,\phi_3]-F_{23}=0$. Presumably,
the extra term arises as a sort of quantum correction in the
renormalization group flow to the infrared.

In the limit that the gauge theory reduces to the two-dimensional
sigma-model, $S$-duality maps a $(B,A,A)$-brane wrapping the fiber
of the Hitchin fibration to a $(B,B,B)$ zerobrane. In the abelian
case, one can verify directly that the boundary conditions that we
have described for these branes are $S$-dual.

We can construct another $(B,A,A)$-brane that seemingly is
likewise supported on a fiber of the Hitchin fibration by
replacing $T'$ with $T''=T'\G_{2367}=-\G_{1467}$.  Imposing
$T''\epsilon=\epsilon$ and $T''\lambda\vert=\lambda\vert$, we get
\eqn\loppo{[\G^{IJ}F_{IJ},T'']\epsilon=0,} which implies again
Dirichlet boundary conditions for $A_6=\phi_2$ and $A_7=\phi_3$
and Neumann boundary conditions for $A_2$ and $A_3$.

Of course, for a given fiber ${\CMF}$ of the Hitchin fibration
$\pi:\MH\to{{\cmmib B}}$, there should be many branes, associated
with the possible flat line bundles on ${\CMF}$.  We have given
two constructions, but because of time-reversal symmetry, each
leads to a trivial Chan-Paton line bundle (or at most one that is
of order two).  It hopefully is possible to modify the gauge
theory construction to make this bundle vary.

\subsec{Branes Of Type $(A,B,A)$}\subseclab\xgoto

Our next goal is to construct from the gauge theory some branes of
type $(A,B,A)$. The most interesting of these is the canonical
coisotropic brane. As a prelude, we introduce the abbreviation
$\Gamma_{\pm I\pm J}=(\pm \G_I\pm \G_J)/\sqrt 2$. Similarly, we
take $F_{\pm I\pm J,K}=(\pm F_{IK}\pm F_{JK})/\sqrt 2$, etc.

Let us define \eqn\jokko{T'=-\Gamma_{15}\Gamma_{2-7}\Gamma_{3+6}.}
As we will see shortly, this choice leads to branes of type
$(A,B,A)$.  One point of this definition is that $T'$ is invariant
under simultaneous rotation of the $x^2-x^3$ and $x^6-x^7$ planes.
Thus in the twisted theory, $T'$ does not depend on the choice of
the coordinates $x^2,x^3$, provided they are orthogonal. For the
same reason, if we consider the twisted theory on $M=\RR_+^2\times
C$, $T'$ depends only on the metric on $C$, not on the specific
coordinate choice. The branes we will construct using $T'$ depend
only on the conformal class of the metric, i.e. the complex
structure on $C$.

We can again construct two kinds of branes, with boundary
conditions $T'\lambda|=\lambda|$ or $T'\lambda|=-\lambda|$.  In
the former case, supersymmetry gives in the usual way
$[\G^{IJ}F_{IJ},T']\epsilon=0$.  This tells us that $F_{IJ}=0$ if
one of $I,J$ is in the space generated by the $1,5,2-7,$ and $3+6$
directions and the other is orthogonal to it. In particular,
$F_{0I}|=0$ for $I=5,2-7,3+6$, and we impose Dirichlet boundary
conditions for such values of $I$.  But $F_{1I}=0$ for
$I=2+7,3-6,4,8,9$, and we impose Neumann boundary conditions on
$A_I$ for such $I$.

In such a brane, $A_2-\phi_3$ and $A_3+\phi_2$ have specified
boundary values, while $A_2+\phi_3$ and $A_3-\phi_2$ are free to
fluctuate.  To understand this result, it helps to recall that in
the complex structure $J$, the holomorphic coordinates are
$2(A_z+i\phi_z)=(A_2-i A_3)+i(\phi_2-i\phi_3)=(A_2+\phi_3)-i(
A_3-\phi_2)$ and $2(A_{\bar z}+i\phi_{\bar z})=(A_2+i
A_3)+i(\phi_2+i\phi_3)=(A_2-\phi_3)+i( A_3+\phi_2)$.  So our
result means that for this type of brane, ${\cal A}_{\bar
z}=A_{\bar z}+i\phi_{\bar z}$ is fixed and ${\cal A}_{ z}=A_{
z}+i\phi_{ z}$ is unconstrained.  This defines a submanifold $W'$
of $\MH$ which is clearly a complex submanifold in the complex
structure $J$.  In addition, $W'$ is Lagrangian from the point of
view of the holomorphic symplectic form
$\Omega_J=(-i/4\pi)\int_C\Tr\, \delta {\cal A}\wedge \delta {\cal
A}$.   Indeed, $W'$ is middle-dimensional, and in addition
$\Omega_J$ vanishes on $W'$, since $W'$ is characterized by
$\delta {\cal A}_{\bar z}=0$. So the brane supported on $W'$, with
a trivial Chan-Paton bundle, is an $(A,B,A)$-brane, as expected.
$W'$ can be characterized as the submanifold of $\MH$ consisting
of flat $G_{{\Bbb{C}}}$ bundles with a specified holomorphic
structure.

For the other type of brane, with $T'\lambda|=-\lambda|$,
supersymmetry requires $\{\G^{IJ}F_{IJ},T'\}\epsilon=0$.  This
leads to two conditions.  The first is that $F_{IJ}=0$ if $I,J$
are in the subspace perpendicular to directions $1,5,2-7,$ and
$3+6$.  In particular, $F_{0I}=0$ if direction $I$ is in this
subspace, and we impose Dirichlet boundary conditions for such
$I$. Thus, in particular, $A_2+\phi_3$ and $A_3-\phi_2$ obey
Dirichlet boundary conditions. The second consequence of
$\{\G^{IJ}F_{IJ},T'\}\epsilon=0$ is that $F$ must be anti-selfdual
when restricted to the subspace generated by directions
$1,5,2-7,3+6$. This leads to modified Neumann boundary conditions
on $A_5,$ $A_{2-7}$, and $A_{3+6}$: \eqn\nono{\eqalign{
F_{15}+F_{2-7,3+6}&=0\cr F_{1,2-7}+F_{3+6,5} & =0\cr
F_{1,3+6}+F_{5,2-7}&=0.\cr}}

For this type of brane, $A_{2+7}$ and $A_{3-6}$ are fixed on the
boundary while $A_{2-7}$ and $A_{3+6}$ are free to fluctuate.  So,
from a low energy point of view, the roles of $\CA_z$ and
$\CA_{\bar z}$ are exchanged, relative to the branes with
$T'\lambda|=\lambda|$.  Thus we get a brane supported on a
submanifold $W''$ of $\MH$ that parametrizes flat $G_{\Bbb{C}}$
bundles with a fixed antiholomorphic structure.  This again is a
brane of type $(A,B,A)$.

There is another and more trivial way to exchange the roles of
$\CA_z$ and $\CA_{\bar z}$: we could reverse the sign of $\phi_2$
and $\phi_3$ or in other words replace $T'$ by
$T''=-\Gamma_{15}\Gamma_{2+7}\Gamma_{3-6}$.  So we really have two
families of branes supported on manifolds parameterizing bundles
with fixed holomorphic or antiholomorphic structure.  The two
families differ by the conditions placed on modes that are massive
away from singularities of $\MH$; they may differ at singularities
of $\MH$.

\bigskip\noindent{\it $S$-Duality And The Canonical Coisotropic Brane}

By applying $S$-duality to the examples that we just constructed,
we can obtain new branes, which will also be branes of type
$(A,B,A)$. We know how to apply $S$-duality to the condition for
unbroken supersymmetry: $T'\epsilon=\epsilon$ is replaced by $\hat
T\epsilon=\epsilon$ where \eqn\nofo{\hat T ={1+\HG\over \sqrt
2}T'{1-\HG\over \sqrt 2}=-{\Gamma_{15}\over
2}\left((\Gamma_{26}+\Gamma_{37})-\HG(\Gamma_{23}+\Gamma_{67})\right).}
To get new $(A,B,A)$-branes in this way, we must construct
boundary conditions that preserve the supersymmetries with $\hat
T\epsilon=\Gamma_{2367}\epsilon=\epsilon$.

We look for a supersymmetric brane with $\hat
T\lambda|=-\lambda|$. After a somewhat lengthy calculation, we
deduce from the condition $\{\G^{IJ}F_{IJ},\hat T\}\epsilon=0$ for
supersymmetry that the boundary conditions on the bosons must be
\eqn\tomgo{\eqalign{F_{15}+F_{26}+F_{37}& = 0 \cr
                    F_{12}+F_{06}-F_{56}& = 0 \cr
                    F_{13}+F_{07}-F_{57}& = 0 \cr
                    F_{16}+F_{52}-F_{02}& = 0 \cr
                    F_{17}+F_{53}-F_{03}& = 0 ,\cr}}
along with $F_{0J}=0$ for $J=4,8,9$.

We set the massive fields $A_J,\,J=4,8,9$ to zero on the boundary.
Otherwise, all fields obey modified Neumann boundary conditions
and are free to fluctuate on the boundary, so the $(A,B,A)$-brane
obtained this way is a space-filling brane whose target space is
all of $\MH$.

We have seen mixed Neumann-Dirichlet boundary conditions before,
but there is a key difference here. In previous examples, the
boundary conditions become purely Dirichlet or purely Neumann if
one sets to zero the fields $A_4,A_5,A_8$, and $A_9$ that play no
role in the low energy theory away from singularities, so they are
irrelevant in the effective two-dimensional field theory of the
brane.  In the present example, this is not the case. After
discarding the ``massive'' fields and choosing the gauge $A_1=0$,
the boundary conditions become \eqn\bdrycc{\eqalign{
D_0\phi_2+\partial_1 A_2&=0,\cr D_0\phi_3+\partial_1 A_3&=0,\cr
F_{02}-\partial_1\phi_2&=0,\cr F_{03}-\partial_1\phi_3&=0.}}

Mixed Dirichlet-Neumann boundary conditions on the sigma-model
fields arise when the gauge field on the brane is not flat;
specifically, the boundary condition is \eqn\zembro{
f_{MN}\partial_0 X^N+g_{MN}\partial_1 X^N=0, } where $f$ is the
curvature of the target-space gauge field, and $g$ is the
target-space metric. Comparing with \bdrycc\ we find \eqn\membro{
f={1\over 2\pi}\int_C |d^2z| \Tr\, \left(\delta \phi_z\wedge\delta
A_\bz+\delta\phi_\bz\wedge \delta A_z\right)=\omega_J. } This is
the target-space gauge field for the canonical coisotropic
$(A,B,A)$-brane in the sigma model with target $\MH(G,C)$. So that
is the brane that we have found.

\bigskip\noindent{\it Boundary Observables for the Canonical Coisotropic Brane}

Now we want to make a simple observation about boundary
observables for the canonical coisotropic brane.  We consider
local operators ${\EUO}$ inserted at a point $P$ of the boundary
of $M$. The boundary, of course, looks like $\Bbb{R}\times C$.  We
will consider the cohomology of the topological supercharge $Q$ of
the $A$-model in complex structure $K$ acting on local operators
inserted at a point $s\times p\in \Bbb{R}\times C$.  The metric of
$C$ is irrelevant mod $\{Q,\dots\}$, so we can assume $C$ to be
flat near $p$.

We can classify local operators by their dimension and also by the
``spin'' with which they transform under rotations of $C$ around
the point $p$.  (The considerations will be local, so it will not
matter that the global structure of $C$ typically breaks the
rotation symmetry.)  The natural dimensions in the twisted theory
are 1 for $A,\phi,\psi,\tilde\psi$, and 2 for
$\chi,\eta,\tilde\eta$.

We consider $Q$-invariant operators of dimension $n$ and spin $n$.
To have these quantum numbers, a gauge-invariant operator must be
constructed only from $\phi_z$, $\psi_z$, $\tilde\psi_z$, and the
covariant derivative $D_z$.  But for an operator to be
$Q$-invariant, $D_z$ cannot appear, since $A_z$ is not
$Q$-invariant even on the boundary.  The boundary condition $\hat
T\psi|=-\psi|$ implies that
\eqn\bdrry{\eqalign{\psi_2|&=-i\tilde\psi_3|\cr
\psi_3|&=i\tilde\psi_2|.\cr}} With $\psi_z=(\psi_2-i\psi_3)/2$,
this leads to $\psi_z|=\tilde\psi_z|$. So $Q$-invariant boundary
observables of dimension $n$ and spin $n$ are functions just of
$\phi_z$ and $\psi_z$, without any derivatives.

 The fields $\phi_z$ and $\psi_z$ have ghost number $\EUK=0$ and
 $\EUK=1$, respectively.  So an operator of this type that also has
 $\EUK=0$ is a gauge-invariant function of $\phi_z$ only, such as
 ${\EUO}=\Tr\,\phi_z^n$ for some $n$.  A typical operator of
 $\EUK=1$ is ${\EUO}'={\rm Tr}\,\phi_z^{n-1}\psi_z$.

 Classically, the boundary conditions ensure that on the boundary
 $[Q,\phi_z\,]=\{Q,\psi_z\,\}=0$, so all operators ${\EUO}$
 and ${\EUO}'$ of the sort just described are nonzero elements
 of the cohomology of $Q$.  What happens quantum mechanically?
 Could there be a quantum correction to the action of $Q$, such that the
 exact quantum formula would be $[Q,{\EUO}]=\epsilon{\EUO}'$
 for some $\epsilon$?  If so, at the quantum level, ${\EUO}$ and
 ${\EUO}'$ would pair up and disappear from the cohomology.

 That this does not occur can be argued using time-reversal
 symmetry.  We define an orientation-reversing symmetry ${\EUT}$
 that reverses the sign of the time coordinate $x^0$ and acts
 trivially on other coordinates of $M$.  Allowing for the minus
 sign with which ${\EUT}$ acts on $\phi$ (see eqn. \yoro), ${\EUT}$
 is a symmetry of the boundary conditions that define the
 canonical coisotropic brane.  ${\EUT}$ commutes with the
 topological supercharge of the $A$-model in complex structure
 $K$, since $t=1$ is a fixed point of ${\EUT}$ (see eqn. \zongor).
 Also, $\phi_z$ is odd under ${\EUT}$, in view of \yoro, but
 \zoro\
 plus the boundary condition $\psi_z|=\tilde\psi_z|$ implies that
 $\psi_z|$ is even under ${\EUT}$.  So the operators ${\EUO} $
 and ${\EUO}'$ cannot pair up under the action of $Q$, and must
 survive in the cohomology.   More generally, consider a local
 boundary operator of dimension and spin $n$ and ghost number $k$
 that is $Q$-invariant at the classical level.  Such an operator
 is determined by a gauge-invariant function $F(\phi_z,\psi_z)$
 that is of degree $n-k$ in $\phi_z$ and degree $k$ in $\psi_z$.
 It transforms as $(-1)^{n-k}$ under ${\EUT}$.   For $Q$ to act
 non-trivially on such an operator at the quantum level, it would
 have to leave $n$ unchanged and increase $k$ by 1.  This would
reverse the eigenvalue of ${\EUT}$, and so is impossible.  Thus,
the cohomology of $Q$ in the space of local operators of dimension
and spin $n$ coincides with the classical result, and is given
simply by the space of gauge-invariant functions
$F(\phi_z,\psi_z)$.

\listrefs
\end